\documentclass[a4paper,fleqn,usenatbib]{mnras}

\usepackage{newtxtext,newtxmath}

\usepackage[T1]{fontenc}
\usepackage{ae,aecompl}

\usepackage{graphicx}	
\usepackage{amsmath}	
\usepackage{amssymb}	
\usepackage{subfig}
\usepackage[labelfont=bf]{caption}
\usepackage{threeparttable}
\usepackage{times,tikz}

\graphicspath{{Figures/}}

\def\kms{km~s$^{-1}$}
\def\Section{\S}
\def\HI{H{\sc i} }
\def\XCO{X$_\text{CO}$ }
\def\FDS{\citealt{Iodice2016,Iodice2017}, Peletier et al., in prep., \citealt{Venhola2017,Venhola2018}}

\title[The ALMA Fornax Cluster Survey]{The ALMA Fornax Cluster Survey I: stirring and stripping of the molecular gas in cluster galaxies}

\author[N. Zabel et al.]{Nikki Zabel,$^{1}$\thanks{E-mail: ZabelNJ@cardiff.ac.uk}
Timothy A. Davis,$^{1}$
Matthew W. L. Smith,$^{1}$
Natasha Maddox,$^{2,3}$ \newauthor
George J. Bendo,$^{4}$
Reynier Peletier,$^{5}$	
Enrichetta Iodice,$^{6}$
Aku Venhola,$^{7}$
Maarten Baes,$^{8}$ \newauthor
Jonathan I. Davies,$^{1}$
Ilse de Looze,$^{9,10,11}$ 
Haley Gomez, $^{1}$
Marco Grossi,$^{12}$ \newauthor
Jeffrey D. P. Kenney,$^{13}$ 
Paolo Serra,$^{14}$ 
Freeke van de Voort,$^{15,16}$
Catherine Vlahakis,$^{17}$ \newauthor
Lisa M. Young$^{18,19}$
\\
$^{1}$School of Physics and Astronomy, Cardiff University, Queen's Building, The Parade, Cardiff, CF24 3AA, Wales, UK\\
$^{2}$ASTRON, the Netherlands Institute for Radio Astronomy, Oude Hoogeveensedijk 4, 7991 PD, Dwingeloo, The Netherlands\\
$^{3}$Faculty of Physics, Ludwig-Maximilians-Universit\"at, Scheinerstr. 1, 81679 Munich, Germany \\
$^{4}$UK ALMA Regional Centre Node, Jodrell Bank Centre for Astrophysics, School of Physics and Astronomy, The University of Manchester, \\ 
Oxford Road, Manchester M13 9PL, UK \\
$^{5}$Kapteyn Astronomical Institute, University of Groningen, PO Box 72, 9700 AB Groningen, The Netherlands\\
$^{6}$INAF-Astronomical Observatory of Capodimonte, via Moiariello 16, Naples, I-80131, Italy\\
$^{7}$Astronomy Research Unit, University of Oulu, 90014, Oulu, Finland\\
$^{8}$Sterrenkundig Observatorium, Department of Physics and Astronomy, Universiteit Gent, Krijgslaan 281 S9, B-9000 Gent, Belgium\\
$^{9}$Department of Physics and Astronomy, University College London, Gower Street, London WC1E 6BT, UK \\
$^{10}$Sterrenkundig Observatorium, Universiteit Gent, Krijgslaan 281 S9, B-9000 Gent, Belgium \\
$^{11}$Institute of Astronomy, University of Cambridge, Madingley Road, Cambridge CB3 0HA, UK \\
$^{12}$Observatório do Valongo, Universidade Federal do Rio de Janeiro, Ladeira Pedro Antônio 43, 20080-090 Rio de Janeiro, RJ, Brazil\\
$^{13}$Yale University Astronomy Department, P.O. Box 208101, New Haven, CT 06520-8101, USA\\
$^{14}$INAF - Osservatorio Astronomico di Cagliari, Via della Scienza 5, I-09047 Selargius (CA), Italy\\
$^{15}$Heidelberg Institute for Theoretical Studies, Schloss-Wolfsbrunnenweg 35, D-69118 Heidelberg, Germany \\
$^{16}$Astronomy Department, Yale University, PO Box 208101, New Haven, CT 06520-8101, USA\\
$^{17}$National Radio Astronomy Observatory, 520 Edgemont Road, Charlottesville, VA 22903-2475, USA\\
$^{18}$Physics Department, New Mexico Tech, 801 Leroy Place, Socorro, NM 87801, USA \\
$^{19}$National Radio Astronomy Observatory, Socorro, NM 87801, USA
}

\date{Accepted 2018 November 26. Received 2018 November 20; in original form 2018 October 11}

\pubyear{2018}

\begin{document}
\label{firstpage}
\pagerange{\pageref{firstpage}--\pageref{lastpage}}
\maketitle

\begin{abstract}
We present the first results of the ALMA Fornax Cluster Survey (AlFoCS): a complete ALMA survey of all members of the Fornax galaxy cluster that were detected in \HI or in the far infrared with \textit{Herschel}. The sample consists of a wide variety of galaxy types, ranging from giant ellipticals to spiral galaxies and dwarfs, located in all (projected) areas of the cluster. It spans a mass range of $10^{\sim 8.5 - 11} M_\odot$. The CO(1-0) line was targeted as a tracer for the cold molecular gas, along with the associated 3 mm continuum. CO was detected in 15 of the 30 galaxies observed. All 8 detected galaxies with stellar masses below $3 \times 10^9 M_\odot$ have disturbed molecular gas reservoirs, only 7 galaxies are regular/undisturbed. This implies that Fornax is still a very active environment, having a significant impact on its members. Both detections and non-detections occur at all projected locations in the cluster. Based on visual inspection, and the detection of molecular gas tails in alignment with the direction of the cluster centre, in some cases ram pressure stripping is a possible candidate for disturbing the molecular gas morphologies and kinematics. Derived gas fractions in almost all galaxies are lower than expected for field objects with the same mass, especially for the galaxies with disturbed molecular gas, with differences of sometimes more than an order of magnitude. The detection of these disturbed molecular gas reservoirs reveals the importance of the cluster environment for even the tightly bound molecular gas phase.
\end{abstract}




\section{Introduction}
\label{sec:intro}
It has long been known that galaxies in cluster environments evolve differently from their counterparts in the field. In particular, the relative number of early-type galaxies in cluster environments is significantly higher than in the field \citep[e.g.][]{Oemler1974,Dressler1980}. In addition, the galaxies that are present in clusters have a smaller atomic gas reservoir than their counterparts in the field \citep{Haynes1984,Cayatte1990,Solanes2001,Gavazzi2005}. The clustering of galaxies generates an extreme environment, that is likely capable of quenching the star formation in galaxies, transforming them from blue, late-type galaxies to red ellipticals.

Over the years, various processes have been proposed as the responsible mechanism for this transformation. Ram pressure stripping (RPS) was first suggested as a candidate by \cite{Gunn1972}, and similarly viscous stripping by \cite{Nulsen1982}, starvation by \cite{Larson1980}, and thermal evaporation by \cite{Cowie1977}. Furthermore there are galaxy-galaxy interactions, such as harassment \citep{Moore1996} and mergers. So called pre-processing, which takes place at higher redshifts when the clusters are first formed and the galaxies' velocities are still relatively low, also plays a role in shaping the galaxies, in the form of minor mergers and tidal interactions (\citealt{Mihos2004,Fujita2004}, see \citealt{Boselli2006} for an extended review). The relative importance of the different mechanisms is still poorly understood. In any case, it is clear that the cluster environment plays a fundamental role in galaxy evolution, especially keeping in mind that $\sim$40\% of galaxies live in groups or clusters \citep[e.g.][]{Robotham2011}, and the majority of the local galaxies live in groups \citep[e.g.][]{Zabludoff1998}.

It is well known that the atomic gas in galaxies is affected by the above-mentioned processes. The situation is more complicated for the molecular gas, because it is more tightly bound to the galaxy and distributed more centrally. The debate about this has therefore been more lively. Early research often concluded that the molecular gas in cluster galaxies is the same as that in field galaxies, and is unaffected by the cluster environment \citep[e.g.][]{Stark1986, Kenney1989, Casoli1991, Boselli1995, Boselli2006}. It was not until more recently that indications of deficiency, i.e. a lower mass than expected based on statistics of similar galaxies in the field, were observed for the molecular gas as well \citep[e.g.][]{Vollmer2008,Fumagalli2009,Boselli2014} and also for dust \citep[e.g.][]{Cortese2010,Cortese2012}. Although, these deficiencies are smaller than for H{\sc i}. On average galaxies which are H{\sc i} deficient by a factor of $\sim$10 are CO deficient by a factor of $\sim$2. \citet{Lee2017} report examples of three galaxies in the Virgo cluster that are ram pressure stripped of their molecular gas as well as their atomic gas. At higher redshifts, evidence of molecular gas stripping and deficiencies in clusters has also recently been observed \citep[e.g.][]{Noble2018,Wang2018,Stach2017}, although cluster galaxies with molecular gas contents similar to \citep[e.g.][]{Rudnick2017} or even higher than \citep[e.g.][]{Hayashi2018} field galaxies are found as well.

Because molecular gas is the direct fuel for star formation, the effects of the cluster environment on this phase of the interstellar medium (ISM) have immediate consequences for the star formation rate of the host galaxy. If it is directly affected by environmental processes, this could have important implications for the quenching of cluster members and therefore for galaxy evolution as a whole.

The goal of this work is to investigate whether the cluster environment indeed affects the molecular gas in galaxies, and if so, attempt to identify which processes are mainly responsible for this. In order to do this, we focus our attention on the Fornax cluster. Fornax is among the two nearest galaxy clusters, together with the Virgo cluster. They are located at 19.95 \citep{Tonry2001} and 16.8 Mpc (NASA/IPAC Extragalactic Database), respectively. Both clusters are therefore ideal laboratories to study the effects of the cluster environment on galaxies at high resolution. Extensive catalogues exist for both clusters, compiled by \cite{Binggeli1985} for Virgo and by \cite{Ferguson1989} for Fornax. Other, more recent studies of the Virgo cluster include the deep optical Next Generation Virgo Survey (NGVS, \citealt{Ferrarese2012}), the \textit{Herschel} Virgo Cluster Survey (HeViCS, \citealt{Davies2010}) in the far infrared, the GALEX Ultraviolet Virgo Cluster Survey (GUViCS, \citealt{Boselli2011}) in the UV, and the blind narrow-band H$\alpha$ + [NII] imaging survey Virgo Environmental Survey Tracing Ionised Gas Emission (VESTIGE, \citealt{Boselli2018}). Located in the southern hemisphere, Fornax has been studied less than its northern counterpart. However, recently more and more studies of the Fornax cluster have appeared. These include the optical Fornax Deep Survey (FDS, \FDS), the Herschel Fornax Cluster Survey (HeFoCS, \citealt{Davies2013}) the integral-field spectroscopic survey Fornax3D \citep{Sarzi2018}, the blind \HI Australia Telescope Compact Array (ATCA) survey \citep{Lee-Waddell2018}, and soon the MeerKAT Fornax HI and radio continuum survey \citep{Serra2016}.

There are some fundamental differences between both clusters that add to the importance of studying the Fornax cluster in addition to the Virgo cluster. First, Fornax is much smaller than Virgo, with Virgo being $\sim$10 times as massive as Fornax \citep{Jordan2007}. It is home to $\sim$2000 galaxies, while Fornax harbours only 350 (detected at the time of the catalogues mentioned above, both complete in magnitudes up to $B_T \approx 18$, and containing members with magnitudes up to $B_T \approx 20$). Despite its smaller size, the Fornax cluster has a number density of roughly three times that of Virgo. Fornax is also more regular and dynamically evolved than Virgo, and has a lower velocity dispersion. Because it is more relaxed, environment and density related effects are easier to identify in Fornax: galaxies in its centre will be more strongly affected by density effects than galaxies in the outskirts. In Virgo these effects are harder to identify, because it is still in the process of assembling. The central hot gas density in Fornax is 4 times lower than in Virgo, and its temperature twice as low \citep{Schindler1999,Paolillo2002,Scharf2005}. These differences suggest that ram pressure stripping plays less of a role in the Fornax cluster, compared to Virgo. According to \cite{Davies2013} ram pressure stripping should be a factor 16 less important in Fornax, based on the equation from \cite{Gunn1972}: $P_r \approx \rho_e v^2$, where $P_r$ is the ram pressure, $\rho_e$ is the intracluster density, and $v$ the velocity of the galaxy. The higher number density in Fornax, on the other hand, suggests that galaxy-galaxy interactions are more important. In this work we turn to a resolved study of the ISM in Fornax galaxies to investigate these processes further.

\cite{Horellou1995} carried out an H{\sc i} and $^{12}$CO(1-0) survey of 21 spirals and lenticulars in the Fornax cluster, using the Nan\c{c}ay radio telescope (France) and the Swedish-ESO Submillimeter Telescope (SEST, \citealt{Booth1989}), respectively. They detected 16 galaxies in HI, and 11 were detected in CO. They found that on average the CO emission of Fornax galaxies is weak: about five times lower than that of spirals in the Virgo cluster. From this it follows that the corresponding molecular gas masses are low as well: they found H$_2$ masses that are about ten times lower than the atomic gas masses. They attribute the decreased molecular gas masses to reduced star-formation activity, and argue that it is in agreement with low far-infrared, radio continuum and H$\alpha$ luminosities. They comment, however, that although the CO emission found for the Fornax galaxies is low compared to that in infrared-selected samples, that may be typical for spirals in optically-selected samples. In this work we revisit the CO($J$=1-0) in the Fornax cluster and investigate whether these observations can be confirmed.

The ALMA Fornax Cluster Survey is a complete survey of the 30 Fornax cluster members that were detected in 3 or more \textit{Herschel} Space Observatory \citep{Pilbratt2010} bands with the \textit{Herschel} Fornax Cluster Survey \citep{Fuller2014} or in H{\sc i} (\citealt{Waugh2002}, Loni et al. in prep. based on ATCA data). The CO(1-0) rotational line (rest frequency: 115.271 GHz) was observed to create spatially resolved maps of the cold molecular gas and its kinematics in these galaxies. The survey covers a range of different galaxy stellar masses and morphologies. 

A full description of the sample can be found in \Section \ref{sec:sample_selection}. The observations, data reduction, and ancillary data are described in \Section \ref{sec:observations}. In \Section \ref{sec:results} we present moment maps of the CO emission of the detected galaxies, as well as their position-velocity diagrams (PVDs) and spectra, and a comparison with optical observations. H$_2$ masses are estimated and compared with the expected H$_2$ masses for field galaxies. In \Section \ref{sec:discussion} we discuss the results, and the morphologies and kinematics of the galaxies in the sample. Various environmental processes are considered as possible candidates for the irregularities observed, and the surprising detection of several dwarf galaxies is discussed. In \Section \ref{sec:conclusions} we summarise the work, and distil conclusions. Although accurate distance measurements are available for some of the AlFoCS galaxies, here we adopt the distance to the Fornax cluster (19.95 Mpc, \citealt{Tonry2001}) as a common distance to all galaxies. 

\section{Sample selection}
\label{sec:sample_selection}

\begin{figure}
	\centering
	\includegraphics[width=0.52\textwidth]{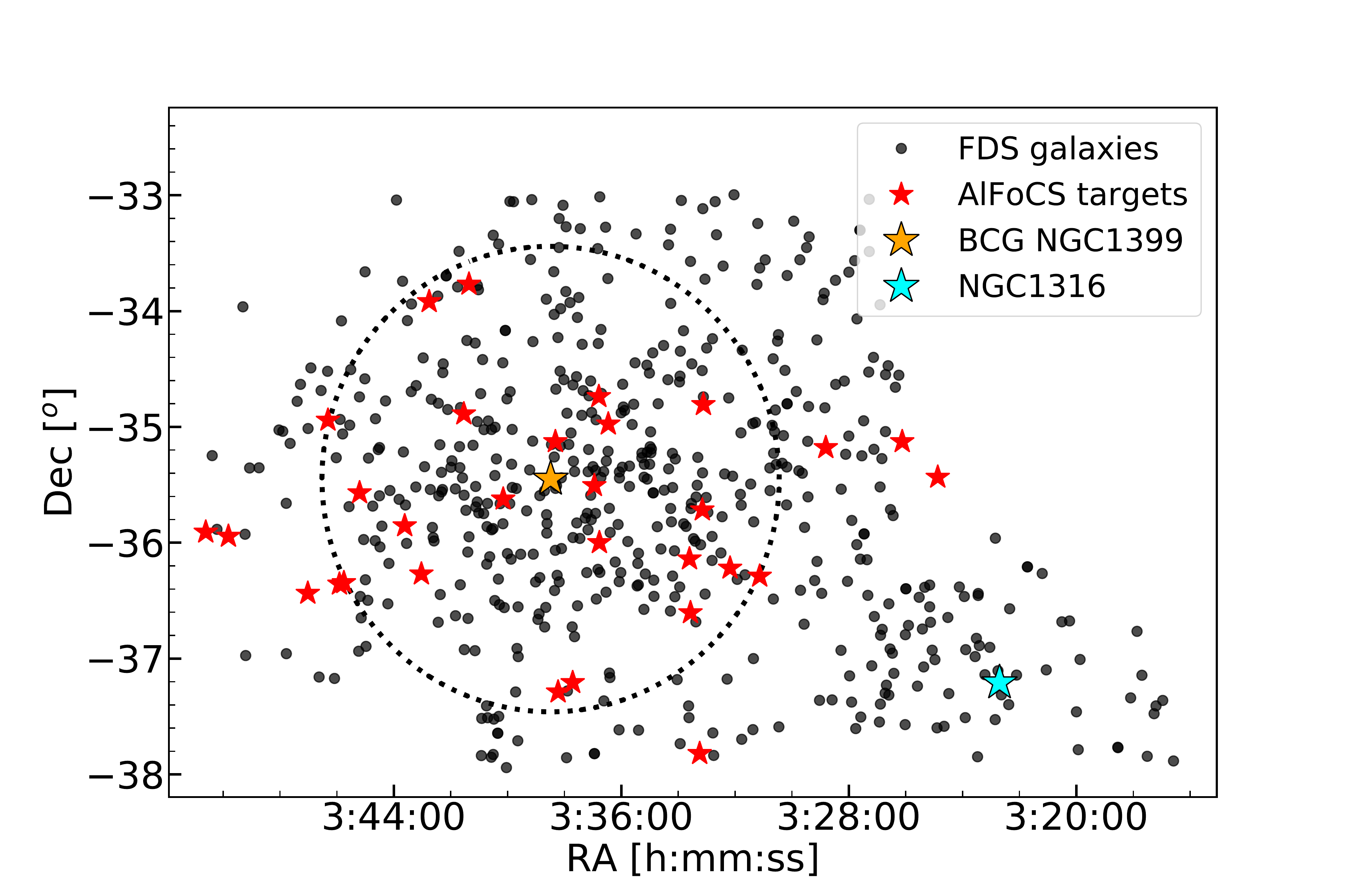}
	\caption{Map of the Fornax cluster. The black dots represent Fornax Deep Survey galaxies from \citet[][see \Section \ref{sub:optical_data}]{Venhola2018}, the red stars represent the AlFoCS sample. The central galaxy, NGC1399, is indicated with a larger yellow star, and the virial radius (located at 0.7 Mpc, \citealt{Drinkwater2001a}) is shown as a dotted line. AlFoCS targets are distributed evenly over the cluster (except for the infalling subgroup, which was not covered by \textit{Herschel}). NGC1316, the central galaxy of the infalling subgroup, is indicated with a cyan star.}
	\label{fig:Fornax}
\end{figure}

\begin{figure}
	\centering
	\includegraphics[width=0.45\textwidth]{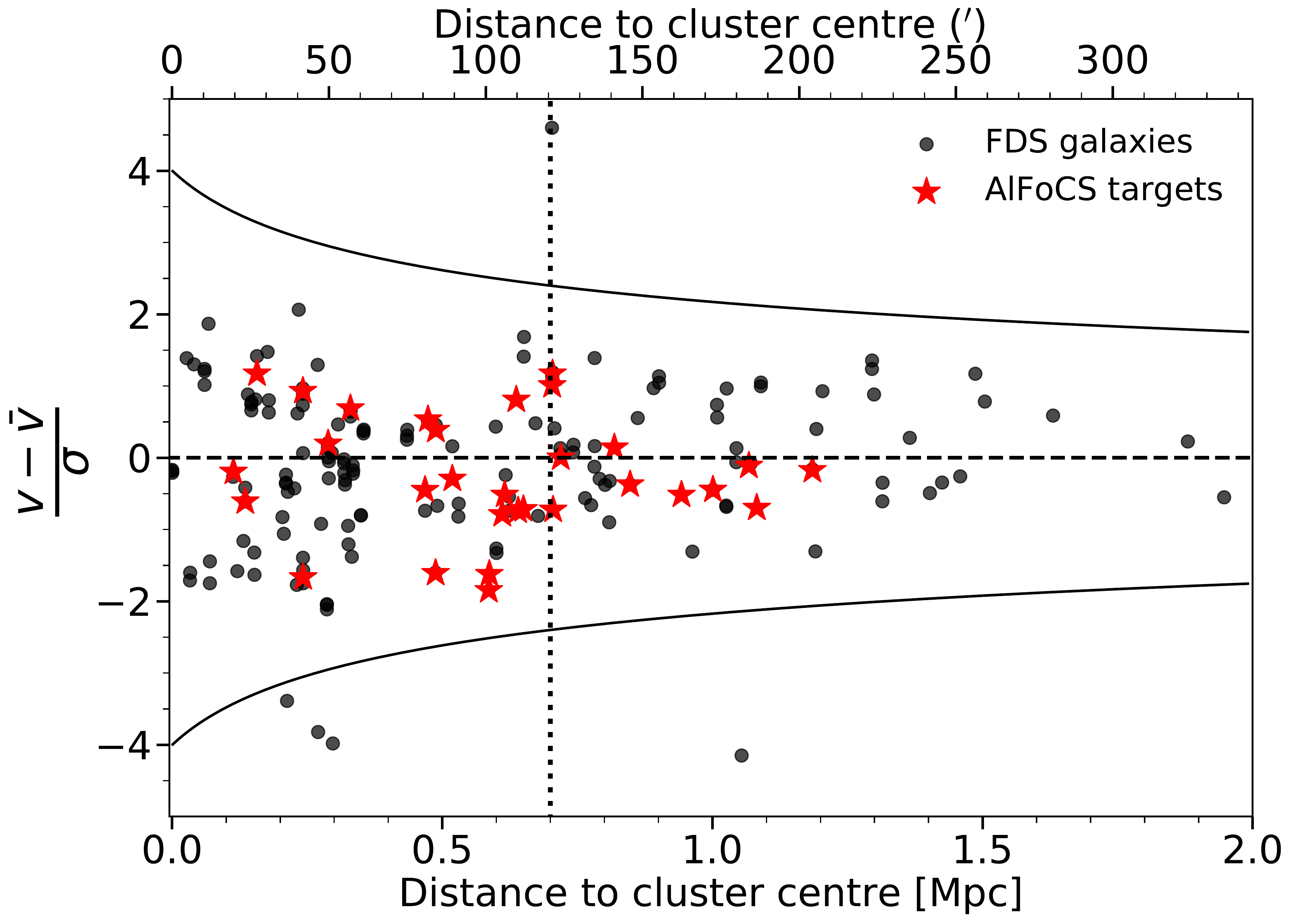}
	\caption{Caustic diagram of the Fornax cluster. The black data points represent the FDS galaxies for which velocity information is available, and the red stars represent the AlFoCS targets. The solid lines represent the escape velocities in the cluster as a function of distance from the cluster centre. The vertical dotted line indicates the virial radius at 0.7 Mpc \citep{Drinkwater2001a}. $\bar{v} = 1493$ \kms and $\sigma = 374$ \kms \citep{Drinkwater2001a}. The AlFoCS targets are distributed evenly in the caustic space.}
	\label{fig:trumpet}
\end{figure}

Our sample is based on the Fornax Cluster Catalogue \citep[FCC,][]{Ferguson1989}. From this catalogue, galaxies with stellar masses $>3 \times 10^8 \text{ M}_\odot$ were selected to ensure high enough metallicity to detect CO. Furthermore, galaxies were selected to contain dust \citep{Fuller2014} or HI down to $\sim3 \times 10^7 M_\odot$ (\citealt{Waugh2002}, Loni et al. in prep. based on ATCA data). This suggests ongoing star formation activity, and therefore the presence of a molecular gas reservoir. Whether a galaxy was selected based on its FIR emission or \HI content is listed in Table \ref{tab:targets}. The application of these criteria on the FCC leads to a sample of 30 galaxies, spanning a wide range of morphological types, varying from giant ellipticals to irregular dwarfs. A wide range of locations within the cluster is covered by the survey. This is shown in Figure \ref{fig:Fornax}, where we compare the locations of the AlFoCS galaxies with the locations of the galaxies in the Fornax Deep Survey (FDS, \FDS). The FDS is a recent optical survey of the Fornax cluster, containing 573 galaxies, and is described in more detail in \Section \ref{sub:optical_data}. The FDS galaxies are shown as black dots, and the galaxies targeted here are shown as red stars. The brightest cluster galaxy (BCG) NGC1399 is shown as a bigger yellow star, and the dotted line represents the virial radius of the cluster according to \citet{Drinkwater2001a}. The central galaxy of the currently infalling subgroup in the south-east of the figure, NGC1316, is indicated with a cyan star. Aside from a slight ($<$10\%) deficiency of galaxies in the innermost ($\sim$350 kpc or 1 degree) radius of the cluster centre (defined as the location of NGC1399), the AlFoCS targets are spread evenly among the cluster galaxies: they are located at all directions from the cluster centre, and both close to the central galaxy, and outside the virial radius. There are no observations in the infalling subgroup around NGC1316, as this area was not covered by \textit{Herschel}.

\begin{table*}
	\centering
	\begin{threeparttable}
	\caption{Key properties of the galaxies in the sample.}
	\label{tab:targets}
	\begin{tabular}{lllllll}
	\hline
	Common name & FCC \# & RA & Dec & $cz$ & $M_\star$ & Selection \\
	- & - & (J2000) & (J2000) & (\kms) & (log($M_{\odot}$)) & - \\
	(1) & (2) & (3) & (4) & (5) & (6) & (7) \\
	\hline
	FCC32 & 32 & 03h24m52.4s & -35d26m08s & 1319 $^\diamond$ & $9.23^{+0.04}_{-0.07}$* & FIR \\
	FCC44 & 44 & 03h26m07.4s & -35d07m39s & 1233 $^\diamond$ & $8.50^{+0.07}_{-0.17}$* & FIR \\
	NGC1351A & 67 & 03h28m48.7s & -35d10m41s & 1354 & 9.45$^\dagger$ & FIR, \HI \\
	MGC-06-08-024 & 90 & 03h31m08.2s & -36d17m25s & 1814 $^\diamond$ & 8.98$^\dagger$ & FIR, \HI \\
	FCC102 & 102 & 03h32m10.7s & -36d13m15s & 1722 $^\ddagger$ & $8.36^{+0.08}_{-0.10}$* & \HI \\
	ESO358-G015 & 113 & 03h33m06.8s & -34d48m29s & 1389 & 8.88$^\dagger$ & FIR, \HI \\
	ESO358-16 & 115 & 03h33m09.2s & -35d43m07s & 1701 $^\diamond$ & $8.32^{+0.07}_{-0.09}$* & \HI \\
	FCC117 & 117 & 03h33m14.6s & -37d49m11s & - & $7.77^{+0.18}_{-0.20}$* & FIR \\
	FCC120 & 120 & 03h33m34.2s & -36d36m21s & 849 $^\ddagger$ & $8.50^{+0.07}_{-0.09}$* & \HI \\
	NGC1365 & 121 & 03h33m36.4s & -36d08m25s & 1638 $^\diamond$ & 11.16$^\dagger$ & FIR, \HI \\
	NGC1380 & 167 & 03h36m27.6s & -34d58m34s & 1878 $^\diamond$ & 10.98$^\dagger$ & FIR \\
	FCC177 & 177 & 03h36m47.5s & -34d44m23s & 1562 $^\diamond$ & $10.4^{+0.01}_{-0.02}$* & FIR \\
	NGC1386 & 179 & 03h36m46.2s & -35d59m58s & 869 $^\diamond$ & 10.5$^\dagger$ & FIR \\	
	NGC1387 & 184 & 03h36m57.0s & -35d30m24s & 1303 $^\diamond$ & 10.77$^\dagger$ & FIR \\
	FCC198 & 198 & 03h37m42.7s & -37d12m30s & - & $8.09^{+0.05}_{-0.07}$* & FIR \\
	FCC206 & 206 & 03h38m13.5s & -37d17m25s & 1403 $^\diamond$ & $9.01^{+0.07}_{-0.10}$* & FIR \\
	FCC207 & 207 & 03h38m19.3s & -35d07m45s & 1421 $^\diamond$ & $8.78^{+0.04}_{-0.05}$* & FIR \\
	NGC1427A & 235 & 03h40m09.3s & -35d37m28s & 2029 $^\diamond$ & 9.78$^\dagger$ & FIR, \HI \\
	FCC261 & 261 & 03h41m21.5s & -33d46m09s & 1710 $^\ddagger$ & 8.58$^\dagger$ & FIR \\
	PGC013571 & 263 & 03h41m32.6s & -34d53m18s & 1725 $^\diamond$ & 9.2$^\dagger$ & FIR, \HI \\
	FCC282 & 282 & 03h42m45.3s & -33d55m14s & 1266 $^\ddagger$ & 9.0$^\dagger$ & FIR \\
	NGC1437A & 285 & 03h43m02.2s & -36d16m24s & 891 $^\ddagger$ & 9.38$^\dagger$ & FIR, \HI \\
	NGC1436 & 290 & 03h43m37.1s & -35d51m11s & 1388 $^\diamond$ & 10.1$^\dagger$ & FIR, \HI \\
	FCC302 & 302 & 03h45m12.1s & -35d34m15s & 816 $^\ddagger$ & $8.48^{+0.09}_{-0.07}$* & \HI \\
	FCC306 & 306 & 03h45m45.4s & -36d20m48s & 891 $^\ddagger$ & 8.68$^\dagger$ & FIR, \HI \\
	NGC1437B & 308 & 03h45m54.8s & -36d21m25s & 1515 $^\ddagger$ & 9.39$^\dagger$ & FIR, \HI \\
	ESO358-G063 & 312 & 03h46m19.0s & -34d56m37s & 1920 $^\ddagger$ & 10.04$^\dagger$ & FIR, \HI \\
	FCC316 & 316 & 03h47m01.5s & -36d26m15s & 1547 $^\diamond$ & $8.64^{+0.07}_{-0.12}$* & FIR \\
	FCC332 & 332 & 03h49m49.0s & -35d56m44s & 1327 $^\diamond$ & 8.63$^\dagger$ & FIR \\
	ESO359-G002 & 335 & 03h50m36.7s & -35d54m34s & 1431 $^\diamond$ & 9.21$^\dagger$ & FIR \\	
	\hline
	\end{tabular}
	\textit{Notes:} 1: Common name of the galaxy; 2: Fornax Cluster Catalogue number of the galaxy; 3: Right ascension; 4: Declination; 5: Velocity (defined as the object's redshift times the speed of light); 6: Stellar mass. $^\ddagger$See \Section \ref{sub:redshifts}; $^\diamond$Redshifts from NASA/IPAC Extragalactic Database; $^\dagger$Stellar masses from \citet{Fuller2014}; *Stellar masses derived from 3.6 $\mu$m images, (see \Section \ref{sub:H2_masses}); 7: Whether the galaxy was selected based on an \HI (\citealt{Waugh2002}, Loni et al. in prep. based on ATCA data) or a FIR \citep{Fuller2014} detection (or both).
	\end{threeparttable}
\end{table*}

To confirm that all targets are indeed cluster members, we create a caustic diagram of all galaxies with known velocities: the (projected) velocities of the galaxies (corrected for the velocity of the cluster and galaxy-to-galaxy velocity dispersion within the cluster) of the cluster as a function of their distance from the cluster centre. This is shown in Figure \ref{fig:trumpet}. The mean velocity and velocity dispersion of the Fornax cluster were taken from \citet{Drinkwater2001a}, and equal 1493 \kms and 374 \kms, respectively. The velocities of the individual galaxies are a combination of velocities from the FCC, the 2dF Galaxy Redshift Survey \citep{Colless2001,Drinkwater1999}, and the 2MASS Redshift Survey \citep{Huchra2012}. Note that velocity information is unavailable for 470 of the 573 FDS galaxies, and these were omitted from the figure. The solid lines represent the escape velocities at each projected distance from the cluster centre, assuming a Navarro-Frenk-White (NFW) density profile for the cluster dark matter distribution \citep{Navarro1997}. They were derived using equation 7 and 16 from \citet{Shull2014}, featuring a dark matter concentration parameter, which was estimated using equation 3 from \citet{Coe2010}. The dotted line again represents the virial radius, and the colours are the same as in Figure \ref{fig:Fornax}. All AlFoCS galaxies shown here have velocities well below the escape velocity at their location, and are distributed evenly in the caustic space.

The locations, velocities, and stellar masses of the galaxies observed are listed in Table \ref{tab:targets}.

\section{Observations and data reduction}
\label{sec:observations}

\subsection{ALMA data}
\label{sub:ALMA_data}

\begin{table*}
	\centering
	\begin{threeparttable}
	\caption{Observational parameters.}
	\label{tab:obspars}
	\begin{tabular}{cccccccccc}
		\hline
		SB & Date & \# ants & TOT & Bandpass cal. & Flux cal. & CF spw 3 & CV spw 3 & BW spw 3 & CF spws 0, 1, 2 \\
		- & - & - & (mins.) & - & - & (GHz) & (\kms) & (\kms) & (GHz, resp.) \\
		(1) & (2) & (3) & (4) & (5) & (6) & (7) & (8) & (9) & (10) \\
		\hline
		Single fields & 07-01-2016 & 42 & 52 & J0336-3616 & J0336-3616  & 114.547 & 1885 & 4898 & 113.001, 100.939, 102.544  \\
		Small mosaics & 11-01-2016 & 46 & 125 & Uranus & J0336-3616 & 114.756 & 1340 & 4907 & 112.818, 100.824, 102.713 \\
		Dwarfs & 12-01-2016 & 43 & 251 & Uranus & J0336-3616 & 114.716 & 1445 & 4900 & 113.161, 101.089, 102.703 \\
		\hline
	\end{tabular}
	\textit{Notes:} 1: Scheduling Block; 2: Date of the observations; 3: Number of antennas used; 4: Total observation time in minutes; 5: Bandpass calibrator; 6: Flux calibrator; 7: Central frequency of spectral window 3 (centred on the $^{12}$CO(1-0) line); 8: Central velocity of spectral window 3 (centred on the $^{12}$CO(1-0) line); 9: Bandwidth of spectral window 3 (centred on the $^{12}$CO(1-0) line) 10: Central frequencies of the remaining spectral windows.
	\end{threeparttable}
\end{table*}

Atacama Large Millimeter/submillimeter Array (ALMA) observations of the $^{12}$CO(1-0) line in 29 AlFoCS targets were carried out under project 2015.1.00497.S (PI: Timothy Davis). ALMA's 12 m configuration was used, which has a primary beam size of $\sim$55'' at $\sim$115 GHz. In cases where the FIR emission of the galaxy extends beyond this scale, multiple pointings are combined into a mosaic to ensure that CO is observed all the way to the outskirts of the galaxy. The largest recoverable scale is 25''. Band 3 observations were performed between the 7th and 12th of January 2016, subdivided in three Scheduling Blocks (SBs) in order to meet the sensitivity requirements of the different targets whilst keeping maximum efficiency: single fields, small mosaics, and dwarfs. The sensitivities achieved are listed in Table \ref{tab:observed_props}. The first Scheduling Block consists of one Execution Block (EB): uid\_\_\_A002\_Xaeaf96\_X515. The small mosaics are divided over two Execution Blocks: uid\_\_\_A002\_Xaec9ef\_X5c0 and uid\_\_\_A002\_Xaec9ef\_X88a. The same is true for the dwarfs, which are divided over Execution Blocks uid\_\_\_A002\_Xaecf7b\_X32d4 and uid\_\_\_A002\_Xaecf7b\_X3943. For each SB one spectral window was centred at 114.756, 114.547, and 114.716 GHz, respectively, to target the $^{12}$CO(1-0) rotational line. The bandwidths are 1.875 GHz, covering 3840 channels. The other spectral windows, covering 128 channels each with total bandwidths of 2 GHz, were used to target the band 3 continuum of the individual galaxies. Their central frequencies, along with other details of the observations, are listed in Table \ref{tab:obspars}. The expected calibration uncertainty of the data is 10\%. Synthesized beam sizes and the sensitivities achieved are listed in Table \ref{tab:observed_props}.

\subsubsection{Data reduction}
\label{subsub:data_reduction}
\indent The data were calibrated manually using the Common Astronomy Software Applications package (CASA, version 5.1.1, \citealt{McMullin2007}), using standard ALMA calibration scripts\footnote{The scripts used can be found on \url{https://github.com/NikkiZabel/AlFoCS_data_reduction_scripts}}. Several antennas were flagged manually, mostly because of high system temperatures or outliers in the data of the flux calibrator. The resulting ``dirty'' images were then ``cleaned'' using the \verb|tCLEAN| algorithm \citep{Hogbom1974} in CASA. In cases where both CO and continuum are detected, a continuum estimate is created using the full line-free bandwidth, and subtracted from the channels containing the CO line using the \verb|uvcontsub| command. Cleaning of the channels containing the CO line was done interactively, using a natural weighting scheme (equivalent to a Briggs weighting scheme \citep{Briggs1995} with a robust parameter of 2). Many of the sources have extended emission, and using natural weighting will help ensure that this is recovered in the data. This choice also maximises the signal-to-noise at the cost of decreased spatial resolution. The channel widths of most final data cubes are 10 \kms, as is usually chosen for this type of data \citep[e.g.][]{Alatalo2013}, and the pixel sizes 0.5 arcseconds. Exceptions are the dwarf galaxies FCC207 and FCC261, for which channel widths of 2 \kms were used, because of their narrow line widths (see Table \ref{tab:observed_props}). The result is a three-dimensional RA-Dec-velocity data cube for each galaxy. Primary beam (PB) corrections are then carried out as a separate step using the \verb|impbcor| command, allowing us to store both PB corrected and non PB corrected data cubes. Beam sizes and sensitivities are listed in Table \ref{tab:observed_props}. Typical rms noise levels are around $\sim$3 mJy/beam. Channel maps of all detected galaxies in the sample can be found in the online data of the journal, or on \url{https://github.com/NikkiZabel/AlFoCS_channel_maps}.

\subsubsection{NGC1365}
\label{subsub:NGC1365}
In order to expand our sample, an already reduced image of NGC1365 was taken from the ALMA archive (project ID: 2015.1.01135.S, PI: Egusa, Fumi). It was observed on 20 March 2016. ALMA's 12m configuration was used, with a primary beam size of $\sim$55'' at $\sim$115 GHz. The mosaic covers an area of $\sim 6.6' \times \sim 4.4'$. The central frequency of spectral window 3 (the window centred on the $^{12}$CO(1-0) line) is 114.848 GHz or 1100 \kms. The bandwidths are 1.875 GHz, covering 3840 channels. The spectral resolution is 2.55 \kms. To obtain the final data cube, the \verb|CLEAN| algorithm in CASA version 4.7.0 was used. A continuum estimated and subtracted from the channels containing the CO line as described in \Section \ref{subsub:data_reduction}. A briggs weighting scheme was adopted \citep{Briggs1995} with a robust parameter of 0.5. The pixel sizes of the final data cube are 0.3 arcseconds, and the channel width 5 \kms. The synthesized beam size and the sensitivity achieved are listed in Table \ref{tab:observed_props}.

Aside from the data reduction, this observation is treated the same as the galaxies observed as part of this survey. 

\subsection{Mopra data}
\label{sub:Mopra}
Additional single-dish observations of Fornax cluster galaxies from the Mopra Fornax Cluster CO-Line Legacy Survey (PI: M.W.L. Smith) are included, a survey of $^{12}$CO(1-0) in 28 galaxies in the Fornax cluster, carried out between the nights of 08-08-2012 and 17-09-2012. The Mopra Spectrometer (MOPS) was used in Wideband Mode, centred at a rest frequency of 115.500 GHz for all targets. Its coverage is 8.3 GHz (or 30,378 \kms), and its spectral resolution 0.915 \kms. The FWHM of the beam is 33 $\pm$ 2'' at 115 GHz \citep{Ladd2005}. The calibration uncertainty is less than 10\% \citep{Ladd2005}, we adopt a conservative value of 10\% here. The data were reduced using the ATNF \verb|LIVEDATA| \citep{Barnes2001} and \verb|GRIDZILLA| \citep{Sault1995} packages. \verb|LIVEDATA| is used to fit baselines and transform the raw datafiles to SDFITS files. We fit linear baselines to all spectra and mask the top and bottom 300 channels. \verb|GRIDZILLA| is then used to combine these files into data cubes. The spatial resolution of these cubes is 0.25' per pixel. We use our own scripts to combine the data from the various pointings into a mosaic. For a few objects only single pointings were required. For these objects the data reduction is done using our own scripts to obtain the quotient spectrum by subtracting and dividing by the obtained reference spectra, perform baseline subtraction, and velocity-bin the data. A ripple in the baseline is present in some of the data. This is a known issue with the Mopra telescope, and attempts to mitigate it here, for example by flagging in Fourier space, were not successful. The noise levels in these data are higher, but the data are still usable for the aims of this work.

\subsection{Optical data}
\label{sub:optical_data}
To allow for a comparison of the distribution of the cold molecular gas with the stellar bodies of the galaxies, and to create three-colour images, \textit{r}-, \textit{g}-, and \textit{u}-band images were obtained from the Fornax Deep Survey (FDS, \FDS) for all galaxies in which CO(1-0) was detected in AlFoCS. The FDS is a new, deep multi-band optical survey of the Fornax cluster, which covers 26 square degrees around the virial radius, including the SW sub-group centered on NGC1316 \citep{Iodice2017}. It has been obtained with the ESO VLT Survey Telescope (VST), which is a 2.6-meter diameter optical survey telescope located at Cerro Paranal, Chile \citep{Schipani2012}. The imaging is done in the \textit{u}', \textit{g}', \textit{r}' and \textit{i}'-bands using the 1$^o \times 1^o$ field of view OmegaCAM instrument \citep{Kuijken2002} attached to VST. The deep images provide data with excellent resolution with mean seeing of 1 arcsec and pixel size of 0.2 arcsec. The quality of the data and the photometry of the galaxies are described in detail in \citet{Venhola2018}. The survey area is covered with homogeneous depth with the 1$\sigma$ limiting surface brightness over 1 pixel area of 26.6, 26.7, 26.1 and 25.5 mag arcsec$^{-2}$ in \textit{u}', \textit{g}', \textit{r}' and \textit{i}', respectively. When averaged over a 1 arcsec$^2$ area, these numbers correspond to 28.3, 28.4, 27.8, 27.2 mag arcsec$^{-2}$ in \textit{u}', \textit{g}', \textit{r}' and \textit{i}', respectively. The photometric calibration errors of the FDS are 0.04, 0.03, 0.03, and 0.04 mag in \textit{u}', \textit{g}', \textit{r}' and \textit{i}', respectively. \citet{Venhola2018} produced S\'{e}rsic model fits for all the dwarf galaxies within the survey area using GALFIT \citep{Peng2002,Peng2010}. In addition, Iodice et al. (2018) have studied all bright ($m_B<15$ mag) ETGs  inside the virial radius of the  cluster (some of them are presented in this work). They released the total magnitudes, effective radii and stellar masses and discussed the structure and colors of the galaxy outskirts.

\subsection{Redshift determinations}
\label{sub:redshifts}
A subset of the AlFoCS objects were observed with the 3.9m Anglo-Australian Telescope at the Siding Spring Observatory as part of a larger programme. The AAOmega spectrograph (\citealt{Sharp2006}; \citealt{Saunders2004}) and the Two-degree Field (2dF; \citealt{Lewis2002}) fibre positioner were used, along with the 580V and 385R gratings, providing spectral coverage over 3740--8850\AA. The spectra were reduced using the 2dFDR software package (\citealt{Croom2004}), and spectral classifications and redshifts were determined using MARZ (\citealt{Hinton2016}). Velocities derived from these redshifts are listed in Table \ref{tab:targets}, indicated with a $\ddagger$.

\subsection{Moment maps}
\label{sub:moment_maps}
\begin{figure*}

	\centering

	\subfloat[]
	{\hspace{-5mm}\includegraphics[height=0.35\textwidth \label{subfig:rgb_reg}]{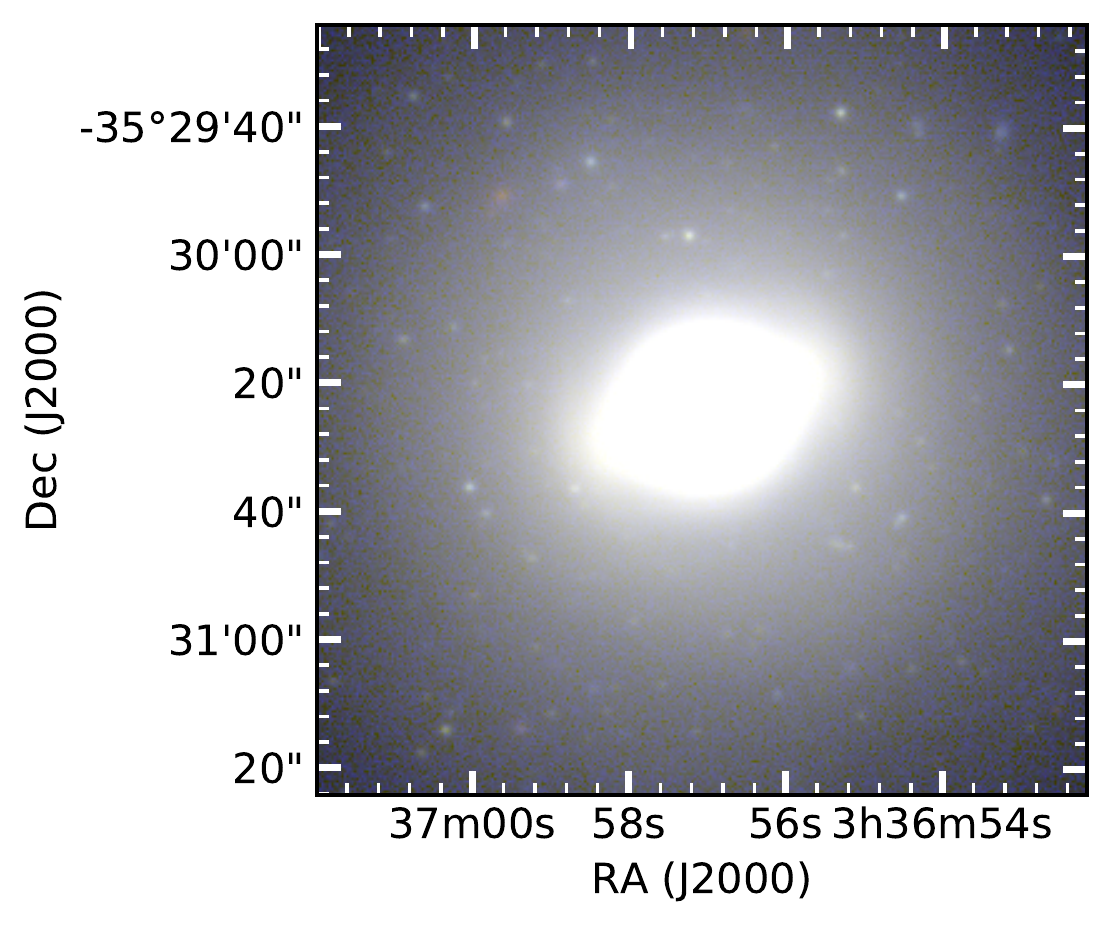}}	
	\hspace{5mm}
	\subfloat[]
		{\includegraphics[height=0.35\textwidth \label{subfig:intens_map_reg}]{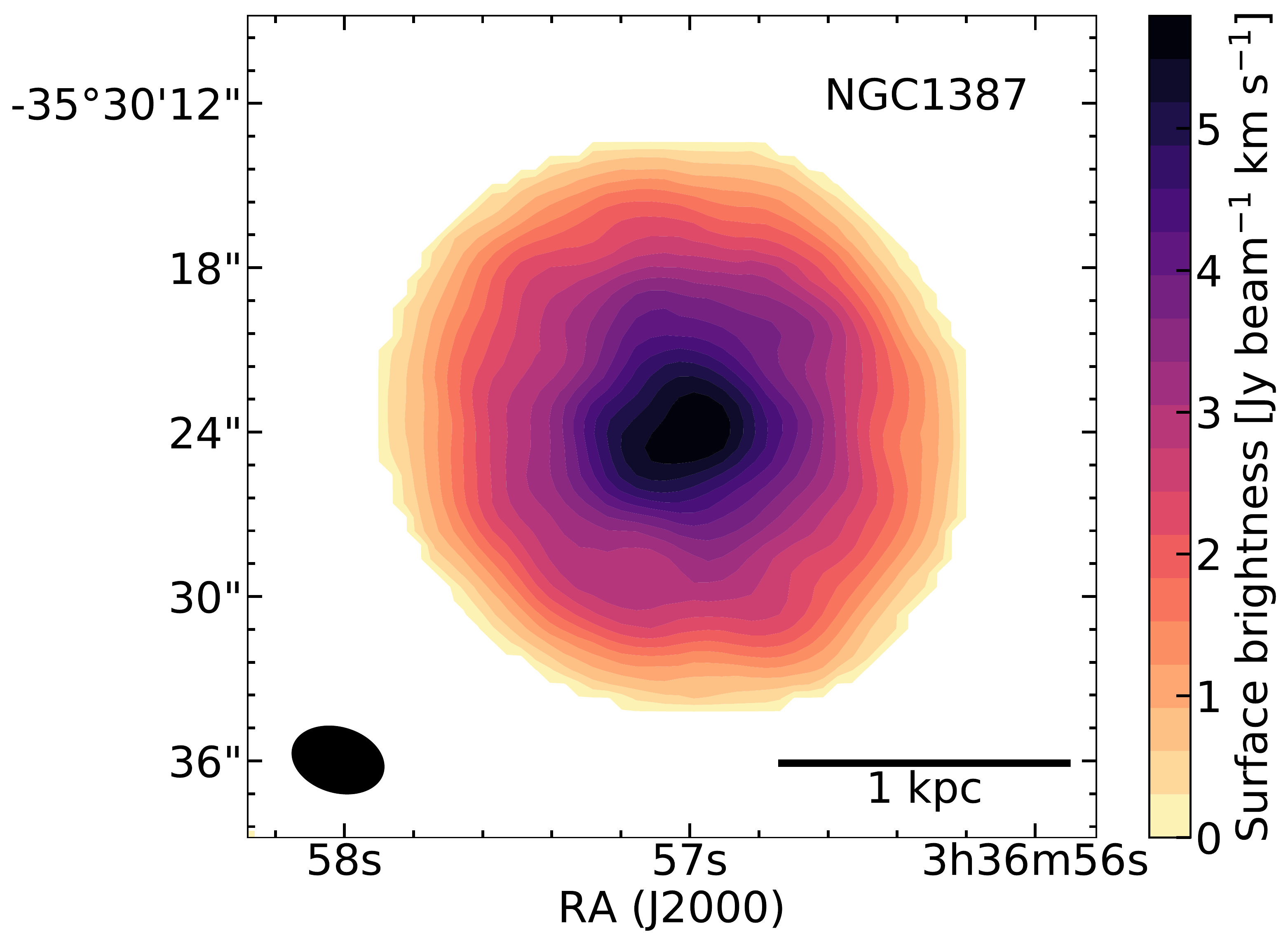}}	
	
	
	\subfloat[]
		{\hspace{0mm}\includegraphics[height=0.35\textwidth \label{subfig:vel_map_reg}]{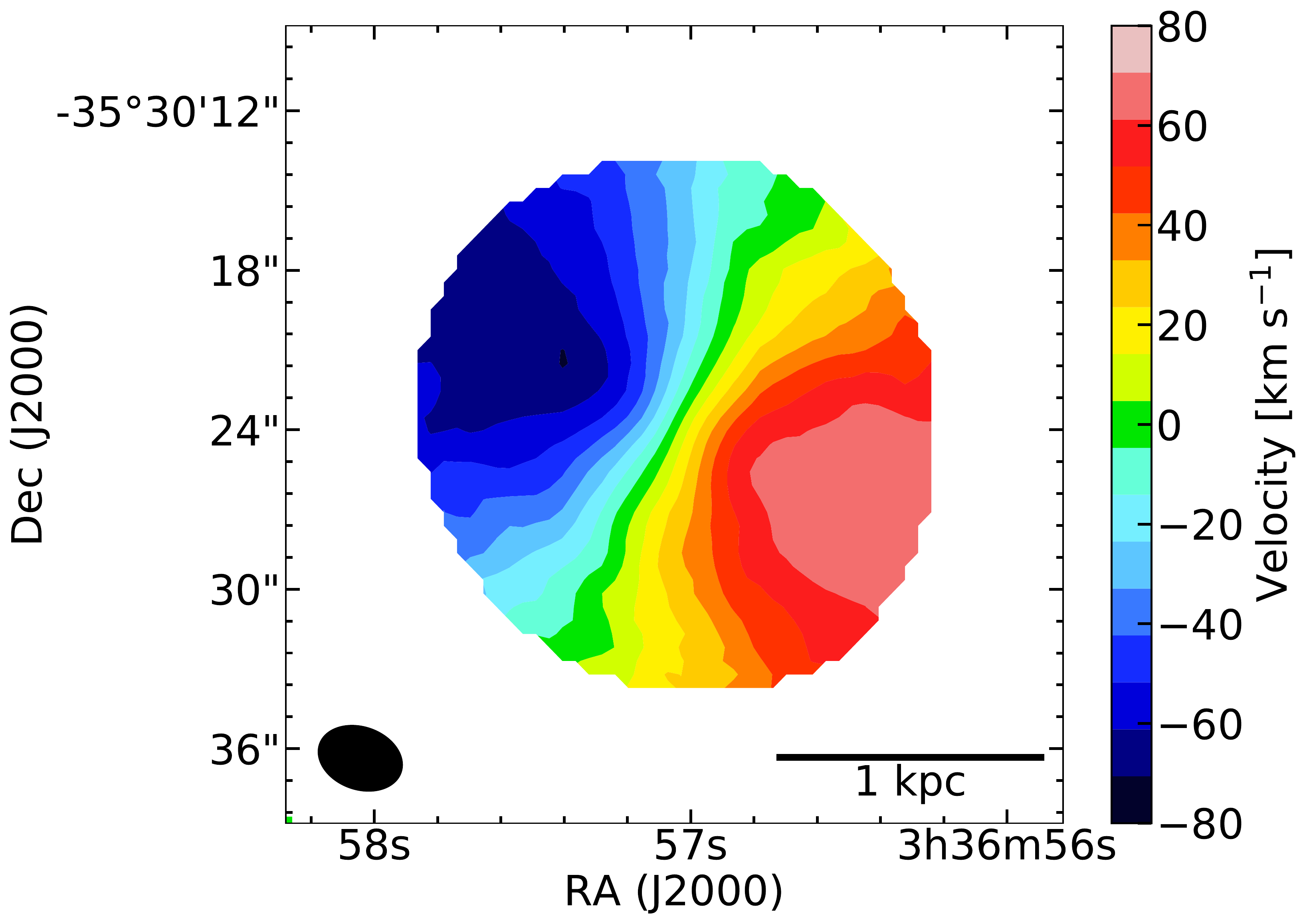}}
	\hspace{10mm}
	\subfloat[]
		{\includegraphics[height=0.35\textwidth \label{subfig:vel_disp_map_reg}]{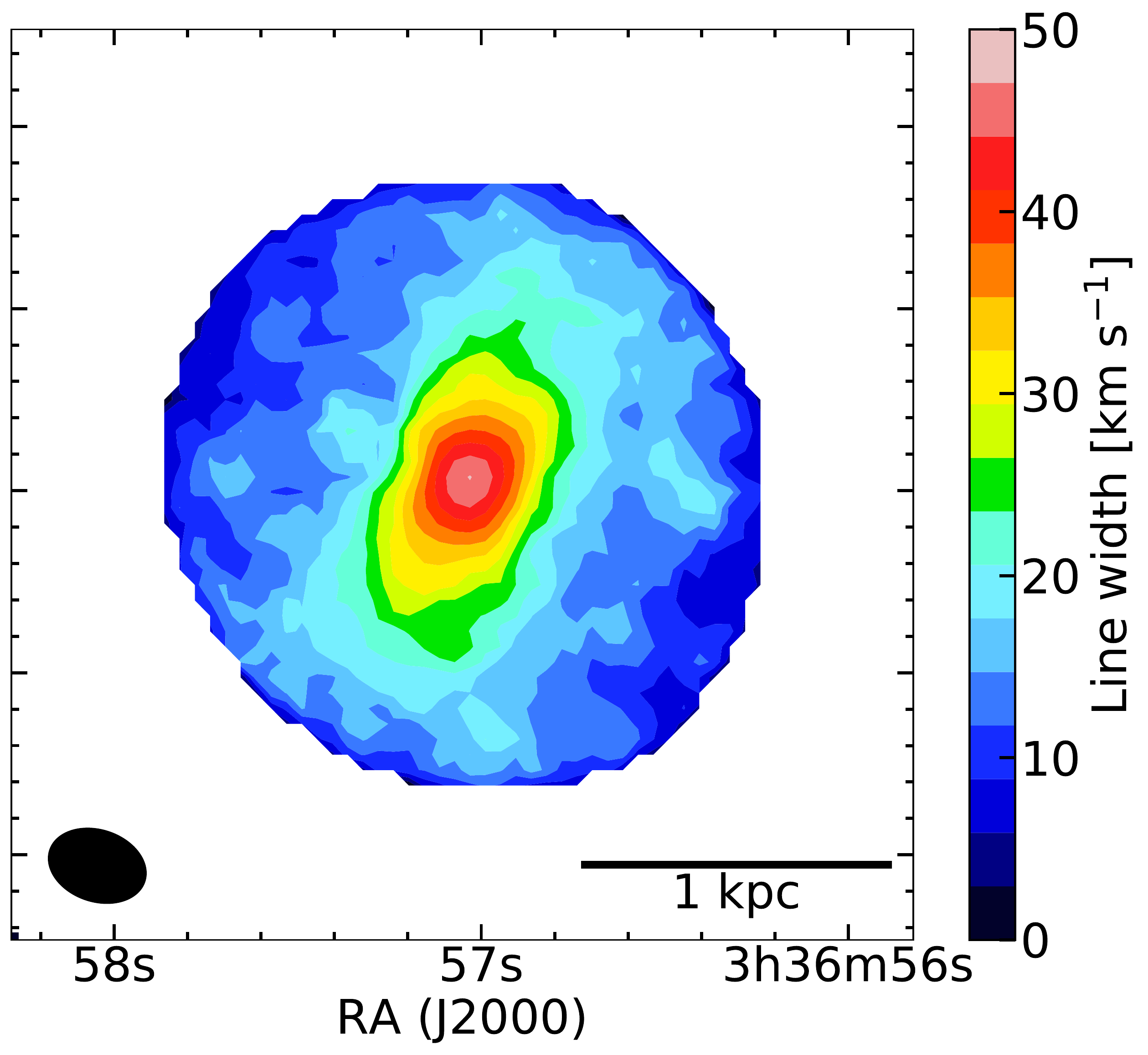}}
		
		
	\subfloat[]
		{\includegraphics[height=0.39\textwidth \label{subfig:PVD_reg}]{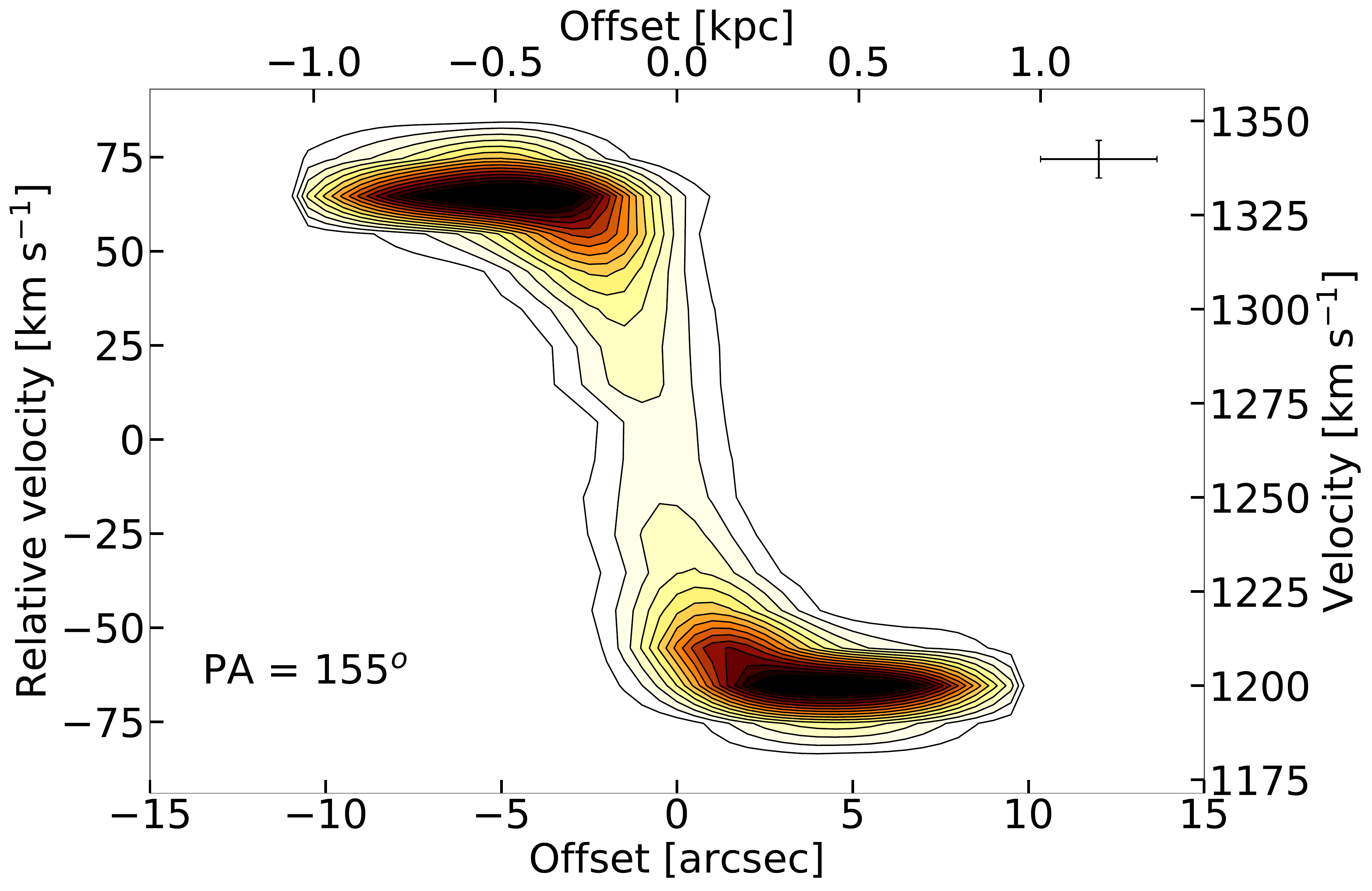}}	
	\hspace{6mm}
	\subfloat[]
		{\includegraphics[height=0.355\textwidth \label{subfig:spectrum_reg}]{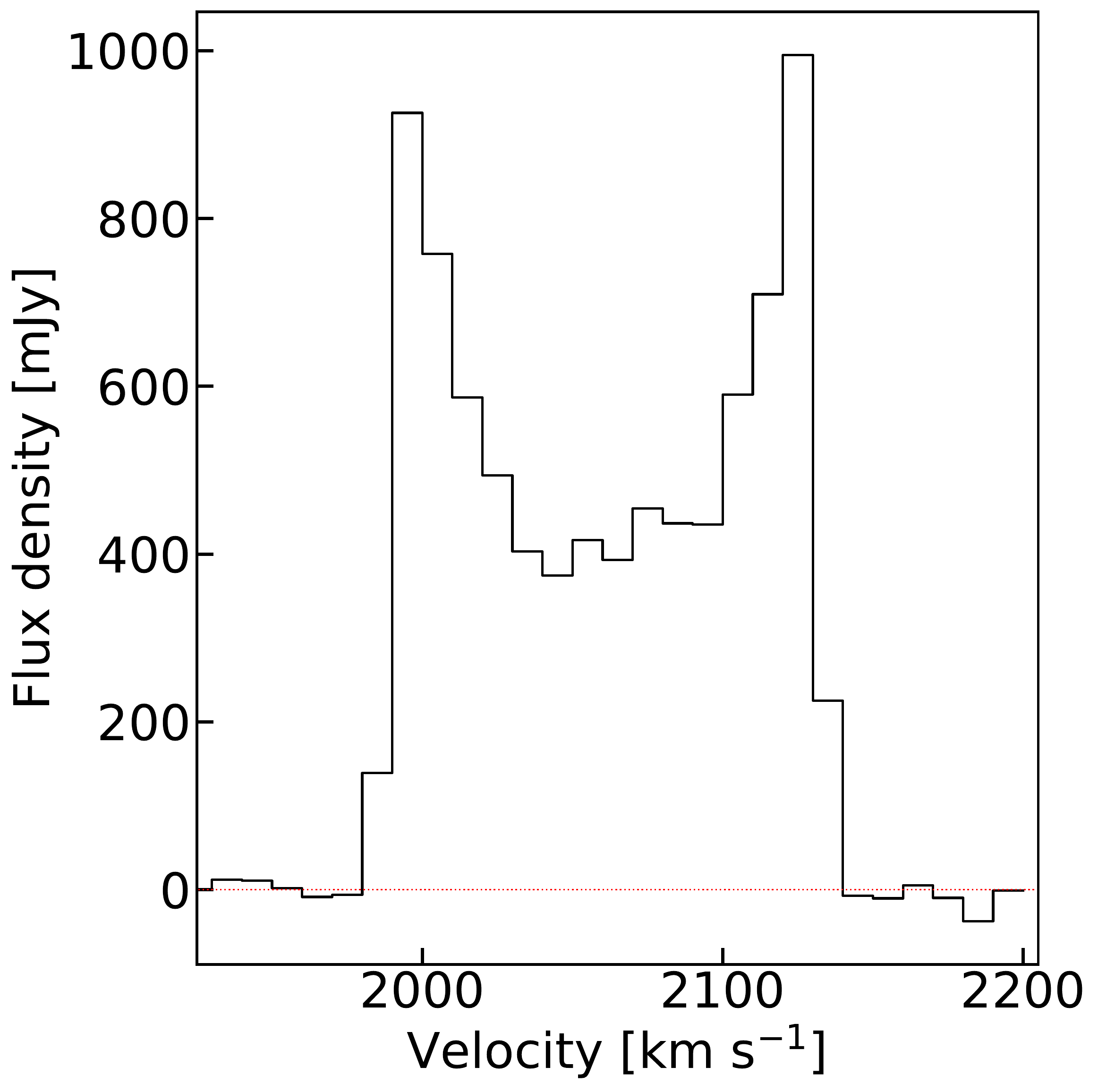}}
		
		
	\caption{a: Three-colour (\textit{r}-\textit{g}-\textit{u}) image of NGC1387. b: Moment zero map: distribution of the cold molecular gas as traced by the ALMA CO data. c: Moment 1 map: velocity map of the cold molecular gas. Each colour represents a 10 \kms velocity channel. d: Moment 2 map: linewidth of the CO integrated spectrum. e: Position-velocity diagram or of the cold molecular gas. The uncertainties in the spatial and velocity directions are indicated in the upper right corner. f: the CO(1-0) line. The beam of the observations is shown in the lower left corners of the moment maps, as well as a 1 kpc scale bar in the lower right corners. NGC1387 is a very regular galaxy with symmetric moment maps.}
	\label{fig:moment-maps_reg}
\end{figure*}

\begin{figure*}
		
	\subfloat[]
	{\hspace{0mm}\includegraphics[height=0.35\textwidth \label{subfig:rgb_irreg}]{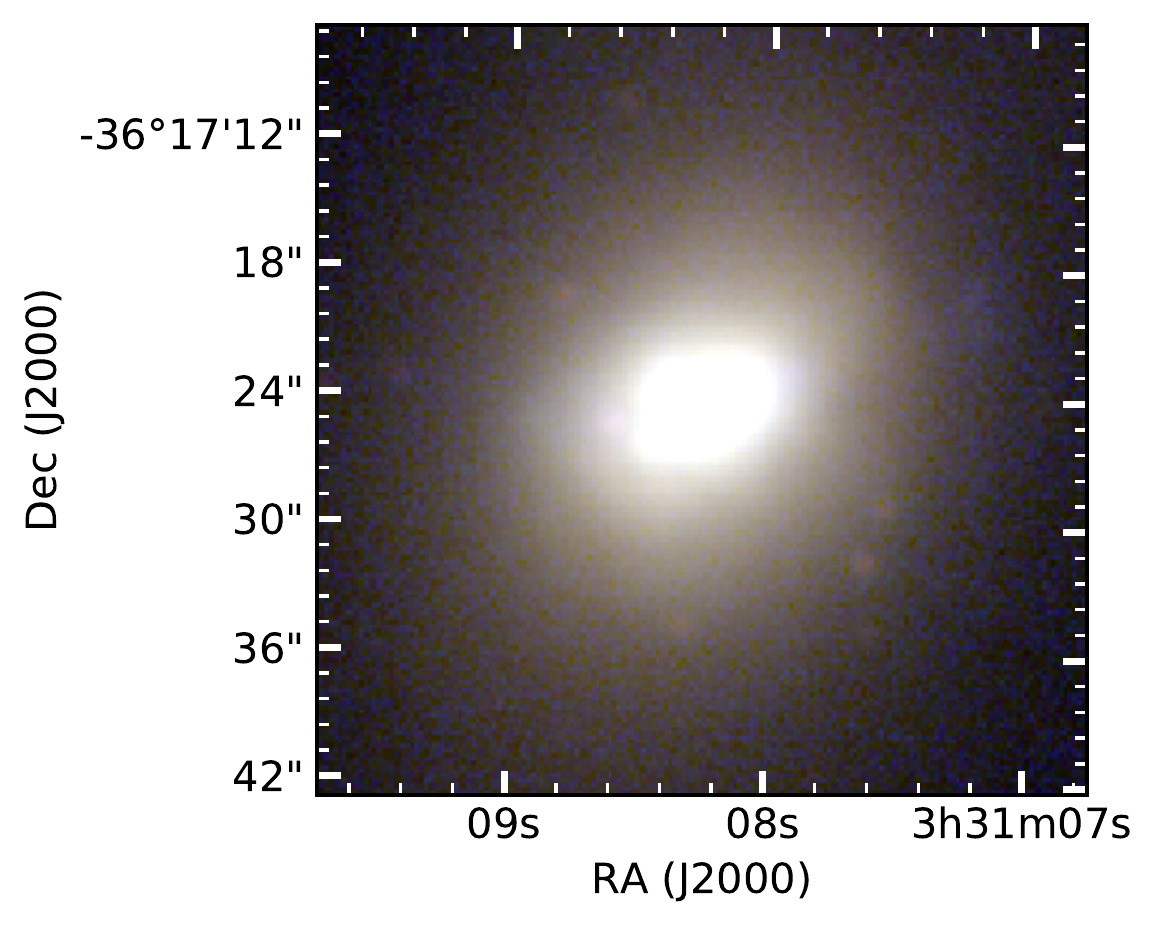}}	
	\hspace{0mm}
	\subfloat[]
		{\includegraphics[height=0.35\textwidth \label{subfig:intens_map_irreg}]{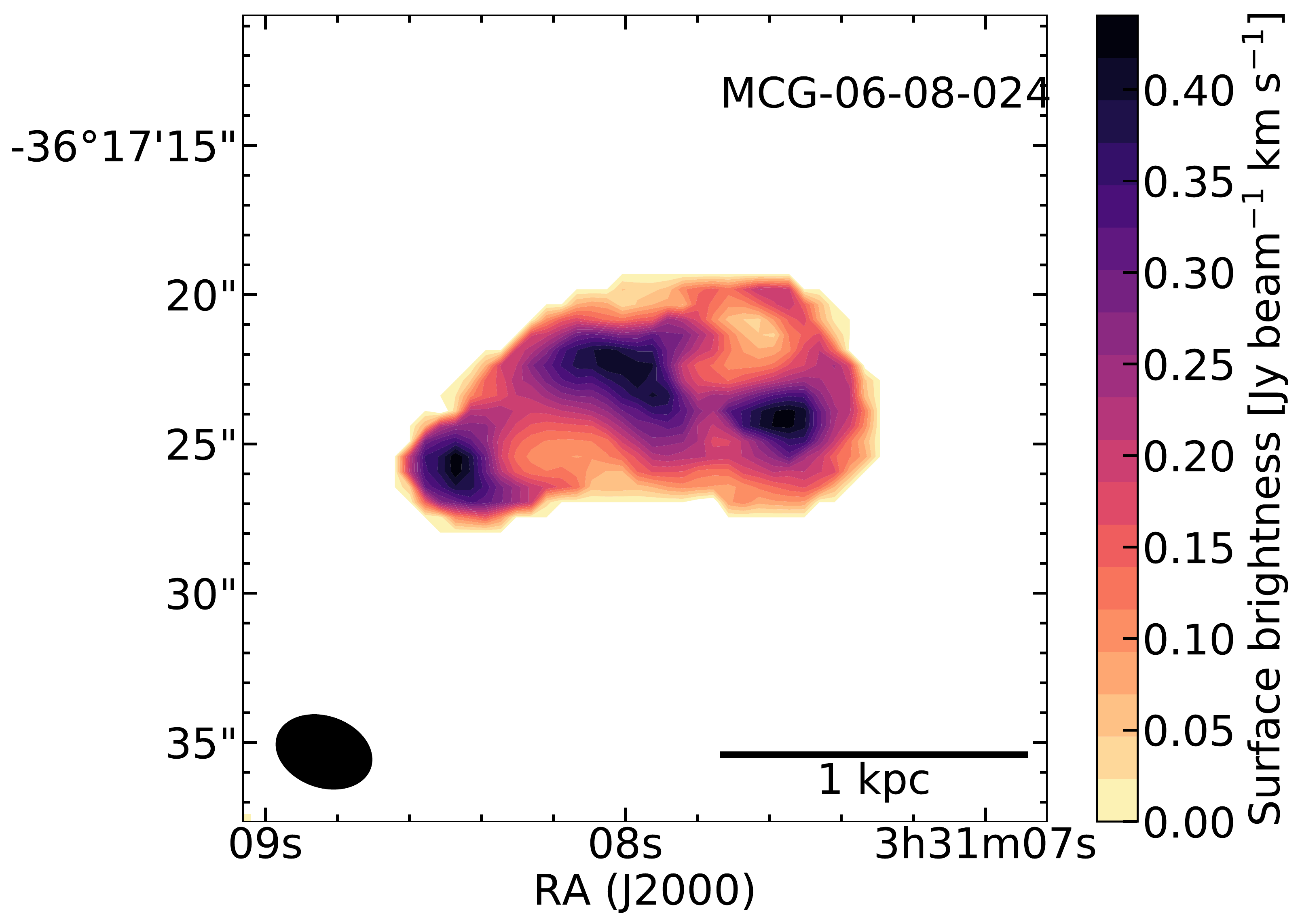}}	
	
	
	\subfloat[]
		{\hspace{-3mm}\includegraphics[height=0.35\textwidth \label{subfig:vel_map_irreg}]{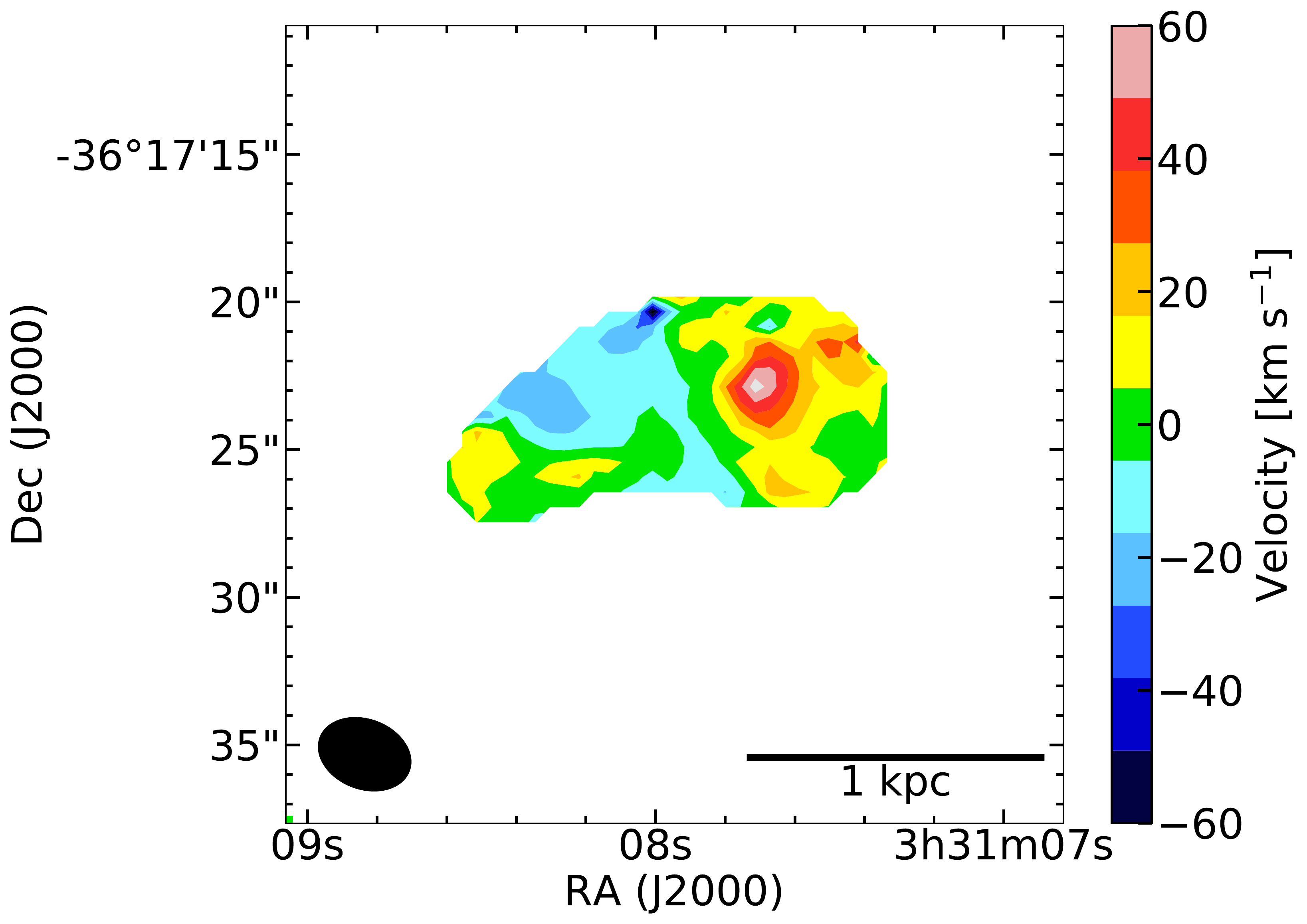}}
	\hspace{3mm}
	\subfloat[]
		{\includegraphics[height=0.35\textwidth \label{subfig:vel_disp_map_irreg}]{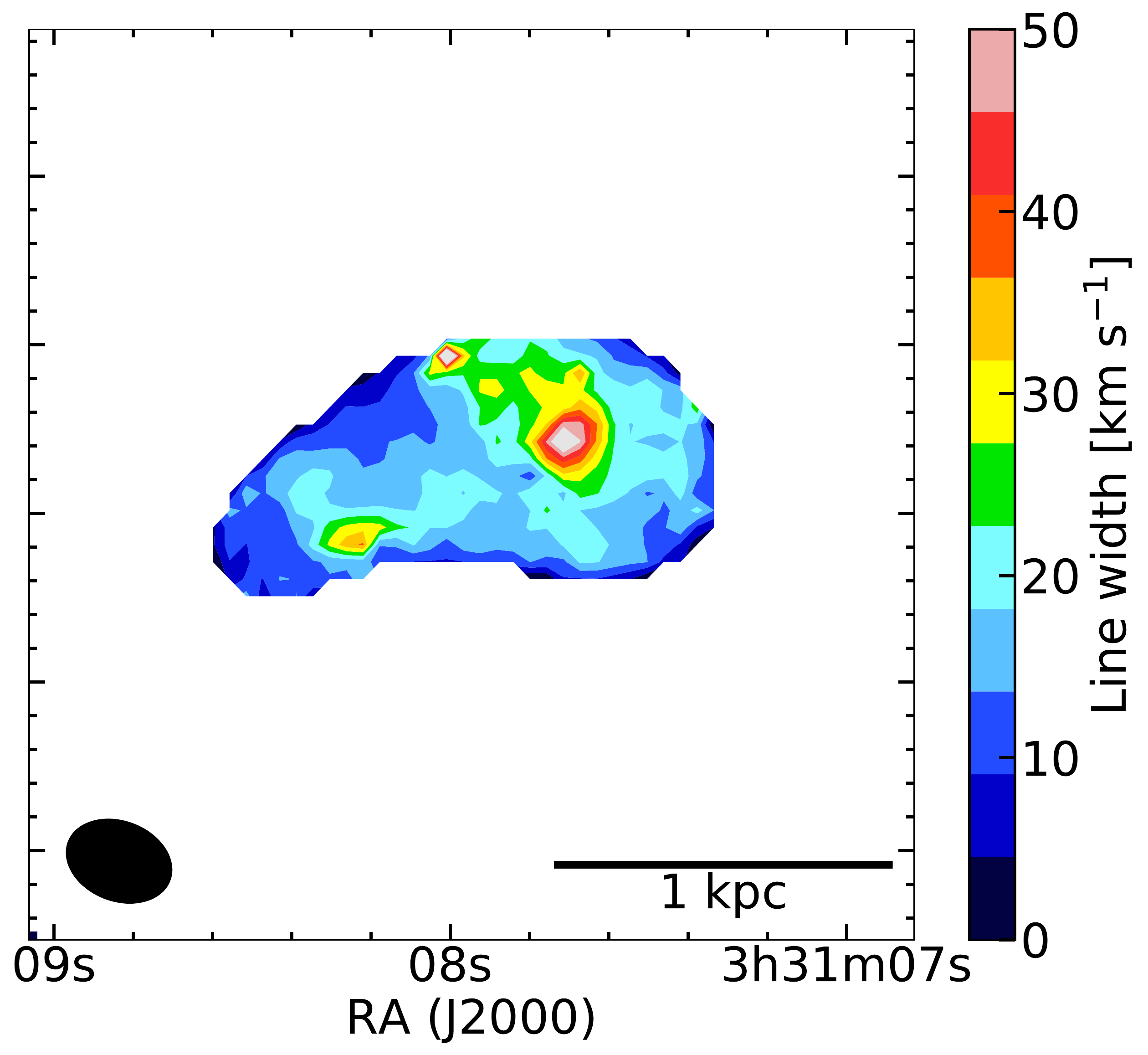}}
		
		
	\subfloat[]
		{\includegraphics[height=0.39\textwidth \label{subfig:PVD_irreg}]{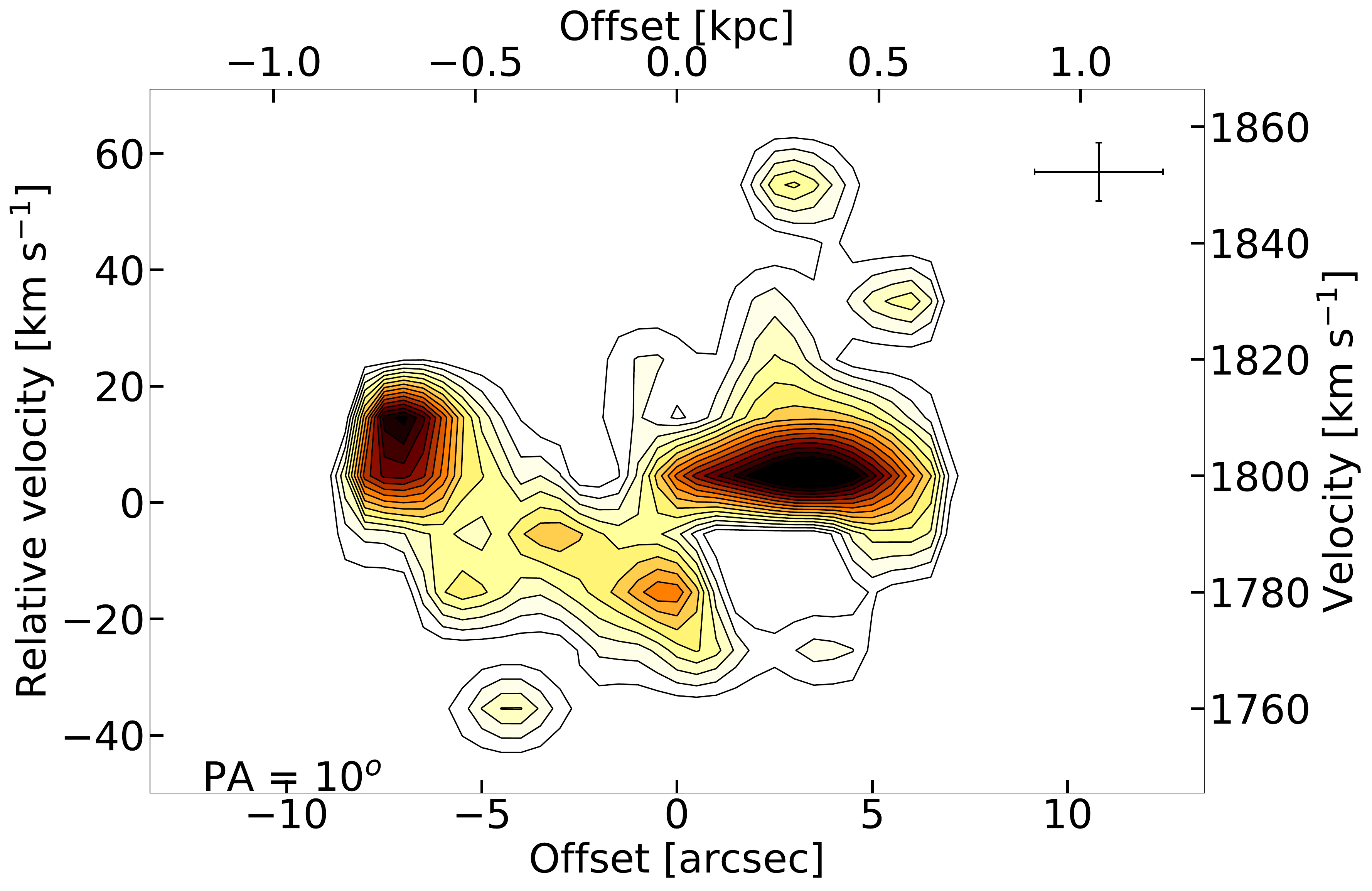}}	
	\hspace{6mm}
	\subfloat[]
		{\includegraphics[height=0.355\textwidth \label{subfig:spectrum_irreg}]{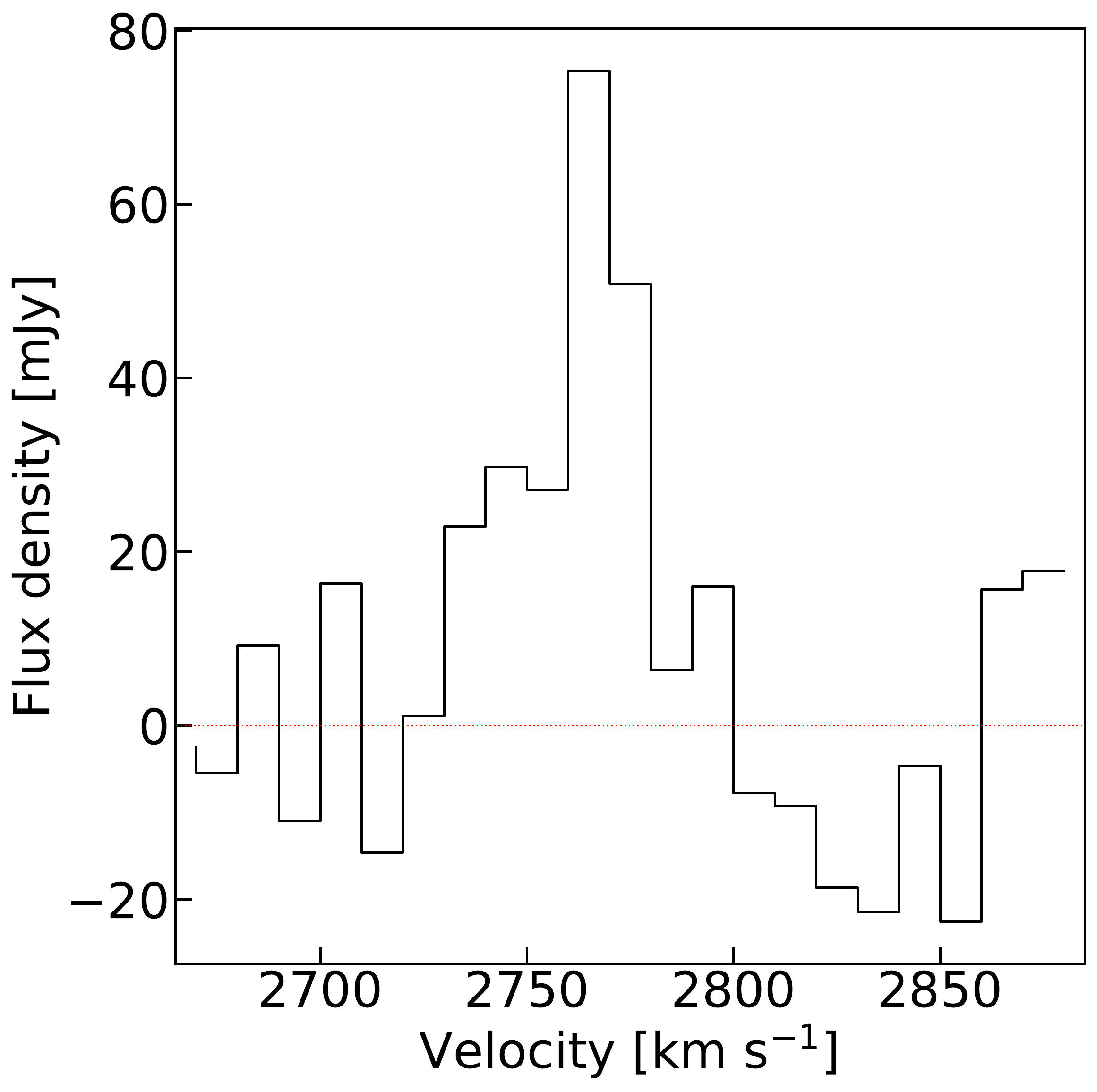}}
		
	\caption{MGC-06-08-024 (FCC90), similar to Figure \ref{fig:moment-maps_reg}. MGC-06-08-024 is galaxy with irregular CO emission, and therefore has irregular moment maps and an irregular position-velocity diagram.}
	\label{fig:moment-maps_irreg}
\end{figure*}

\indent Cleaned data cubes were used to produce moment maps of the CO(1-0) line emission, using the masked moment method \citep{Dame2011}. While PB corrected images are used in the remainder of this work, for the purpose of clarity uncorrected maps are presented in Figure \ref{fig:moment-maps_reg}, Figure \ref{fig:moment-maps_irreg}, and Appendix \ref{app:moment_maps}. In order to create the mask, a Gaussian smoothing was applied to a copy of the data cube, in both spatial axes as well as the velocity axis, with a full width at half the maximum (FWHM) of 1.5 times the beam's major axis for the spatial axes, and 4 channels (proven to be optimal from previous experience) for the velocity axis. Using this smoothed copy as a mask, the data cubes were then ``clipped'' to the $x\sigma$ level, which means that all spaxels below this value are set to zero, where $x$ is chosen to give the best visual result, and equals 3 or 4.

\begin{figure*}
	
	\centering
	
	\subfloat[]
	{\includegraphics[height=0.4\textwidth \label{subfig:overplot-MCG}]{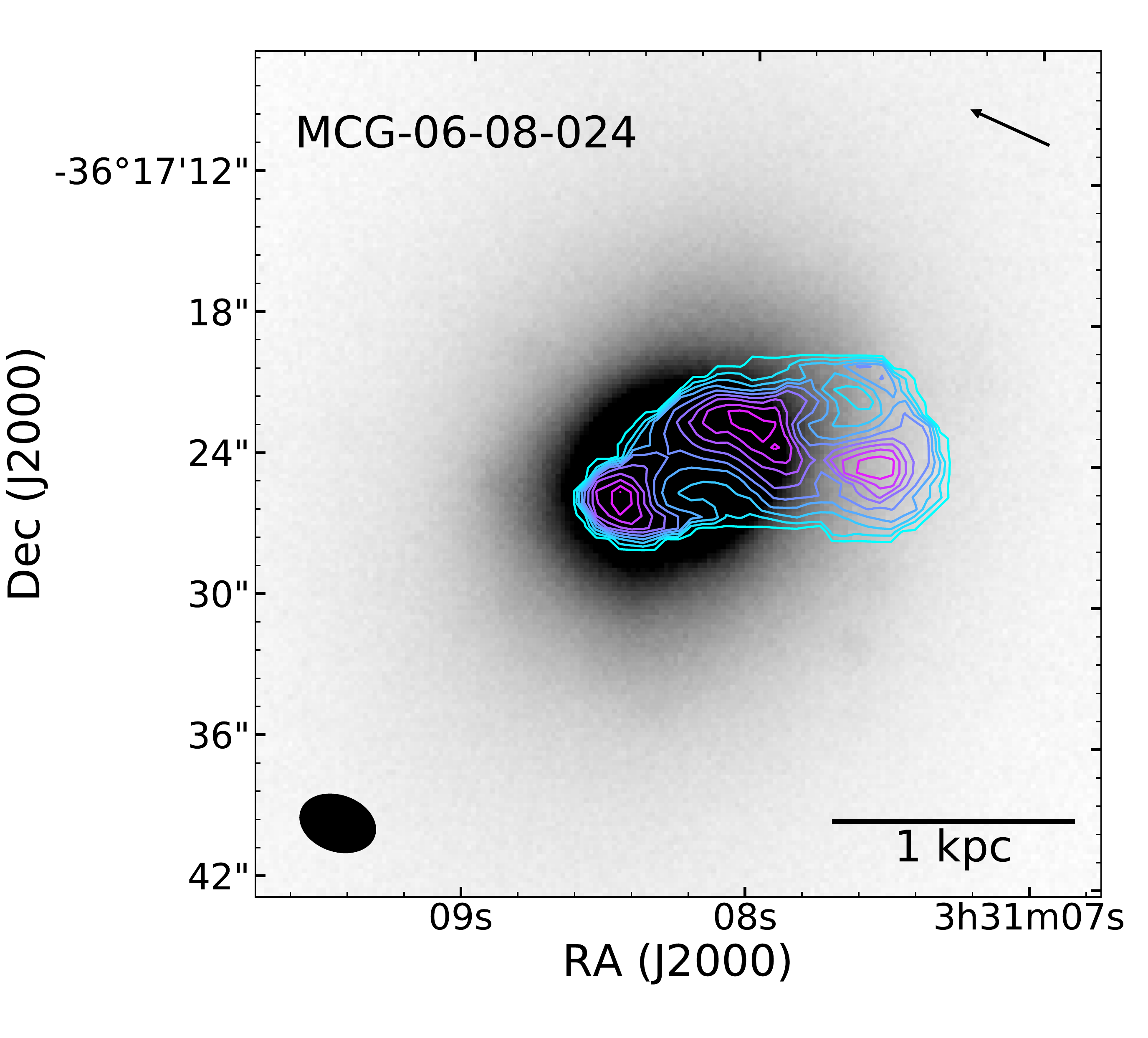}}	
	\subfloat[]
	{\includegraphics[height=0.4\textwidth \label{subfig:overplot-G002}]{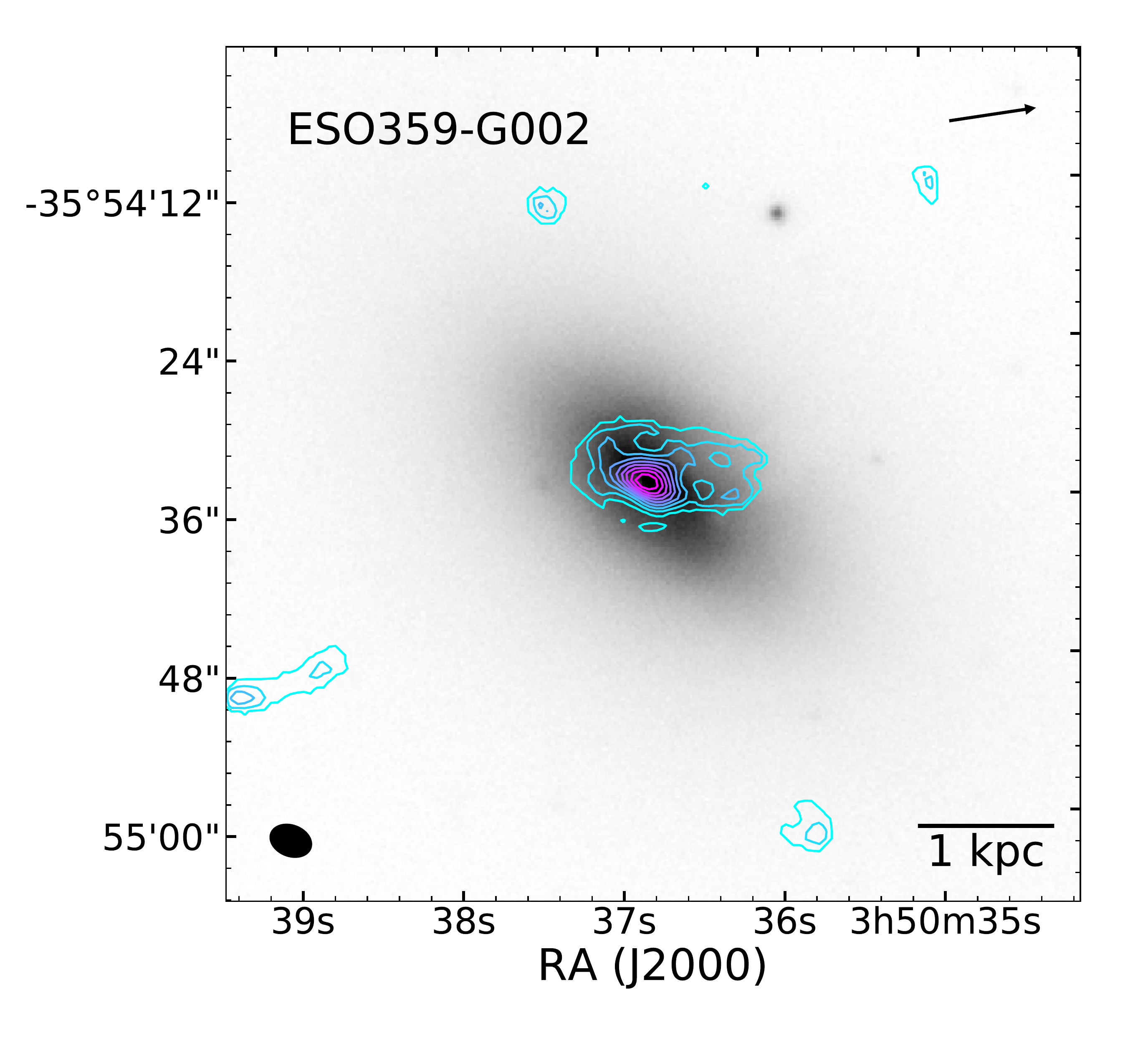}}
	
	\hspace{1mm}\subfloat[]
	{\includegraphics[height=0.4\textwidth \label{subfig:overplot-282}]{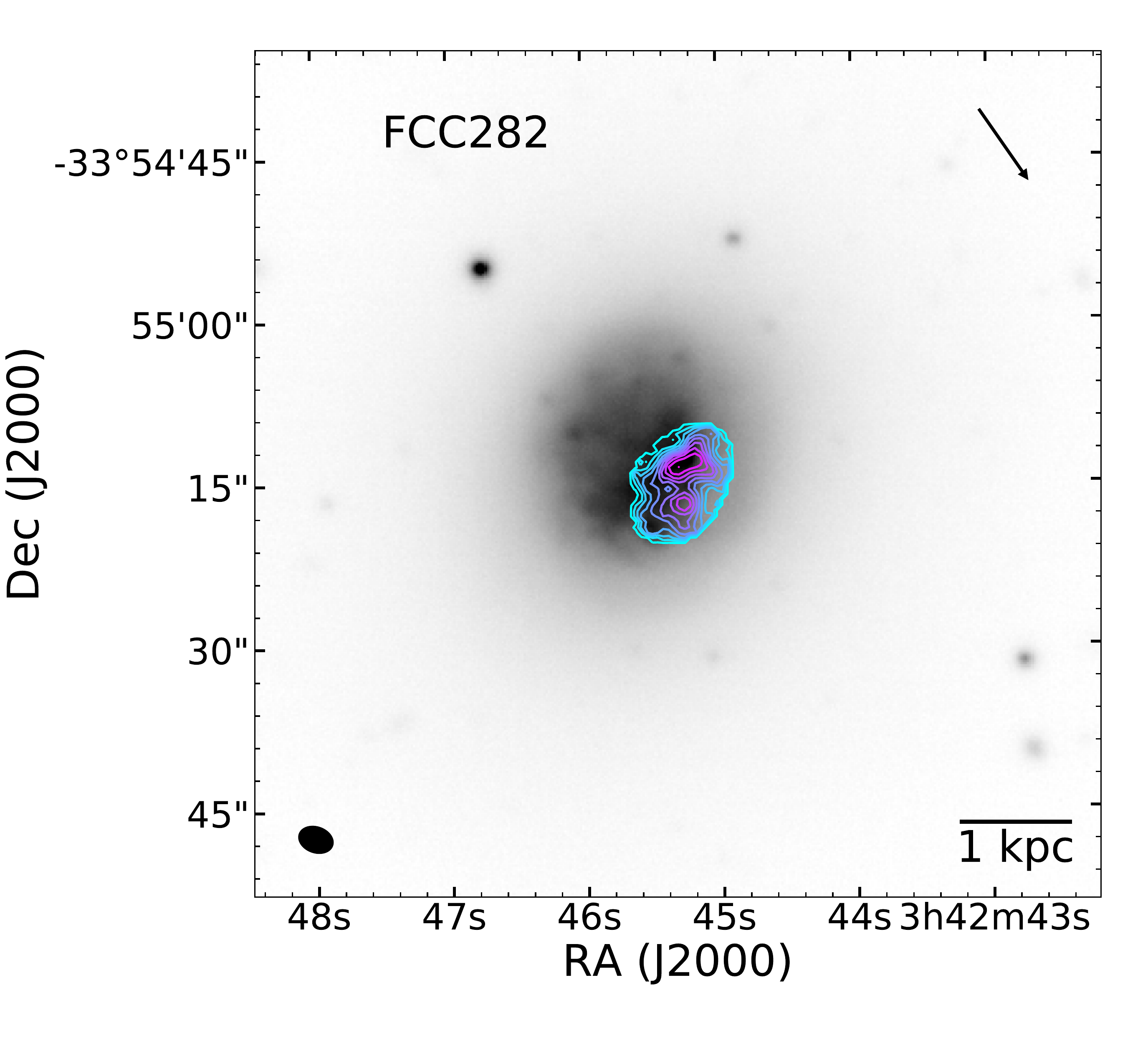}}	
	\hspace{2mm}
	\subfloat[]
	{\includegraphics[height=0.4\textwidth \label{subfig:overplot-332}]{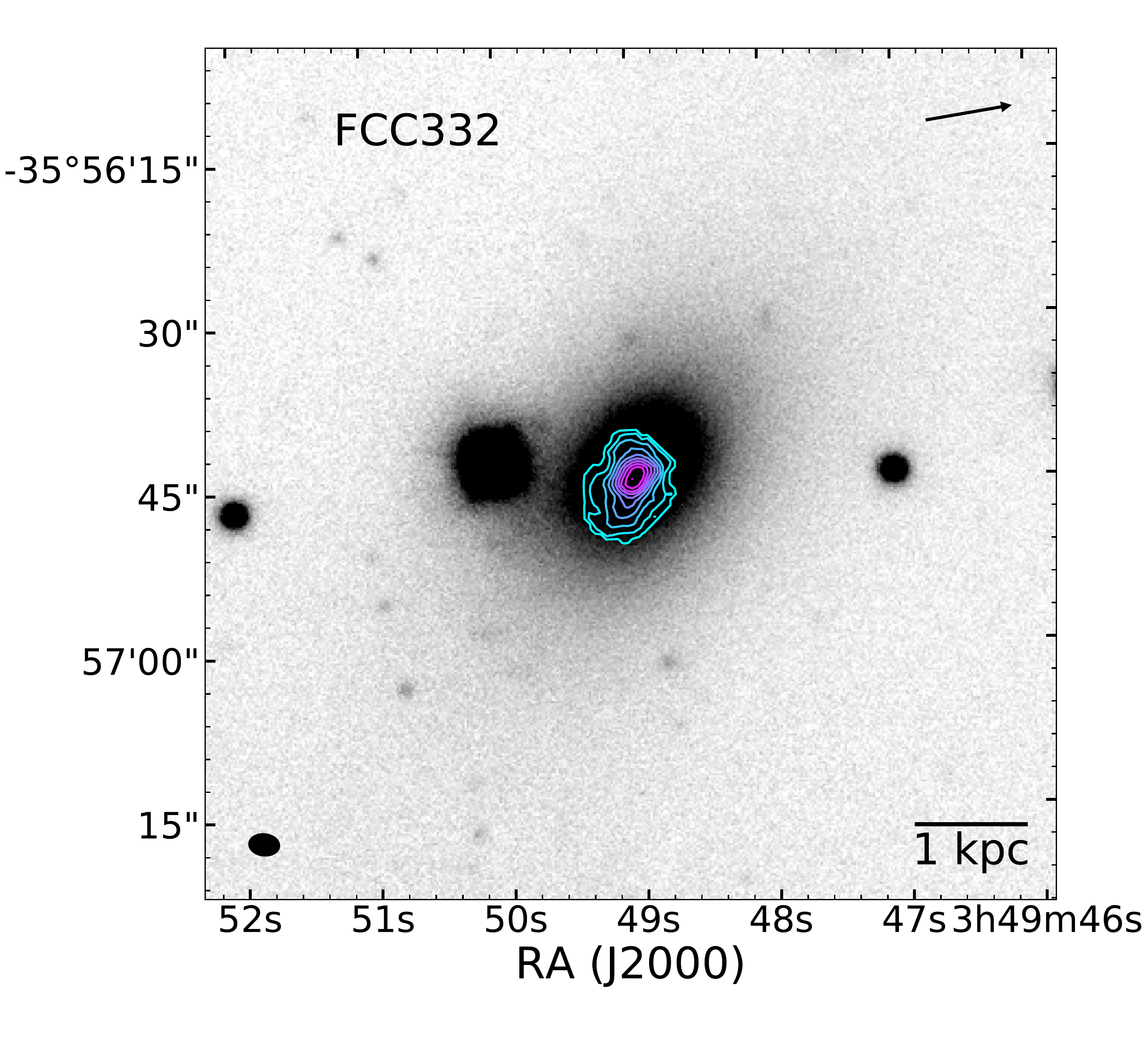}}	
	
	\caption{CO(1-0) emission overplotted on optical (\textit{g}-band) images from the FDS (\FDS). The CO emission is shown as 10 coloured contours, the outer contour is set equal to 3 or 4 $\sigma$, while the level of the innermost contour depends on the highest signal measured in the galaxy. The beam is shown in the lower left corners, and a 1 kpc scalebar in the lower right corners. The arrow in the upper right corners indicates the direction towards the cluster centre. The molecular gas in these galaxies is asymmetric with respect to their stellar bodies. In \ref{subfig:overplot-MCG} and \ref{subfig:overplot-G002} it extends beyond the stellar body and forms a tail aligned with the direction of the cluster centre.}
	\label{fig:overplots}
\end{figure*}

In Figure \ref{fig:moment-maps_reg} and Figure \ref{fig:moment-maps_irreg} moment maps of NGC1387 and MCG-06-08-024 are shown, serving as examples of the regular and disturbed galaxies, respectively (see \Section \ref{sub:morphologies} for more details). The top left panel of each of these figures is a three-colour image, constructed using the \textit{r}-, \textit{g}-, and \textit{u}- band images from the Fornax Deep Survey (\FDS, see \Section\ref{sub:optical_data}). 
The top right panels are intensity or moment zero maps of the cold molecular gas as traced by the ALMA CO data, showing its spatial distribution. The black ellipse in the lower left corner shows the beam of the observations, and a 1 kpc scalebar is shown in the lower right corner. This is the same in the other two moment maps.
The middle left panels are velocity or moment one maps of the galaxies. Each of the colours represents a 10 \kms (2 \kms for FCC207 and FCC261, see \Section \ref{subsub:data_reduction}) velocity channel. The warm colours represent the positive, redshifted velocities, and the cold colours represent the negative, blueshifted velocities. 
Middle right figures are moment two maps, representing the linewidth. 

The bottom left figures are position-velocity diagrams (PVDs), which reveal the motion of gas along the major axes of the galaxies. They are obtained by defining a slit the size of the beam along the major axis of the galaxy in the data cube, and collapsing it along the minor axis. The errorbars in the upper right corner indicate the PSF FWHM (horizontal) and channel width (vertical). 
The bottom right figures show the part of the galaxy's spectrum containing the CO(1-0) line. The spectrum was obtained by defining a rectangular aperture around the detected emission in the spatial directions, large enough to contain all its CO emission, and then collapsing the data cube along both spatial axes. 

In NGC1387 (Figure \ref{subfig:intens_map_reg}) the gas is distributed as an almost face-on disk, with the projected intensity decreasing radially. Its velocities vary between -80 and +80 \kms relative to the systemic velocity, which is determined by taking the mean of the moment one map shown here. The line is widest in a band along the kinematic minor axis, due to beam smeared rotation. The PVD of NGC1387 (Figure \ref{subfig:PVD_reg}) is very regular, showing a smooth and symmetric ``rotation curve'', which reaches its maximum very quickly. The double-peaked line profile, typical for a disk, is clearly visible in its spectrum (Figure \ref{subfig:spectrum_reg}). 

In MCG-06-08-024 the molecular gas is distributed irregularly, around three different maxima. The velocities of the gas are between -60 and +60 \kms relative to the systemic velocity. The PVD of MCG-06-08-024 has a very irregular shape.%

Similar images of the remaining 14 detected galaxies were created in the same way, and can be found in Appendix \ref{app:moment_maps}.

\subsection{Comparison to optical morphology}
\label{subsub:optical}
Figure \ref{fig:overplots} overplots the CO integrated intensity contours on top of optical images of the galaxies (\textit{g}-band images of the FDS were used, see \Section \ref{sub:optical_data}). The CO emission is shown as 10 coloured contours, the outer contour being equal to 3-4$\sigma$, while the innermost contour depends on the highest signal measured in the galaxy in question. The arrows in the upper right corners point towards the cluster centre (here defined as the location of the BCG NGC1399). Similar plots for the remaining galaxies can be found in Appendix \ref{app:overplots}. 

The galaxies in Figure \ref{fig:overplots} are all examples of galaxies with irregular CO emission, asymmetric compared to the galaxy's stellar body. In MCG-06-08-024 (Figure \ref{subfig:overplot-MCG}) and ESO359-G002 (Figure \ref{subfig:overplot-G002}) the molecular gas forms a tail that extends beyond the stellar body. These galaxies are discussed further in \Section \ref{sub:RPS}. Other examples of galaxies with asymmetric CO emission are FCC207 (Figure \ref{subfig:overplot_FCC207}) and FCC261 (Figure \ref{subfig:overplot_FCC261}). In the cases of the regular galaxies, such as ESO358-G063 (Figure \ref{subfig:overplot_ESO358-G063}), NGC1386 (Figure \ref{subfig:overplot_NGC1386}), NGC1387 (Figure \ref{subfig:overplot_NGC1387}), and NGC1351A (Figure \ref{subfig:overplot_NGC1351A}), the CO emission follows the optical shape of the galaxy. The CO emission in NGC1380 (Figure \ref{subfig:overplot_NGC1380}) is very compact compared to its stellar body in our images, but has been shown to be distributed in a regular disk by \citet{Boizelle2017}.

It would be interesting to compare the CO morphologies to \HI morphologies, especially for the galaxies that exhibit asymmetric CO emission or gas tails. This would show us whether these galaxies also have \HI tails, which is expected if ram pressure stripping is at play. The current \HI observations available are not of sufficient resolution to do this. However, in the future we will be able to use data from the MeerKAT Fornax Survey for this purpose.

\begin{table*}
	\centering
	\begin{threeparttable}
	\caption{Observed and derived properties of the AlFoCS targets.}
	\label{tab:observed_props}
	\begin{tabular}{llllllllll}
	\hline
	Common name & FCC \# & Reg./dist. & Gauss/box & $b_{\text{maj}}; b_{\text{min}}; b_{\text{PA}}$ & rms & $\Delta v$ & $\int S_\nu d \nu$ & log$_{10}$($M_{\text{H}_{2}}$) & Deficiency \\
	- & - & - & - & (''; ''; $^\text{o}$) & (mJy beam$^{-1}$) & (\kms) & (Jy \kms) & ($M_{\odot}$) & (dex) \\
	(1) & (2) & (3) & (4) & (5) & (6) & (7) & (8) & (9) & (10) \\
	\hline
	FCC32 & 32 & - & G & 2.4; 1.8; 84 & 2.6 & 50 & $\leq$ 2.1 & $\leq$ 8.02 & $\leq$ 0.01 \\
	FCC44 & 44 & - & G & 2.8; 2.0; 85 & 2.4 & 50 & $\leq$ 1.9 & $\leq$ 8.48 & $\leq$ 1.62 \\
	NGC1351A & 67 & R & B & 2.6; 2.0; 61 & 3.6 & 250 $\pm$ 20 & 19.9 $\pm$ 2.0 & 7.83 $\pm$ 0.07 & -0.54 $\mp$ 0.01 \\
	MGC-06-08-024 & 90 & D & G & 3.3; 2.4; 71 & 3.0 & 31 $\pm$ 7 & 1.71 $\pm$ 0.22 & 6.97 $\pm$ 0.07 & -1.11 $\mp$ 0.01 \\
	FCC102 & 102 & - & G & 2.8; 2.0; 85 & 2.4 & 50 & $\leq$ 1.9 & $\leq$ 8.00 & $\leq$ 0.77 \\
	ESO358-G015 & 113 & - & G & 3.1; 2.1; 71 & 3.2 & 50 & $\leq$ 2.5 & $\leq$ 7.61 & $\leq$ -0.36 \\
	ESO358-16 & 115 & - & G & 3.3; 2.3; 73 & 3.3 & 50 & $\leq$ 2.6 & $\leq$ 7.66 & $\leq$ 1.36 \\
	FCC117 & 117 & - & G & 2.8; 2.0; 84 & 2.4 & 50 & $\leq$ 1.9 & $\leq$ 7.53 & $\leq$ 1.04 \\
	FCC120 & 120 & - & G & 2.8; 2.0; 84 & 2.4 & 50 & $\leq$ 1.9 & $\leq$ 8.32 & $\leq$ 0.44 \\
	NGC1365 & 121 & R & B & 2.4; 2.0; 12 & 12 & 940 & 1221 $\pm$ 20 & 9.49 $\pm$ 0.04 & 0.53 $\pm$ 0.01 \\
	NGC1380 & 167 & R & B & 2.6; 2.0; 80 & 3.6 & 660 $\pm$ 20 & 18.1 $\pm$ 1.8 & 7.67 $\pm$ 0.06 & -1.39 $\mp$ 0.01 \\
	FCC177 & 177 & - & G & 3.3; 2.4; 73 & 3.3 & 50 & $\leq$ 2.6 & $\leq$ 8.14 & $\leq$ -0.99 \\
	NGC1386 & 179 & R & B & 3.3; 2.4; 72 & 2.9 & 540 $\pm$ 20 & 88.9 $\pm$ 8.9 & 8.37 $\pm$ 0.04 & -0.61 $\mp$ 0.01 \\
	NGC1387 & 184 & R & B & 3.3; 2.4; 72 & 3.0 & 200 $\pm$ 20 & 83.3 $\pm$ 8.3 & 8.33 $\pm$ 0.04 & -0.74 $\mp$ 0.01 \\
	FCC198 & 198 & - & G & 2.8; 2.0; 84 & 2.4 & 50 & $\leq$ 1.9 & $\leq$ 7.82 & $\leq$ 2.17 \\
	FCC206 & 206 & - & G & 2.8; 2.0; 83 & 2.5 & 50 & $\leq$ 2.0 & $\leq$ 7.34 & $\leq$ 0.71 \\
	FCC207 & 207 & D & G & 2.8; 2.0; 83 & 2.6 & 11 $\pm$ 3 & 0.6 $\pm$ 0.3 & 6.54 $\pm$ 0.22 & -1.33 $\mp$ 0.01 \\
	NGC1427A & 235 & - & G & 2.9; 2.3; 80 & 2.2 & 50 & $\leq$ 1.7 & $\leq$ 7.42 & $\leq$ -1.21 \\
	FCC261 & 261 & D & G & 2.9; 2.0; 84 & 2.6 & 9.5 $\pm$ 3.9 & 0.27 $\pm$ 0.55 & 6.27 $\pm$ 0.88 & -1.47 $\mp$ 0.01 \\
	PGC013571 & 263 & D & G & 3.3; 2.4; 72 & 3.1 & 54 $\pm$ 10 & 7.0 $\pm$ 0.71 & 7.22 $\pm$ 0.05 & -1.02 $\mp$ 0.01 \\
	FCC282 & 282 & D & G & 3.2; 2.4; 70 & 3.1 & 36 $\pm$ 4 & 3.0 $\pm$ 0.34 & 7.15 $\pm$ 0.05 & -0.95 $\mp$ 0.01 \\
	NGC1437A & 285 & - & G & 3.0; 2.1; 70 & 3.0 & 50 & $\leq$ 2.3 & $\leq$ 7.83 & $\leq$ -0.85 \\
	NGC1436 & 290 & R & B & 2.6; 2.0; 79 & 3.2 & 260 $\pm$ 20 & 97.6 $\pm$ 9.8 & 8.44 $\pm$ 0.05 & -0.44 $\mp$ 0.01 \\
	FCC302 & 302 & - & G & 2.8; 2.0; 83 & 2.5 & 50 & $\leq$ 2.0 & $\leq$ 8.82 & $\leq$ 0.78 \\
	FCC306 & 306 & - & G & 2.8; 2.0; 84 & 2.4 & 50 & $\leq$ 1.9 & $\leq$ 8.10 & $\leq$ -0.11 \\
	NGC1437B & 308 & D & G & 3.2; 2.4; 69 & 3.1 & 91 $\pm$ 14 & 17 $\pm$ 1.67 & 7.76 $\pm$ 0.04 & -0.59 $\mp$ 0.01 \\
	ESO358-G063 & 312 & R & B & 2.6; 2.0; 80 & 3.3 & 380 $\pm$ 20 & 131.5 $\pm$ 13.2 & 8.57 $\pm$ 0.05  & -0.34 $\mp$  0.01 \\
	FCC316 & 316 & - & G & 2.8; 2.0; 82 & 2.7 & 50 & $\leq$ 2.1 & $\leq$ 7.31 & $\leq$ 0.53 \\
	FCC332 & 332 & D & G & 2.8; 2.0; 84 & 2.3 & 30 $\pm$ 5 & 2.0 $\pm$ 0.25 & 7.18 $\pm$ 0.06 & -0.61 $\mp$ 0.01 \\
	ESO359-G002 & 335 & D & G & 3.2; 2.4; 69 & 3.1 & 37 $\pm$ 5 & 2.0 $\pm$ 0.24 & 6.92 $\pm$ 0.05 & -1.33 $\mp$ 0.01 \\
	\hline
	\end{tabular}
	\textit{Notes:} 1: Common name of the galaxy; 2: Fornax Cluster Catalogue number of the galaxy; 3: Whether the morphology and kinematics of the molecular gas in the galaxy are regular (R) or disturbed (D) (see \Section \ref{sub:morphologies}); 4: Whether the line profile of the CO(1-0) line is best described by a Gaussian (G) or a boxy (B) profile (see \Section \ref{sub:H2_masses}); 4/7: Upper limits were determined assuming a Gaussian line profile with a FWHM of 50 \kms (see \Section \ref{subsub:upper_limits}); 5: Beam major axis, minor axis and position angle; 6: the typical rms in a single channel in the line-free channels of the data cube; 7: the width of the CO integrated spectrum (see \Section \ref{sub:H2_masses}); 8: the total CO emission; 9: total M$_{\text{H}_2}$ mass derived from the CO emission (see \Section \ref{sub:H2_masses}); 10: H$_2$ deficiency, defined as $M_{\text{H}_2, \text{observed}} - M_{\text{H}_2, \text{expected}}$ (see \Section \ref{sub:H2_masses}).
	\end{threeparttable}
\end{table*}

\section{Results}
\label{sec:results}

\begin{figure*}
	\centering
	\includegraphics[width=0.8\textwidth]{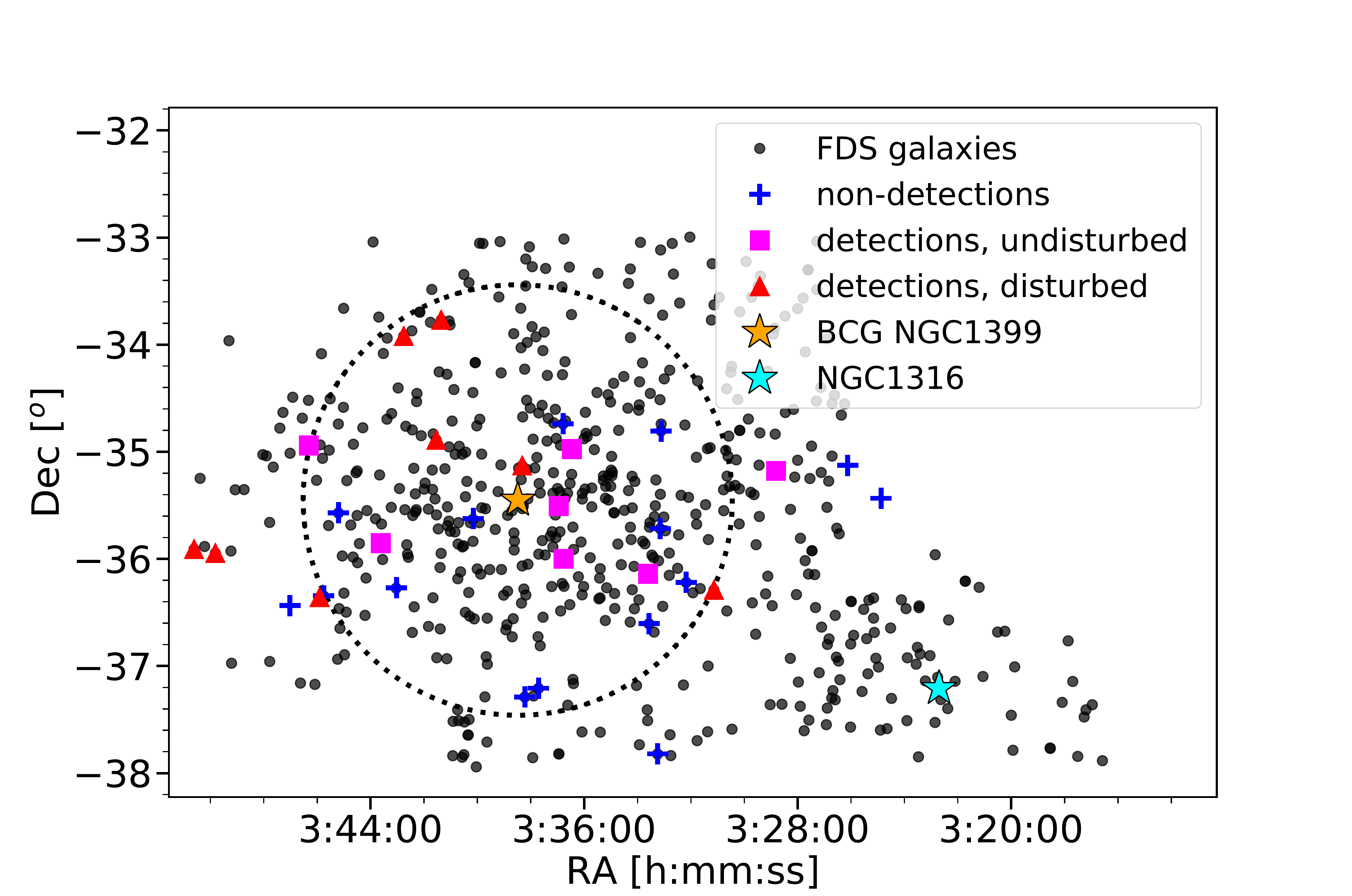}
	\caption{Map of the Fornax cluster, similar to Figure \ref{fig:Fornax}. The coloured symbols represent the AlFoCS targets, their shape and colour indicate whether CO was detected and if so, whether it is morphologically and kinematically disturbed or regular, as indicated in the legend. The central galaxy, NGC1399, is indicated with a yellow star, and the virial radius with a dotted line. The central galaxy of the infalling group in the lower right corner of the figure, NGC1316, is indicated with a cyan star. Non-detections, disturbed galaxies, and regular galaxies are distributed evenly over the cluster.}
	\label{fig:Fornax_coloured}
\end{figure*}

CO was detected (at $> 3 \sigma$) in 15 of the 30 galaxies observed. In Figure \ref{fig:Fornax_coloured} the (projected) locations of the detections and non-detections within the cluster are shown, and morphologically and kinematically regular and disturbed galaxies are highlighted. All FDS (see \Section \ref{sub:optical_data}) galaxies are shown as black dots, the AlFoCS galaxies are shown in colour. Non-detections are shown as blue plus signs, the pink squares are galaxies in which CO is detected and morphologically and kinematically regular or undisturbed, and the red triangles represent galaxies in which CO is detected and morphologically and kinematically disturbed (see \Section \ref{sub:morphologies} for more details). Both detections and non-detections are distributed evenly over the cluster. At first glance it looks like there are slightly more non-detections south of the cluster centre, however this is not statistically significant. Galaxies with disturbed molecular gas reservoirs seem to be mainly located close to or outside the virial radius.

\subsection{Marginal detections}
\label{sub:marginal}
In ESO358-G015, FCC32, and NGC1437A CO is detected, but only marginally. In ESO358-G015 and NGC1437A these are 4 - 5 $\sigma$ detections, but the emission comes from small features away from the galactic centre, and it is not clear whether this emission is related to the galaxy observed. For FCC32 we find a tentative 2 $\sigma$ peak at the centre of the galaxy. These features are likely noise, and for these reasons we do not consider these observations further in this work.

\subsection{Continuum detections}
\label{sub:cont_detections}

\begin{figure*}
	\centering
	\subfloat[NGC1380]
		{\includegraphics[height=0.29\textwidth]{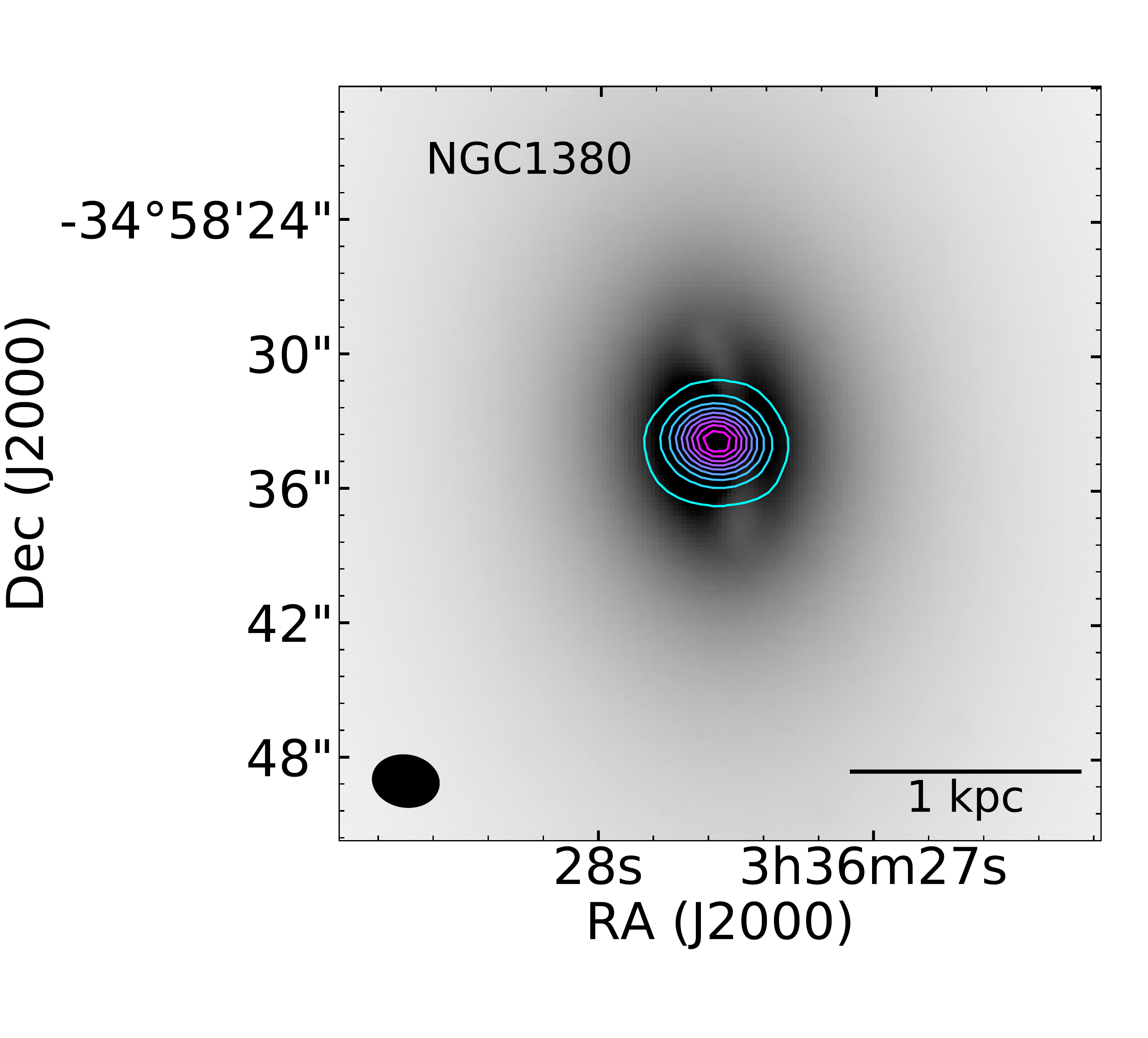}}
	\subfloat[NGC1386]
		{\includegraphics[height=0.28\textwidth]{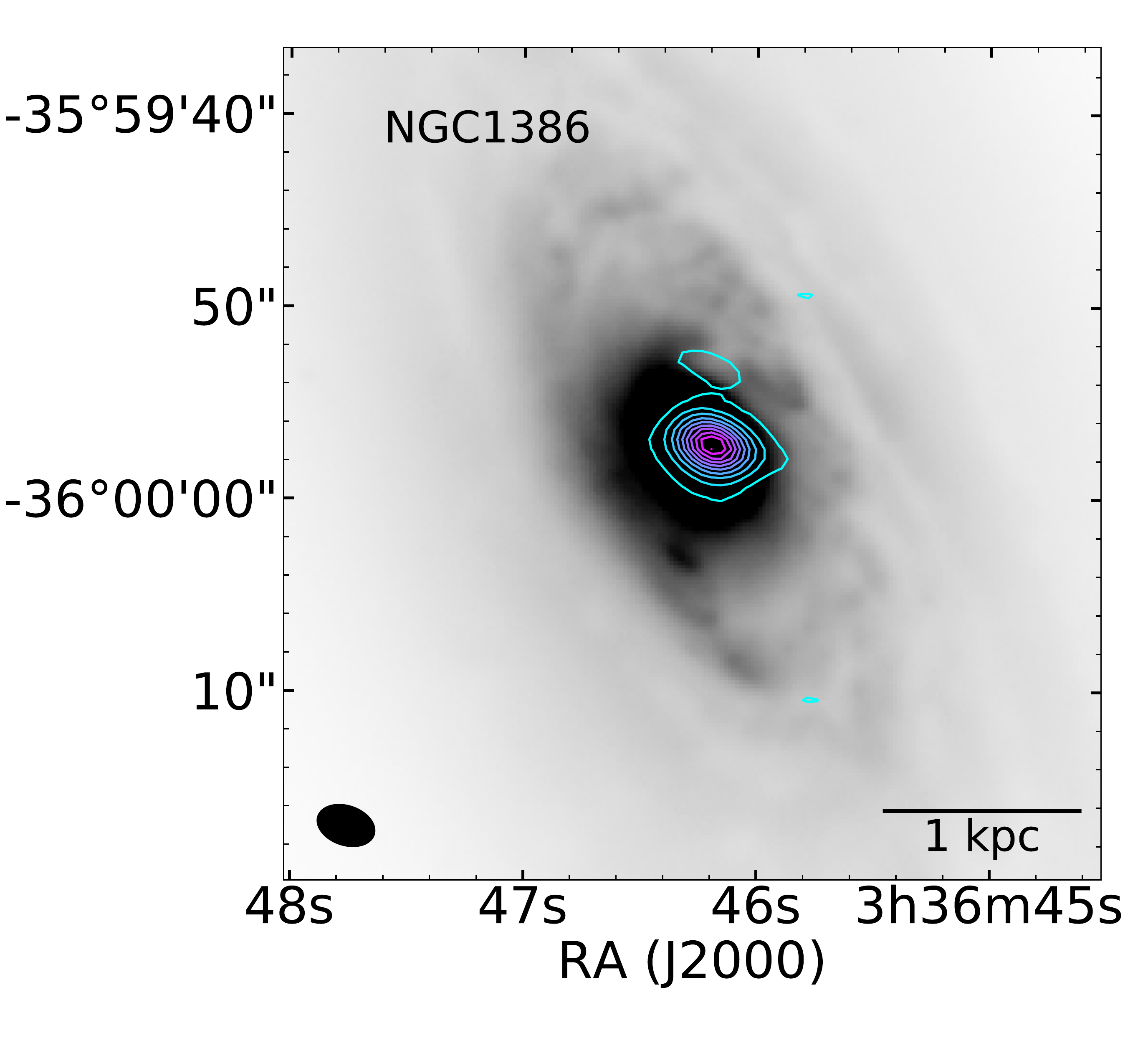}}	
	\subfloat[NGC1387]
		{\includegraphics[height=0.28\textwidth]{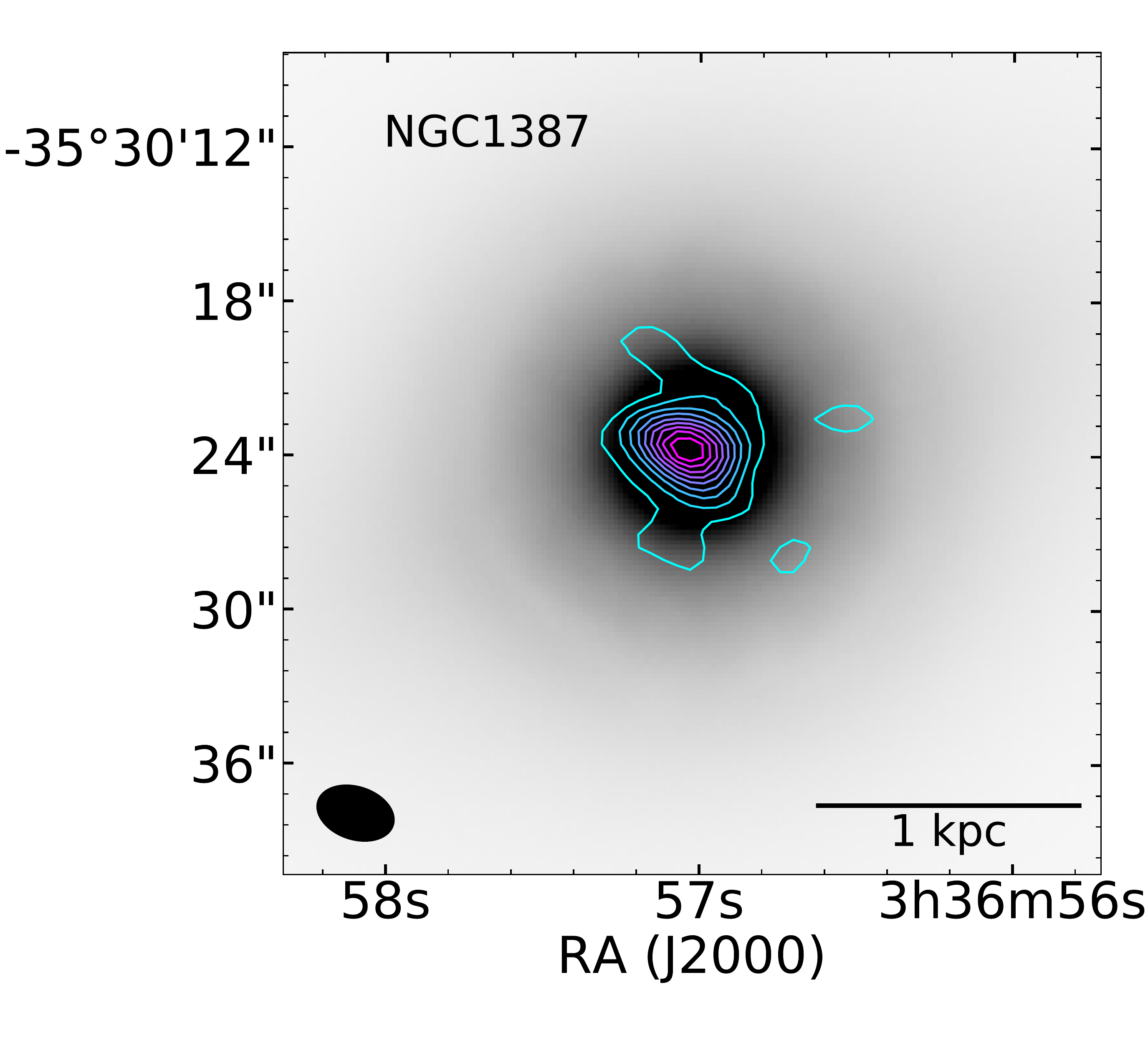}}
	\caption{3 mm continuum emission overplotted on optical (\textit{g}-band) images from the FDS (see \Section \ref{sub:optical_data}), similar to Figure \ref{fig:overplots}. The emission is shown as 10 coloured contours; the lower (outer) contour level equals 5$\sigma$. The emission originates from the galaxies' centres, and is likely due to a combination of AGN activity and thermal emission from dust.}
	\label{fig:continua}
\end{figure*}

Continuum (3 mm) was detected in NGC1380, NGC1386, NGC1387, and NGC1427A. In Figure \ref{fig:continua} the continuum maps of NGC1380, NGC1386, and NGC1387 are shown as coloured contours overplotted on the \textit{g}-band images from the FDS, similar to Figure \ref{fig:overplots}. In all three cases the continuum emission originates from the galactic centre. Two galaxies, NGC1380 and NGC1386, are known to harbour active galactic nuclei (AGN, e.g. \citealt{Boizelle2017,Lena2015,Rodriguez-Ardila2017}). The emission we detect is an unresolved point source at the galactic centre, but has a positive spectral index (see Table \ref{tab:continuum}). It is possible that both thermal and non-thermal emission is contributing the observed emission in these sources.

The 3 mm continuum emission in NGC1387 has a point-like morphology in the lower sideband, but when imaged at the higher frequencies several additional point sources are also detected, in the region where we know dust and molecular gas are present. This additional emission leads to the very large spectral index measured for this source (see Table \ref{tab:continuum}).
Given this, the detected 3 mm emission is again likely due to a mix of AGN activity and thermal emission from dust. 

In the case of NGC1427A the emission originates from a small source at the edge of the galaxy. This is shown and discussed separately in \Section \ref{subsub:NGC1427A}.  

\subsection{H$_2$ masses}
\label{sub:H2_masses}
H$_2$ masses for all detected galaxies were estimated using the following equation:
\begin{equation}
	\label{eq:H2mass}
	M_{\text{H}_2} = 2m_\text{H} \ D^2 \ X_{\text{CO}} \ \frac{\lambda^2}{2\ k_\text{B}} \int S_\nu\ d \nu\ ,
\end{equation}
where $m_\text{H}$ is the mass of a hydrogen atom, $D$ is the distance to the galaxy, $X_{\text{CO}}$ is the CO-to-H$_2$ mass conversion factor, $\lambda$ is the rest wavelength of the line observed, $k_\text{B}$ is the Boltzmann constant, and $\int S_\nu\ d\nu$ the total flux of the line observed. 

We use the metallicity-dependent mass conversion factor derived from \citet[eqn. 25]{Accurso2017}:
\begin{multline}
\text{log}\ \alpha_{\text{CO}} = 14.752 - 1.623 \left[ 12 + \text{log} \left( \text{O/H} \right) \right] \\ 
+ 0.062\ \text{log}\ \Delta \left( \text{MS} \right),
\label{eq:accurso}
\end{multline}
where $12 + \text{log} \left( \text{O/H} \right)$ is the metallicity and $\text{log}\ \Delta \left( \text{MS} \right)$ the distance from the main sequence, discussed below. The $1 \sigma$ spread in $\text{log}\ \alpha_{\text{CO}}$ from this relation is 0.165 dex. It is multiplied by $2.14 \times 10^{20}$ to obtain \XCO \citep{Bolatto2013}. For reference, this equation gives a conversion factor of $2.08 \pm 0.02 \times 10^{20}$ for solar metallicity ($12 + \text{log} \left( \text{O/H} \right)$ = 8.69, \citealt{Asplund2009}). Since we do not have independent metallicity measurements for each object, metallicities were derived directly from the stellar masses of the galaxies, using the mass-metallicity relation from \citet{Sanchez2017}, which uses the calibration from \citet{Pettini2004}. Stellar masses ($M_\star$) are listed in Table \ref{tab:targets}. They were taken from \cite{Fuller2014} where possible (see Table \ref{tab:targets}). Alternatively, they were obtained from aperture photometry on archival Wide-field Infrared Survey Explorer (WISE, \citealt{Wright2010}) band 1 (3.6 $\mu$m) images, assuming a mass-to-light ratio of 1 (see Table \ref{tab:targets}). Apertures were chosen using the effective radii determined by \citet[][see \Section \ref{sub:optical_data}]{Venhola2018} if available, and alternatively chosen by eye. Uncertainties on the stellar mass in these cases are a combination of the uncertainty in the effective radius and the rms in the image.

\begin{table*}
	\begin{threeparttable}
	\centering
	\caption{Properties of the detected 3mm continuum emission.}
	\label{tab:continuum}
	\begin{tabular}{llllllll}
	\hline
	Galaxy & Frequency & Flux density & Frequency USB & Flux density USB & Frequency LSB & Flux density LSB & Spectral index \\
	- & (GHz) & (mJy) & (GHz) & (mJy) & (GHz) & (mJy) & - \\
	(1) & (2) & (3) & (4) & (5) & (6) & (7) & (8) \\
	\hline 
	NGC1380 & 107.765 & 4.18 $\pm$ 0.04 & 113.763 & 4.65 $\pm 0.08$ & 101.775 & 4.12 $\pm 0.04$ & 1.1 $\pm$ 0.2 \\
	NGC1386 & 107.718 & 3.69 $\pm$ 0.05 & 113.750 & 3.99 $\pm 0.07$ & 101.748 & 3.64 $\pm 0.07$ & 0.8 $\pm$ 0.2 \\
	NGC1387 & 107.718 & 1.85 $\pm$ 0.06 & 113.750 & 4.3 $\pm 0.1$ & 101.748 & 1.04 $\pm$ 0.06 & 12.7$^{+0.6}_{-0.5}$ \\
	NGC1427A & 107.765 & 0.16 $\pm$ 0.03 & 113.763 & 0.20 $\pm 0.06$ & 101.775 & 0.16 $\pm 0.03$ & 2.0$^{+3.0}_{-3.5}$ \\
	\hline
	\end{tabular}
	\textit{Notes:} 1: Name of the galaxy; 2: Central frequency of the 3 mm continuum; 3: Flux density of the 3 mm continuum emission; 4: Central frequency of the upper sideband; 5: Flux density of the continuum in the upper sideband; 6: Central frequency of the lower sideband; 7: Flux density of the continuum in the lower sideband; 8: Spectral index of the continuum emission.
	\end{threeparttable}
\end{table*}

The $X_{\text{CO}}$ calibration from \cite{Accurso2017} requires a distance from the main sequence \citep[e.g.][]{Brinchmann2004,Noeske2007,Elbaz2007}. Here we assume a distance from the main sequence $\Delta$MS = 0 for all galaxies. It is a second order parameter, so varying this does not strongly affect our results. Equation \ref{eq:accurso} is valid for values of -0.8 $<$ $\Delta$MS $<$ 1.3. Varying $\Delta$MS over this range results in a maximum error of 0.08 in $\alpha_{\text{CO}}$, which is indeed small compared to the other errors.

To make sure we include all the CO emission, while minimising the inclusion of noise, galaxies were subdivided into two groups: a group whose line profiles are best described by a Gaussian profile (mostly dwarf galaxies with narrow CO lines), and another group whose line profiles are best described by a box profile (mostly larger galaxies). Which profile best describes a galaxy is listed in Table \ref{tab:observed_props}. The widths of the CO integrated spectra are given. For boxy line profiles an uncertainty of 20 \kms (the equivalent of two channels) is adopted, for Gaussian profiles the formal fitting errors on the linewidth are quoted. For the first group we fit a Gaussian to the CO(1-0) line and integrate this fit to obtain the total line flux. For the second group, we integrate directly under the line observed. In this case the boundaries of the line are determined using the PVDs. Uncertainties are a combination of the error on the total integrated line emission $\int S_\nu d\nu$ and an adopted 10\% calibration error, and are often dominated by the latter. For galaxies with a boxy profile, the error in the integrated line emission is estimated according to the following equation, adapted from equation 1 from \citet{Young2011}:
 
\begin{equation}
\sigma_I^2 = \left( \Delta v \right) ^2 \sigma ^ 2 N_l,
\end{equation}
where $N_l$ is the number of channels that is summed over, $\Delta v$ the width of each channel, and $\sigma$ the rms noise level in the line free part of the spectrum. For galaxies with an approximately Gaussian line profile, the error on the total integrated line emission is estimated by combining the formal fitting errors on the parameters of the fit. The resulting molecular gas masses are listed in Table \ref{tab:observed_props}.

\begin{table*}
	\begin{threeparttable}
	\caption{Key properties and derived quantities of the Mopra targets included in this work.}
	\label{tab:mopra}
	\centering
	\begin{tabular}{lllllll}
	\hline
	Name & RA & Dec & Stellar mass & rms & log$_{10}(M_{\text{H}_2}$) & Deficiency \\
	- & (J2000) & (J2000) & (log($M_{\odot}$)) & (mJy \kms) & ($M_\odot$) & (dex) \\
	(1) & (2) & (3) & (4) & (5) & (6) & (7) \\
	\hline
	NGC1316 & 03h22m41.718s & -37d12m29.62s & 10.0$^\dagger$ & 273 & $\leq$ 8.27 & $\leq$ -0.52 \\
	NGC1317 & 03h22m44.286s & -37d06m13.28s & 9.98$^\dagger$ & 192 & 8.69 $\pm$ 0.04 & -0.08 $\pm$ 0.01 \\
	NGC1350 & 03h31m08.12s & -33d37m43.1s & 10.71$^\dagger$ & 602 & $\leq$ 8.61 & $\leq$ -0.38 \\
	ESO359 G3 & 03h52m00.92s & -33d28m03.5s & $10.11^{+0.01\ \ddagger}_{-0.02}$ & 130 & $\leq$ 7.95 & $\leq$ 0.05 \\
	FCCB857/858* & 03h33m19.49s & -35d20m41.4s & $9.31^{+0.01\ \ddagger}_{-0.04}$ & 34 & $\leq$ 7.36 & $\leq$ -0.07 \\
	FCCB950 & 03h34m31.65s & -36d52m20.7s & $9.47^{+0.02\ \ddagger}_{-0.04}$ & 31 & $\leq$ 7.32 & $\leq$ -0.64 \\
	FCCB990 & 03h35m11.38s & -33d22m25.6s & $9.39^{+0.02\ \ddagger}_{-0.04}$ & 97 & $\leq$ 7.82 & $\leq$ 0.04\\
	FCCB713** & 03h31m20.94s & -35d29m29.9s & - & 55 & $\leq$ 7.57 & $\leq$ -0.59 \\
	FCCB792** & 03h32m25.95s & -38d05m33.8s & - & 21 & $\leq$ 7.16 & $\leq$ -0.48 \\
	FCCB1317** & 03h39m11.70s & -33d31m56.0s & - & 97 & $\leq$ 7.81 & $\leq$ 0.10 \\
	\hline
	\end{tabular}
	\textit{Notes:} 1: Name of the galaxy observed; 2: Right ascension; 3: Declination; 4: Stellar mass (see \Section \ref{sub:H2_masses}); 5: rms in the spectrum; 6: Derived molecular gas mass (see \Section \ref{sub:H2_masses}); 7: H$_2$ deficiency (see \Section \ref{sub:gas_fractions}); *FCCB857 and FCCB858 are close to each other on the sky and were therefore contained within one beam. The stellar mass quoted here is the addition of the stellar masses of both galaxies. The coordinates of FCCB858 are quoted here. **These galaxies were observed but later found to be background objects. They are therefore omitted in Figures \ref{fig:gas_fraction} and \ref{fig:mass_distance}, and stellar masses were therefore not determined for them; $^\dagger$Stellar masses from \citet{Fuller2014}; $^\dagger$Stellar masses derived from 3.6 $\mu$m images, (see \Section \ref{sub:H2_masses}).
	\end{threeparttable}
\end{table*}

\subsubsection{Upper limits}
\label{subsub:upper_limits}
For non-detections, 3$\sigma$ upper limits were determined using the rms in the (spatial) inner area of the PB corrected data cubes. Since all non-detections can be considered dwarf galaxies, we assume Gaussian line profiles with FWHM of 50 km s$^{-1}$. This is slightly broader than the profiles of the dwarf galaxies detected here, and therefore a conservative assumption. The maximum of the assumed line profile was set to 3 times the rms in the corresponding data cube. We use stellar mass dependent CO-to-H$_2$ conversion factors, as described above in \Section \ref{sub:H2_masses}. We then use Equation \ref{eq:H2mass} to obtain the upper limits for the H$_2$ mass listed in Table \ref{tab:observed_props}.

\subsubsection{Mopra}
\label{subsub:Mopra}

Of the 28 galaxies observed with Mopra, CO was detected in one additional galaxy which was not observed with ALMA; NGC1317. After removing data affected by bad weather, we were able to obtain this one additional H$_2$ mass measurement and 8 additional upper limits. Due to a problem with the observations, we only have single, central pointing observations of NGC1317. Since the molecular gas is usually centrally located, however, we expect this to cover most if not all of its CO emission. Upper limits are 3$\sigma$ upper limits, estimated as described above. Despite the rather prominent baseline ripple in some of the observations, a known issue with the Mopra Telescope (see \Section \ref{sec:observations}), these upper limits provide reasonably good constraints. The resulting upper limits, as well as the estimated H$_2$ mass of NGC1317, are listed in Table \ref{tab:mopra}.

\subsection{Gas fractions \& deficiencies}
\label{sub:gas_fractions}
\begin{figure*}
	\hspace{3mm}	
	\centering
	\includegraphics[width=0.8\textwidth]{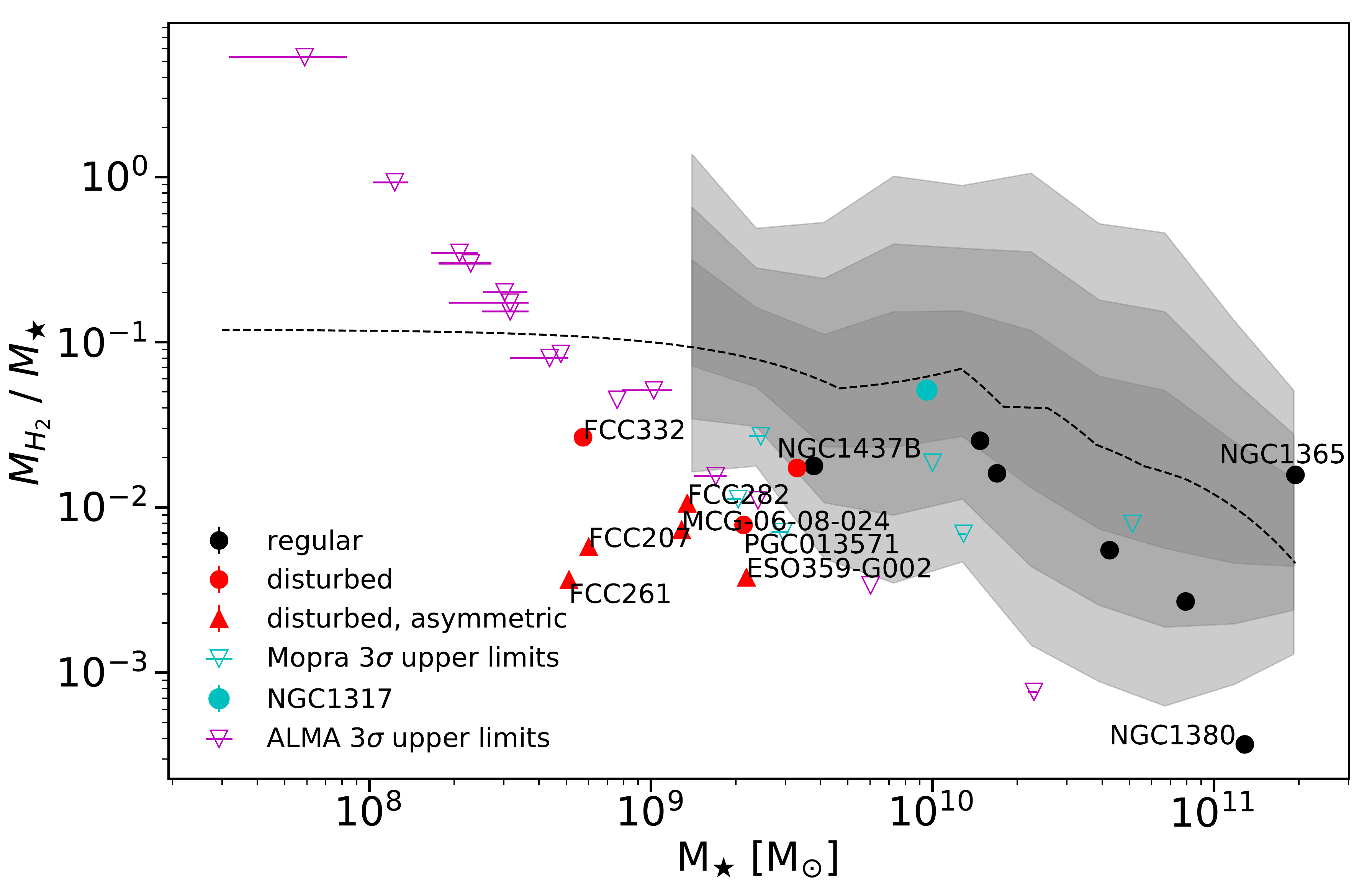}
	\caption{Molecular gas fraction, as a function of stellar mass. Black dots are regular galaxies, red markers are disturbed galaxies. The shape of the marker indicates whether the galaxy may be undergoing ram pressure stripping, based on visual inspection. ALMA upper limits are shown as magenta open triangles, and Mopra upper limits as cyan open triangles. The Mopra detection of NGC1317 is shown as a cyan dot. Within the shaded area, the dashed line represents the expected gas fraction based on \citet{Saintonge2017}. The three shades of grey indicate the 1, 2, and 3 $\sigma$ levels (from the inside out) of the xCOLD GASS data. Outside the shaded area the dashed line is based on linear extrapolation (in log space). Galaxies with high deficiencies and the galaxies that were classified as disturbed are labelled. There is a discrepancy between the expected gas fractions and the gas fractions observed, especially the disturbed galaxies are H$_2$ deficient compared to field galaxies.}
	\label{fig:gas_fraction}
\end{figure*}

In Figure \ref{fig:gas_fraction} the galaxies' molecular-to-stellar mass ratios are shown as a function of their stellar mass (see \Section \ref{sub:H2_masses} for more details about the stellar masses used here), and compared with those of field control galaxies with the same stellar masses. The molecular gas fraction is given by $\left( \frac{M_{\text{H}_2}}{M_{\text{total}}} \right)$. Since $M_{\text{H{\sc i}}}$ and $M_{\text{H}_2}$ are relatively small contributions to the total mass of the galaxy compared to the stellar mass, for convenience and consistency with the definition in \citet{Saintonge2017} (see below), we define the gas fraction here as $\left( \frac{M_{\text{H}_2}}{M_\star} \right)$. We use the extended CO Legacy Database for \textit{GALEX} Arecibo SDSS Survey (xCOLD GASS, \citealt{Saintonge2017}) as a field galaxy control sample. xCOLD GASS is a survey of molecular gas in the local universe, built upon its predecessor COLD GASS \citep{Saintonge2011}. It is a mass-selected ($M_\star > 10^9 M_\odot$) survey of galaxies in the redshift interval 0.01 $< z < 0.05$ from the SDSS, and is therefore representative of the local galaxy population within this mass range. We use the relation based on the median values that they obtained by subdividing the sample in bins based on their stellar mass \citep[see][their Figure 10]{Saintonge2017}, and interpolate linearly (in log space) to obtain the relation represented by the dashed line. Since xCOLD GASS galaxies were selected to have stellar masses $M_\star > 10^9 M_\odot$, the first stellar mass bin is located at $M_\star = 9.388 M_\odot$. Below this stellar mass, the dashed line is obtained using linear extrapolation (in log space). Expected mass fractions for galaxies in this mass range, 5 detections and 11 upper limits, should be treated with caution. The shaded areas represent the 1, 2, and 3 sigma levels in the xCOLD GASS data, from dark to lighter. Galaxies with disturbed molecular gas are shown in red, and galaxies with regular, undisturbed molecular gas are shown in black (see \Section \ref{sub:morphologies} for the definitions). Galaxies that have clear gas tails that extend beyond their optical emission, or otherwise asymmetric CO emission, are indicated with red triangles. ALMA upper limits for the H$_2$ mass are shown as magenta open triangles, and Mopra upper limits as cyan open triangles. NGC1317, the only Mopra detection included here, is shown as a cyan dot. There is a systematic offset between the xCOLD GASS H$_2$ mass fractions for field galaxies and our values of up to about $\sim$1 dex. This offset is not very significant at an individual level for regular galaxies, whose offset is mostly within or close to the 1$\sigma$ scatter in the xCOLD GASS data. With exception of NGC1437B and FCC332, all disturbed galaxies lie below 3$\sigma$ (for FCC261 and FCC207 we cannot be certain because they lie below the mass range of the xCOLD GASS data, but based on this figure it seems plausible to assume they would fall below 3$\sigma$ as well). In particular the galaxies with asymmetric CO emission have low gas fractions. 

We define H$_2$ deficiencies here as log($M_{H_{2}, \text{observed}}$) - log($ M_{H_{2}, \text{expected}}$). Estimates of the H$_2$ deficiency for each galaxy are listed in Table \ref{tab:observed_props}. Galaxies with regular CO emission have an average H$_2$ deficiency of -0.50 dex, and galaxies with disturbed CO emission have an average H$_2$ deficiency of -1.1 dex. In Figure \ref{fig:mass_distance} H$_2$ deficiencies are plotted as a function of the (projected) distance between the galaxy and the cluster centre. Markers and colours are the same as in Figure \ref{fig:gas_fraction}. It seems like galaxies within a (projected) radius of 0.4 Mpc from the cluster centre are slightly more deficient than galaxies outside this radius. However, Kolmogorov-Smirnov and Mann-Whitney \textit{U} tests are unable to reject the null hypothesis that both groups of galaxies are drawn from the same distribution at more than $\sim$2$\sigma$. Possible explanations for this and a further discussion of this figure can be found in \Section \ref{sub:RPS}. 

\section{Discussion}
\label{sec:discussion}

\subsection{Gas morphologies \& kinematics}
\label{sub:morphologies}
The galaxies detected here can be divided into two categories: galaxies with disturbed molecular gas morphologies and regular systems. Whether a galaxy is morphologically disturbed or regular is determined by visual inspection of the moment 0 and 1 maps (see Figures \ref{fig:moment-maps_reg}, \ref{fig:moment-maps_irreg}, Appendix \ref{app:moment_maps}, and Table \ref{tab:observed_props}). Non-disturbed galaxies have molecular gas that is concentrated symmetrically around the galactic centre, whereas galaxies with disturbed morphologies contain molecular gas that is asymmetric with respect to the (optical) centre of the galaxy. It sometimes has a very irregular shape, and, in some cases, even extends beyond the galaxy's stellar body (see \Section \ref{subsub:rps_gals}). Of the galaxies detected here, eight are classified as disturbed galaxies, and seven have regular molecular gas morphologies. 

The galaxies with morphologically disturbed molecular gas reservoirs also have disturbed molecular gas kinematics. Looking at the velocity maps in Figure \ref{fig:moment-maps_reg} and the regular galaxies in Appendix \ref{app:moment_maps}, regular galaxies follow a standard ``spider diagram'' shape, indicative of a regular rotation. Disturbed galaxies, on the other hand, have irregular velocity maps, indicating the presence of non-circular motions. In some cases rotation is still present (in NGC1437B and PGC013571, for example, Figures \ref{fig:NGC1437B} and \ref{fig:PGC013571}, respectively), in other cases no rotation can be identified (for example, ESO359-G002 and FCC332, Figures \ref{fig:ESO359-G002} and \ref{fig:FCC332}, respectively). This is also reflected in the PVDs, which look like smooth rotation curves for the regular galaxies, but have very asymmetric and irregular shapes for the disturbed galaxies. Maps of the CO(1-0) linewidth of the regular galaxies often reveal symmetric structures such as rings or spiral arms (see, for example, Figures \ref{fig:ESO358-G063}). For disturbed galaxies this is, again, much more irregular (for example, Figure \ref{fig:NGC1437B}). A further discussion of each galaxy in detail can be found in Appendix \ref{sub:individual}.

\subsection{Stripping and gas stirring in Fornax in comparison with the field}
\label{sub:things}
There is a clear mass split between galaxies with regular and disturbed molecular gas morphologies, where all galaxies with stellar masses below $3 \times 10^9 M_\odot$ have disturbed molecular gas (see Figure \ref{fig:gas_fraction}). In the absence of a comparable field sample tracing molecular gas at these stellar masses, we compare this result to the Local Irregulars That Trace Luminosity Extremes, The H{\sc i} Nearby Galaxy Survey (LITTLE THINGS, \citealt{Hunter2012}). LITTLE THINGS is a multi-wavelength survey of 37 dwarf irregular and 4 blue compact nearby ($\leq$ 10.3 Mpc) (field) dwarf galaxies that is centred around H{\sc i}-line data, obtained with the National Radio Astronomy Observatory (NRAO) Very Large Array (VLA). It has high sensitivity ($\leq$1.1 mJy beam$^{-1}$ per channel), high spectral resolution ($\leq$2.6 \kms), and high angular resolution ($\sim$6''), resulting in detailed intensity and velocity maps. If the molecular gas in a galaxy is disturbed, we expect their atomic gas to be disturbed as well. Therefore this comparison, although not ideal, is still meaningful. Categorising the LITTLE THINGS dwarfs in the same way as the AlFoCS galaxies (see above, \Section \ref{sub:morphologies}), only about half of these dwarf galaxies show disturbed \HI kinematics and morphologies. Since all AlFoCS galaxies with stellar masses lower than $3 \times 10^9 M_\odot$ have disturbed morphologies and kinematics, this indicates that these low mass galaxies are more disturbed than their counterparts in the field. This suggests that Fornax is still a very active environment, having significant effects on its members. Furthermore, it implies that less massive galaxies are more susceptible to the effects of the cluster environment, likely because of their shallower potential wells. This difference in gas deficiency between massive and less massive galaxies is also seen in simulations \citep[e.g.][]{Voort2017}, and is likely driven by their shallower potential wells.

\begin{figure}
	\centering
	\includegraphics[width=0.48\textwidth]{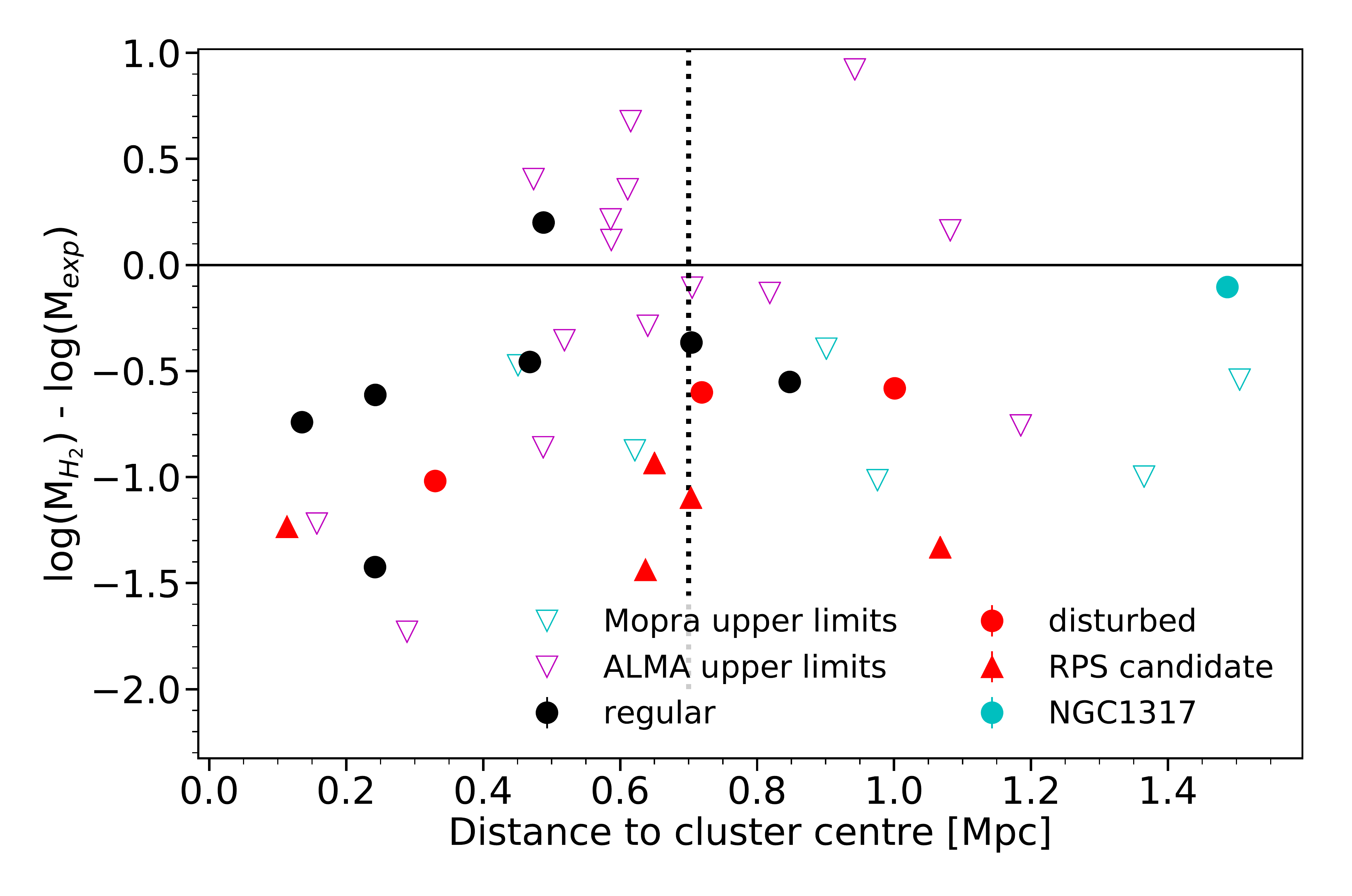}
	\caption{Molecular gas mass deficiencies (see \Section \ref{sub:gas_fractions}) as a function of the (projected) distance to the cluster centre (defined as the location of NGC1399). Marker shapes and colours are the same as in Figure \ref{fig:gas_fraction}. There is no clear correlation between a galaxy's H$_2$ deficiency and its distance from the cluster centre.}
	\label{fig:mass_distance}
\end{figure}

\subsection{Ram pressure stripping or galaxy-galaxy interactions?}
\label{sub:RPS}
The AlFoCS galaxies with disturbed molecular gas reservoirs are H$_2$ deficient compared to field galaxies (see Figure \ref{fig:gas_fraction} and \Section \ref{sub:gas_fractions}). This confirms the result from \citet{Horellou1995}, who find that the CO emission in Fornax cluster galaxies is relatively weak, and the H$_2$ masses relatively low. AlFoCS galaxies have deficiencies up to -1.1 dex (see \Section \ref{sub:gas_fractions}). These deficiences are higher than those found in \citet{Boselli2014}, who find H$_2$ deficiencies of a factor $\sim$2 for the most \HI deficient galaxies in the Virgo cluster. The molecular gas in the most deficient AlFoCS galaxies is centrally located and asymmetric. Mechanisms that are possibly responsible for this include ram pressure stripping and galaxy-galaxy interactions. 

Two of the irregular galaxies, MCG-06-08-024 and ESO359-G002, show molecular gas tails that extend well beyond the brightest parts of the galaxy's stellar body (see \Section \ref{subsub:rps_gals}). Together with the dwarfs FCC207 and FCC261, they have the lowest gas fractions of the disturbed galaxies (see Figure \ref{fig:gas_fraction} and Table \ref{tab:observed_props}). In both cases, this tail is aligned with the direction of the cluster centre (see \Section \ref{subsub:rps_gals}). This, in combination with their low gas fractions, can be interpreted as a sign of ongoing ram pressure stripping. This is striking, since RPS is not thought to affect the molecular gas much, as it is bound much more tightly to the galaxy than the atomic gas. Moreover, RPS is thought to be less important in the Fornax cluster than in, for example, the Virgo cluster, given its relatively small size and large density of galaxies (see \Section \ref{sec:intro}). 

The fact that the gas tails align with the direction of the cluster centre is, however, not necessarily proof that ram pressure stripping is in play. There are confirmed RPS tails pointing in all directions, even nearly perpendicular to the direction of the cluster centre \cite[e.g.][]{Kenney2014}. This is also seen in simulations \citep[e.g.][]{Yun2018}. Moreover, the kinematics of these galaxies are more irregular than expected based on RPS alone, which suggests that a past galaxy-galaxy interaction may be (co-)responsible for this. In deep FDS images (Iodice et al. 2018, submitted to A\&A), MCG-06-08-024 shows a very disturbed morphology in the outskirts, which could indicate a past galaxy-galaxy interaction. Furthermore, these RPS candidates are not necessarily close to the cluster centre, nor do they have particularly high velocities, as one might expect for galaxies that are undergoing RPS. However, \cite{Jaffe2018}, recently found galaxies undergoing RPS all over the cluster, and also in a wide variety of locations in the velocity phase-space. Simulations by \citet{Yun2018} show that ram pressure stripped galaxies are more common beyond half the virial radius, where most of the AlFoCS galaxies with disturbed molecular gas are located. Both galaxies discussed here have relatively low masses and shallow potential wells, so they are expected to be susceptible to ram pressure stripping. \citet{Yun2018} also find that galaxies with shallow potential wells can experience extended stripping due to weak ram pressure. 
Based on these data alone, it is difficult to say whether it is ram pressure affecting these galaxies. The combination with additional data, for example a study of the stellar kinematics of these galaxies, would allow us to distinguish between galaxy-galaxy interactions and ram pressure stripping with more certainty. 

Several other galaxies, such as FCC282 and FCC332, also show asymmetric molecular gas reservoirs, and were therefore labelled as possible RPS candidates. Asymmetric molecular gas distributions and molecular gas tails can, however, also be the result of galaxy-galaxy interactions. Other galaxies, such as NGC1437B and FCC261, have relatively massive neighbours that are close to them on the sky, which could mean that they are experiencing tidal forces. NGC1437B is the least H$_2$ deficient of the disturbed galaxies (see Figure \ref{fig:gas_fraction}). If we look at its velocity map and PVD (see Figure \ref{fig:NGC1437B}), we can see that it has maintained its rotation and still shows a coherent structure, but it appears to be influenced by a pull on its south side. Although a second tail at the north side is missing, this could be an indication of an ongoing tidal interaction. Although the extension of the molecular gas on the south side of the galaxy does not align with the direction of the cluster centre, it is also possible that this asymmetry is caused by RPS, depending on the galaxy's orbit through the cluster (see above). It is currently still relatively far out, located approximately at the virial radius on the sky.

In Figure \ref{fig:mass_distance}, there appears to be no correlation between a galaxy's H$_2$ mass deficiency and its distance from the cluster centre. Although we suffer from small number statistics, there are a few other possible explanations for this:

\begin{itemize}
\item We are looking at a 2D projection of the cluster, the positions of the galaxies along the line of sight are not taken into account. 
\item Lower mass galaxies end up more H$_2$ deficient than their higher mass counterparts, because of their shallower potential wells. The total H$_2$ mass per galaxy is therefore more a function of their intrinsic mass than of their location in the cluster.
\item The responsible mechanism is galaxy-galaxy interactions. While RPS is much more effective in the cluster centre, depending quadratically on the density of the hot halo, galaxy-galaxy interactions are, relatively, more common at the outskirts of the cluster. If the latter play a role, we would expect less of a trend in the gas deficiencies as we move away from the cluster centre.
\item The galaxies are moving through the cluster, so if they experienced RPS when they were near its centre, they can have moved to the outskirts of the cluster since then.
\item The galaxies were selected to have FIR emission, and therefore galaxies that lost all their gas are excluded from the sample. 
\end{itemize}

\subsection{Dwarfs}
\label{sub:dwarfs}
Among the detections are several galaxies with low stellar masses, that can be classified as early-type dwarfs. Four of these have stellar masses $M_\star \leq 1.0 \times 10^9 M_\odot$. It was long thought that early-type dwarf galaxies in cluster environments would not have a molecular ISM, due to their expected short stripping timescales and shallow potentials. 

The currently accepted theory is that these galaxies are the remnants of low mass late-type galaxies that have fallen into the cluster. This hypothesis is supported by the presence of visible structures such as spiral arms, bars, disks, and nuclei and cores \citep{Lisker2006a,Lisker2006b,Jerjen2000,Barazza2002,DeRijcke2003}, the rotational support of their stellar kinematics \citep{Pedraz2002,Rys2013}, and the detection of significant amounts of gas and dust in some of them \citep{Conselice2002,Serego2007,DeLooze2010,DeLooze2013,Serego2013}. 

It is also possible that they were already early-type dwarf galaxies to begin with, but re-accreted material through tidal interactions either with the intracluster medium or with another galaxy, which could also trigger new star formation. Inconsistencies in the observations supporting the infalling-spiral scenario \citep{Miller1998,Sanchez2012} may add to the favourability of this idea. However, the limited spatial resolution of most of these studies to date make it hard to draw strong conclusions. 

\citet{DeLooze2013} observed ``transition-type dwarf galaxies'' (TTDs) in the Virgo cluster. These galaxies are dwarfs that have an apparent early-type morphology, but still show dust emission and thus evidence of a cold ISM and star formation. They posit that TTDs are in the process of having their ISM removed by the cluster environment, transforming them from late-type dwarfs to quiescent ones (see \citealt{Boselli2008, Koleva2013} for a more detailed description of the definition and identification of TTDs). They find that many of the dust properties of these objects lie in between what is expected for early-type galaxies and for late-type galaxies, supporting the hypothesis that they are infalling low mass spirals that are in the process of being quenched. The presence of central cores and dust concentrations are additional evidence in favour of this outside-in gas removal theory. 

We suspect that the dwarf galaxies we observe here are TTDs moving through the cluster and being stripped of their gas and thus in the process of being quenched. Each dwarf galaxy observed has a very disturbed and irregular molecular ISM, both morphologically and kinematically, suggesting that they are being stripped by the hot intracluster gas (for example MCG-06-08-024 and ESO359-G002, see Figures \ref{fig:moment-maps_irreg} and \ref{fig:ESO359-G002}), or being torn apart by tidal forces. This is in favour of the hypothesis that they are the remnants of infalling gas-rich galaxies. The observation that galaxies with disturbed molecular gas reservoirs seem to favour locations around the virial radius (see \Section \ref{sec:results}) supports the idea that these dwarfs are starting their first passage through the cluster, or have just crossed it for the first time.

\subsection{NGC1427A}
\label{subsub:NGC1427A}

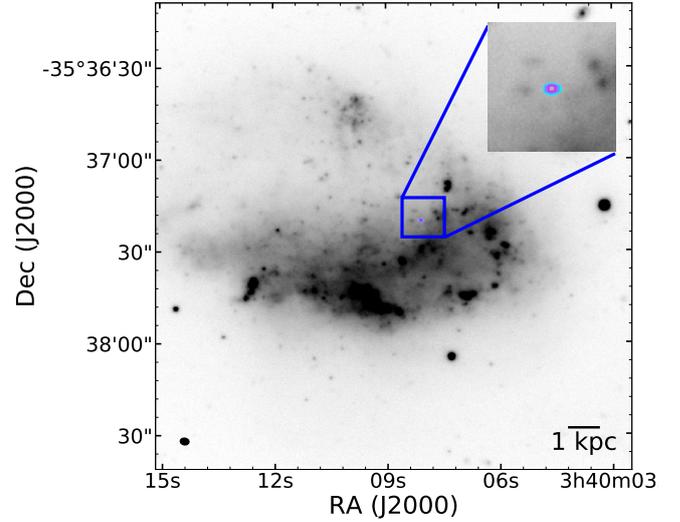
\begin{figure}
\centering
\begin{tikzpicture}
    \node[anchor=south west] (image) at (0,0) {\includegraphics[width=0.48\textwidth]{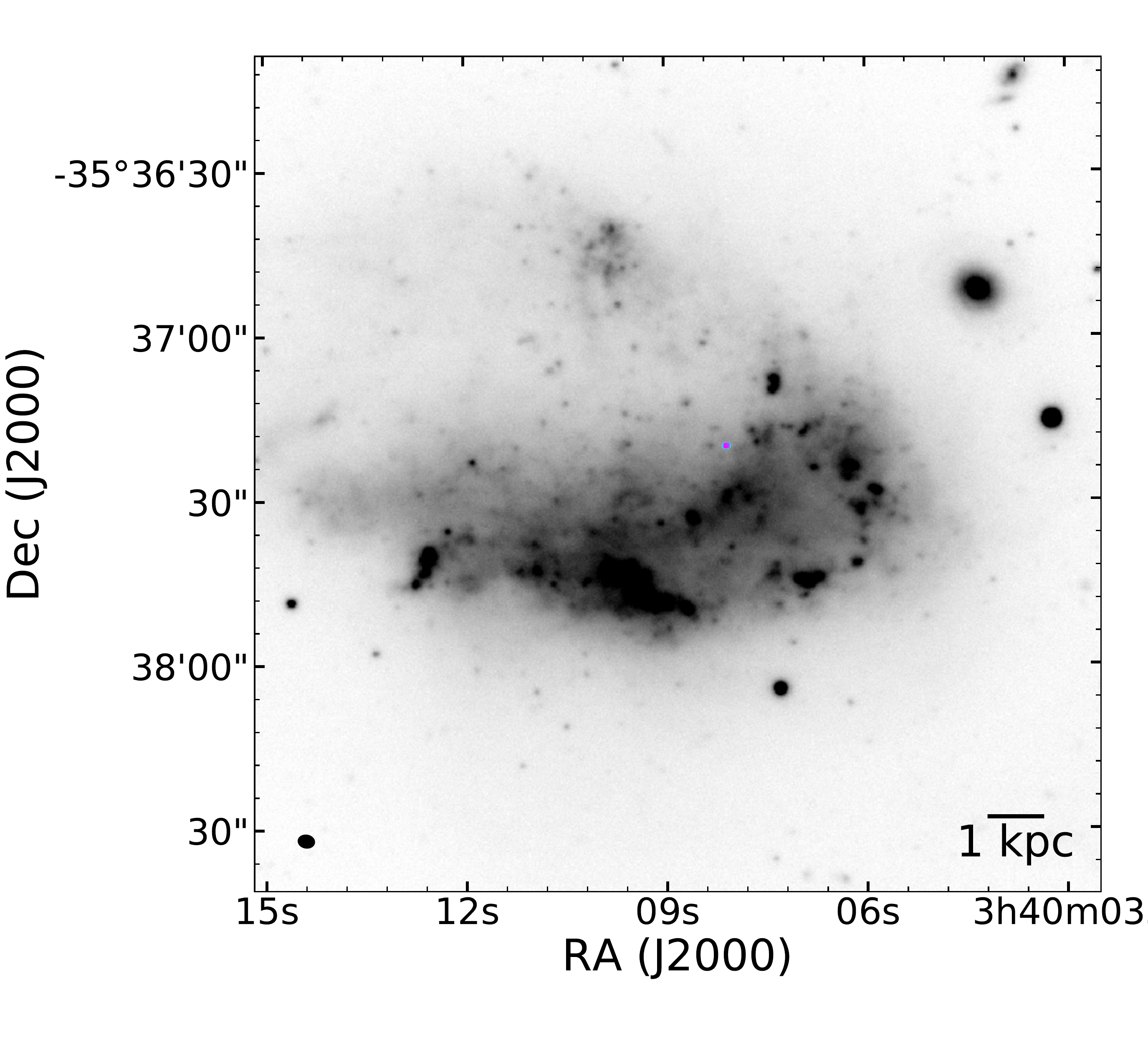}};
    
    	\begin{scope}[x={(image.south east)},y={(image.north west)}]
        \draw[blue,very thick] (0.601,0.541) rectangle (0.665,0.607);
           \end{scope}    
    
    	\begin{scope}[x={(image.south east)},y={(image.north west)}]
        	\draw[blue,very thick] (0.601,0.607) -- (0.731,0.896);	
	\draw[blue,very thick] (0.665,0.540) -- (0.923,0.681);
           \end{scope}    
    
    \node[anchor=south west] (image) at (6.26,5.3) {\includegraphics[width=0.095\textwidth]{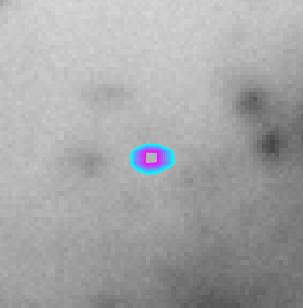}};
\end{tikzpicture}

	\caption{3 mm continuum emission in the observation of NGC1427A, overplotted on an optical (\textit{g}-band) image from the FDS (see \Section \ref{sub:optical_data}). The emission is shown as 10 coloured contours; the lower (outer) contour level equals 4$\sigma$, the inner one 5.6$\sigma$. The emission likely originates from a background source.}
	\label{fig:NGC1427A}
\end{figure}
	
NGC1427A has been proposed to be undergoing ram pressure stripping \citep{Chaname2000,Mora2015}, although recent H{\sc i} observations by \citet{Lee-Waddell2018} suggest that previous tidal interactions are responsible for the galaxy's irregular shape, and for star formation triggering in the disk. It was detected in all five \textit{Herschel} bands \citep{Fuller2014}. Based on the above observations, the expectation was to detect CO in this galaxy. However, none was detected. 

NGC1427A is vigorously star forming. It is expected to have approximately solar metallicity, based on both its stellar mass and analysis of the colours of its star clusters \citep{Mora2015}. Its atomic hydrogen mass is $M_{\text{\HI}} = 2.1 \times 10^9 M_\odot$ \citep{Lee-Waddell2018}, we find an upper limit on its molecular hydrogen mass of $M_{\text{H}_2} = 2.63 \times 10^7$, and its SFR is 0.05 $\pm$ 0.03 \citep{Mora2015}. This leads to the surprisingly high atomic-to-molecular gas ratio $\frac{M_{\text{\HI}}}{M_{\text{H}_2}} > 79$ and short depletion time $\frac{M_{\text{H}_2}}{\text{SFR}} <$1.3 Gyr.

It is possible that the flux is resolved out, since the galaxy extends well beyond the largest recoverable scale of 25'' (see Figure \ref{fig:NGC1427A}). However, as CO emission is broken between velocity channels due to its motion within the galaxy, in our mosaic we would still expect to see some emission.

If the CO emission has not been resolved out, this makes the non-detection of CO surprising. A natural explanation for this would be if the galaxy underwent a merger or other event which has diluted its gas phase metallicity. We test this by deriving the gas-phase metallicity from the measured gas-to-dust ratio. This is done using the empirical relation between the gas-to-dust ratio and metallicity from \citet[][Figure 4 and Table 1]{RemyRuyer2014} for galaxies in the low metallicity regime, where we expect the galaxy to be. Using the relation for the higher metallicity regime instead would result in an even lower metallicity, thus only amplifying the analysis below. We use the relation for a metallicity-dependent \XCO. Rewriting for the metallicity gives:
\begin{equation}
12 + \text{log}(\text{O/H}) = \frac{- \left( \text{log} \left( \frac{\text{G}}{\text{D}} \right) - b \right)}{\alpha_\text{L}} + x_\odot
\end{equation}
where $b$ = 0.96, $\alpha_\text{L} = 3.10 \pm 1.33$, and $x_\odot$ = 8.69 the solar metallicity \citep{Asplund2009}. The gas-to-dust ratio is derived using the \HI and H$_2$ masses stated above, and a dust mass of $4.2 \times 10^{6} M_\odot$ from \citet{Fuller2014}. Since our upper limit on the H$_2$ mass depends on X$_\text{CO}$, we combine the relation above and that between \XCO  and metallicity (Equation \ref{eq:accurso}) to estimate the metallicity (and thus the resulting X$_\text{CO}$) in this object. The resulting metallicity, assuming these relations hold, is $12 + \text{log}(\text{O/H})$ = 8.12 (or 0.27 $Z_\odot$). This is significantly lower than the metallicity derived from the stellar mass ($12 + \text{log}(\text{O/H})$ = 8.71; 1.05 $Z_\odot$, see \Section \ref{sub:H2_masses}), and that found for young star clusters by \citet{Mora2015}, who find values of $\sim$0.4-1 $Z_\odot$. It implies an \XCO of $\sim18 \times 10^{20}$ cm$^{-2}$ (K \kms)$^{-1}$ (see \Section \ref{sub:H2_masses}), in which case the H$_2$ mass limit we can set in this object increases to $\sim 2.1 \times 10^{8}$ $M_\odot$. This revised values for the atomic-to-molecular gas ratio and the molecular gas depletion time yield $>$10 and $<$4.2 Gyr respectively, much more consistent with canonical values.

We do detect a 3 mm continuum source in our NGC1427A observations. This is shown in Figure \ref{fig:NGC1427A}, where the continuum is overplotted as coloured contours on a \textit{g}-band image from the FDS, similar to Figure \ref{fig:continua}. The emission originates from a source towards the north of NGC1427A, slightly to the east side relative to its centre. There are two possible candidates for its origin: either the emission comes from a background source, or from an AGN associated with NGC1427A, that has been moved off-centre as a result of a galaxy-galaxy interaction. The latter interpretation seems quite speculative. However based on the information currently available and the galaxy's turbulent history, it is a possibility. In case the continuum originates from a background source, we would expect to see an optical counterpart in the optical FDS image. These images are quite deep, and can detect point sources down to $\sim$25-26 mag. However, no optical counterpart is detected. Moreover, no emission lines are detected in archival MUSE observations of this area. Thus, if it is indeed a background object, it must be heavily obscured by dust.

\section{Conclusions / summary}
\label{sec:conclusions}
We have presented the first results from the ALMA Fornax Cluster Survey (AlFoCS), a complete survey of the CO(1-0) in all Fornax cluster galaxies above $>3 \times 10^8 \text{ M}_\odot$ that contain dust \citep{Fuller2014} and/or HI (\citealt{Waugh2002}, Loni et al. in prep. based on ATCA data). The goal of this survey is to study the effects of the cluster environment on the cold molecular gas inside galaxies. We present moment zero, one, and two maps, as well as position-velocity diagrams, spectra, and comparisons with optical images for all galaxies detected. Furthermore we estimate H$_2$ masses and derive the corresponding H$_2$ deficiencies compared to field galaxies. The main conclusions from this initial analysis are:

\begin{itemize}
\item The cold molecular gas in galaxies is indeed affected by the cluster environment. All galaxies with stellar masses below $3 \times 10^9 M_\odot$ (8 out of 15 detected galaxies) have morphologically and kinematically disturbed gas reservoirs. ``Disturbed'' means that their molecular gas is distributed asymmetrically with respect to the optical centre of the galaxy, sometimes with irregular shapes or large tails. The moment one maps and PVDs show irregular motions, and in most cases no rotation can be identified. This suggests that Fornax is still a very active environment, having a significant impact on its members. More massive galaxies are probably experiencing the same interactions, however the molecular gas may not be affected in the same way, because of their deeper potential wells. \\

\item Continuum was detected in four of the galaxies observed. In three of them it is likely associated with AGN activity. In one case (NGC1427A) the emission does not originate from the centre of the galaxy, but rather from the galaxy's edge. It is unclear whether the source of this emission is a background object or an AGN that was moved off-centre due to a recent galaxy-galaxy interaction. \\

\item Galaxies with regular CO emission have an average H$_2$ deficiency of -0.50 dex, and galaxies with disturbed CO emission have an average H$_2$ deficiency of -1.1 dex. AlFoCS galaxies with disturbed molecular gas reservoirs are therefore, with few exceptions, significantly deficient in H$_2$ compared to their counterparts in the field (as probed by the xCOLD GASS sample). \\

\item Whether a galaxy has a molecular gas reservoir, and whether that reservoir is disturbed or regular, appears to be independent of the galaxy's location within the cluster. However, the sample size is small. \\

\item Two molecular gas tails were detected, that extend beyond the galaxy's stellar body and align with the direction of the cluster centre. These galaxies are possibly undergoing ram pressure stripping. Relatively high H$_2$ deficiencies support this explanation. Several other galaxies have molecular gas reservoirs that are asymmetric with respect to their stellar bodies as well, and are therefore possible RPS candidates, however it is difficult to draw definite conclusions from these data alone. \\

\end{itemize}

To be able to really distinguish between ram pressure stripping and galaxy-galaxy interactions, we need to compare the stellar kinematics (derived from e.g. MUSE observations: \citealt{Sarzi2018}) of the galaxies in question with the kinematics of their disturbed molecular gas. If both the molecular gas and stars show similar kinematics, ram pressure stripping can be ruled out. If only the molecular gas kinematics are disturbed, on the other hand, we can confirm that ram pressure stripping plays a role. This will be the aim of a future work. Similarly, we will compare our data with those of the \textit{Herschel} Fornax Cluster Survey \citep{Fuller2014} to compare the molecular gas and dust distributions, and derive gas-to-dust ratios for the galaxies in our sample. \\

In conclusion, the detection of a relatively high number of galaxies with disturbed molecular gas reservoirs and H$_2$ deficiencies of sometimes more than an order of magnitude reveal the importance of the cluster environment for even the tightly bound molecular gas phase, and motivate further study of environmental effects on molecular gas in nearby clusters. 

\section*{Acknowledgements}
This publication has received funding from the European Union's Horizon 2020 research and innovation programme under grant agreement No 730562 [RadioNet].

This project has received funding from the European Research Council (ERC) under the European Union's Horizon 2020 research and innovation programme (grant agreement no. 679627; project name FORNAX).

NZ acknowledges support from the European Research Council (ERC) in the form of Consolidator Grant CosmicDust (ERC-2014-CoG-647939)

TAD acknowledges support from a Science and Technology Facilities Council Ernest Rutherford Fellowship.

FvdV is supported by the Klaus Tschira Foundation.

R.F.P. acknowledges financial support from the European Union's Horizon 2020 research and innovation program under the Marie Skłodowska-Curie grant agreement No. 721463 to the SUNDIAL ITN network. 

This paper makes use of the following ALMA data: ADS/JAO.ALMA\#2015.1.01135.S. ALMA is a partnership of ESO (representing its member states), NSF (USA) and NINS (Japan), together with NRC (Canada) and NSC and ASIAA (Taiwan) and KASI (Republic of Korea), in cooperation with the Republic of Chile. The Joint ALMA Observatory is operated by ESO, AUI/NRAO and NAOJ. 

The Mopra radio telescope is part of the Australia Telescope National Facility which is funded by the Australian Government for operation as a National Facility managed by CSIRO.

This publication makes use of data products from the Wide-field Infrared Survey Explorer, which is a joint project of the University of California, Los Angeles, and the Jet Propulsion Laboratory/California Institute of Technology, funded by the National Aeronautics and Space Administration.

This research has made use of the NASA/IPAC Extragalactic Database (NED), which is operated by the Jet Propulsion Laboratory, California Institute of Technology, under contract with the National Aeronautics and Space Administration.




\bibliographystyle{mnras}
\bibliography{References} 

\begin{thebibliography}{}
\makeatletter
\relax
\def\mn@urlcharsother{\let\do\@makeother \do\$\do\&\do\#\do\^\do\_\do\%\do\~}
\def\mn@doi{\begingroup\mn@urlcharsother \@ifnextchar [ {\mn@doi@}
  {\mn@doi@[]}}
\def\mn@doi@[#1]#2{\def\@tempa{#1}\ifx\@tempa\@empty \href
  {http://dx.doi.org/#2} {doi:#2}\else \href {http://dx.doi.org/#2} {#1}\fi
  \endgroup}
\def\mn@eprint#1#2{\mn@eprint@#1:#2::\@nil}
\def\mn@eprint@arXiv#1{\href {http://arxiv.org/abs/#1} {{\tt arXiv:#1}}}
\def\mn@eprint@dblp#1{\href {http://dblp.uni-trier.de/rec/bibtex/#1.xml}
  {dblp:#1}}
\def\mn@eprint@#1:#2:#3:#4\@nil{\def\@tempa {#1}\def\@tempb {#2}\def\@tempc
  {#3}\ifx \@tempc \@empty \let \@tempc \@tempb \let \@tempb \@tempa \fi \ifx
  \@tempb \@empty \def\@tempb {arXiv}\fi \@ifundefined
  {mn@eprint@\@tempb}{\@tempb:\@tempc}{\expandafter \expandafter \csname
  mn@eprint@\@tempb\endcsname \expandafter{\@tempc}}}

\bibitem[\protect\citeauthoryear{{Accurso} et~al.,}{{Accurso}
  et~al.}{2017}]{Accurso2017}
{Accurso} G.,  et~al., 2017, \mn@doi [\mnras] {10.1093/mnras/stx1556}, \href
  {http://adsabs.harvard.edu/abs/2017MNRAS.470.4750A} {470, 4750}

\bibitem[\protect\citeauthoryear{{Alatalo} et~al.,}{{Alatalo}
  et~al.}{2013}]{Alatalo2013}
{Alatalo} K.,  et~al., 2013, \mn@doi [\mnras] {10.1093/mnras/sts299}, \href
  {http://adsabs.harvard.edu/abs/2013MNRAS.432.1796A} {432, 1796}

\bibitem[\protect\citeauthoryear{{Asplund}, {Grevesse}, {Sauval}  \&
  {Scott}}{{Asplund} et~al.}{2009}]{Asplund2009}
{Asplund} M.,  {Grevesse} N.,  {Sauval} A.~J.,   {Scott} P.,  2009, \mn@doi
  [\araa] {10.1146/annurev.astro.46.060407.145222}, \href
  {http://adsabs.harvard.edu/abs/2009ARA%26A..47..481A} {47, 481}

\bibitem[\protect\citeauthoryear{{Barazza}, {Binggeli}  \& {Jerjen}}{{Barazza}
  et~al.}{2002}]{Barazza2002}
{Barazza} F.~D.,  {Binggeli} B.,   {Jerjen} H.,  2002, \mn@doi [\aap]
  {10.1051/0004-6361:20020875}, \href
  {http://adsabs.harvard.edu/abs/2002A%26A...391..823B} {391, 823}

\bibitem[\protect\citeauthoryear{{Barnes} et~al.,}{{Barnes}
  et~al.}{2001}]{Barnes2001}
{Barnes} D.~G.,  et~al., 2001, \mn@doi [\mnras]
  {10.1046/j.1365-8711.2001.04102.x}, \href
  {http://adsabs.harvard.edu/abs/2001MNRAS.322..486B} {322, 486}

\bibitem[\protect\citeauthoryear{{Binggeli}, {Sandage}  \&
  {Tammann}}{{Binggeli} et~al.}{1985}]{Binggeli1985}
{Binggeli} B.,  {Sandage} A.,   {Tammann} G.~A.,  1985, \mn@doi [\aj]
  {10.1086/113874}, \href {http://adsabs.harvard.edu/abs/1985AJ.....90.1681B}
  {90, 1681}

\bibitem[\protect\citeauthoryear{{Boizelle}, {Barth}, {Darling}, {Baker},
  {Buote}, {Ho}  \& {Walsh}}{{Boizelle} et~al.}{2017}]{Boizelle2017}
{Boizelle} B.~D.,  {Barth} A.~J.,  {Darling} J.,  {Baker} A.~J.,  {Buote}
  D.~A.,  {Ho} L.~C.,   {Walsh} J.~L.,  2017, \mn@doi [\apj]
  {10.3847/1538-4357/aa8266}, \href
  {http://adsabs.harvard.edu/abs/2017ApJ...845..170B} {845, 170}

\bibitem[\protect\citeauthoryear{{Bolatto}, {Wolfire}  \& {Leroy}}{{Bolatto}
  et~al.}{2013}]{Bolatto2013}
{Bolatto} A.~D.,  {Wolfire} M.,   {Leroy} A.~K.,  2013, \mn@doi [\araa]
  {10.1146/annurev-astro-082812-140944}, \href
  {http://adsabs.harvard.edu/abs/2013ARA%26A..51..207B} {51, 207}

\bibitem[\protect\citeauthoryear{{Booth} et~al.,}{{Booth}
  et~al.}{1989}]{Booth1989}
{Booth} R.~S.,  et~al., 1989, \aap, \href
  {http://adsabs.harvard.edu/abs/1989A%26A...216..315B} {216, 315}

\bibitem[\protect\citeauthoryear{{Boselli} \& {Gavazzi}}{{Boselli} \&
  {Gavazzi}}{2006}]{Boselli2006}
{Boselli} A.,  {Gavazzi} G.,  2006, \mn@doi [\pasp] {10.1086/500691}, \href
  {http://adsabs.harvard.edu/abs/2006PASP..118..517B} {118, 517}

\bibitem[\protect\citeauthoryear{{Boselli}, {Casoli}  \& {Lequeux}}{{Boselli}
  et~al.}{1995}]{Boselli1995}
{Boselli} A.,  {Casoli} F.,   {Lequeux} J.,  1995, \aaps, \href
  {http://adsabs.harvard.edu/abs/1995A%26AS..110..521B} {110, 521}

\bibitem[\protect\citeauthoryear{{Boselli}, {Boissier}, {Cortese}  \&
  {Gavazzi}}{{Boselli} et~al.}{2008}]{Boselli2008}
{Boselli} A.,  {Boissier} S.,  {Cortese} L.,   {Gavazzi} G.,  2008, \mn@doi
  [\aap] {10.1051/0004-6361:200809546}, \href
  {http://adsabs.harvard.edu/abs/2008A%26A...489.1015B} {489, 1015}

\bibitem[\protect\citeauthoryear{{Boselli} et~al.,}{{Boselli}
  et~al.}{2011}]{Boselli2011}
{Boselli} A.,  et~al., 2011, \mn@doi [\aap] {10.1051/0004-6361/201016389},
  \href {http://adsabs.harvard.edu/abs/2011A%26A...528A.107B} {528, A107}

\bibitem[\protect\citeauthoryear{{Boselli}, {Cortese}, {Boquien}, {Boissier},
  {Catinella}, {Gavazzi}, {Lagos}  \& {Saintonge}}{{Boselli}
  et~al.}{2014}]{Boselli2014}
{Boselli} A.,  {Cortese} L.,  {Boquien} M.,  {Boissier} S.,  {Catinella} B.,
  {Gavazzi} G.,  {Lagos} C.,   {Saintonge} A.,  2014, \mn@doi [\aap]
  {10.1051/0004-6361/201322313}, \href
  {http://adsabs.harvard.edu/abs/2014A%26A...564A..67B} {564, A67}

\bibitem[\protect\citeauthoryear{{Boselli} et~al.,}{{Boselli}
  et~al.}{2018}]{Boselli2018}
{Boselli} A.,  et~al., 2018, \mn@doi [\aap] {10.1051/0004-6361/201732407},
  \href {http://adsabs.harvard.edu/abs/2018A%26A...614A..56B} {614, A56}

\bibitem[\protect\citeauthoryear{{Briggs}}{{Briggs}}{1995}]{Briggs1995}
{Briggs} D.~S.,  1995, in American Astronomical Society Meeting Abstracts.
  p.~1444

\bibitem[\protect\citeauthoryear{{Brinchmann}, {Charlot}, {White}, {Tremonti},
  {Kauffmann}, {Heckman}  \& {Brinkmann}}{{Brinchmann}
  et~al.}{2004}]{Brinchmann2004}
{Brinchmann} J.,  {Charlot} S.,  {White} S.~D.~M.,  {Tremonti} C.,  {Kauffmann}
  G.,  {Heckman} T.,   {Brinkmann} J.,  2004, \mn@doi [\mnras]
  {10.1111/j.1365-2966.2004.07881.x}, \href
  {http://adsabs.harvard.edu/abs/2004MNRAS.351.1151B} {351, 1151}

\bibitem[\protect\citeauthoryear{{Casoli}, {Boisse}, {Combes}  \&
  {Dupraz}}{{Casoli} et~al.}{1991}]{Casoli1991}
{Casoli} F.,  {Boisse} P.,  {Combes} F.,   {Dupraz} C.,  1991, \aap, \href
  {http://adsabs.harvard.edu/abs/1991A%26A...249..359C} {249, 359}

\bibitem[\protect\citeauthoryear{{Cayatte}, {van Gorkom}, {Balkowski}  \&
  {Kotanyi}}{{Cayatte} et~al.}{1990}]{Cayatte1990}
{Cayatte} V.,  {van Gorkom} J.~H.,  {Balkowski} C.,   {Kotanyi} C.,  1990,
  \mn@doi [\aj] {10.1086/115545}, \href
  {http://adsabs.harvard.edu/abs/1990AJ....100..604C} {100, 604}

\bibitem[\protect\citeauthoryear{{Chanam{\'e}}, {Infante}  \&
  {Reisenegger}}{{Chanam{\'e}} et~al.}{2000}]{Chaname2000}
{Chanam{\'e}} J.,  {Infante} L.,   {Reisenegger} A.,  2000, \mn@doi [\apj]
  {10.1086/308364}, \href {http://adsabs.harvard.edu/abs/2000ApJ...530...96C}
  {530, 96}

\bibitem[\protect\citeauthoryear{{Coe}}{{Coe}}{2010}]{Coe2010}
{Coe} D.,  2010, preprint, \href
  {http://adsabs.harvard.edu/abs/2010arXiv1005.0411C} {} (\mn@eprint {arXiv}
  {1005.0411})

\bibitem[\protect\citeauthoryear{{Colless} et~al.,}{{Colless}
  et~al.}{2001}]{Colless2001}
{Colless} M.,  et~al., 2001, \mn@doi [\mnras]
  {10.1046/j.1365-8711.2001.04902.x}, \href
  {http://adsabs.harvard.edu/abs/2001MNRAS.328.1039C} {328, 1039}

\bibitem[\protect\citeauthoryear{{Conselice}}{{Conselice}}{2002}]{Conselice2002}
{Conselice} C.~J.,  2002, \mn@doi [\apjl] {10.1086/341878}, \href
  {http://adsabs.harvard.edu/abs/2002ApJ...573L...5C} {573, L5}

\bibitem[\protect\citeauthoryear{{Cort{\'e}s}, {Kenney}  \&
  {Hardy}}{{Cort{\'e}s} et~al.}{2006}]{Cortes2006}
{Cort{\'e}s} J.~R.,  {Kenney} J.~D.~P.,   {Hardy} E.,  2006, \mn@doi [\aj]
  {10.1086/499075}, \href {http://adsabs.harvard.edu/abs/2006AJ....131..747C}
  {131, 747}

\bibitem[\protect\citeauthoryear{{Cortese} et~al.,}{{Cortese}
  et~al.}{2010}]{Cortese2010}
{Cortese} L.,  et~al., 2010, \mn@doi [\aap] {10.1051/0004-6361/201014550},
  \href {http://adsabs.harvard.edu/abs/2010A%26A...518L..49C} {518, L49}

\bibitem[\protect\citeauthoryear{{Cortese} et~al.,}{{Cortese}
  et~al.}{2012}]{Cortese2012}
{Cortese} L.,  et~al., 2012, \mn@doi [\aap] {10.1051/0004-6361/201118499},
  \href {http://adsabs.harvard.edu/abs/2012A%26A...540A..52C} {540, A52}

\bibitem[\protect\citeauthoryear{{Cowie} \& {Songaila}}{{Cowie} \&
  {Songaila}}{1977}]{Cowie1977}
{Cowie} L.~L.,  {Songaila} A.,  1977, \mn@doi [\nat] {10.1038/266501a0}, \href
  {http://adsabs.harvard.edu/abs/1977Natur.266..501C} {266, 501}

\bibitem[\protect\citeauthoryear{{Croom}, {Saunders}  \& {Heald}}{{Croom}
  et~al.}{2004}]{Croom2004}
{Croom} S.,  {Saunders} W.,   {Heald} R.,  2004, Anglo-Australian Observatory
  Epping Newsletter, \href {http://adsabs.harvard.edu/abs/2004AAONw.106...12C}
  {106, 12}

\bibitem[\protect\citeauthoryear{{Dame}}{{Dame}}{2011}]{Dame2011}
{Dame} T.~M.,  2011, preprint, \href
  {https://ui.adsabs.harvard.edu/#abs/2011arXiv1101.1499D} {p. arXiv:1101.1499}
  (\mn@eprint {arXiv} {1101.1499})

\bibitem[\protect\citeauthoryear{{Davies} et~al.,}{{Davies}
  et~al.}{2010}]{Davies2010}
{Davies} J.~I.,  et~al., 2010, \mn@doi [\aap] {10.1051/0004-6361/201014571},
  \href {http://adsabs.harvard.edu/abs/2010A%26A...518L..48D} {518, L48}

\bibitem[\protect\citeauthoryear{{Davies} et~al.,}{{Davies}
  et~al.}{2013}]{Davies2013}
{Davies} J.~I.,  et~al., 2013, \mn@doi [\mnras] {10.1093/mnras/sts082}, \href
  {https://ui.adsabs.harvard.edu/#abs/2013MNRAS.428..834D} {428, 834}

\bibitem[\protect\citeauthoryear{{De Looze} et~al.,}{{De Looze}
  et~al.}{2010}]{DeLooze2010}
{De Looze} I.,  et~al., 2010, \mn@doi [\aap] {10.1051/0004-6361/201014647},
  \href {http://adsabs.harvard.edu/abs/2010A%26A...518L..54D} {518, L54}

\bibitem[\protect\citeauthoryear{{De Looze} et~al.,}{{De Looze}
  et~al.}{2013}]{DeLooze2013}
{De Looze} I.,  et~al., 2013, \mn@doi [\mnras] {10.1093/mnras/stt1626}, \href
  {http://adsabs.harvard.edu/abs/2013MNRAS.436.1057D} {436, 1057}

\bibitem[\protect\citeauthoryear{{De Rijcke}, {Dejonghe}, {Zeilinger}  \&
  {Hau}}{{De Rijcke} et~al.}{2003}]{DeRijcke2003}
{De Rijcke} S.,  {Dejonghe} H.,  {Zeilinger} W.~W.,   {Hau} G.~K.~T.,  2003,
  \mn@doi [\aap] {10.1051/0004-6361:20021866}, \href
  {http://adsabs.harvard.edu/abs/2003A%26A...400..119D} {400, 119}

\bibitem[\protect\citeauthoryear{{Dressler}}{{Dressler}}{1980}]{Dressler1980}
{Dressler} A.,  1980, \mn@doi [\apj] {10.1086/157753}, \href
  {http://adsabs.harvard.edu/abs/1980ApJ...236..351D} {236, 351}

\bibitem[\protect\citeauthoryear{{Drinkwater}, {Sadler}, {Davies}, {Dickens},
  {Gregg}, {Parker}, {Phillipps}  \& {Smith}}{{Drinkwater}
  et~al.}{1999}]{Drinkwater1999}
{Drinkwater} M.~J.,  {Sadler} E.~M.,  {Davies} J.~I.,  {Dickens} R.~J.,
  {Gregg} M.~D.,  {Parker} Q.~A.,  {Phillipps} S.,   {Smith} R.~M.,  1999, in
  {Morganti} R.,  {Couch} W.~J.,  eds, Looking Deep in the Southern Sky. p.~21
  (\mn@eprint {} {astro-ph/9802095})

\bibitem[\protect\citeauthoryear{{Drinkwater}, {Gregg}  \&
  {Colless}}{{Drinkwater} et~al.}{2001}]{Drinkwater2001a}
{Drinkwater} M.~J.,  {Gregg} M.~D.,   {Colless} M.,  2001, \mn@doi [\apjl]
  {10.1086/319113}, \href {http://adsabs.harvard.edu/abs/2001ApJ...548L.139D}
  {548, L139}

\bibitem[\protect\citeauthoryear{{Elbaz} et~al.,}{{Elbaz}
  et~al.}{2007}]{Elbaz2007}
{Elbaz} D.,  et~al., 2007, \mn@doi [\aap] {10.1051/0004-6361:20077525}, \href
  {http://adsabs.harvard.edu/abs/2007A%26A...468...33E} {468, 33}

\bibitem[\protect\citeauthoryear{{Ferguson}}{{Ferguson}}{1989}]{Ferguson1989}
{Ferguson} H.~C.,  1989, \mn@doi [\aj] {10.1086/115152}, \href
  {http://adsabs.harvard.edu/abs/1989AJ.....98..367F} {98, 367}

\bibitem[\protect\citeauthoryear{{Ferrarese} et~al.,}{{Ferrarese}
  et~al.}{2012}]{Ferrarese2012}
{Ferrarese} L.,  et~al., 2012, \mn@doi [\apjs] {10.1088/0067-0049/200/1/4},
  \href {http://adsabs.harvard.edu/abs/2012ApJS..200....4F} {200, 4}

\bibitem[\protect\citeauthoryear{{Fujita}}{{Fujita}}{2004}]{Fujita2004}
{Fujita} Y.,  2004, \mn@doi [\pasj] {10.1093/pasj/56.1.29}, \href
  {http://adsabs.harvard.edu/abs/2004PASJ...56...29F} {56, 29}

\bibitem[\protect\citeauthoryear{{Fuller} et~al.,}{{Fuller}
  et~al.}{2014}]{Fuller2014}
{Fuller} C.,  et~al., 2014, \mn@doi [\mnras] {10.1093/mnras/stu369}, \href
  {http://adsabs.harvard.edu/abs/2014MNRAS.440.1571F} {440, 1571}

\bibitem[\protect\citeauthoryear{{Fumagalli}, {Krumholz}, {Prochaska},
  {Gavazzi}  \& {Boselli}}{{Fumagalli} et~al.}{2009}]{Fumagalli2009}
{Fumagalli} M.,  {Krumholz} M.~R.,  {Prochaska} J.~X.,  {Gavazzi} G.,
  {Boselli} A.,  2009, \mn@doi [\apj] {10.1088/0004-637X/697/2/1811}, \href
  {http://adsabs.harvard.edu/abs/2009ApJ...697.1811F} {697, 1811}

\bibitem[\protect\citeauthoryear{{Gavazzi}, {Boselli}, {van Driel}  \&
  {O'Neil}}{{Gavazzi} et~al.}{2005}]{Gavazzi2005}
{Gavazzi} G.,  {Boselli} A.,  {van Driel} W.,   {O'Neil} K.,  2005, \mn@doi
  [\aap] {10.1051/0004-6361:20041678}, \href
  {http://adsabs.harvard.edu/abs/2005A%26A...429..439G} {429, 439}

\bibitem[\protect\citeauthoryear{{Gunn} \& {Gott}}{{Gunn} \&
  {Gott}}{1972}]{Gunn1972}
{Gunn} J.~E.,  {Gott} III J.~R.,  1972, \mn@doi [\apj] {10.1086/151605}, \href
  {http://adsabs.harvard.edu/abs/1972ApJ...176....1G} {176, 1}

\bibitem[\protect\citeauthoryear{{Hayashi} et~al.,}{{Hayashi}
  et~al.}{2018}]{Hayashi2018}
{Hayashi} M.,  et~al., 2018, \mn@doi [\apj] {10.3847/1538-4357/aab3e7}, \href
  {http://adsabs.harvard.edu/abs/2018ApJ...856..118H} {856, 118}

\bibitem[\protect\citeauthoryear{{Haynes}, {Giovanelli}  \&
  {Chincarini}}{{Haynes} et~al.}{1984}]{Haynes1984}
{Haynes} M.~P.,  {Giovanelli} R.,   {Chincarini} G.~L.,  1984, \mn@doi [\araa]
  {10.1146/annurev.aa.22.090184.002305}, \href
  {http://adsabs.harvard.edu/abs/1984ARA%26A..22..445H} {22, 445}

\bibitem[\protect\citeauthoryear{{Hinton}, {Davis}, {Lidman}, {Glazebrook}  \&
  {Lewis}}{{Hinton} et~al.}{2016}]{Hinton2016}
{Hinton} S.~R.,  {Davis} T.~M.,  {Lidman} C.,  {Glazebrook} K.,   {Lewis}
  G.~F.,  2016, \mn@doi [Astronomy and Computing]
  {10.1016/j.ascom.2016.03.001}, \href
  {http://adsabs.harvard.edu/abs/2016A%26C....15...61H} {15, 61}

\bibitem[\protect\citeauthoryear{{H{\"o}gbom}}{{H{\"o}gbom}}{1974}]{Hogbom1974}
{H{\"o}gbom} J.~A.,  1974, \aaps, \href
  {http://adsabs.harvard.edu/abs/1974A%26AS...15..417H} {15, 417}

\bibitem[\protect\citeauthoryear{{Horellou}, {Casoli}  \& {Dupraz}}{{Horellou}
  et~al.}{1995}]{Horellou1995}
{Horellou} C.,  {Casoli} F.,   {Dupraz} C.,  1995, \aap, \href
  {http://adsabs.harvard.edu/abs/1995A%26A...303..361H} {303, 361}

\bibitem[\protect\citeauthoryear{{Huchra} et~al.,}{{Huchra}
  et~al.}{2012}]{Huchra2012}
{Huchra} J.~P.,  et~al., 2012, \mn@doi [\apjs] {10.1088/0067-0049/199/2/26},
  \href {http://adsabs.harvard.edu/abs/2012ApJS..199...26H} {199, 26}

\bibitem[\protect\citeauthoryear{{Hunter} et~al.,}{{Hunter}
  et~al.}{2012}]{Hunter2012}
{Hunter} D.~A.,  et~al., 2012, \mn@doi [\aj] {10.1088/0004-6256/144/5/134},
  \href {http://adsabs.harvard.edu/abs/2012AJ....144..134H} {144, 134}

\bibitem[\protect\citeauthoryear{{Iodice} et~al.,}{{Iodice}
  et~al.}{2016}]{Iodice2016}
{Iodice} E.,  et~al., 2016, \mn@doi [\apj] {10.3847/0004-637X/820/1/42}, \href
  {http://adsabs.harvard.edu/abs/2016ApJ...820...42I} {820, 42}

\bibitem[\protect\citeauthoryear{{Iodice} et~al.,}{{Iodice}
  et~al.}{2017}]{Iodice2017}
{Iodice} E.,  et~al., 2017, \mn@doi [\apj] {10.3847/1538-4357/aa6846}, \href
  {http://adsabs.harvard.edu/abs/2017ApJ...839...21I} {839, 21}

\bibitem[\protect\citeauthoryear{{Jaff{\'e}} et~al.,}{{Jaff{\'e}}
  et~al.}{2018}]{Jaffe2018}
{Jaff{\'e}} Y.~L.,  et~al., 2018, \mn@doi [\mnras] {10.1093/mnras/sty500},
  \href {http://adsabs.harvard.edu/abs/2018MNRAS.476.4753J} {476, 4753}

\bibitem[\protect\citeauthoryear{{Jerjen}, {Kalnajs}  \& {Binggeli}}{{Jerjen}
  et~al.}{2000}]{Jerjen2000}
{Jerjen} H.,  {Kalnajs} A.,   {Binggeli} B.,  2000, \aap, \href
  {http://adsabs.harvard.edu/abs/2000A%26A...358..845J} {358, 845}

\bibitem[\protect\citeauthoryear{{Jord{\'a}n} et~al.,}{{Jord{\'a}n}
  et~al.}{2007}]{Jordan2007}
{Jord{\'a}n} A.,  et~al., 2007, \mn@doi [\apjs] {10.1086/516840}, \href
  {http://adsabs.harvard.edu/abs/2007ApJS..171..101J} {171, 101}

\bibitem[\protect\citeauthoryear{{Kenney} \& {Young}}{{Kenney} \&
  {Young}}{1989}]{Kenney1989}
{Kenney} J.~D.~P.,  {Young} J.~S.,  1989, \mn@doi [\apj] {10.1086/167787},
  \href {http://adsabs.harvard.edu/abs/1989ApJ...344..171K} {344, 171}

\bibitem[\protect\citeauthoryear{{Kenney}, {Geha}, {J{\'a}chym}, {Crowl},
  {Dague}, {Chung}, {van Gorkom}  \& {Vollmer}}{{Kenney}
  et~al.}{2014}]{Kenney2014}
{Kenney} J.~D.~P.,  {Geha} M.,  {J{\'a}chym} P.,  {Crowl} H.~H.,  {Dague} W.,
  {Chung} A.,  {van Gorkom} J.,   {Vollmer} B.,  2014, \mn@doi [\apj]
  {10.1088/0004-637X/780/2/119}, \href
  {http://adsabs.harvard.edu/abs/2014ApJ...780..119K} {780, 119}

\bibitem[\protect\citeauthoryear{{Koleva}, {Bouchard}, {Prugniel}, {De Rijcke}
  \& {Vauglin}}{{Koleva} et~al.}{2013}]{Koleva2013}
{Koleva} M.,  {Bouchard} A.,  {Prugniel} P.,  {De Rijcke} S.,   {Vauglin} I.,
  2013, \mn@doi [\mnras] {10.1093/mnras/sts238}, \href
  {http://adsabs.harvard.edu/abs/2013MNRAS.428.2949K} {428, 2949}

\bibitem[\protect\citeauthoryear{{Kuijken} et~al.,}{{Kuijken}
  et~al.}{2002}]{Kuijken2002}
{Kuijken} K.,  et~al., 2002, The Messenger, \href
  {http://adsabs.harvard.edu/abs/2002Msngr.110...15K} {110, 15}

\bibitem[\protect\citeauthoryear{{Ladd}, {Purcell}, {Wong}  \&
  {Robertson}}{{Ladd} et~al.}{2005}]{Ladd2005}
{Ladd} N.,  {Purcell} C.,  {Wong} T.,   {Robertson} S.,  2005, \mn@doi [\pasa]
  {10.1071/AS04068}, \href {http://adsabs.harvard.edu/abs/2005PASA...22...62L}
  {22, 62}

\bibitem[\protect\citeauthoryear{{Larson}, {Tinsley}  \& {Caldwell}}{{Larson}
  et~al.}{1980}]{Larson1980}
{Larson} R.~B.,  {Tinsley} B.~M.,   {Caldwell} C.~N.,  1980, \mn@doi [\apj]
  {10.1086/157917}, \href {http://adsabs.harvard.edu/abs/1980ApJ...237..692L}
  {237, 692}

\bibitem[\protect\citeauthoryear{{Laurikainen}, {Salo}, {Buta}, {Knapen},
  {Speltincx}  \& {Block}}{{Laurikainen} et~al.}{2006}]{Laurikainen2006}
{Laurikainen} E.,  {Salo} H.,  {Buta} R.,  {Knapen} J.,  {Speltincx} T.,
  {Block} D.,  2006, \mn@doi [\aj] {10.1086/508810}, \href
  {http://adsabs.harvard.edu/abs/2006AJ....132.2634L} {132, 2634}

\bibitem[\protect\citeauthoryear{{Lee-Waddell} et~al.,}{{Lee-Waddell}
  et~al.}{2018}]{Lee-Waddell2018}
{Lee-Waddell} K.,  et~al., 2018, \mn@doi [\mnras] {10.1093/mnras/stx2808},
  \href {http://adsabs.harvard.edu/abs/2018MNRAS.474.1108L} {474, 1108}

\bibitem[\protect\citeauthoryear{{Lee} et~al.,}{{Lee} et~al.}{2017}]{Lee2017}
{Lee} B.,  et~al., 2017, \mn@doi [\mnras] {10.1093/mnras/stw3162}, \href
  {http://adsabs.harvard.edu/abs/2017MNRAS.466.1382L} {466, 1382}

\bibitem[\protect\citeauthoryear{{Lena} et~al.,}{{Lena}
  et~al.}{2015}]{Lena2015}
{Lena} D.,  et~al., 2015, \mn@doi [\apj] {10.1088/0004-637X/806/1/84}, \href
  {http://adsabs.harvard.edu/abs/2015ApJ...806...84L} {806, 84}

\bibitem[\protect\citeauthoryear{{Lewis} et~al.,}{{Lewis}
  et~al.}{2002}]{Lewis2002}
{Lewis} I.~J.,  et~al., 2002, \mn@doi [\mnras]
  {10.1046/j.1365-8711.2002.05333.x}, \href
  {http://adsabs.harvard.edu/abs/2002MNRAS.333..279L} {333, 279}

\bibitem[\protect\citeauthoryear{{Lisker}, {Grebel}  \& {Binggeli}}{{Lisker}
  et~al.}{2006a}]{Lisker2006a}
{Lisker} T.,  {Grebel} E.~K.,   {Binggeli} B.,  2006a, \mn@doi [\aj]
  {10.1086/505045}, \href {http://adsabs.harvard.edu/abs/2006AJ....132..497L}
  {132, 497}

\bibitem[\protect\citeauthoryear{{Lisker}, {Glatt}, {Westera}  \&
  {Grebel}}{{Lisker} et~al.}{2006b}]{Lisker2006b}
{Lisker} T.,  {Glatt} K.,  {Westera} P.,   {Grebel} E.~K.,  2006b, \mn@doi
  [\aj] {10.1086/508414}, \href
  {http://adsabs.harvard.edu/abs/2006AJ....132.2432L} {132, 2432}

\bibitem[\protect\citeauthoryear{{McMullin}, {Waters}, {Schiebel}, {Young}  \&
  {Golap}}{{McMullin} et~al.}{2007}]{McMullin2007}
{McMullin} J.~P.,  {Waters} B.,  {Schiebel} D.,  {Young} W.,   {Golap} K.,
  2007, in {Shaw} R.~A.,  {Hill} F.,   {Bell} D.~J.,  eds,  Astronomical
  Society of the Pacific Conference Series Vol. 376, Astronomical Data Analysis
  Software and Systems XVI. p.~127

\bibitem[\protect\citeauthoryear{{Mihos}}{{Mihos}}{2004}]{Mihos2004}
{Mihos} J.~C.,  2004, Clusters of Galaxies: Probes of Cosmological Structure
  and Galaxy Evolution, \href
  {http://adsabs.harvard.edu/abs/2004cgpc.symp..277M} {p.~277}

\bibitem[\protect\citeauthoryear{{Miller}, {Lamb}  \& {Cook}}{{Miller}
  et~al.}{1998}]{Miller1998}
{Miller} M.~C.,  {Lamb} F.~K.,   {Cook} G.~B.,  1998, \mn@doi [\apj]
  {10.1086/306533}, \href {http://adsabs.harvard.edu/abs/1998ApJ...509..793M}
  {509, 793}

\bibitem[\protect\citeauthoryear{{Moore}, {Katz}, {Lake}, {Dressler}  \&
  {Oemler}}{{Moore} et~al.}{1996}]{Moore1996}
{Moore} B.,  {Katz} N.,  {Lake} G.,  {Dressler} A.,   {Oemler} A.,  1996,
  \mn@doi [\nat] {10.1038/379613a0}, \href
  {http://adsabs.harvard.edu/abs/1996Natur.379..613M} {379, 613}

\bibitem[\protect\citeauthoryear{{Mora}, {Chanam{\'e}}  \& {Puzia}}{{Mora}
  et~al.}{2015}]{Mora2015}
{Mora} M.~D.,  {Chanam{\'e}} J.,   {Puzia} T.~H.,  2015, \mn@doi [\aj]
  {10.1088/0004-6256/150/3/93}, \href
  {http://adsabs.harvard.edu/abs/2015AJ....150...93M} {150, 93}

\bibitem[\protect\citeauthoryear{{Navarro}, {Frenk}  \& {White}}{{Navarro}
  et~al.}{1997}]{Navarro1997}
{Navarro} J.~F.,  {Frenk} C.~S.,   {White} S.~D.~M.,  1997, \mn@doi [\apj]
  {10.1086/304888}, \href {http://adsabs.harvard.edu/abs/1997ApJ...490..493N}
  {490, 493}

\bibitem[\protect\citeauthoryear{{Noble} et~al.,}{{Noble}
  et~al.}{2018}]{Noble2018}
{Noble} A.~G.,  et~al., 2018, preprint, \href
  {http://adsabs.harvard.edu/abs/2018arXiv180903514N} {} (\mn@eprint {arXiv}
  {1809.03514})

\bibitem[\protect\citeauthoryear{{Noeske} et~al.,}{{Noeske}
  et~al.}{2007}]{Noeske2007}
{Noeske} K.~G.,  et~al., 2007, \mn@doi [\apjl] {10.1086/517926}, \href
  {http://adsabs.harvard.edu/abs/2007ApJ...660L..43N} {660, L43}

\bibitem[\protect\citeauthoryear{{Nulsen}}{{Nulsen}}{1982}]{Nulsen1982}
{Nulsen} P.~E.~J.,  1982, \mn@doi [\mnras] {10.1093/mnras/198.4.1007}, \href
  {http://adsabs.harvard.edu/abs/1982MNRAS.198.1007N} {198, 1007}

\bibitem[\protect\citeauthoryear{{Oemler}}{{Oemler}}{1974}]{Oemler1974}
{Oemler} Jr. A.,  1974, \mn@doi [\apj] {10.1086/153216}, \href
  {http://adsabs.harvard.edu/abs/1974ApJ...194....1O} {194, 1}

\bibitem[\protect\citeauthoryear{{Paolillo}, {Fabbiano}, {Peres}  \&
  {Kim}}{{Paolillo} et~al.}{2002}]{Paolillo2002}
{Paolillo} M.,  {Fabbiano} G.,  {Peres} G.,   {Kim} D.-W.,  2002, \mn@doi
  [\apj] {10.1086/337919}, \href
  {http://adsabs.harvard.edu/abs/2002ApJ...565..883P} {565, 883}

\bibitem[\protect\citeauthoryear{{Pedraz}, {Gorgas}, {Cardiel},
  {S{\'a}nchez-Bl{\'a}zquez}  \& {Guzm{\'a}n}}{{Pedraz}
  et~al.}{2002}]{Pedraz2002}
{Pedraz} S.,  {Gorgas} J.,  {Cardiel} N.,  {S{\'a}nchez-Bl{\'a}zquez} P.,
  {Guzm{\'a}n} R.,  2002, \mn@doi [\mnras] {10.1046/j.1365-8711.2002.05565.x},
  \href {http://adsabs.harvard.edu/abs/2002MNRAS.332L..59P} {332, L59}

\bibitem[\protect\citeauthoryear{{Peng}, {Ho}, {Impey}  \& {Rix}}{{Peng}
  et~al.}{2002}]{Peng2002}
{Peng} C.~Y.,  {Ho} L.~C.,  {Impey} C.~D.,   {Rix} H.-W.,  2002, \mn@doi [\aj]
  {10.1086/340952}, \href {http://adsabs.harvard.edu/abs/2002AJ....124..266P}
  {124, 266}

\bibitem[\protect\citeauthoryear{{Peng}, {Ho}, {Impey}  \& {Rix}}{{Peng}
  et~al.}{2010}]{Peng2010}
{Peng} C.~Y.,  {Ho} L.~C.,  {Impey} C.~D.,   {Rix} H.-W.,  2010, \mn@doi [\aj]
  {10.1088/0004-6256/139/6/2097}, \href
  {http://adsabs.harvard.edu/abs/2010AJ....139.2097P} {139, 2097}

\bibitem[\protect\citeauthoryear{{Pettini} \& {Pagel}}{{Pettini} \&
  {Pagel}}{2004}]{Pettini2004}
{Pettini} M.,  {Pagel} B.~E.~J.,  2004, \mn@doi [\mnras]
  {10.1111/j.1365-2966.2004.07591.x}, \href
  {http://adsabs.harvard.edu/abs/2004MNRAS.348L..59P} {348, L59}

\bibitem[\protect\citeauthoryear{{Pilbratt} et~al.,}{{Pilbratt}
  et~al.}{2010}]{Pilbratt2010}
{Pilbratt} G.~L.,  et~al., 2010, \mn@doi [\aap] {10.1051/0004-6361/201014759},
  \href {http://adsabs.harvard.edu/abs/2010A%26A...518L...1P} {518, L1}

\bibitem[\protect\citeauthoryear{{R{\'e}my-Ruyer} et~al.,}{{R{\'e}my-Ruyer}
  et~al.}{2014}]{RemyRuyer2014}
{R{\'e}my-Ruyer} A.,  et~al., 2014, \mn@doi [\aap]
  {10.1051/0004-6361/201322803}, \href
  {http://adsabs.harvard.edu/abs/2014A%26A...563A..31R} {563, A31}

\bibitem[\protect\citeauthoryear{{Robotham} et~al.,}{{Robotham}
  et~al.}{2011}]{Robotham2011}
{Robotham} A.~S.~G.,  et~al., 2011, \mn@doi [\mnras]
  {10.1111/j.1365-2966.2011.19217.x}, \href
  {http://adsabs.harvard.edu/abs/2011MNRAS.416.2640R} {416, 2640}

\bibitem[\protect\citeauthoryear{{Rodr{\'{\i}}guez-Ardila}, {Prieto},
  {Mazzalay}, {Fern{\'a}ndez-Ontiveros}, {Luque}  \&
  {M{\"u}ller-S{\'a}nchez}}{{Rodr{\'{\i}}guez-Ardila}
  et~al.}{2017}]{Rodriguez-Ardila2017}
{Rodr{\'{\i}}guez-Ardila} A.,  {Prieto} M.~A.,  {Mazzalay} X.,
  {Fern{\'a}ndez-Ontiveros} J.~A.,  {Luque} R.,   {M{\"u}ller-S{\'a}nchez} F.,
  2017, \mn@doi [\mnras] {10.1093/mnras/stx1401}, \href
  {http://adsabs.harvard.edu/abs/2017MNRAS.470.2845R} {470, 2845}

\bibitem[\protect\citeauthoryear{{Rudnick} et~al.,}{{Rudnick}
  et~al.}{2017}]{Rudnick2017}
{Rudnick} G.,  et~al., 2017, \mn@doi [\apj] {10.3847/1538-4357/aa87b2}, \href
  {http://adsabs.harvard.edu/abs/2017ApJ...849...27R} {849, 27}

\bibitem[\protect\citeauthoryear{{Ry{\'s}}, {Falc{\'o}n-Barroso}  \& {van de
  Ven}}{{Ry{\'s}} et~al.}{2013}]{Rys2013}
{Ry{\'s}} A.,  {Falc{\'o}n-Barroso} J.,   {van de Ven} G.,  2013, \mn@doi
  [\mnras] {10.1093/mnras/sts245}, \href
  {http://adsabs.harvard.edu/abs/2013MNRAS.428.2980R} {428, 2980}

\bibitem[\protect\citeauthoryear{{Saintonge} et~al.,}{{Saintonge}
  et~al.}{2011}]{Saintonge2011}
{Saintonge} A.,  et~al., 2011, \mn@doi [\mnras]
  {10.1111/j.1365-2966.2011.18677.x}, \href
  {http://adsabs.harvard.edu/abs/2011MNRAS.415...32S} {415, 32}

\bibitem[\protect\citeauthoryear{{Saintonge} et~al.,}{{Saintonge}
  et~al.}{2017}]{Saintonge2017}
{Saintonge} A.,  et~al., 2017, \mn@doi [\apjs] {10.3847/1538-4365/aa97e0},
  \href {http://adsabs.harvard.edu/abs/2017ApJS..233...22S} {233, 22}

\bibitem[\protect\citeauthoryear{{S{\'a}nchez-Janssen} \&
  {Aguerri}}{{S{\'a}nchez-Janssen} \& {Aguerri}}{2012}]{Sanchez2012}
{S{\'a}nchez-Janssen} R.,  {Aguerri} J.~A.~L.,  2012, \mn@doi [\mnras]
  {10.1111/j.1365-2966.2012.21301.x}, \href
  {http://adsabs.harvard.edu/abs/2012MNRAS.424.2614S} {424, 2614}

\bibitem[\protect\citeauthoryear{{S{\'a}nchez} et~al.,}{{S{\'a}nchez}
  et~al.}{2017}]{Sanchez2017}
{S{\'a}nchez} S.~F.,  et~al., 2017, \mn@doi [\mnras] {10.1093/mnras/stx808},
  \href {http://adsabs.harvard.edu/abs/2017MNRAS.469.2121S} {469, 2121}

\bibitem[\protect\citeauthoryear{{Sarzi} et~al.,}{{Sarzi}
  et~al.}{2018}]{Sarzi2018}
{Sarzi} M.,  et~al., 2018, preprint, \href
  {https://ui.adsabs.harvard.edu/#abs/2018arXiv180406795S} {p.
  arXiv:1804.06795} (\mn@eprint {arXiv} {1804.06795})

\bibitem[\protect\citeauthoryear{{Sault}, {Teuben}  \& {Wright}}{{Sault}
  et~al.}{1995}]{Sault1995}
{Sault} R.~J.,  {Teuben} P.~J.,   {Wright} M.~C.~H.,  1995, in {Shaw} R.~A.,
  {Payne} H.~E.,   {Hayes} J.~J.~E.,  eds,  Astronomical Society of the Pacific
  Conference Series Vol. 77, Astronomical Data Analysis Software and Systems
  IV. p.~433 (\mn@eprint {} {astro-ph/0612759})

\bibitem[\protect\citeauthoryear{{Saunders} et~al.,}{{Saunders}
  et~al.}{2004}]{Saunders2004}
{Saunders} W.,  et~al., 2004, in {Moorwood} A.~F.~M.,  {Iye} M.,  eds,
  \procspie Vol. 5492, Ground-based Instrumentation for Astronomy. pp 389--400,
  \mn@doi{10.1117/12.550871}

\bibitem[\protect\citeauthoryear{{Scharf}, {Zurek}  \& {Bureau}}{{Scharf}
  et~al.}{2005}]{Scharf2005}
{Scharf} C.~A.,  {Zurek} D.~R.,   {Bureau} M.,  2005, \mn@doi [\apj]
  {10.1086/444531}, \href {http://adsabs.harvard.edu/abs/2005ApJ...633..154S}
  {633, 154}

\bibitem[\protect\citeauthoryear{{Schindler}, {Binggeli}  \&
  {B{\"o}hringer}}{{Schindler} et~al.}{1999}]{Schindler1999}
{Schindler} S.,  {Binggeli} B.,   {B{\"o}hringer} H.,  1999, \aap, \href
  {http://adsabs.harvard.edu/abs/1999A\%26A...343..420S} {343, 420}

\bibitem[\protect\citeauthoryear{{Schipani} et~al.,}{{Schipani}
  et~al.}{2012}]{Schipani2012}
{Schipani} P.,  et~al., 2012, Memorie della Societa Astronomica Italiana
  Supplementi, \href {http://adsabs.harvard.edu/abs/2012MSAIS..19..393S} {19,
  393}

\bibitem[\protect\citeauthoryear{{Serra} et~al.,}{{Serra}
  et~al.}{2016}]{Serra2016}
{Serra} P.,  et~al., 2016, in Proceedings of MeerKAT Science: On the Pathway to
  the SKA. 25-27 May, 2016 Stellenbosch, South Africa (MeerKAT2016). Online at
  <A
  href=``href=''>href=``https://pos.sissa.it/cgi-bin/reader/conf.cgi?confid=277</A>,
  id.8. p.~8 (\mn@eprint {arXiv} {1709.01289})

\bibitem[\protect\citeauthoryear{{Sharp} et~al.,}{{Sharp}
  et~al.}{2006}]{Sharp2006}
{Sharp} R.,  et~al., 2006, in Society of Photo-Optical Instrumentation
  Engineers (SPIE) Conference Series. p. 62690G (\mn@eprint {}
  {astro-ph/0606137}), \mn@doi{10.1117/12.671022}

\bibitem[\protect\citeauthoryear{{Shull}}{{Shull}}{2014}]{Shull2014}
{Shull} J.~M.,  2014, \mn@doi [\apj] {10.1088/0004-637X/784/2/142}, \href
  {http://adsabs.harvard.edu/abs/2014ApJ...784..142S} {784, 142}

\bibitem[\protect\citeauthoryear{{Solanes}, {Manrique},
  {Garc{\'{\i}}a-G{\'o}mez}, {Gonz{\'a}lez-Casado}, {Giovanelli}  \&
  {Haynes}}{{Solanes} et~al.}{2001}]{Solanes2001}
{Solanes} J.~M.,  {Manrique} A.,  {Garc{\'{\i}}a-G{\'o}mez} C.,
  {Gonz{\'a}lez-Casado} G.,  {Giovanelli} R.,   {Haynes} M.~P.,  2001, \mn@doi
  [\apj] {10.1086/318672}, \href
  {http://adsabs.harvard.edu/abs/2001ApJ...548...97S} {548, 97}

\bibitem[\protect\citeauthoryear{{Stach}, {Swinbank}, {Smail}, {Hilton},
  {Simpson}  \& {Cooke}}{{Stach} et~al.}{2017}]{Stach2017}
{Stach} S.~M.,  {Swinbank} A.~M.,  {Smail} I.,  {Hilton} M.,  {Simpson} J.~M.,
   {Cooke} E.~A.,  2017, \mn@doi [\apj] {10.3847/1538-4357/aa93f6}, \href
  {http://adsabs.harvard.edu/abs/2017ApJ...849..154S} {849, 154}

\bibitem[\protect\citeauthoryear{{Stark}, {Knapp}, {Bally}, {Wilson}, {Penzias}
   \& {Rowe}}{{Stark} et~al.}{1986}]{Stark1986}
{Stark} A.~A.,  {Knapp} G.~R.,  {Bally} J.,  {Wilson} R.~W.,  {Penzias} A.~A.,
   {Rowe} H.~E.,  1986, \mn@doi [\apj] {10.1086/164717}, \href
  {http://adsabs.harvard.edu/abs/1986ApJ...310..660S} {310, 660}

\bibitem[\protect\citeauthoryear{{Tonry}, {Dressler}, {Blakeslee}, {Ajhar},
  {Fletcher}, {Luppino}, {Metzger}  \& {Moore}}{{Tonry}
  et~al.}{2001}]{Tonry2001}
{Tonry} J.~L.,  {Dressler} A.,  {Blakeslee} J.~P.,  {Ajhar} E.~A.,  {Fletcher}
  A.~B.,  {Luppino} G.~A.,  {Metzger} M.~R.,   {Moore} C.~B.,  2001, \mn@doi
  [\apj] {10.1086/318301}, \href
  {https://ui.adsabs.harvard.edu/#abs/2001ApJ...546..681T} {546, 681}

\bibitem[\protect\citeauthoryear{{Venhola} et~al.,}{{Venhola}
  et~al.}{2017}]{Venhola2017}
{Venhola} A.,  et~al., 2017, \mn@doi [\aap] {10.1051/0004-6361/201730696},
  \href {http://adsabs.harvard.edu/abs/2017A%26A...608A.142V} {608, A142}

\bibitem[\protect\citeauthoryear{{Venhola} et~al.,}{{Venhola}
  et~al.}{2018}]{Venhola2018}
{Venhola} A.,  et~al., 2018, preprint, \href
  {http://adsabs.harvard.edu/abs/2018arXiv181000550V} {} (\mn@eprint {arXiv}
  {1810.00550})

\bibitem[\protect\citeauthoryear{{Vollmer}, {Braine}, {Pappalardo}  \&
  {Hily-Blant}}{{Vollmer} et~al.}{2008}]{Vollmer2008}
{Vollmer} B.,  {Braine} J.,  {Pappalardo} C.,   {Hily-Blant} P.,  2008, \mn@doi
  [\aap] {10.1051/0004-6361:200810432}, \href
  {http://adsabs.harvard.edu/abs/2008A%26A...491..455V} {491, 455}

\bibitem[\protect\citeauthoryear{{Wang} et~al.,}{{Wang}
  et~al.}{2018}]{Wang2018}
{Wang} T.,  et~al., 2018, preprint, \href
  {http://adsabs.harvard.edu/abs/2018arXiv181010558W} {} (\mn@eprint {arXiv}
  {1810.10558})

\bibitem[\protect\citeauthoryear{{Waugh} et~al.,}{{Waugh}
  et~al.}{2002}]{Waugh2002}
{Waugh} M.,  et~al., 2002, \mn@doi [\mnras] {10.1046/j.1365-8711.2002.05942.x},
  \href {http://adsabs.harvard.edu/abs/2002MNRAS.337..641W} {337, 641}

\bibitem[\protect\citeauthoryear{{Wright} et~al.,}{{Wright}
  et~al.}{2010}]{Wright2010}
{Wright} E.~L.,  et~al., 2010, \mn@doi [\aj] {10.1088/0004-6256/140/6/1868},
  \href {http://adsabs.harvard.edu/abs/2010AJ....140.1868W} {140, 1868}

\bibitem[\protect\citeauthoryear{{Young} et~al.,}{{Young}
  et~al.}{2011}]{Young2011}
{Young} L.~M.,  et~al., 2011, \mn@doi [\mnras]
  {10.1111/j.1365-2966.2011.18561.x}, \href
  {http://adsabs.harvard.edu/abs/2011MNRAS.414..940Y} {414, 940}

\bibitem[\protect\citeauthoryear{{Yun} et~al.,}{{Yun} et~al.}{2018}]{Yun2018}
{Yun} K.,  et~al., 2018, preprint, \href
  {http://adsabs.harvard.edu/abs/2018arXiv181000005Y} {} (\mn@eprint {arXiv}
  {1810.00005})

\bibitem[\protect\citeauthoryear{{Zabludoff} \& {Mulchaey}}{{Zabludoff} \&
  {Mulchaey}}{1998}]{Zabludoff1998}
{Zabludoff} A.~I.,  {Mulchaey} J.~S.,  1998, \mn@doi [\apjl] {10.1086/311312},
  \href {http://adsabs.harvard.edu/abs/1998ApJ...498L...5Z} {498, L5}

\bibitem[\protect\citeauthoryear{{di Serego Alighieri} et~al.,}{{di Serego
  Alighieri} et~al.}{2007}]{Serego2007}
{di Serego Alighieri} S.,  et~al., 2007, \mn@doi [\aap]
  {10.1051/0004-6361:20078205}, \href
  {http://adsabs.harvard.edu/abs/2007A%26A...474..851D} {474, 851}

\bibitem[\protect\citeauthoryear{{di Serego Alighieri} et~al.,}{{di Serego
  Alighieri} et~al.}{2013}]{Serego2013}
{di Serego Alighieri} S.,  et~al., 2013, \mn@doi [\aap]
  {10.1051/0004-6361/201220551}, \href
  {http://adsabs.harvard.edu/abs/2013A%26A...552A...8D} {552, A8}

\bibitem[\protect\citeauthoryear{{van de Voort}, {Bah{\'e}}, {Bower}, {Correa},
  {Crain}, {Schaye}  \& {Theuns}}{{van de Voort} et~al.}{2017}]{Voort2017}
{van de Voort} F.,  {Bah{\'e}} Y.~M.,  {Bower} R.~G.,  {Correa} C.~A.,  {Crain}
  R.~A.,  {Schaye} J.,   {Theuns} T.,  2017, \mn@doi [\mnras]
  {10.1093/mnras/stw3356}, \href
  {http://adsabs.harvard.edu/abs/2017MNRAS.466.3460V} {466, 3460}

\makeatother
\end{thebibliography}



\clearpage

\appendix

\section{Discussion of individual detections}
\label{sub:individual}

\subsubsection{NGC1351A (FCC67)}
\label{subsub:NGC1351A}
\begin{figure*}

	\centering

	\subfloat[\label{subfig:1351A_rgb}]
	{\hspace{-4mm}\includegraphics[height=0.35\textwidth ]{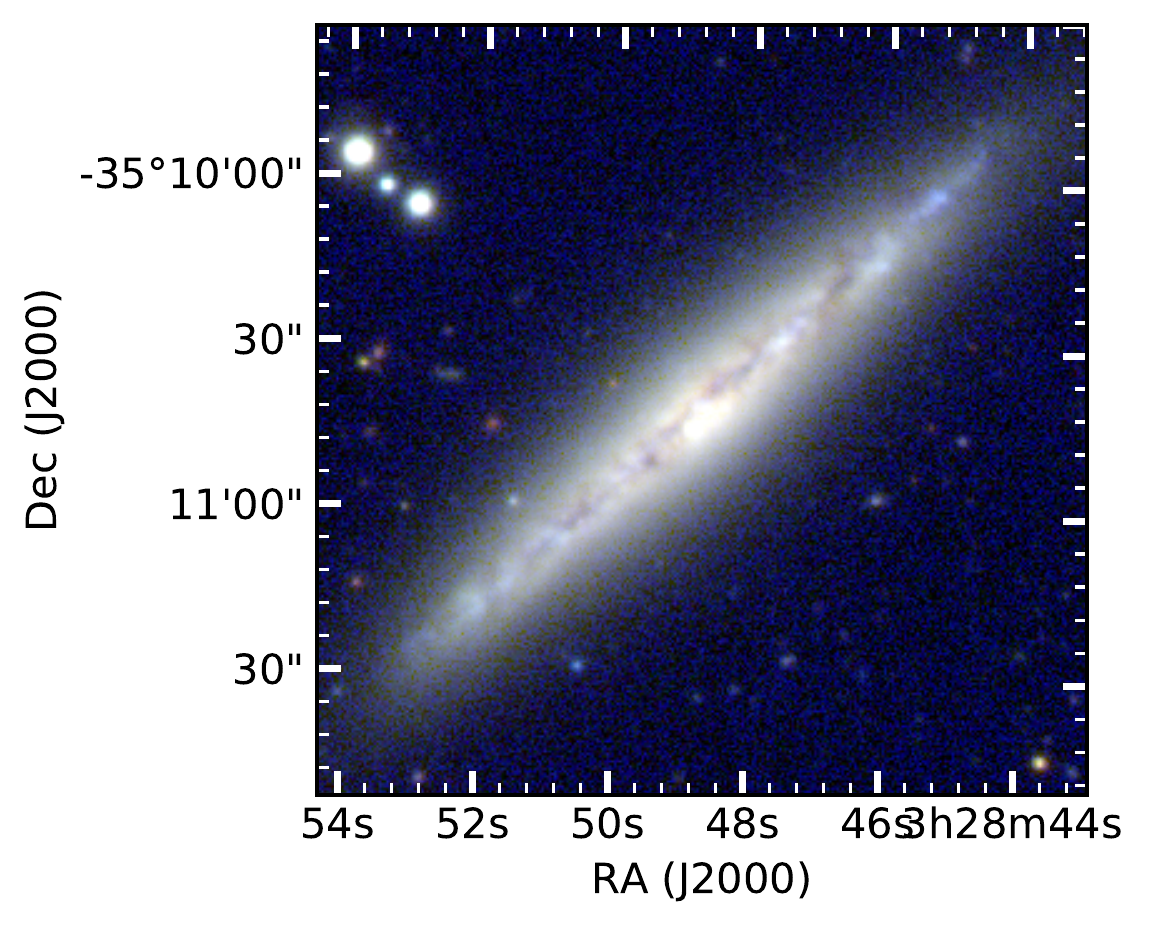}}	
	\hspace{2mm}
	\subfloat[\label{subfig:1351A_mom0}]
		{\includegraphics[height=0.35\textwidth]{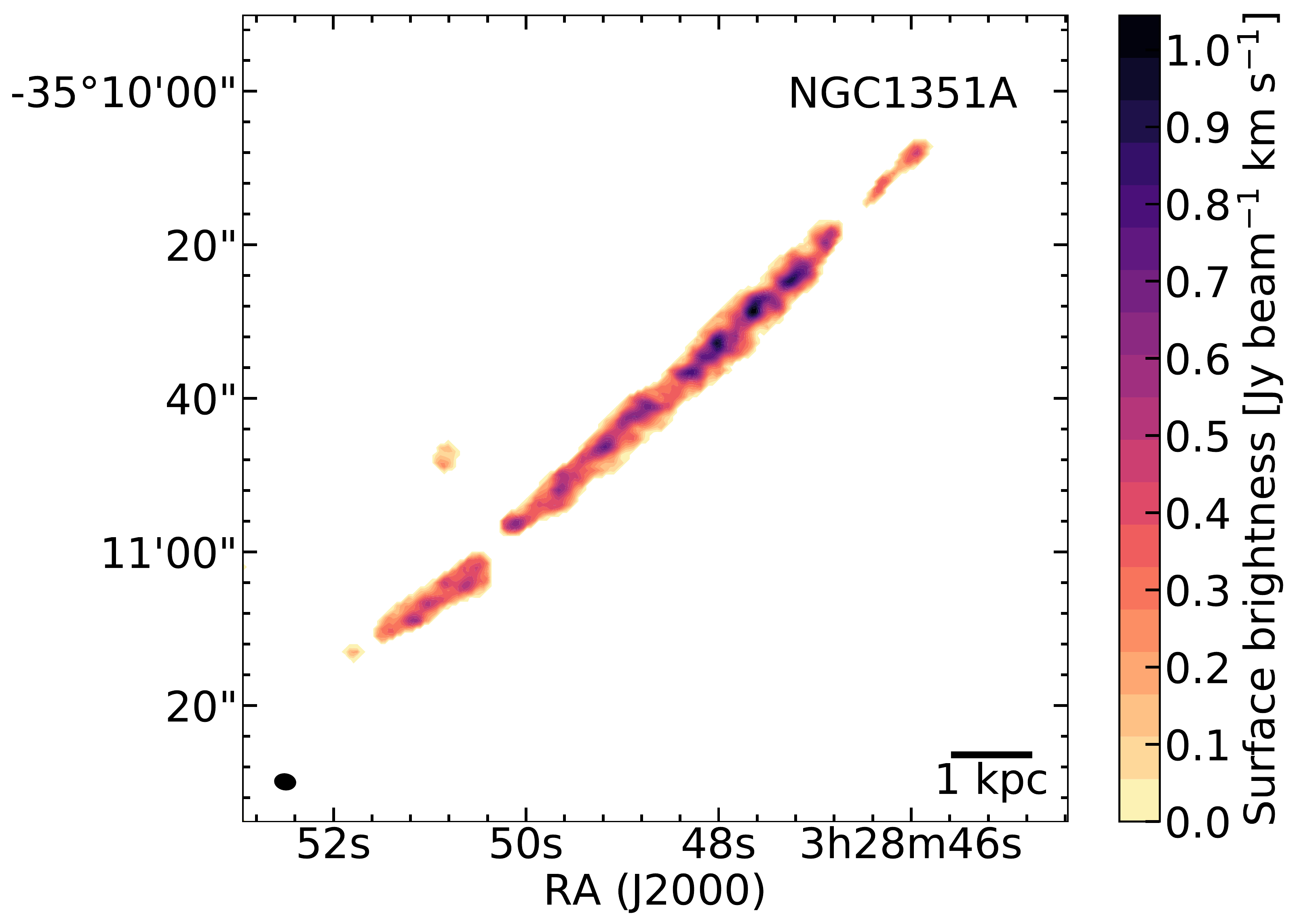}}	
	
	
	\subfloat[\label{subfig:1351A_mom1}]
		{\hspace{-3mm}\includegraphics[height=0.35\textwidth]{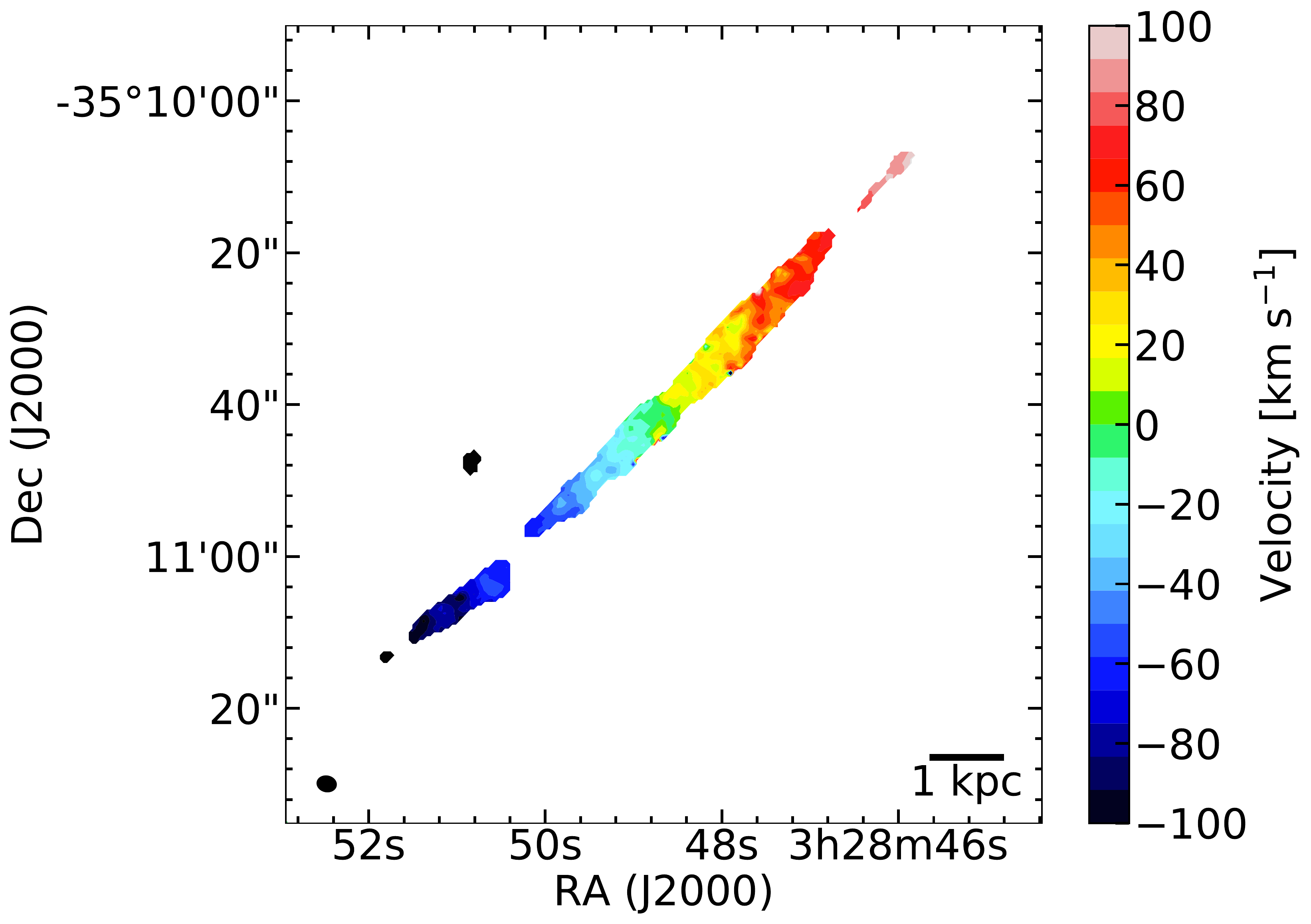}}
	\hspace{7mm}
	\subfloat[\label{subfig:1351A_mom2}]
		{\includegraphics[height=0.35\textwidth]{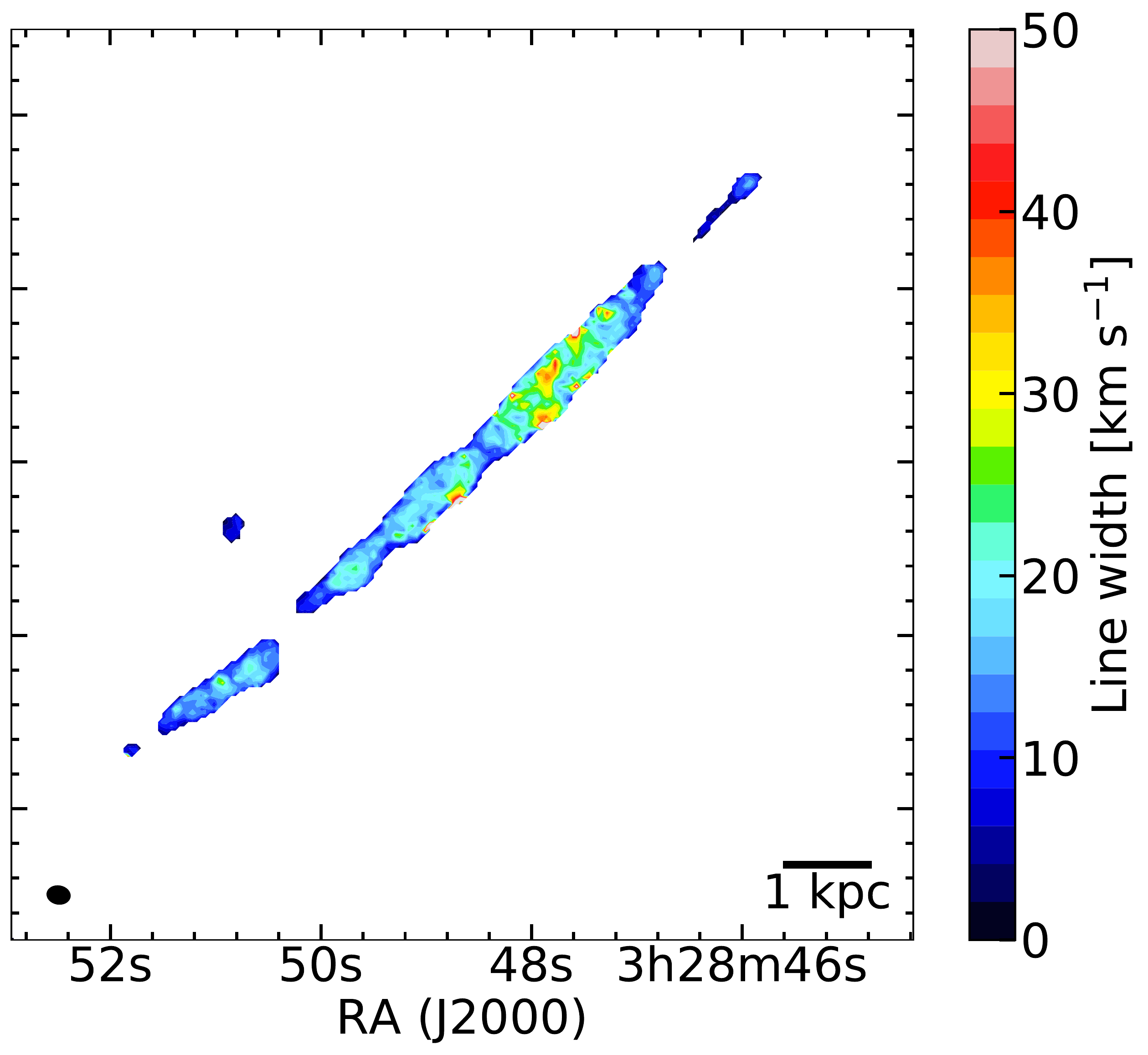}}
		
		
	\subfloat[\label{subfig:1351A_pvd}]
		{\includegraphics[height=0.39\textwidth]{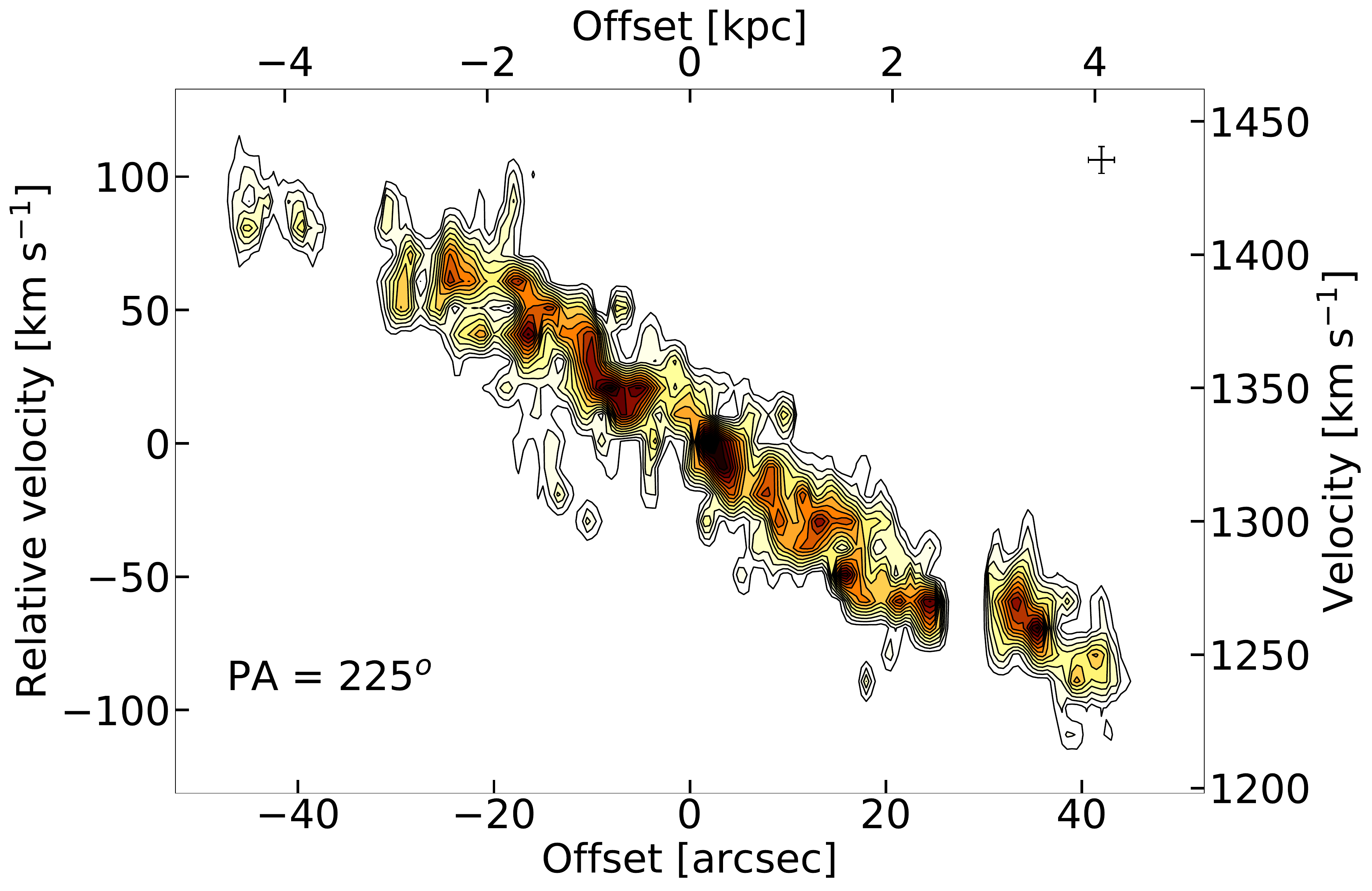}}	
	\hspace{6mm}
	\subfloat[\label{subfig:1351A_spectrum}]
		{\includegraphics[height=0.355\textwidth]{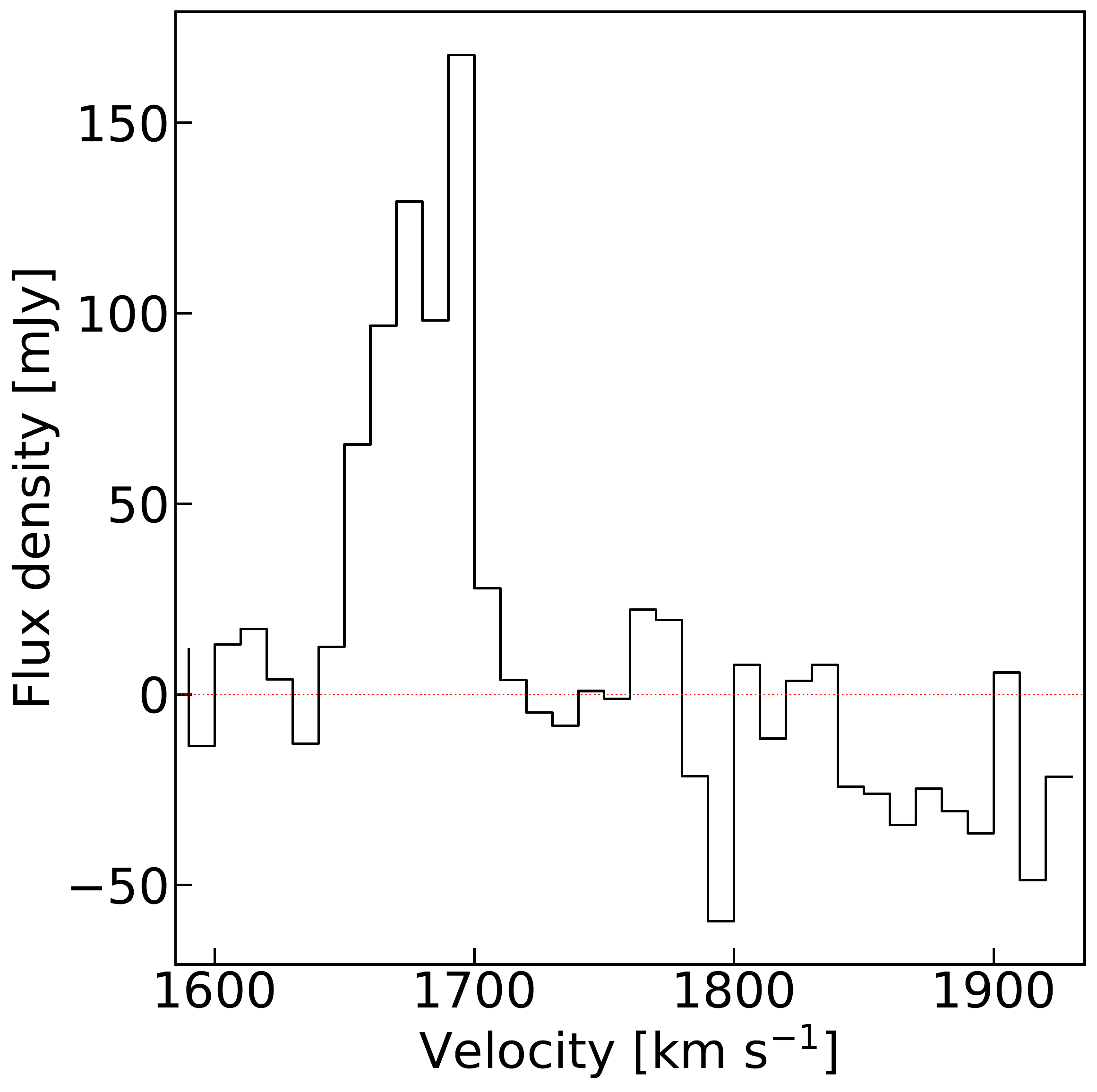}}
		
		
	\caption{a: three-colour image of ESO358-G063. b: moment zero map: distribution of the cold molecular gas as traced by the ALMA CO data. c: moment 1 map: velocity map of the cold molecular gas. Each colour represents a 10 km s$^{-1}$ velocity channel. d: moment 2 map: line width of the CO integrated spectrum. e: Position-velocity diagram of the cold molecular gas. The uncertainties in the spatial and velocity directions are indicated in the upper right corner. f: CO(1 - 0) line. The beam of the observations is shown in the lower left corners of the moment maps, as well as a 1 kpc scale bar in the lower right corners. The beam of the observations is shown in the lower left corners of the moment maps, as well as a 1 kpc scale bar in the lower right corners.}
	\label{fig:NGC1351A}
\end{figure*}

NGC1351A (Figures \ref{fig:NGC1351A} and \ref{subfig:overplot_NGC1351A}) is a good example of an edge-on galaxy. The CO gas shows regular rotation. The detached emission visible from both extremes probably consist of giant molecular associations (GMAs), that are separated from the galaxy's main body by lower density or atomic gas. 

\subsubsection{MCG-06-08-024 (FCC90) \& ESO359-G002 (FCC335)}
\label{subsub:rps_gals}
\begin{figure*}
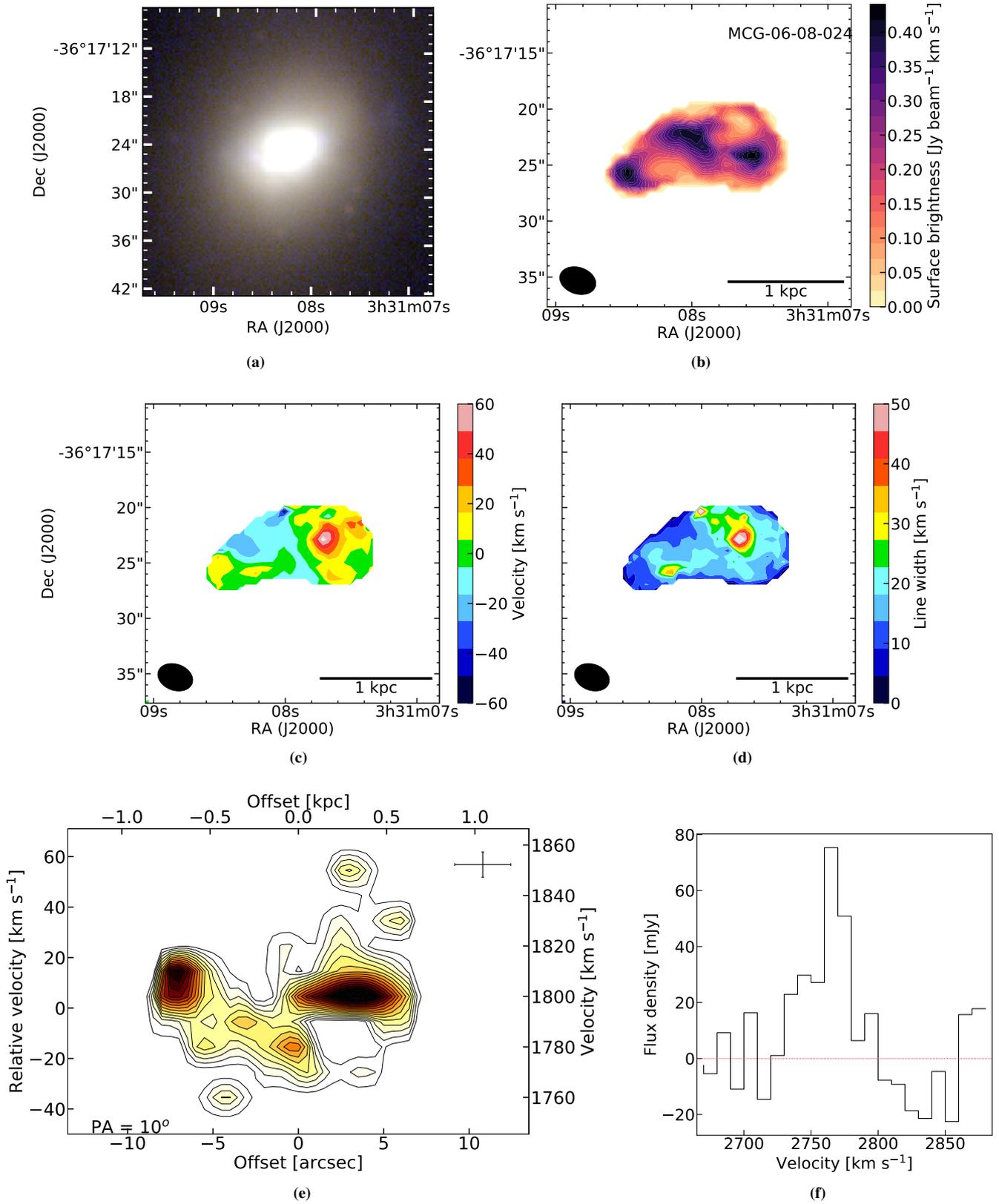

		
	\subfloat[]
	{\hspace{-5mm}\includegraphics[height=0.35\textwidth]{MCG-06-08-024/RGB_image.pdf}}	
	\hspace{-1.5mm}
	\subfloat[]
		{\includegraphics[height=0.35\textwidth]{MCG-06-08-024/zeroth.pdf}}	
	
	
	\subfloat[]
		{\hspace{-5mm}\includegraphics[height=0.35\textwidth]{MCG-06-08-024/first.pdf}}
	\hspace{3mm}
	\subfloat[]
		{\includegraphics[height=0.35\textwidth]{MCG-06-08-024/second.pdf}}
		
		
	\subfloat[]
		{\includegraphics[height=0.39\textwidth]{MCG-06-08-024/PVD.pdf}}	
	\hspace{6mm}
	\subfloat[]
		{\includegraphics[height=0.355\textwidth]{MCG-06-08-024/spectrum.pdf}}
		
	\caption{MGC-06-08-024, similar to Figure \ref{fig:NGC1351A}.}
	\label{fig:MCG-06-08-024}
\end{figure*}

In MCG-06-08-024 (Figures \ref{fig:MCG-06-08-024} and \ref{subfig:overplot-MCG}) and ESO359-G002 (Figures \ref{fig:ESO359-G002} and \ref{subfig:overplot_ESO359-G002}), the CO emission extends in a tail beyond the stellar bodies of the galaxies. The three-colour image of ESO359-G002 shows some dust in the shape of a tail that is similar to the molecular gas. If we compare the directions of these tails to the direction of the cluster centre, we can see that in the case of MCG-06-08-024 the tail points away from the cluster centre, and in the case of ESO359-G002 it points towards the cluster centre. In both cases this could indicate ram pressure stripping: either the galaxy is being stripped of its molecular gas as it falls into the cluster (MCG-06-08-024), or it has already passed through the cluster centre, and is stripped as it is moving towards the other side of the cluster (ESO359-G002). Without detailed information on the orbits of these objects, this is hard to confirm. Ram pressure as a potential candidate for the stirring and stripping of the molecular gas in the Fornax cluster galaxies is discussed in \Section \ref{sub:RPS}.

\subsubsection{NGC1365 (FCC121)}
\label{subsub:NGC1365}
\begin{figure*}

	\centering

	\subfloat[]
	{\hspace{0mm}\includegraphics[height=0.35\textwidth ]{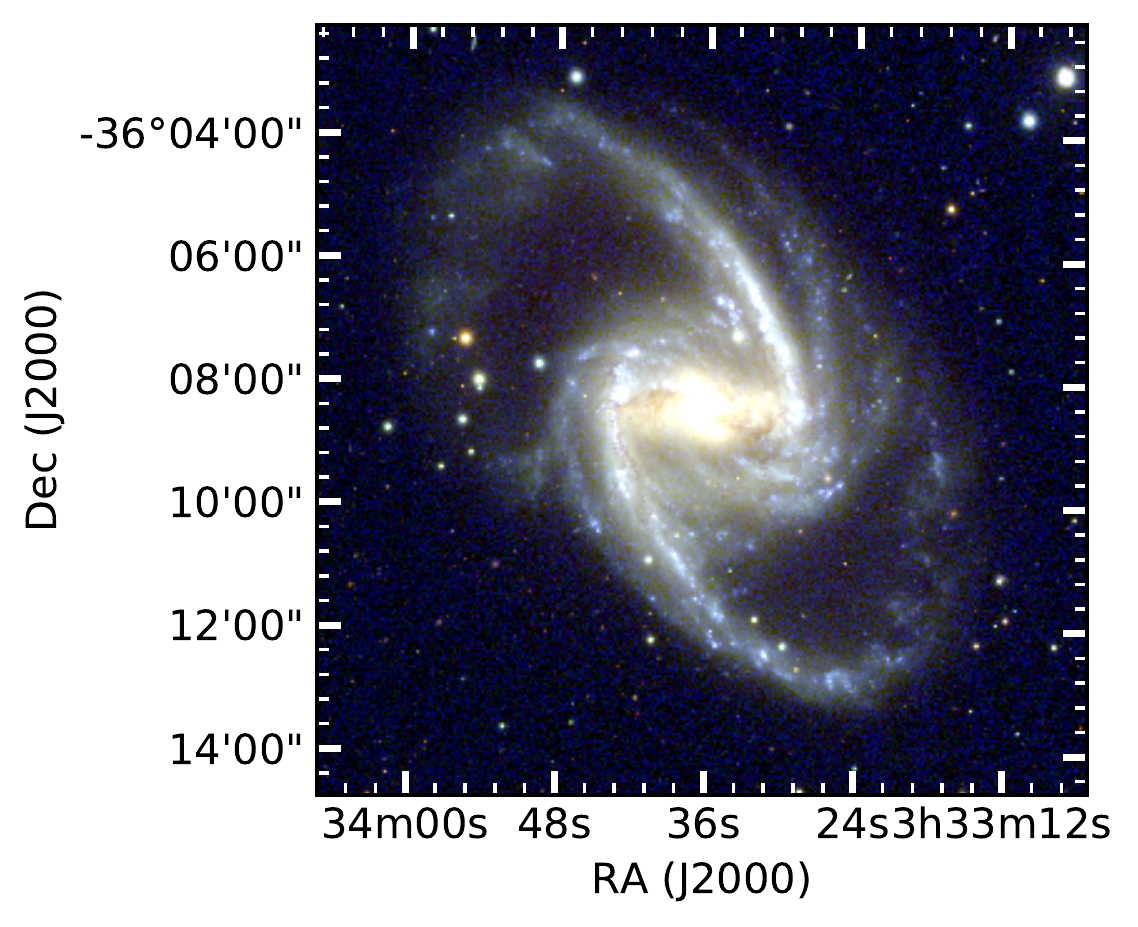}}	
	\hspace{0mm}
	\subfloat[]
		{\includegraphics[height=0.35\textwidth]{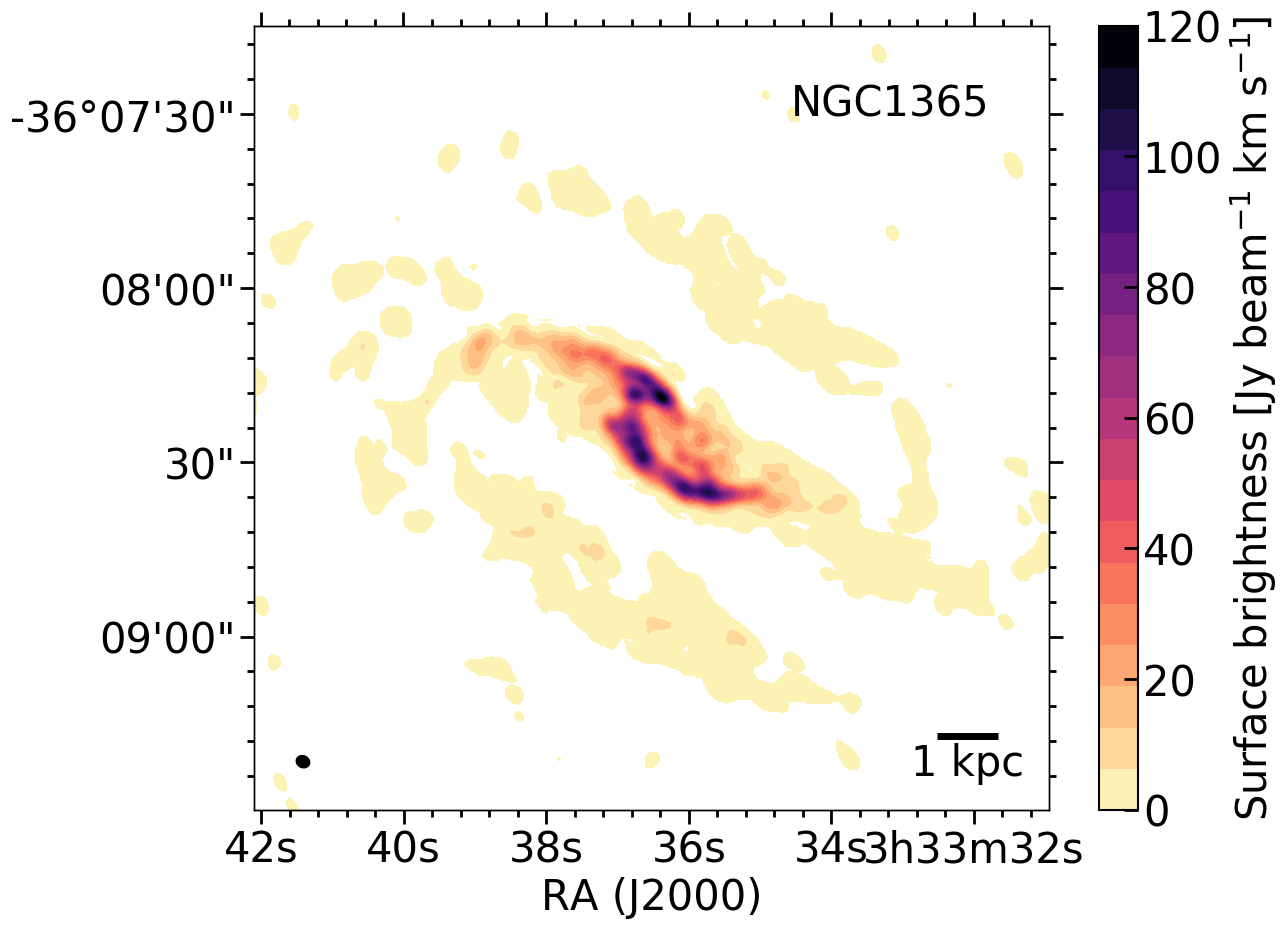}}	
	
	
	\subfloat[]
		{\hspace{0mm}\includegraphics[height=0.35\textwidth]{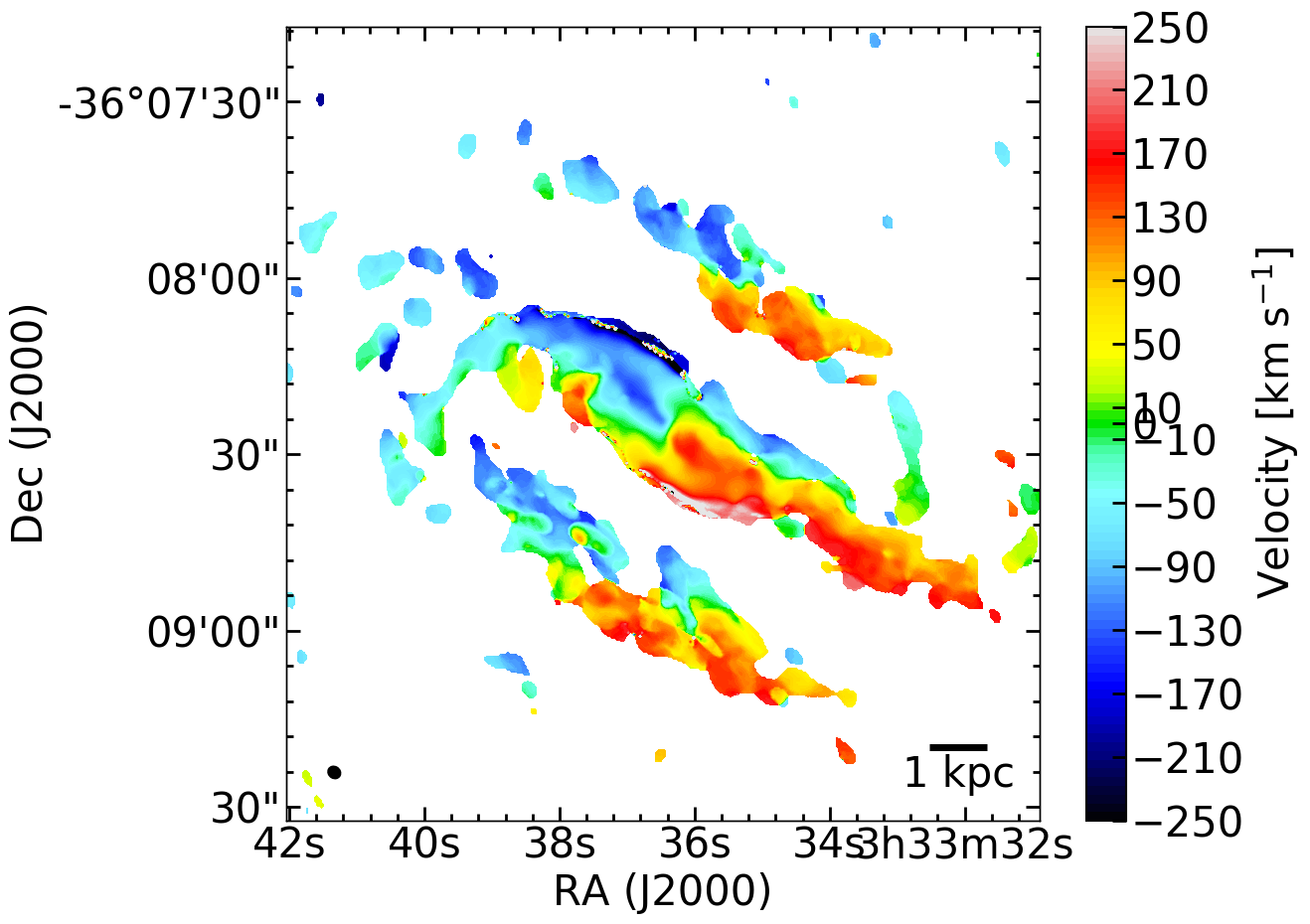}}
	\hspace{3mm}
	\subfloat[]
		{\includegraphics[height=0.35\textwidth]{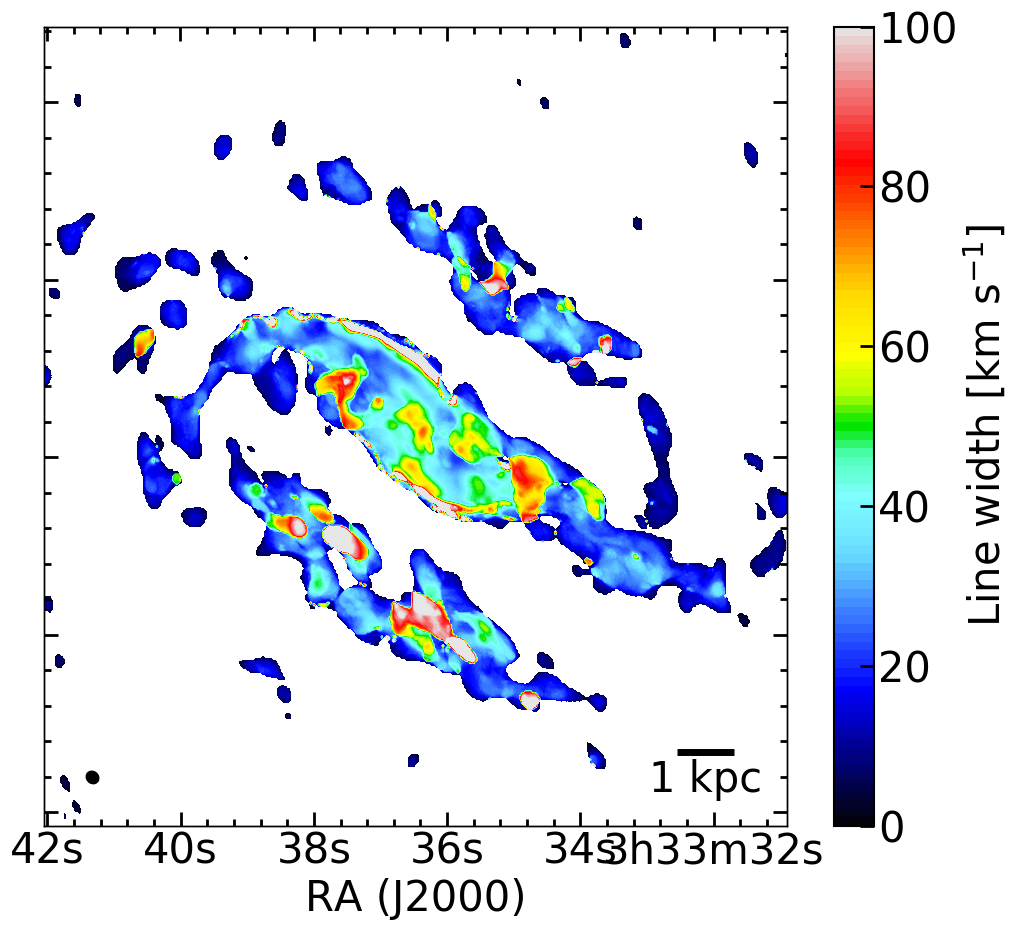}}
		
		
	\subfloat[]
		{\includegraphics[height=0.39\textwidth]{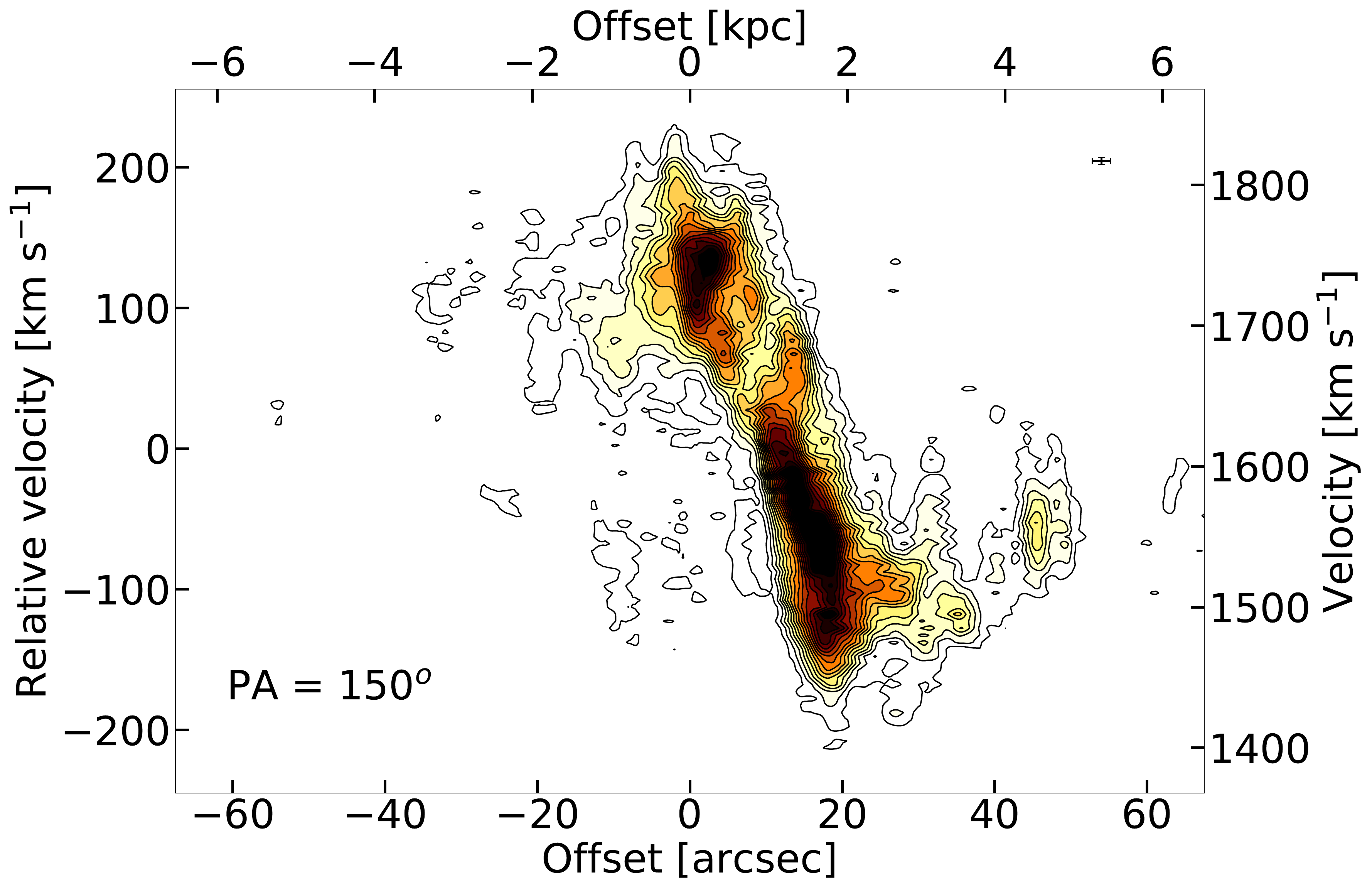}}	
	\hspace{6mm}
	\subfloat[]
		{\includegraphics[height=0.355\textwidth]{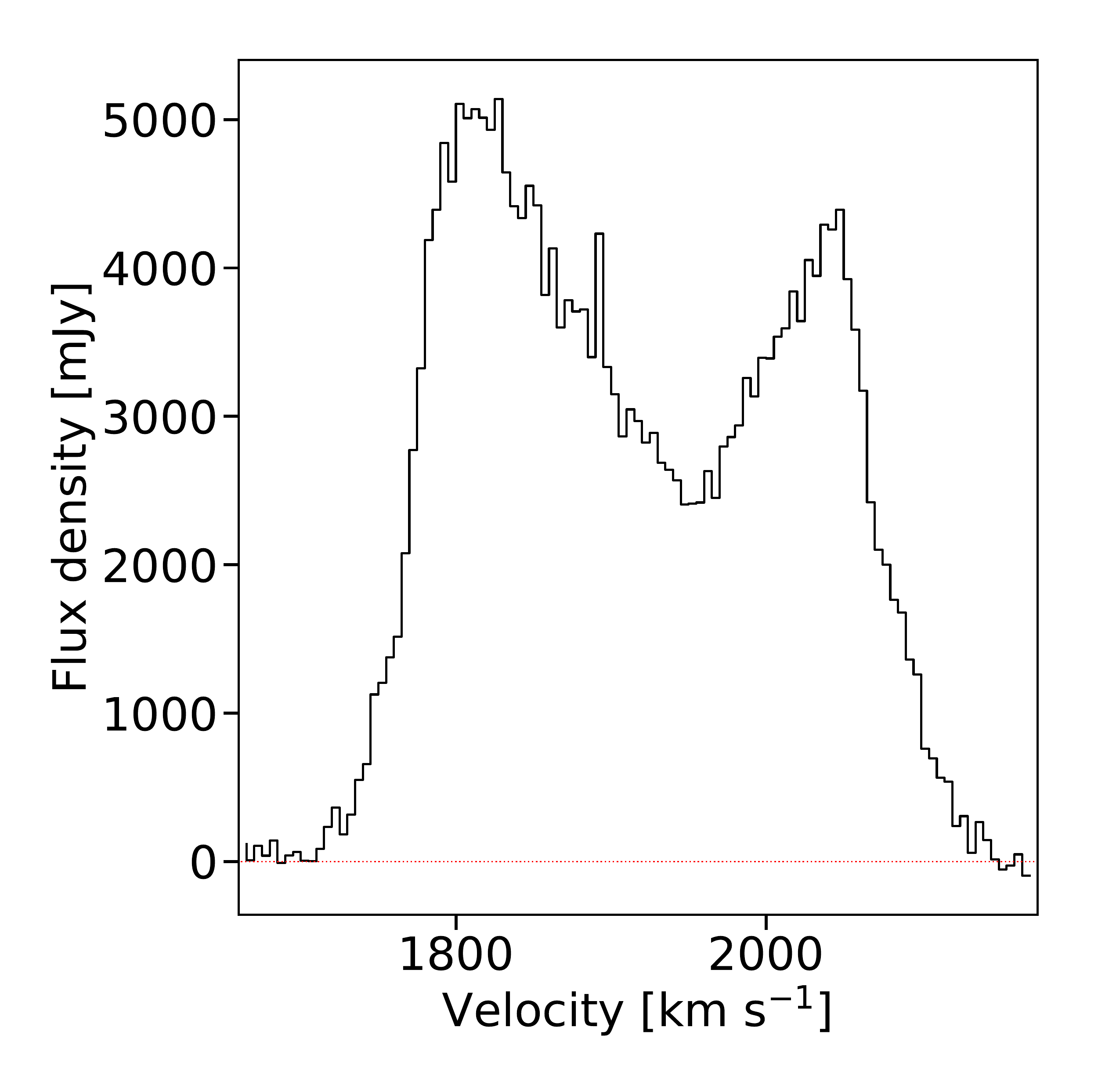}}
		
		
	\caption{NGC1365, similar to Figure \ref{fig:NGC1351A}.}
	\label{fig:NGC1365}
\end{figure*}

NGC1365 (Figures \ref{fig:NGC1365} and  is a giant spiral galaxy, and the only galaxy in our sample that is not H$_2$ deficient compared to the field. The molecular gas shows regular rotation, although in the centre it is warped by the strong bar in this source.

\subsubsection{NGC1380 (FCC167)}
\label{subsub:NGC1380}
\begin{figure*}

	\centering

	\subfloat[]
	{\hspace{-6mm}\includegraphics[height=0.35\textwidth ]{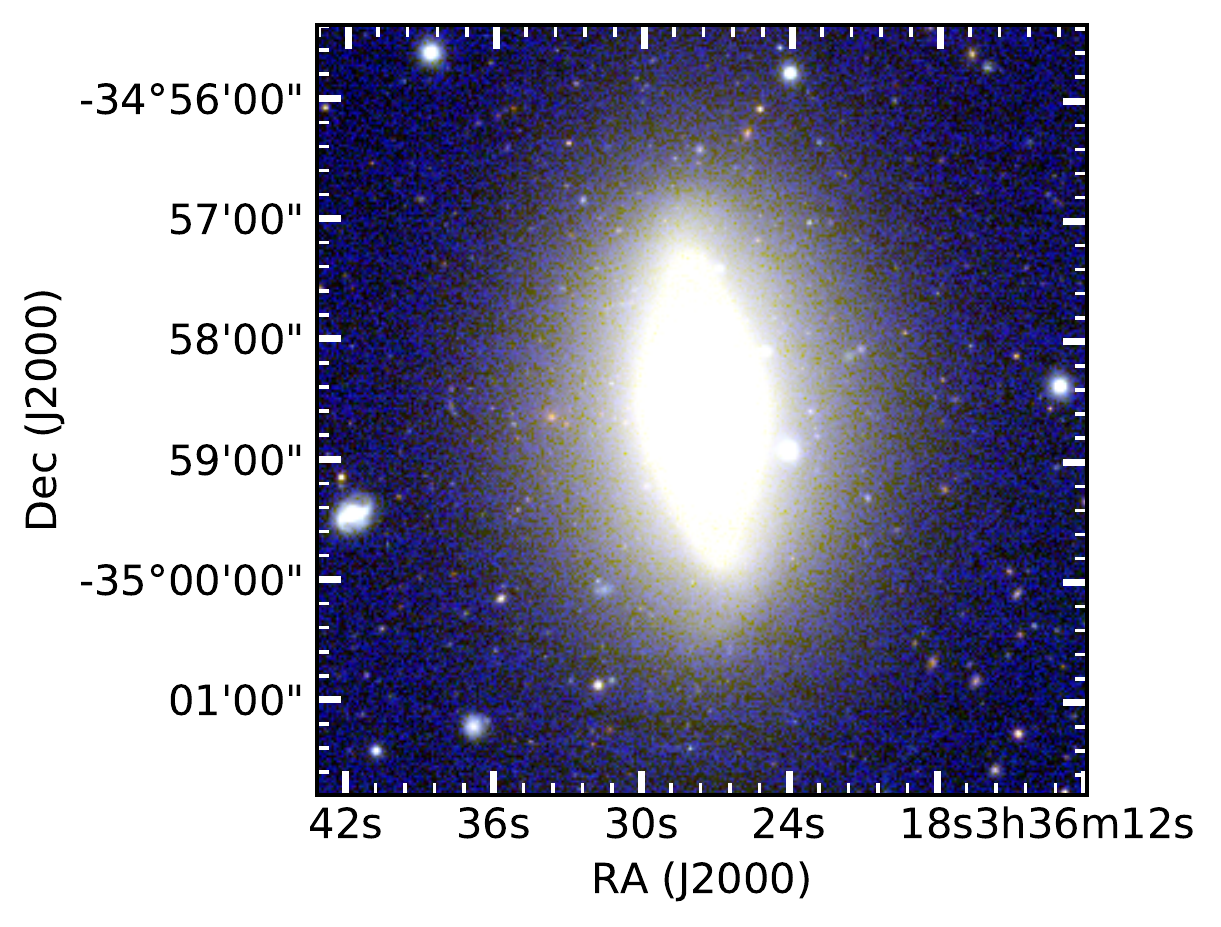}}	
	\hspace{-5mm}
	\subfloat[]
		{\includegraphics[height=0.35\textwidth]{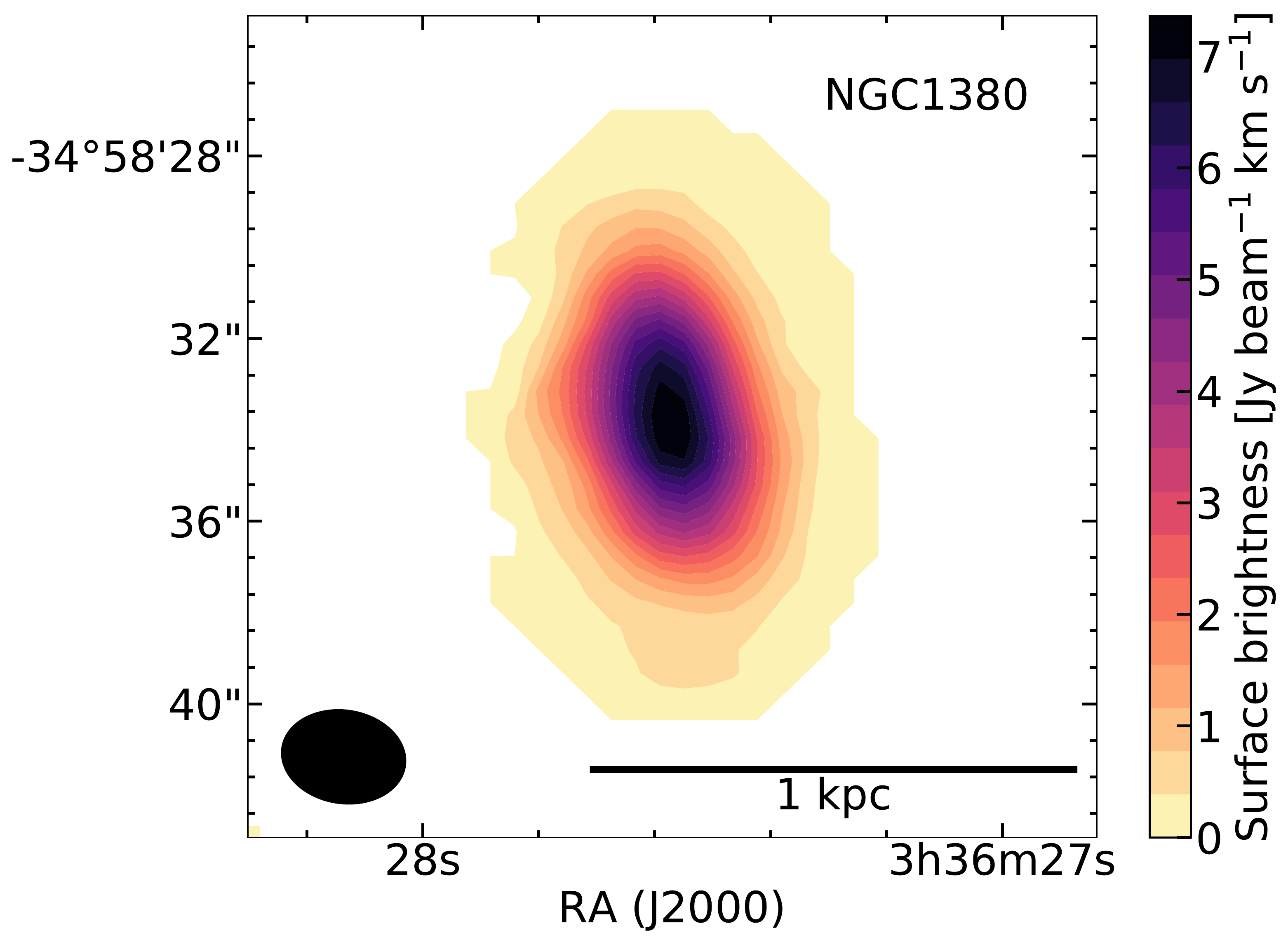}}	
	
	
	\subfloat[]
		{\hspace{-2mm}\includegraphics[height=0.35\textwidth]{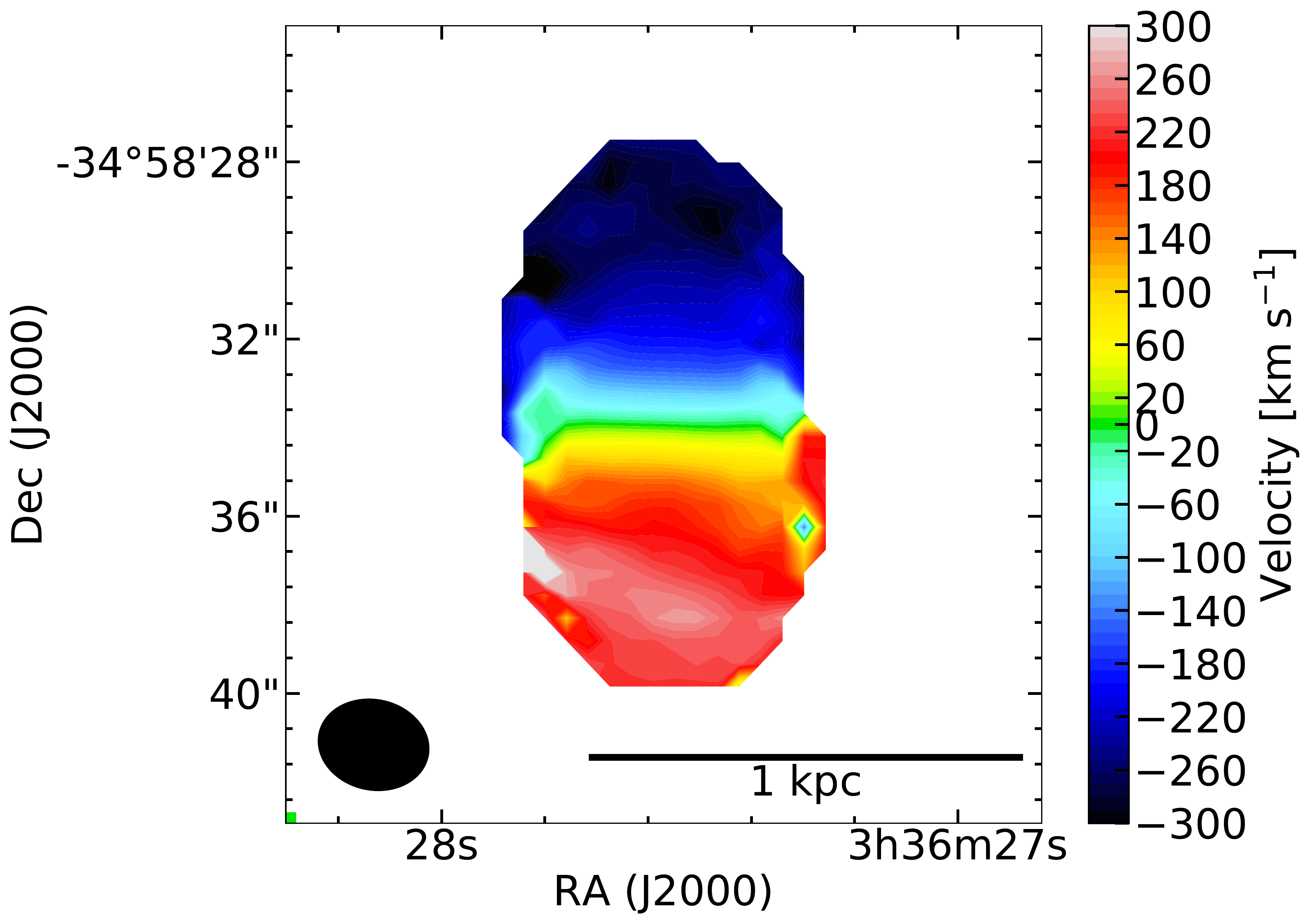}}
	\hspace{7mm}
	\subfloat[]
		{\includegraphics[height=0.35\textwidth]{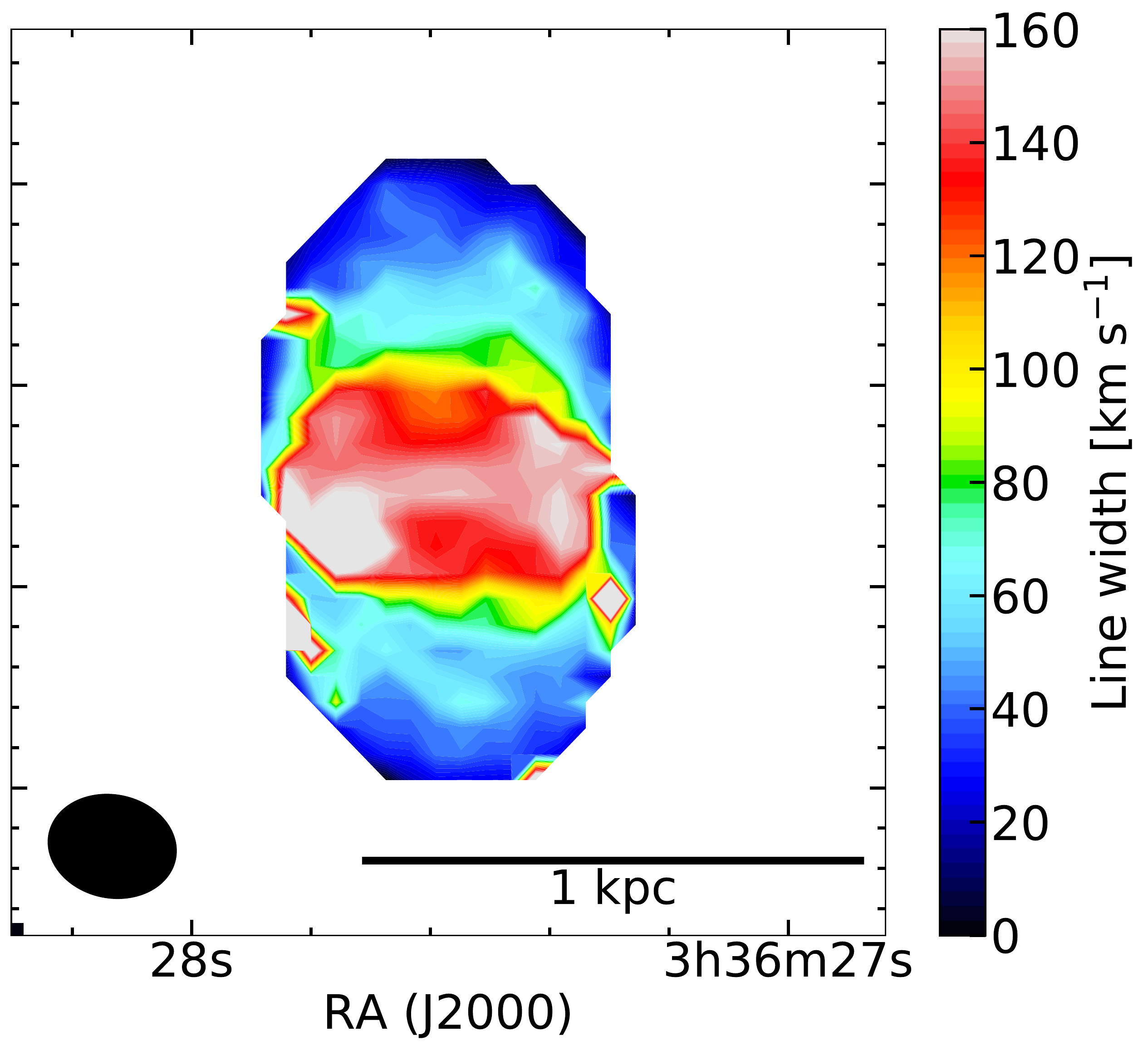}}
		
		
	\subfloat[]
		{\includegraphics[height=0.39\textwidth]{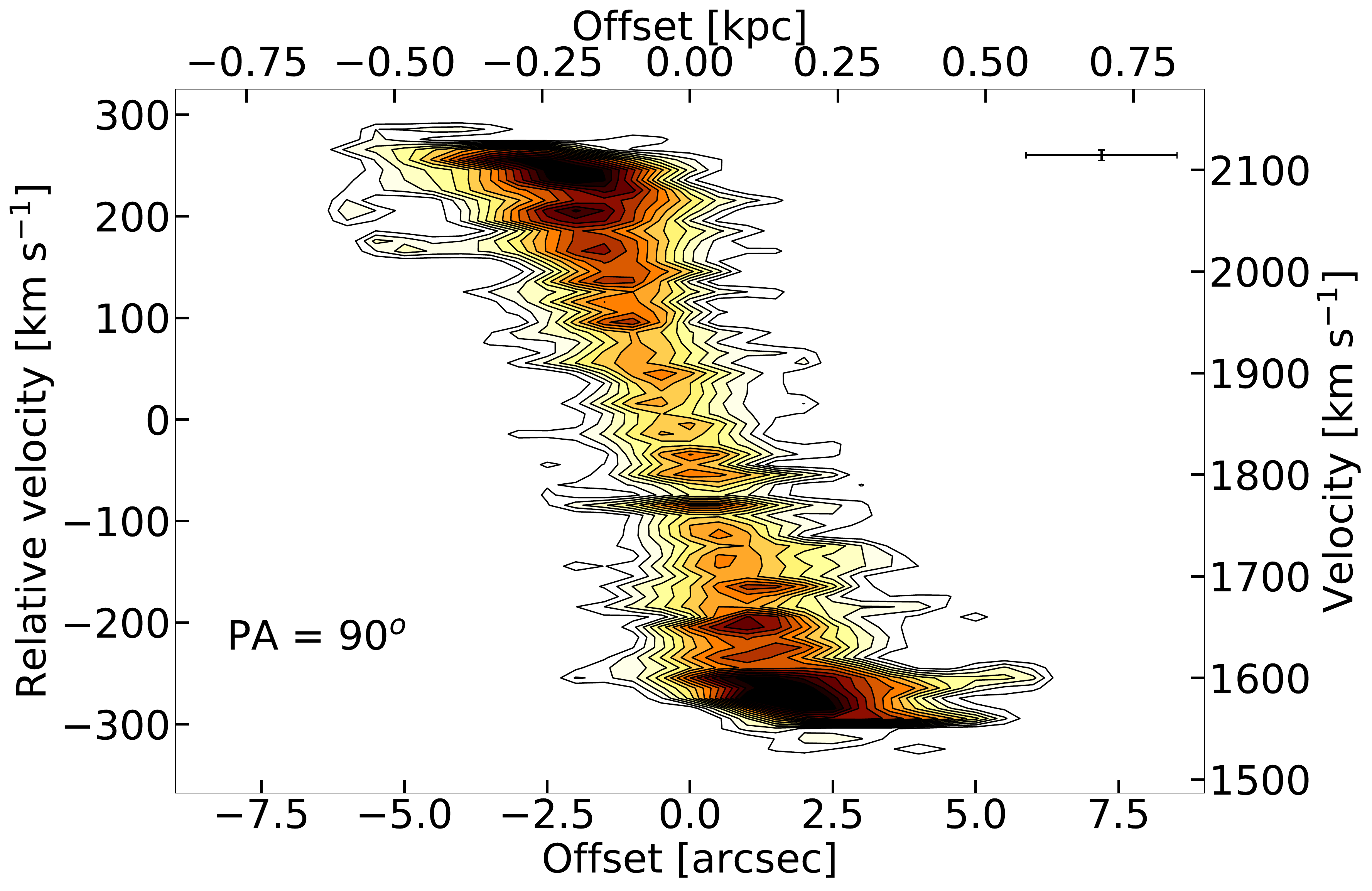}}	
	\hspace{6mm}
	\subfloat[]
		{\includegraphics[height=0.355\textwidth]{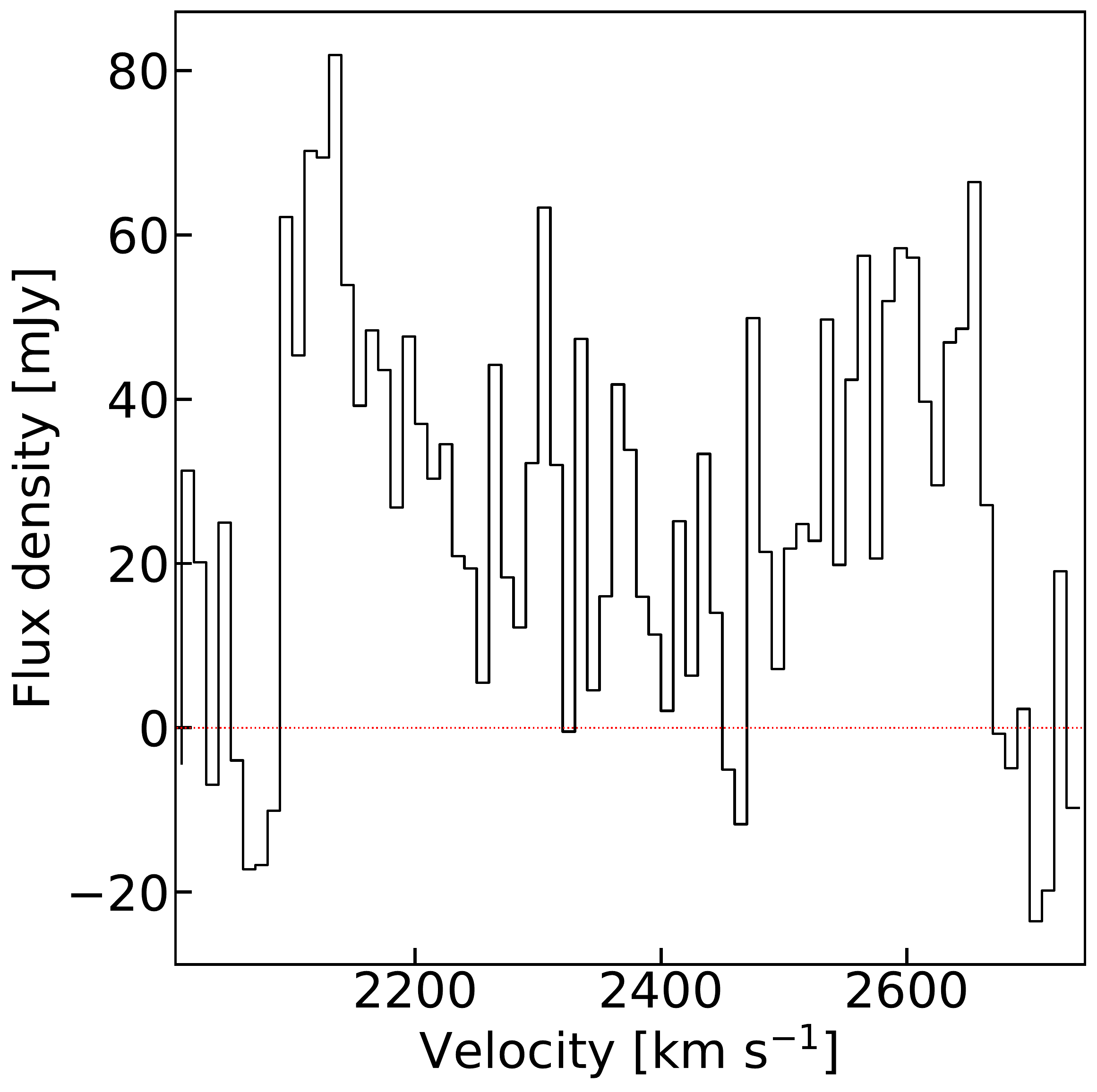}}
		
		
	\caption{NGC1380, similar to Figure \ref{fig:NGC1351A}.}
	\label{fig:NGC1380}
\end{figure*}

NGC1380 (Figures \ref{fig:NGC1380} and \ref{subfig:overplot_NGC1380}) has very centrally located molecular gas in these images, but has been shown to be distributed in a regular disk by \citet{Boizelle2017}. HST images of this galaxy have revealed that the dust in this galaxy is distributed in a ring around the galactic centre. This ring of dust appears quite evidently in the \textit{g}-\textit{i} colour map of the FDS (Iodice et al. 2018, submitted to A\&A). Viaene et al. (in prep.) find that the molecular gas follows a distribution with a similar shape, but at smaller radii. This source will be discussed in detail in Viaene et al. in prep.

\subsubsection{NGC1386 (FCC179)}
\label{subsub:NGC1386}
\begin{figure*}

	\centering

	\subfloat[]
	{\hspace{-6mm}\includegraphics[height=0.35\textwidth ]{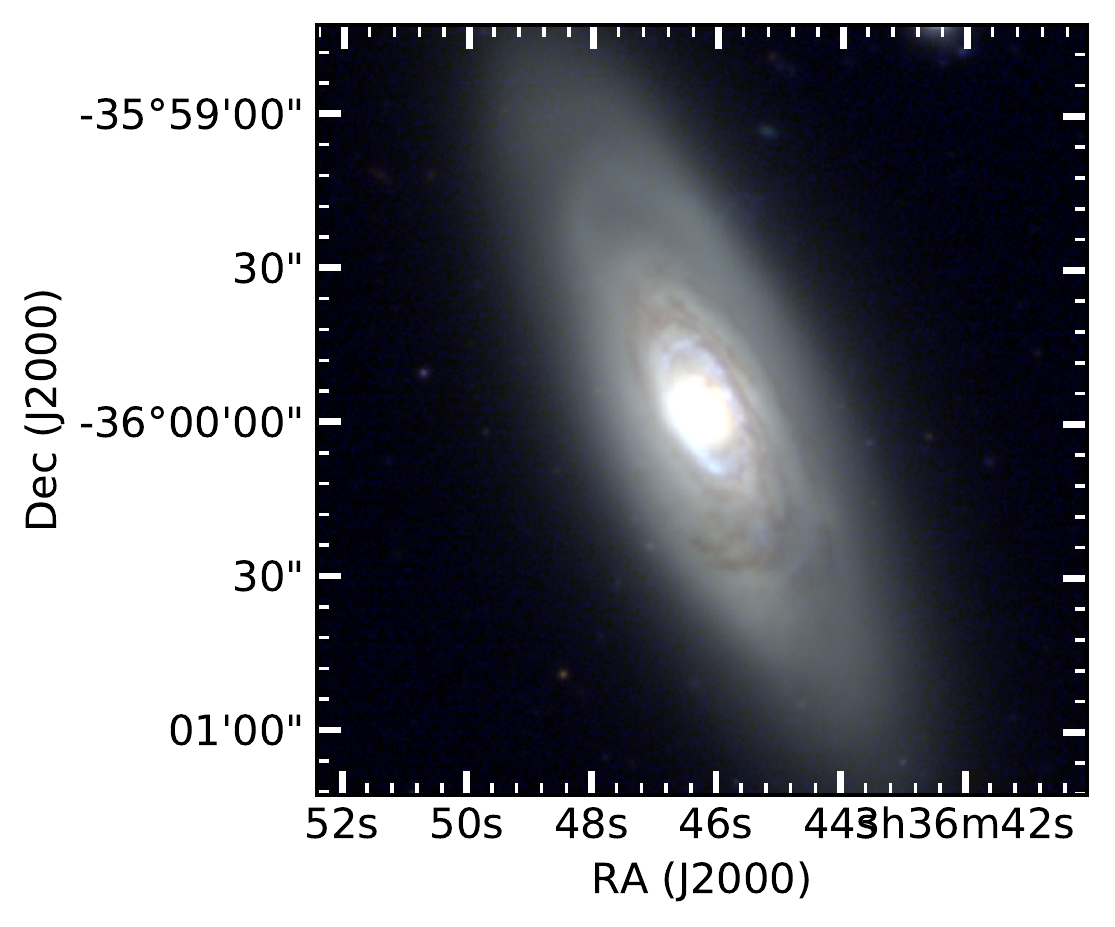}}	
	\hspace{0mm}
	\subfloat[]
		{\includegraphics[height=0.35\textwidth]{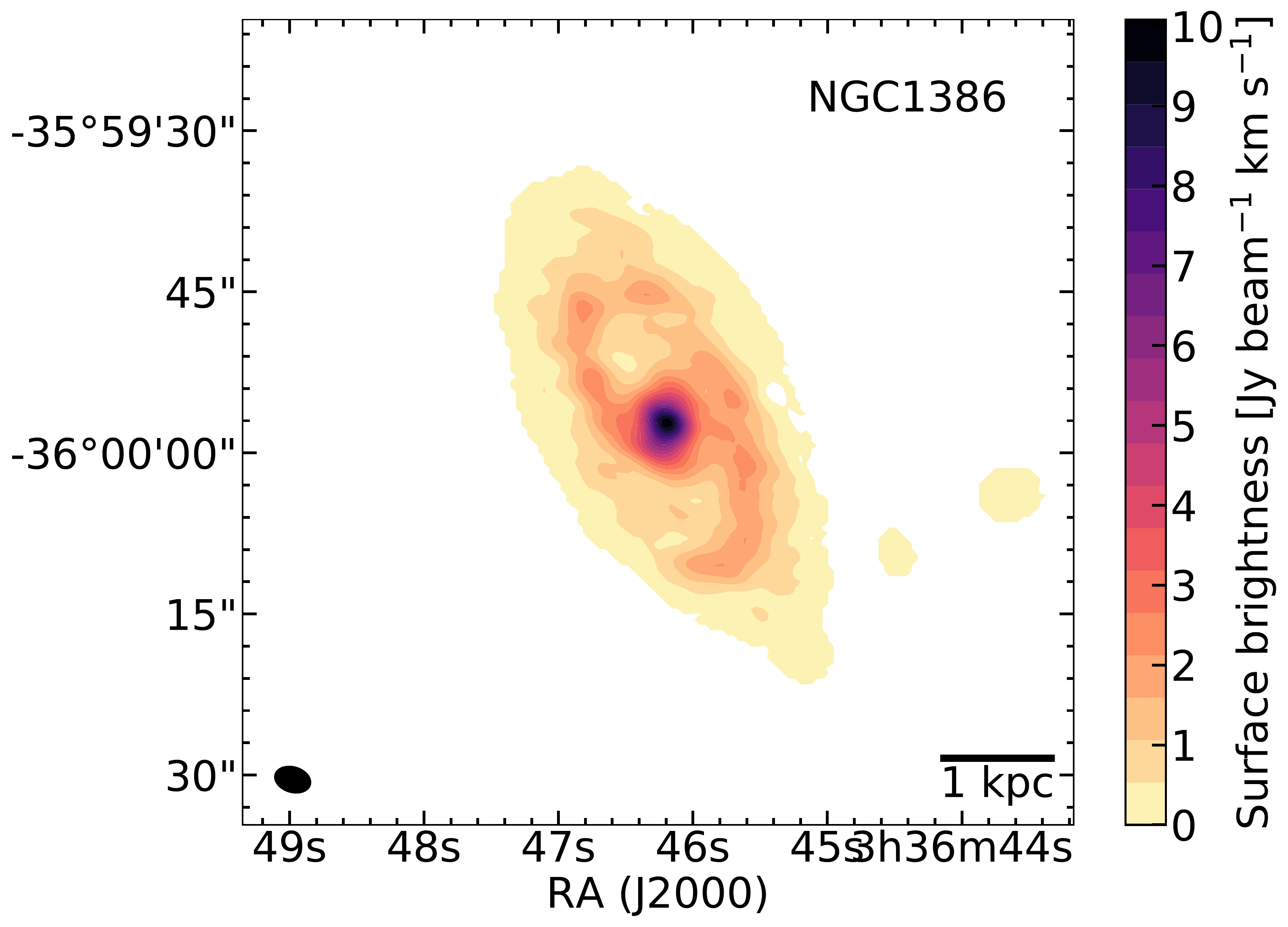}}	
	
	
	\subfloat[]
		{\hspace{-3mm}\includegraphics[height=0.35\textwidth]{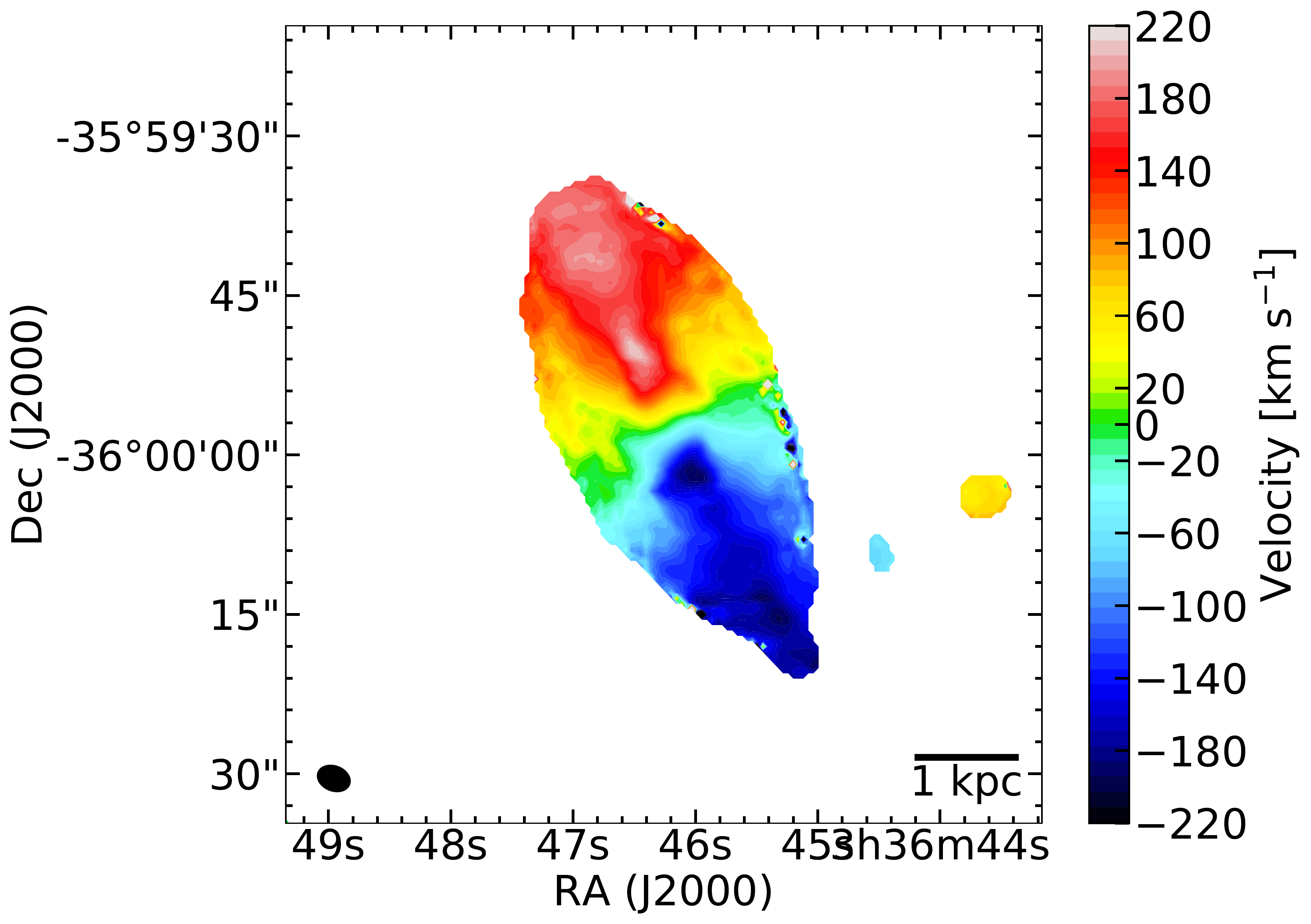}}
	\hspace{7mm}
	\subfloat[]
		{\includegraphics[height=0.35\textwidth]{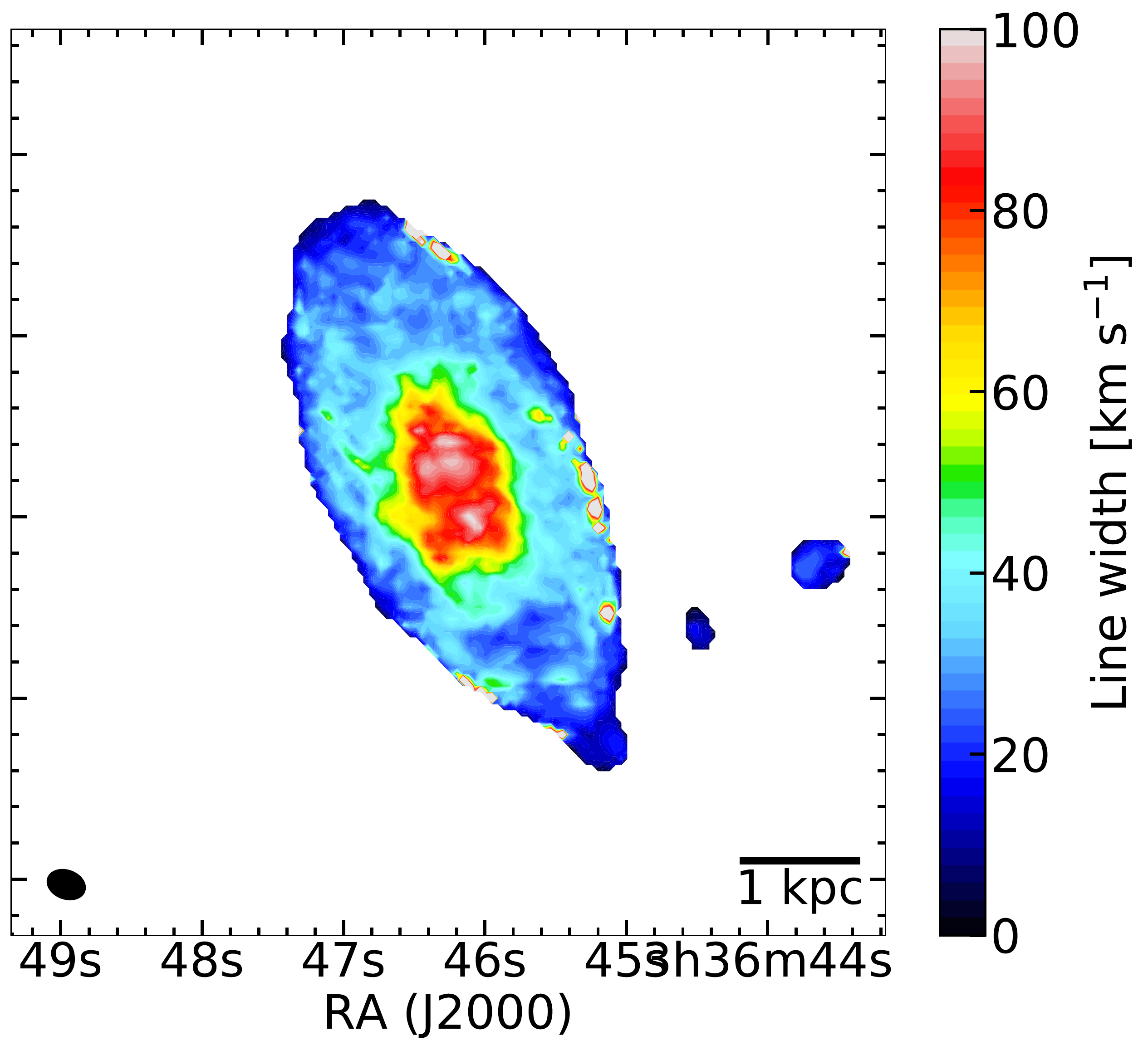}}
		
		
	\subfloat[]
		{\includegraphics[height=0.39\textwidth]{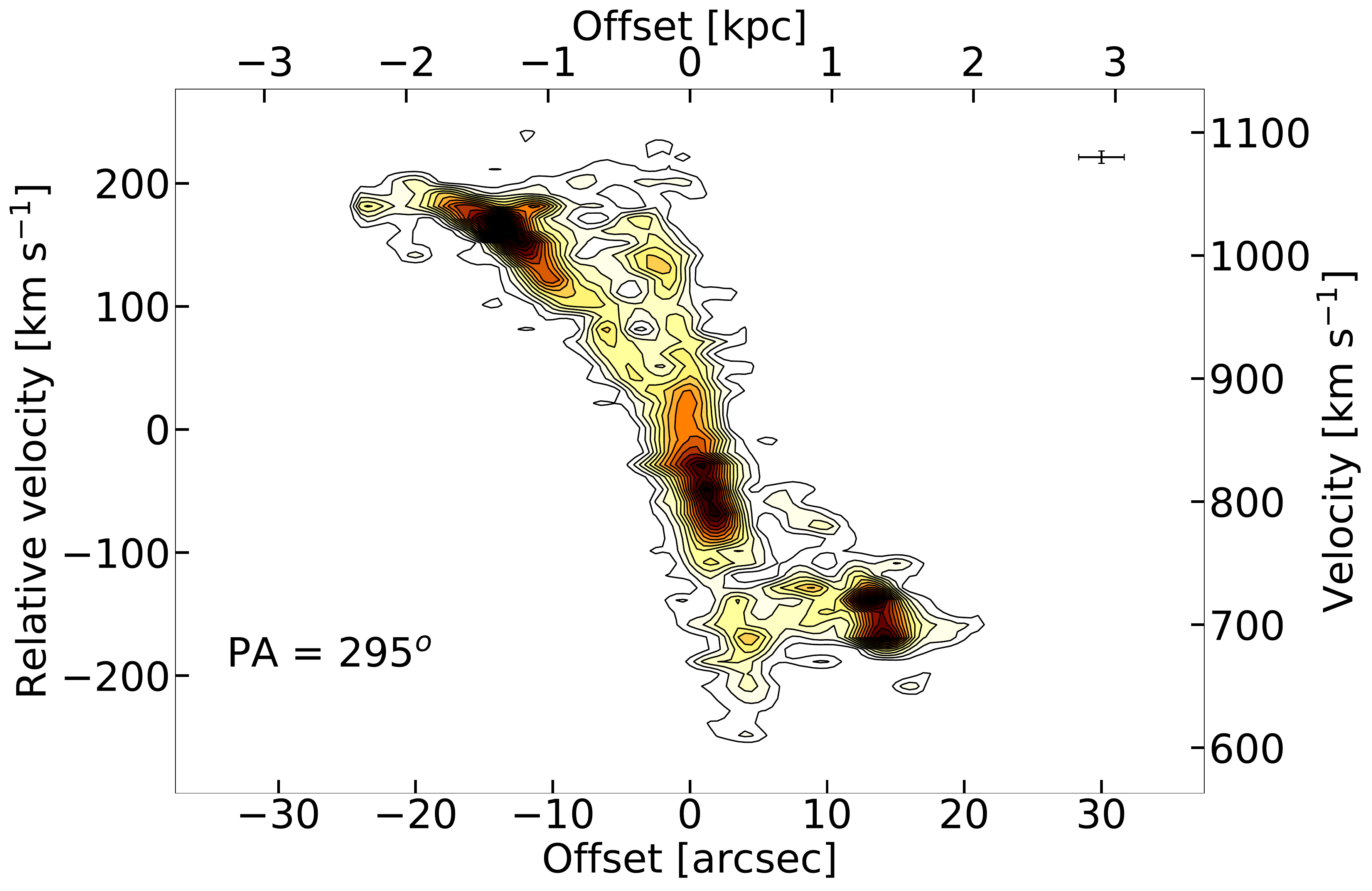}}	
	\hspace{6mm}
	\subfloat[]
		{\includegraphics[height=0.355\textwidth]{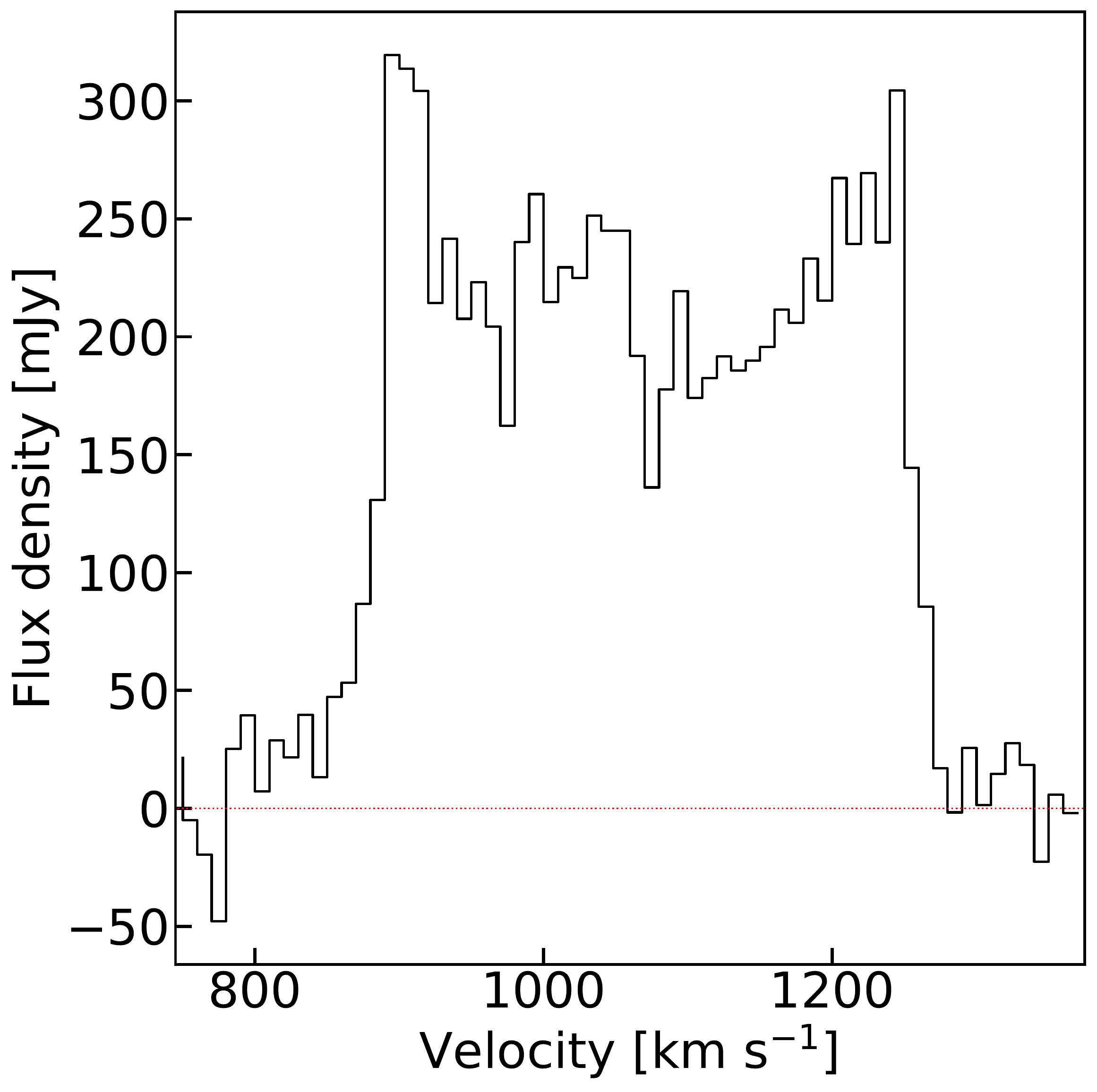}}
		
		
	\caption{NGC1386, similar to Figure \ref{fig:NGC1351A}.}
	\label{fig:NGC1386}
\end{figure*}

NGC1386 (Figures \ref{fig:NGC1386} and \ref{subfig:overplot_NGC1386}) is a good example of a spiral galaxy. The spiral arms are traced in the CO emission. It shows a regular rotation, with slightly increased velocities close to the galactic centre. The X-shape visible in the PVD probably indicates the presence of a bar. 3 mm continuum emission from the centre (see \Section \ref{sub:cont_detections}) is associated with a known AGN.

\subsubsection{NGC1387}
\label{subsub:NGC1387}
\begin{figure*}

	\centering

	\subfloat[]
	{\hspace{-5mm}\includegraphics[height=0.35\textwidth]{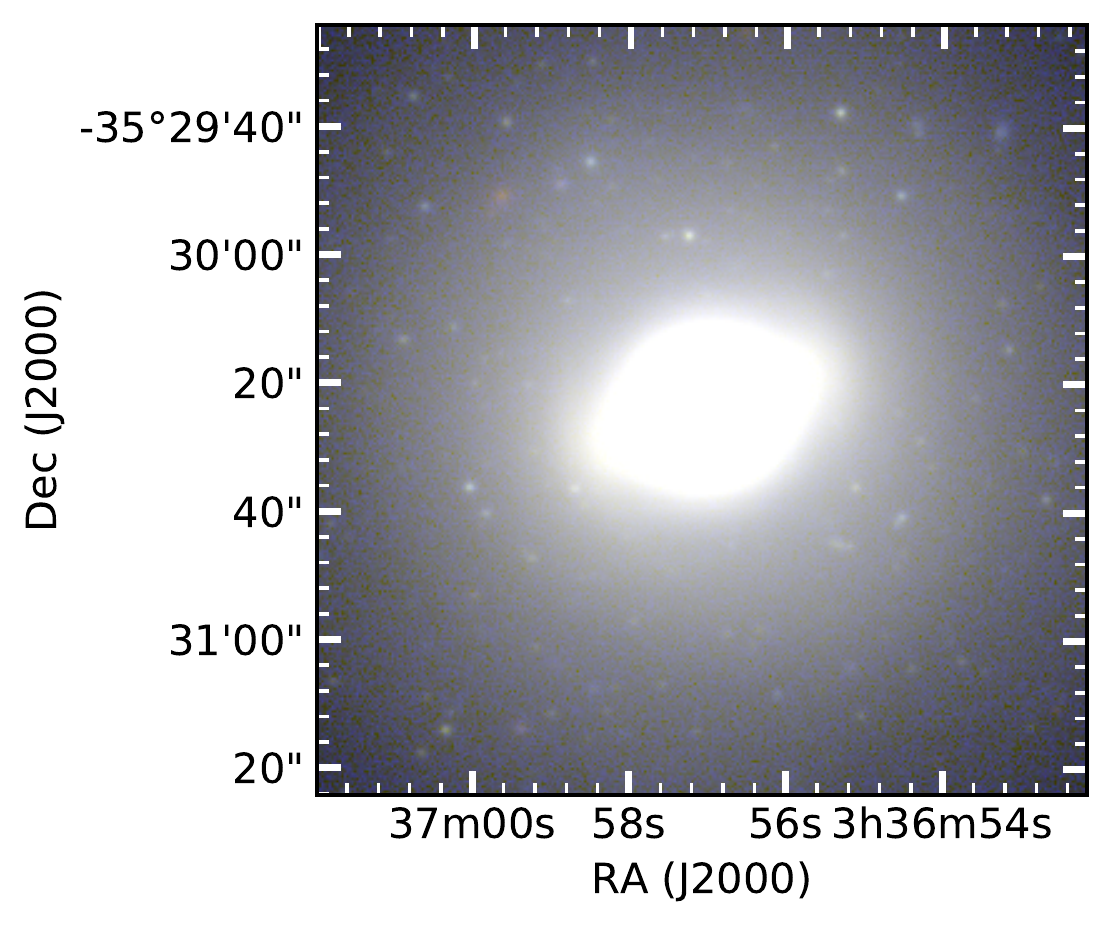}}	
	\hspace{5mm}
	\subfloat[]
		{\includegraphics[height=0.35\textwidth]{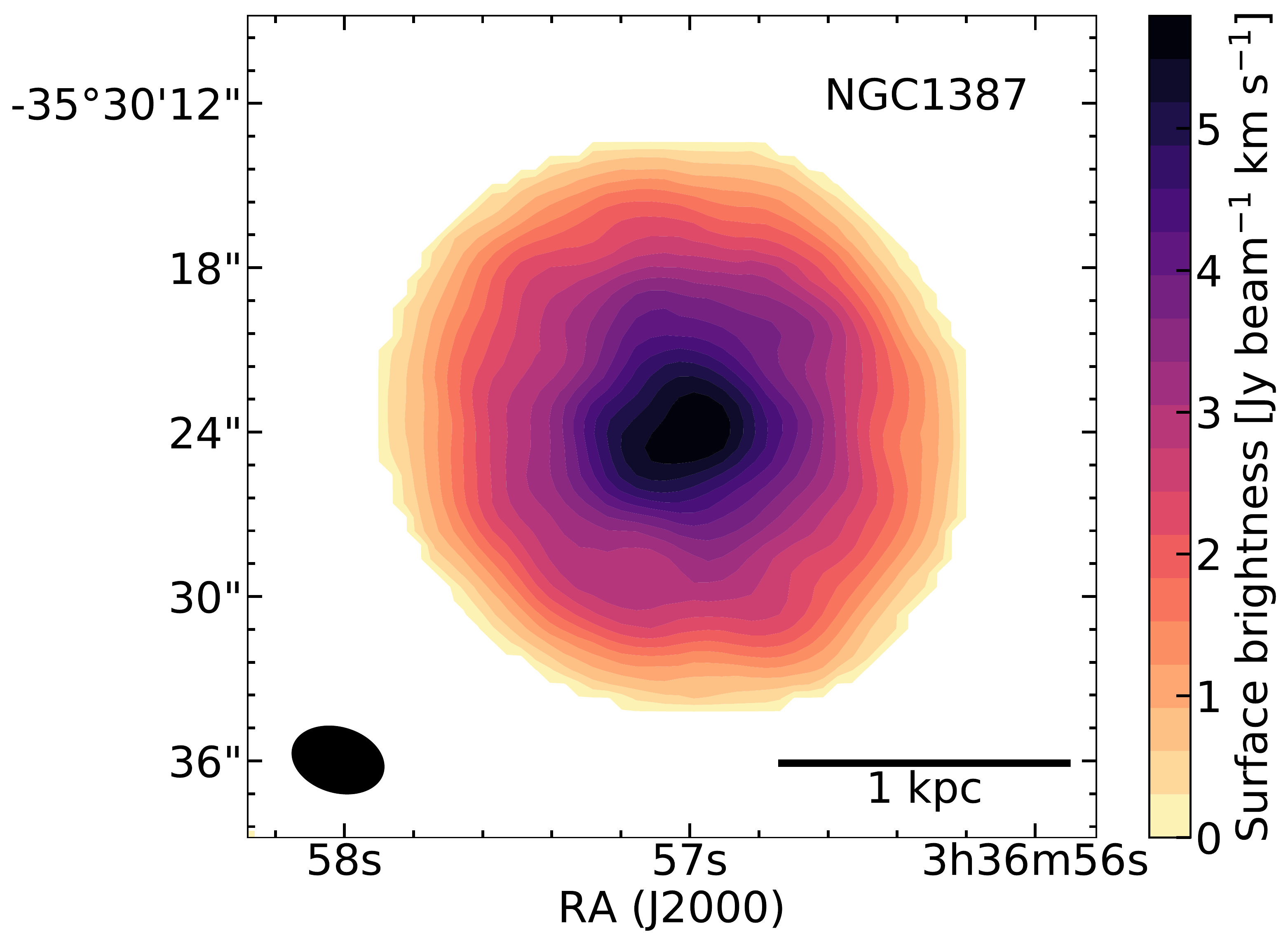}}	
	
	
	\subfloat[]
		{\hspace{-4mm}\includegraphics[height=0.35\textwidth]{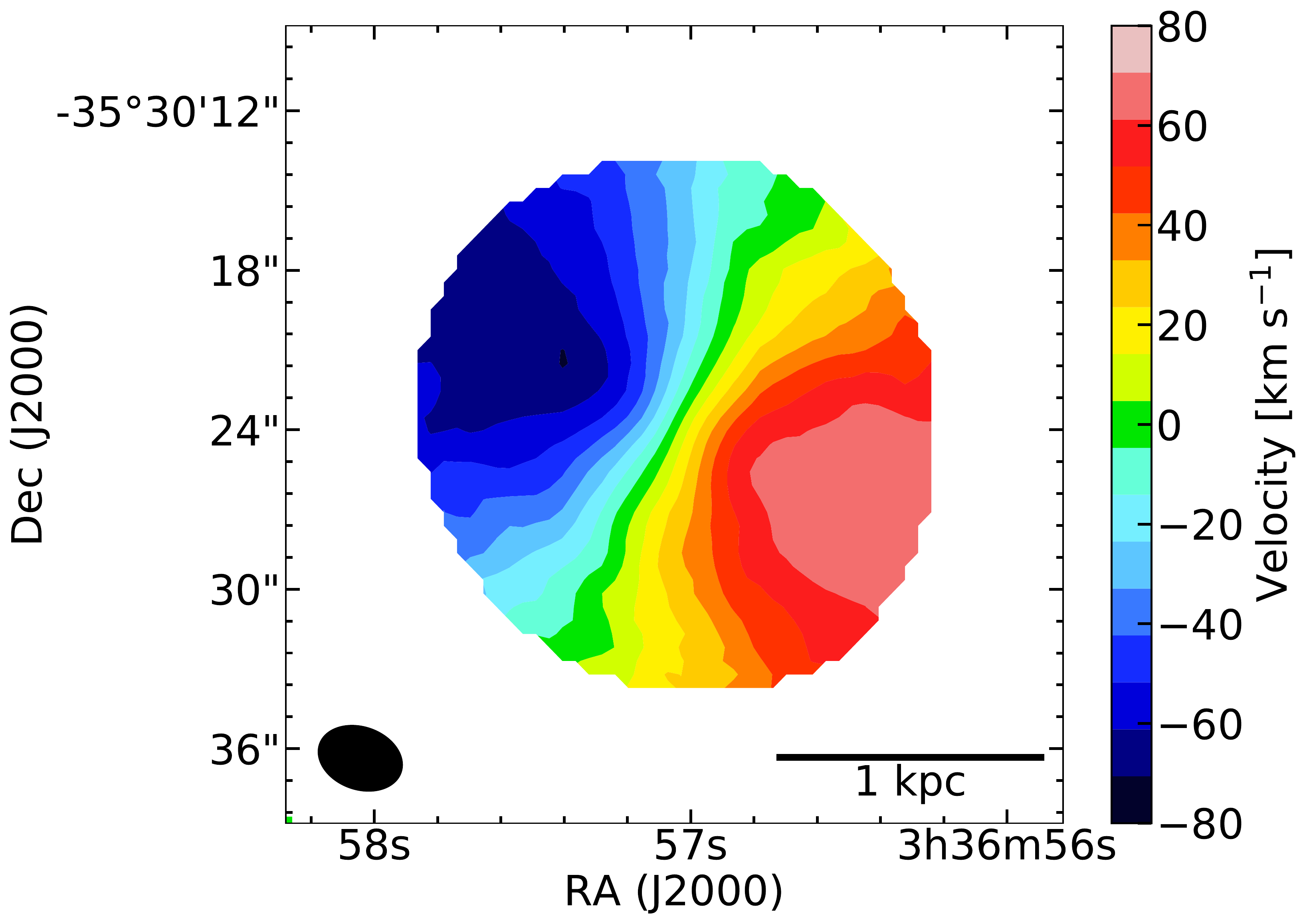}}
	\hspace{7mm}
	\subfloat[]
		{\includegraphics[height=0.35\textwidth]{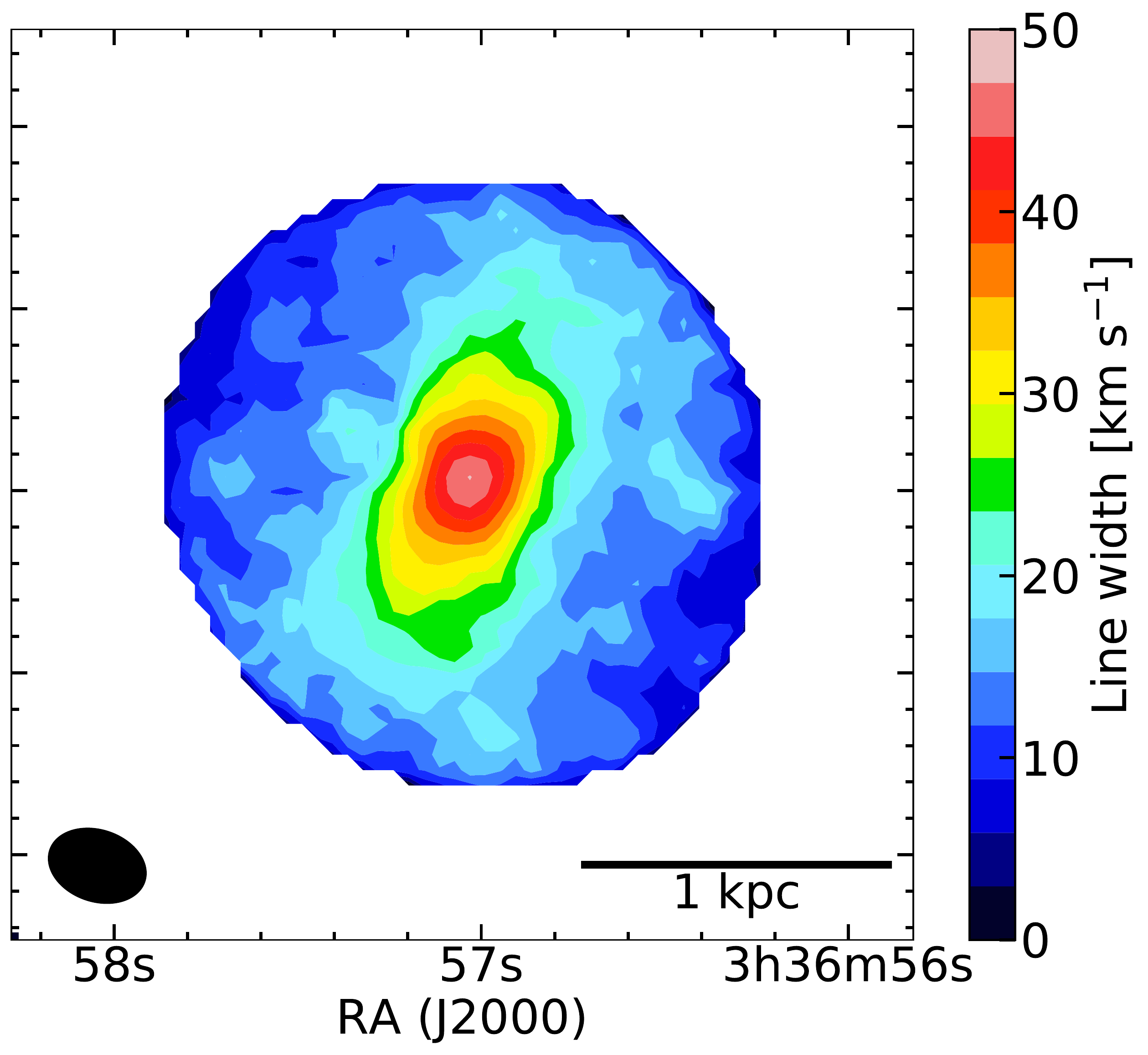}}
		
		
	\subfloat[]
		{\includegraphics[height=0.39\textwidth]{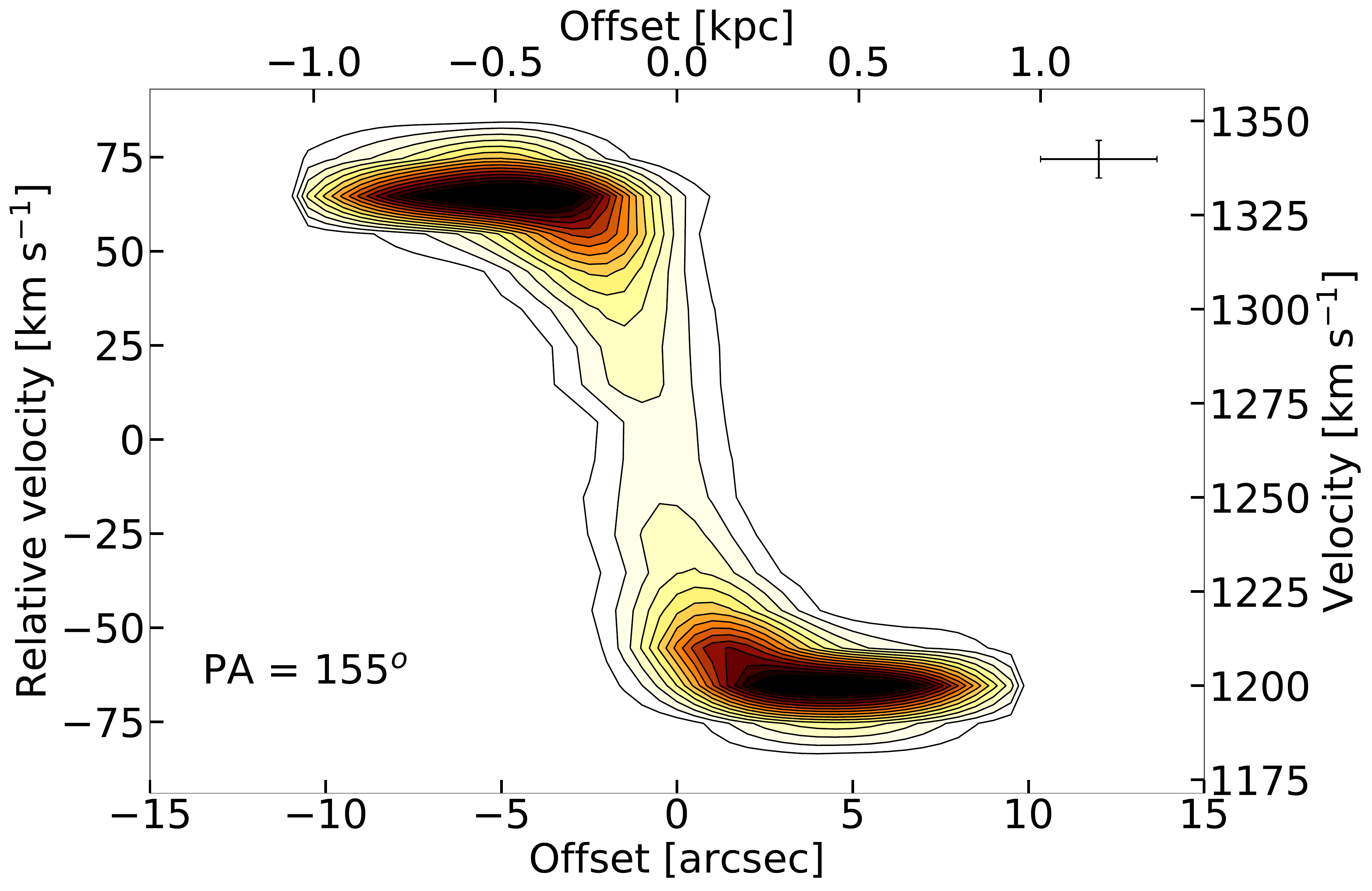}}	
	\hspace{6mm}
	\subfloat[]
		{\includegraphics[height=0.355\textwidth]{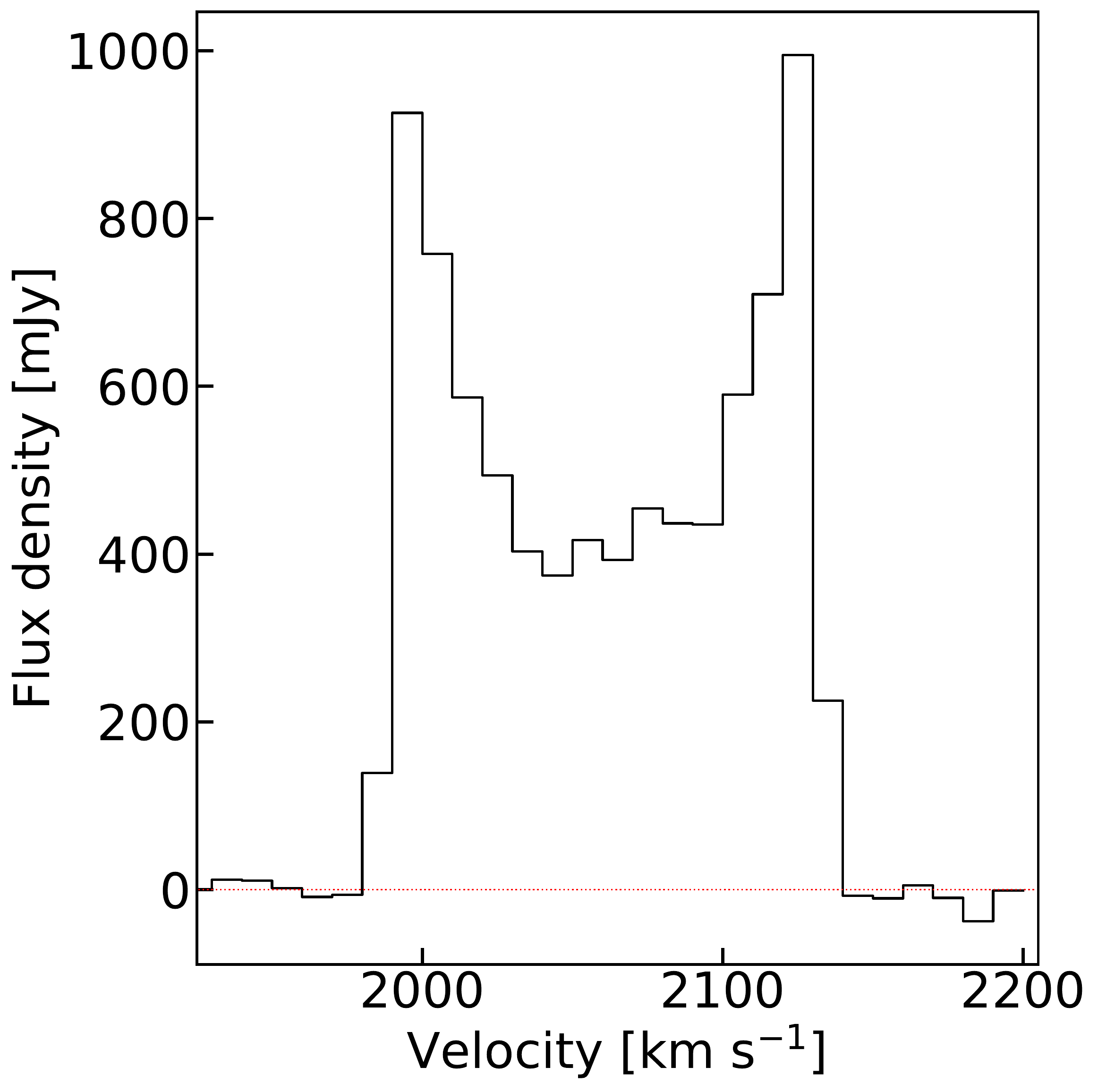}}
		
		
	\caption{NGC1387, similar to Figure \ref{fig:NGC1351A}.}
	\label{fig:NGC1387}
\end{figure*}

NGC1387 (Figures \ref{fig:NGC1387} and \ref{subfig:overplot_NGC1387}) is one of the biggest and brightest early-type galaxies close to the core of the cluster. It is one of the most (if not the most) regular galaxies in the sample, showing centrally distributed CO emission, decreasing radially outward (see Figure \ref{fig:moment-maps_reg}). This gas-disk has an inclination of $\sim18^o$ (Davis et al., in prep.). The galaxy shows a regular rotation, and a double-peaked line profile. Archival HST imaging shows that these features are likely associated with a dust disk, which is compatible with the detected CO emission. The galaxy has a strong bar, which is possibly the result of tidal forces: it is interacting with the BCG NGC1399 \citep [e.g.][]{Iodice2016}. The molecular gas probably remains relatively unaffected by its interaction with NGC1399 because this galaxy is much more massive than the disturbed galaxies in the sample ($M_\star = 5.89 \times 10^{10} M_\odot$), and its gas is deep in its gravitational potential within the X$_2$ orbits of the bar. In deep FDS images (Iodice et al. 2018, submitted to A\&A) it shows a nuclear disk/ring of dust (\textit{g}-\textit{i} = 1.4). A ring of $\sim$6'' was also found in near-infrared images in the \textit{Ks}-band \citep{Laurikainen2006}. 

\subsubsection{FCC207}
\label{subsub:FCC207}
\begin{figure*}

	\centering

	\subfloat[]
	{\hspace{-7mm}\includegraphics[height=0.35\textwidth ]{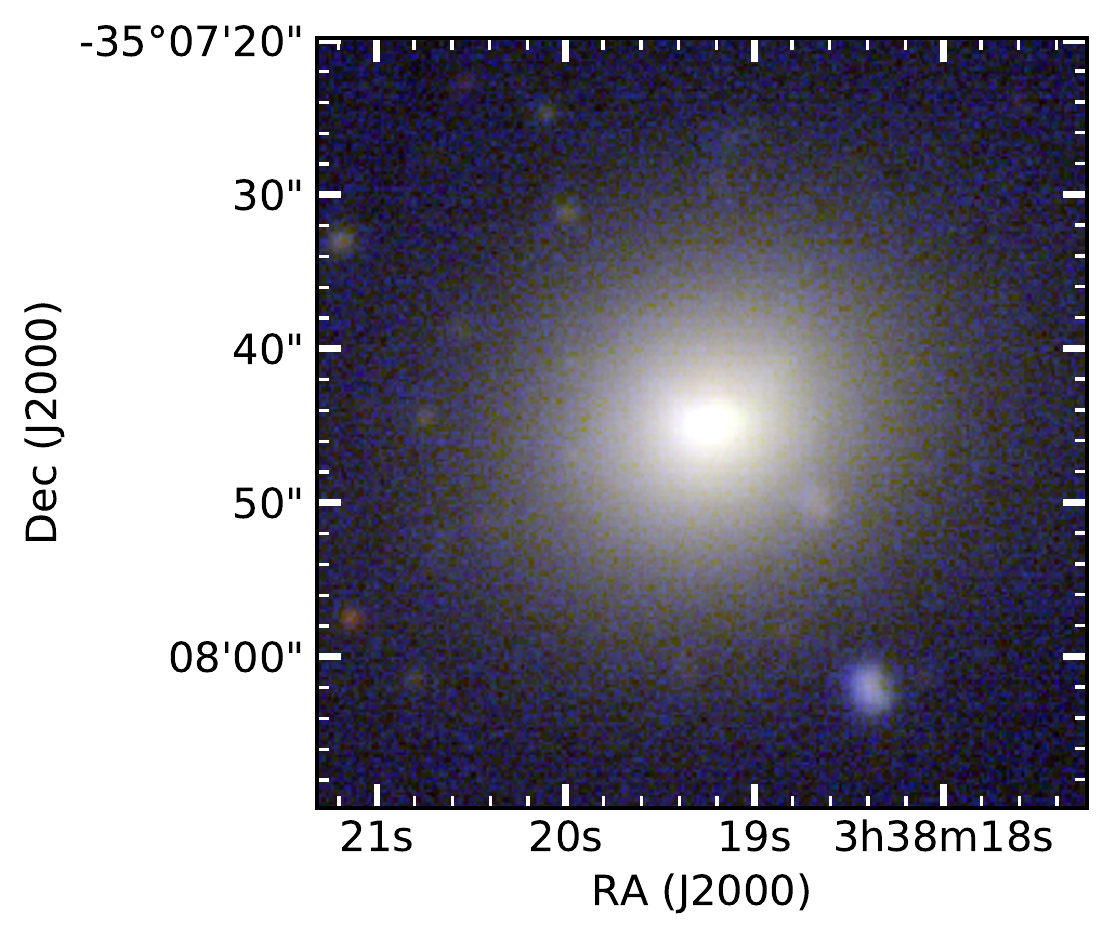}}	
	\hspace{3mm}
	\subfloat[]
		{\includegraphics[height=0.35\textwidth]{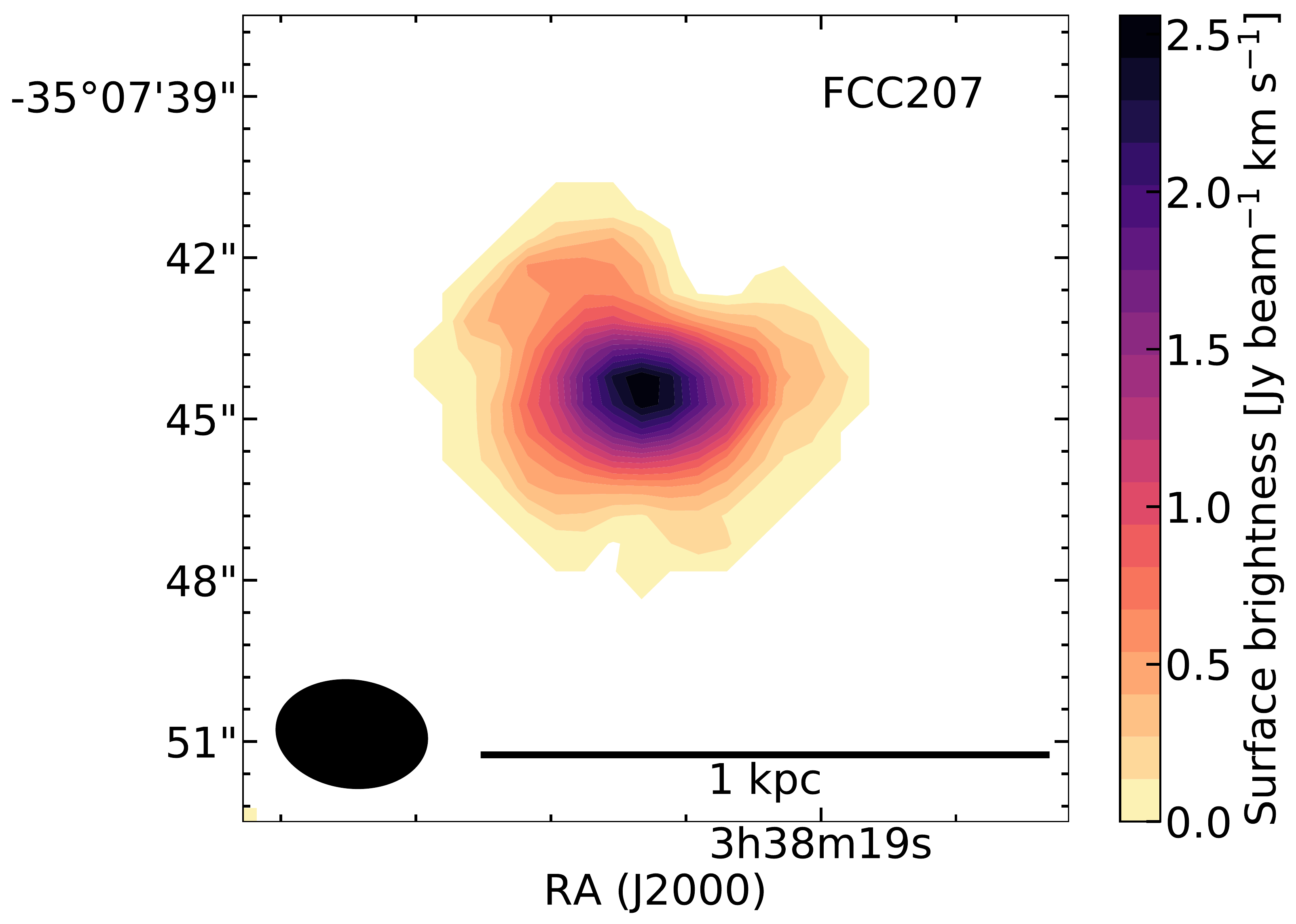}}	
	
	
	\subfloat[]
		{\hspace{-5mm}\includegraphics[height=0.35\textwidth]{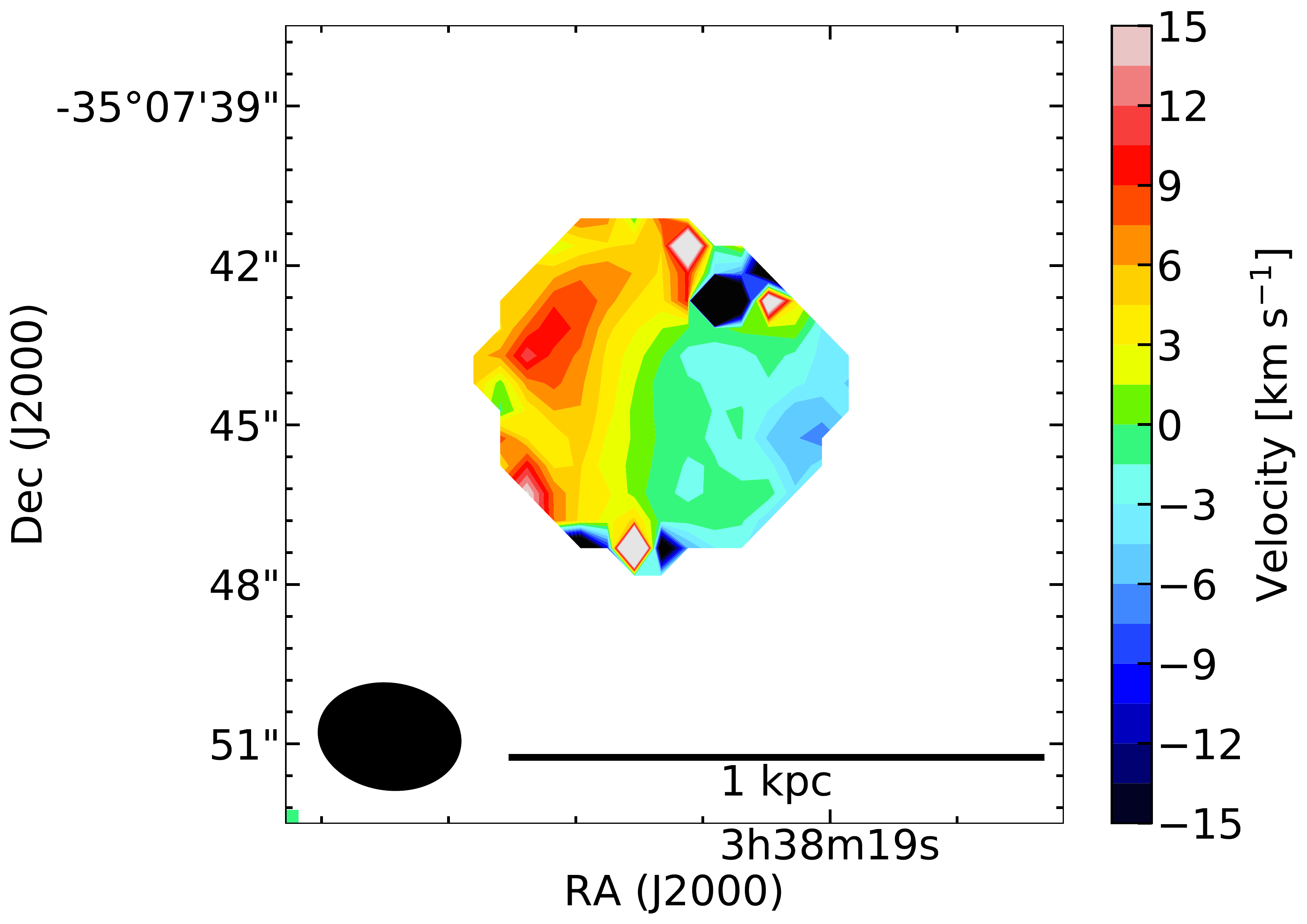}}
	\hspace{7mm}
	\subfloat[]
		{\includegraphics[height=0.35\textwidth]{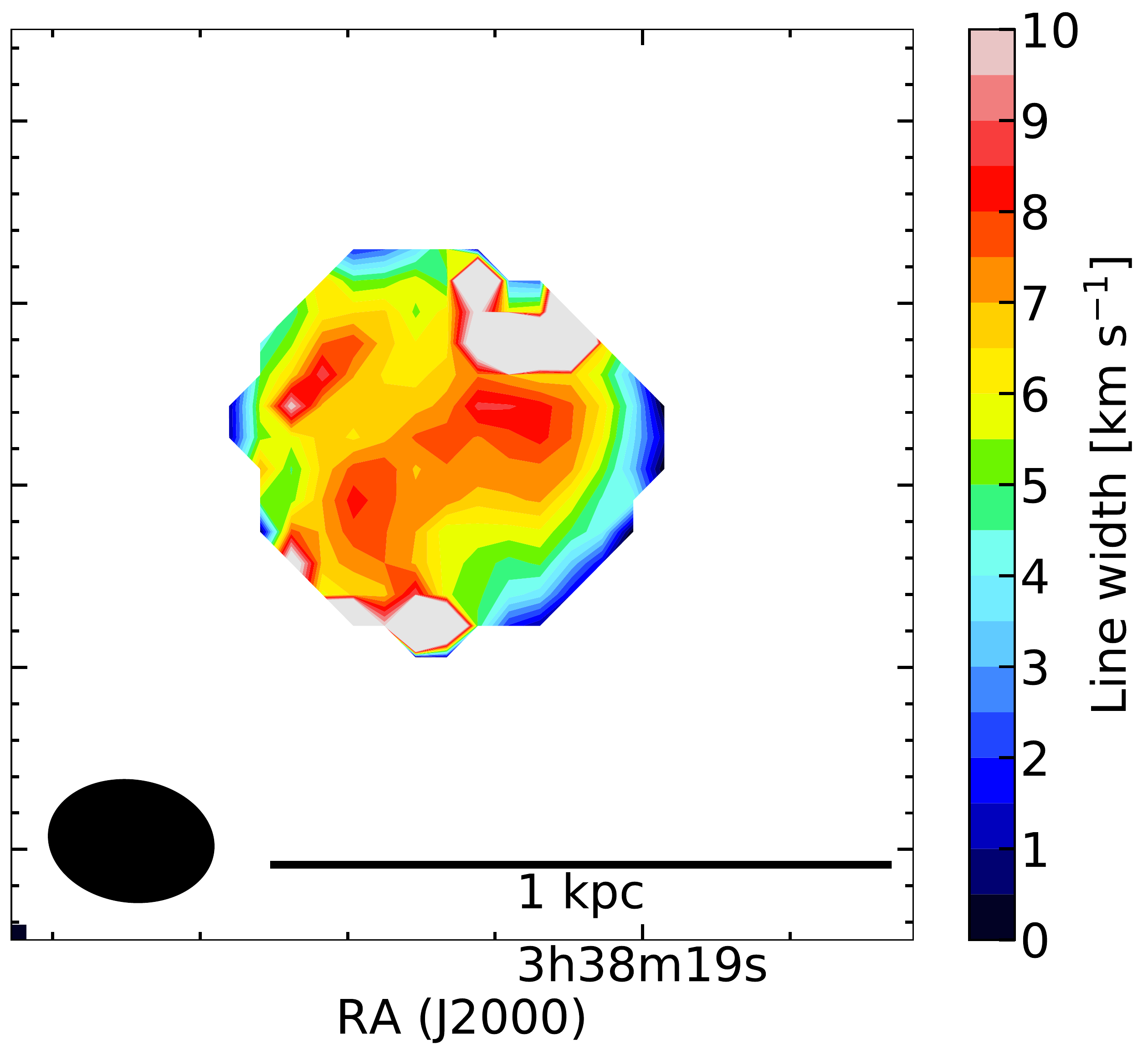}}
		
		
	\subfloat[]
		{\includegraphics[height=0.39\textwidth]{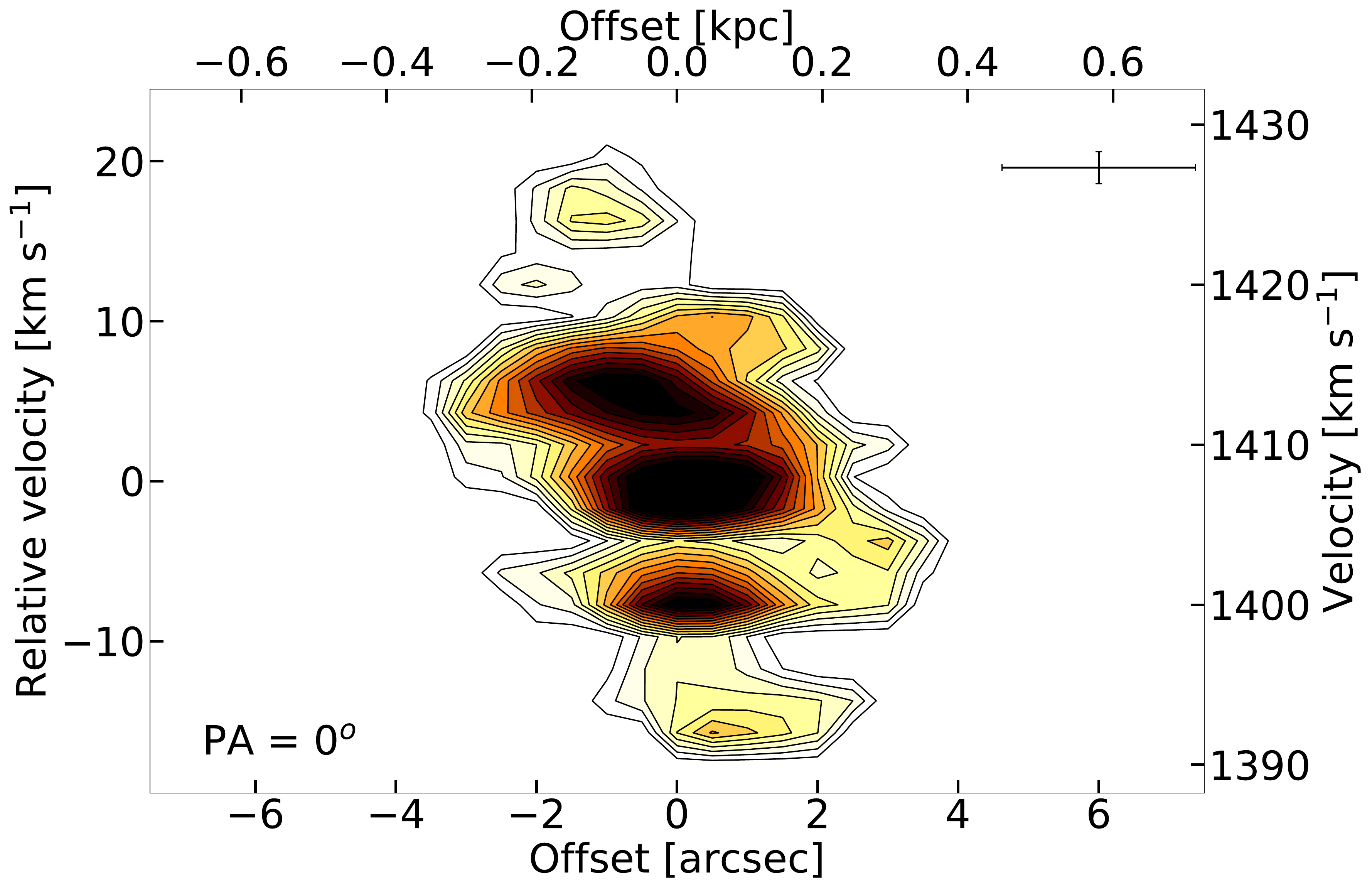}}	
	\hspace{6mm}
	\subfloat[]
		{\includegraphics[height=0.355\textwidth]{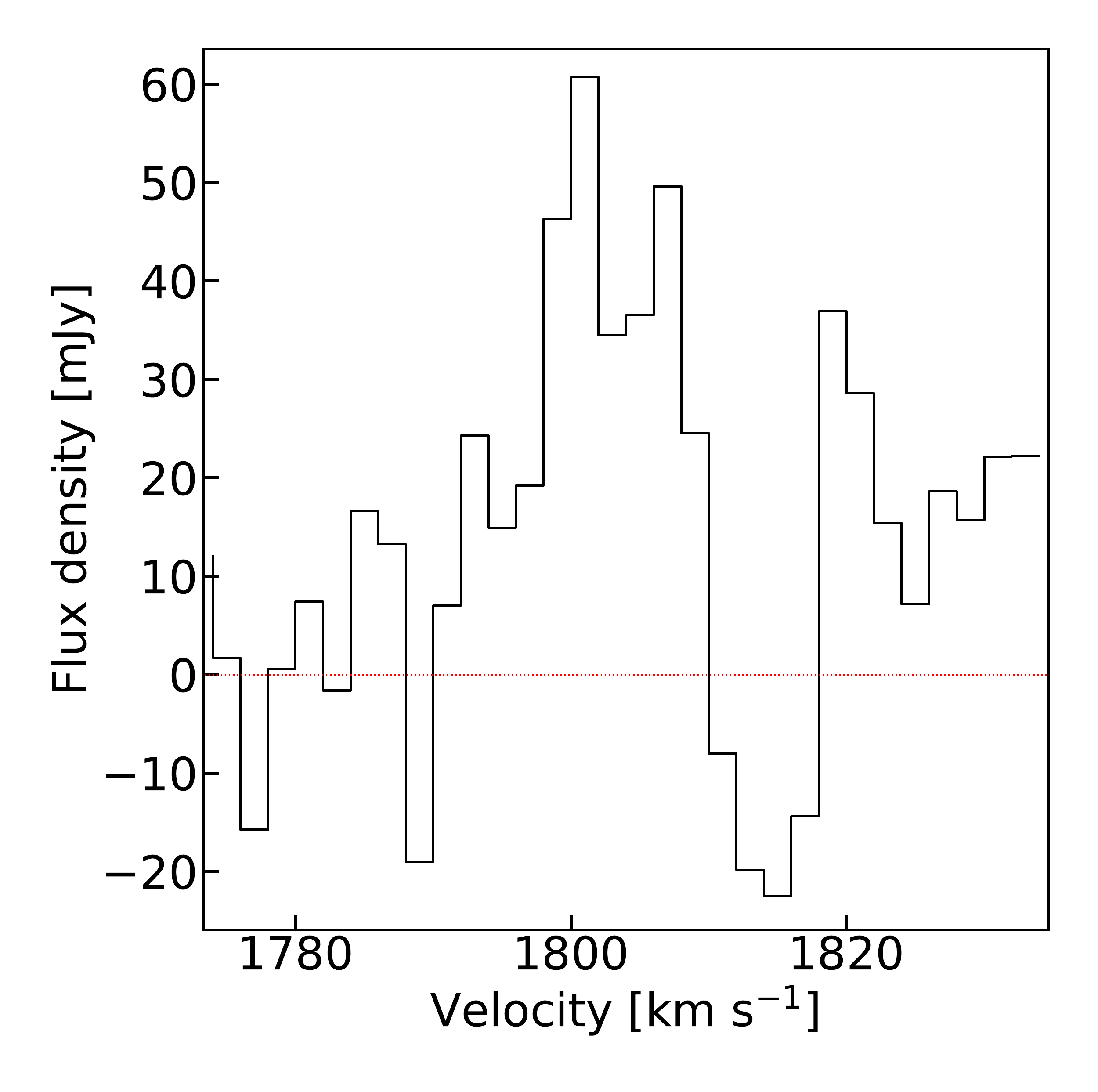}}
		
		
	\caption{FCC207, similar to Figure \ref{fig:NGC1351A}, except that the velocity channels are 2 \kms wide.}
	\label{fig:FCC207}
\end{figure*}

FCC207 (Figures \ref{fig:FCC207} and \ref{subfig:overplot_FCC207}) is a small galaxy with disturbed molecular gas that is off-centre. Some rotation can be identified in the velocity maps, but the kinematics are clearly disturbed. Its stellar body looks relaxed (at the resolution of the data used here).

\subsubsection{FCC261, FCC282, and FCC332}
\label{subsub:FCC261_FCC282_FCC332}
\begin{figure*}

	\centering

	\subfloat[]
	{\hspace{-5mm}\includegraphics[height=0.35\textwidth ]{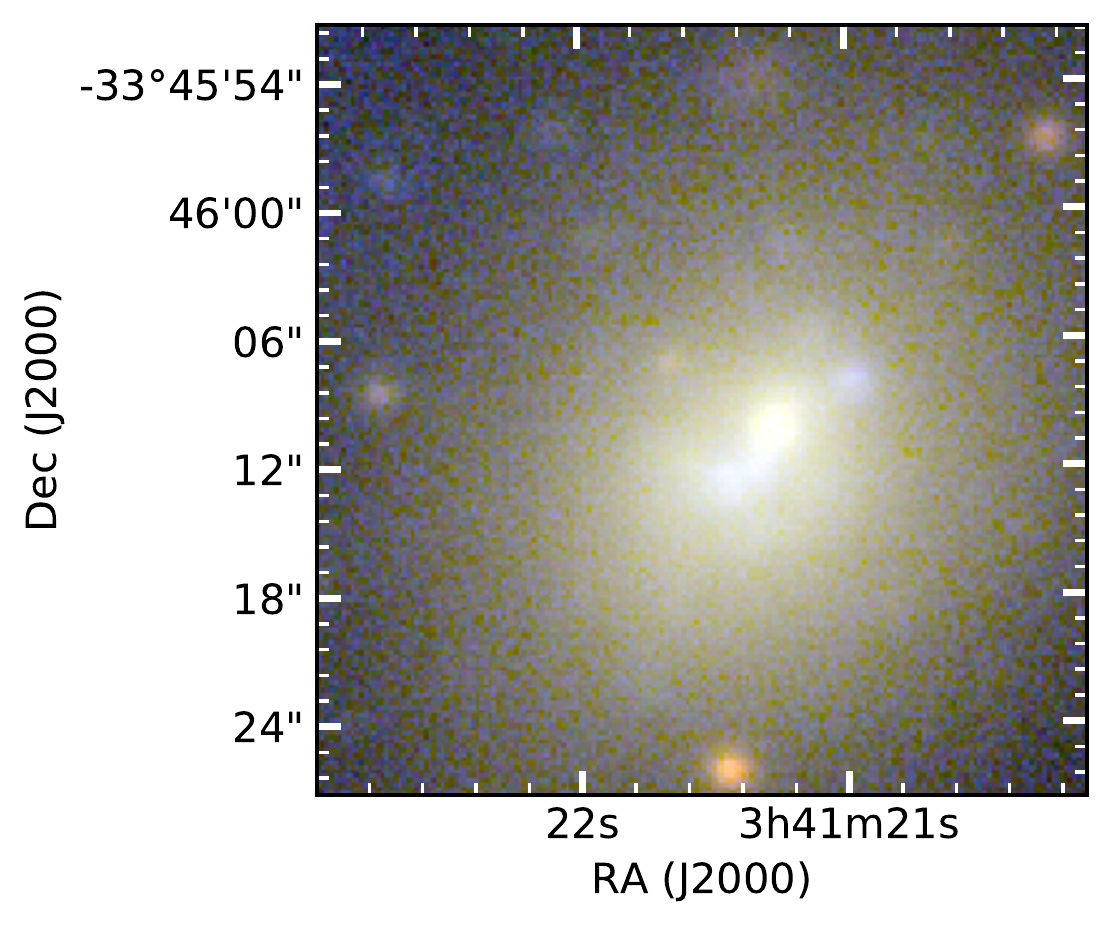}}	
	\hspace{4mm}
	\subfloat[]
		{\includegraphics[height=0.35\textwidth]{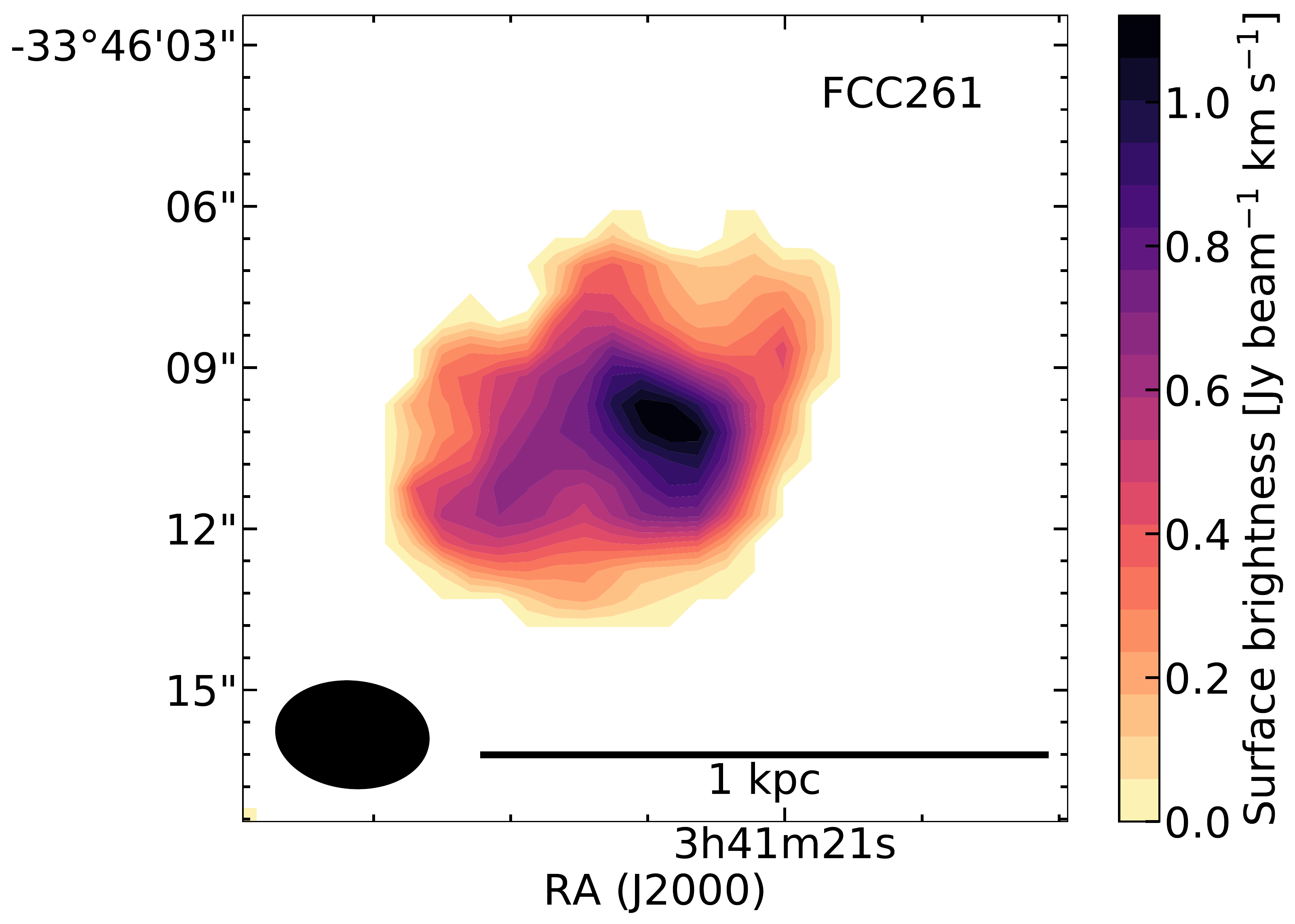}}	
	
	
	\subfloat[]
		{\hspace{-5mm}\includegraphics[height=0.35\textwidth]{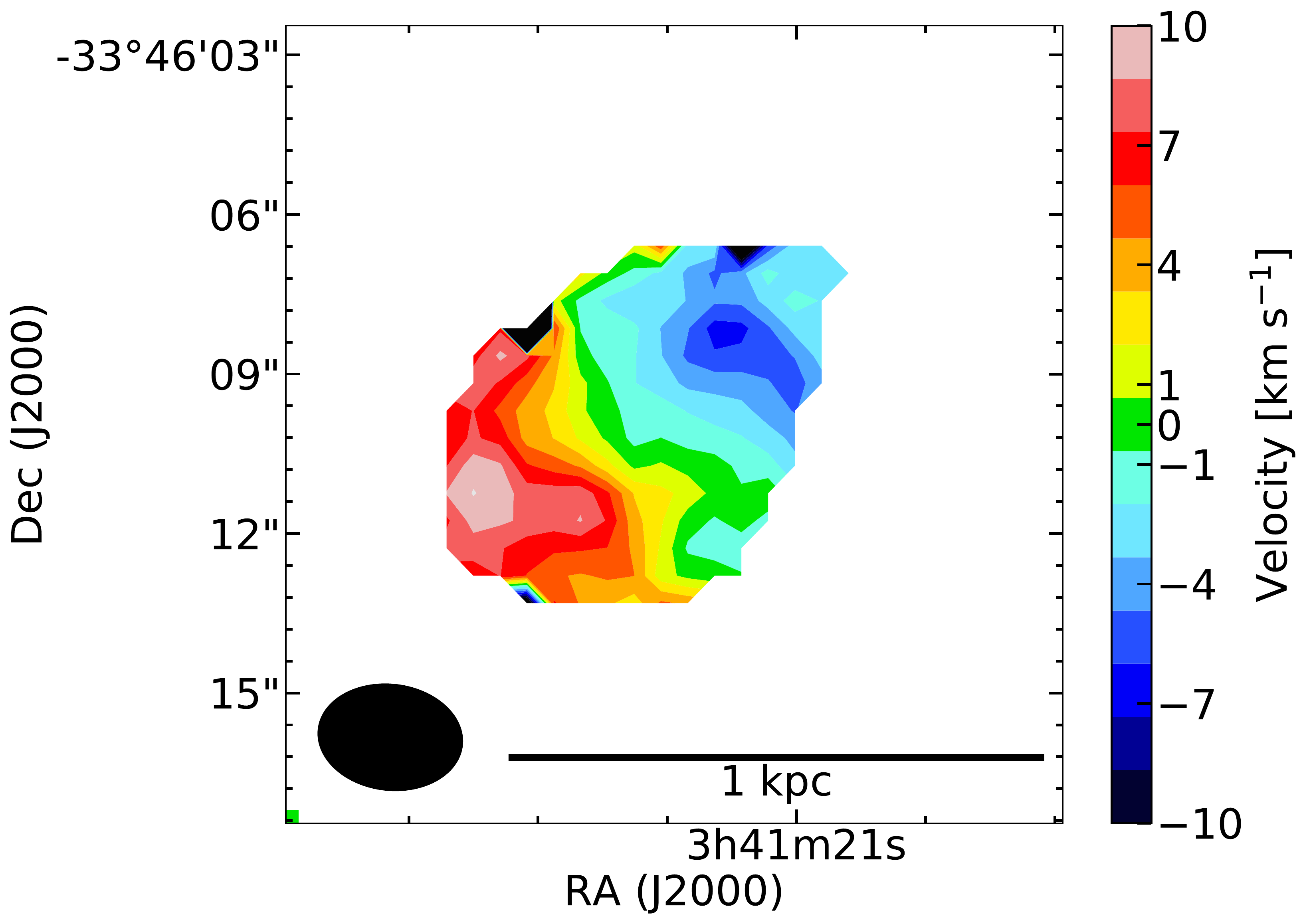}}
	\hspace{7mm}
	\subfloat[]
		{\includegraphics[height=0.35\textwidth]{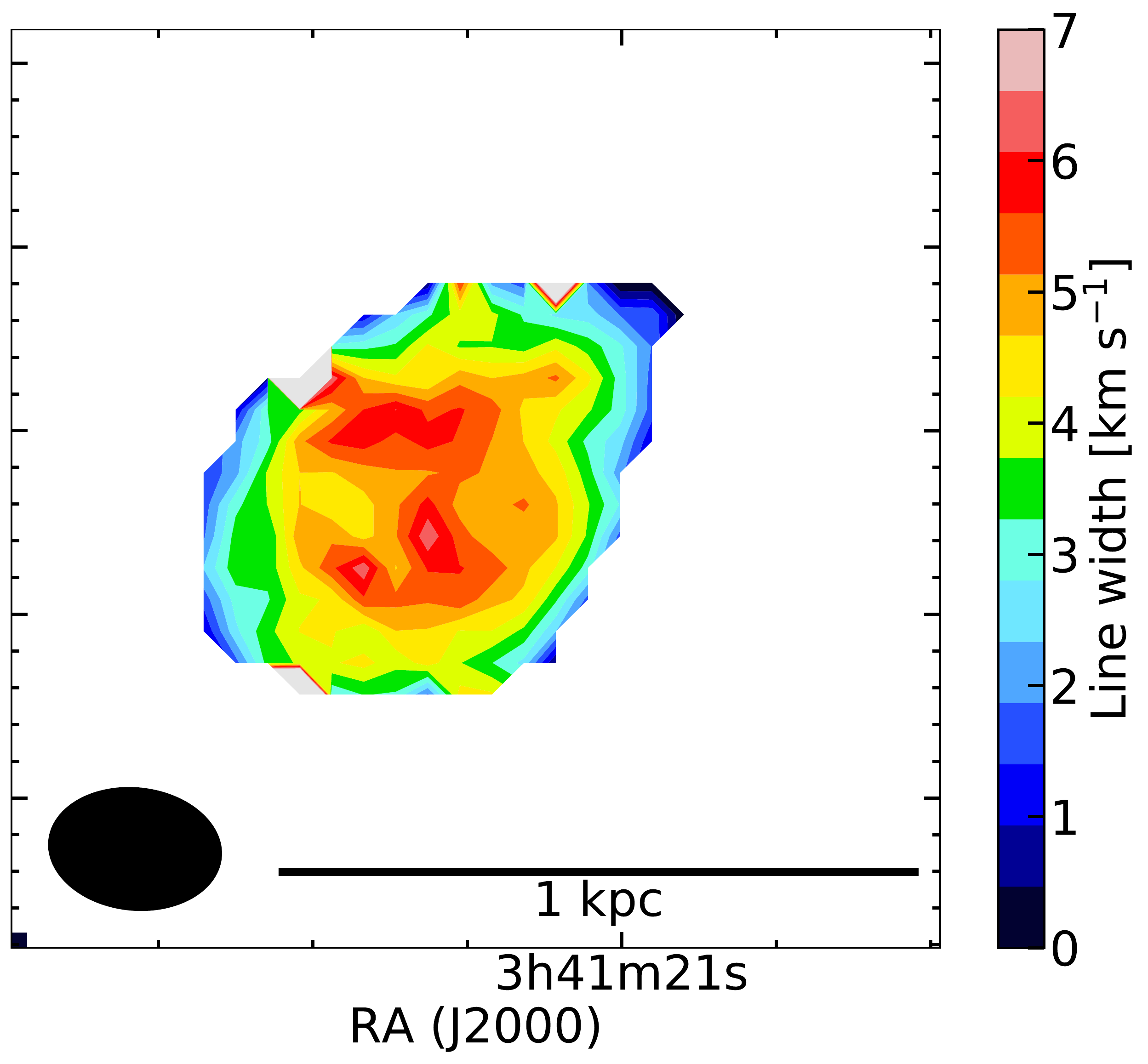}}
		
		
	\subfloat[]
		{\includegraphics[height=0.39\textwidth]{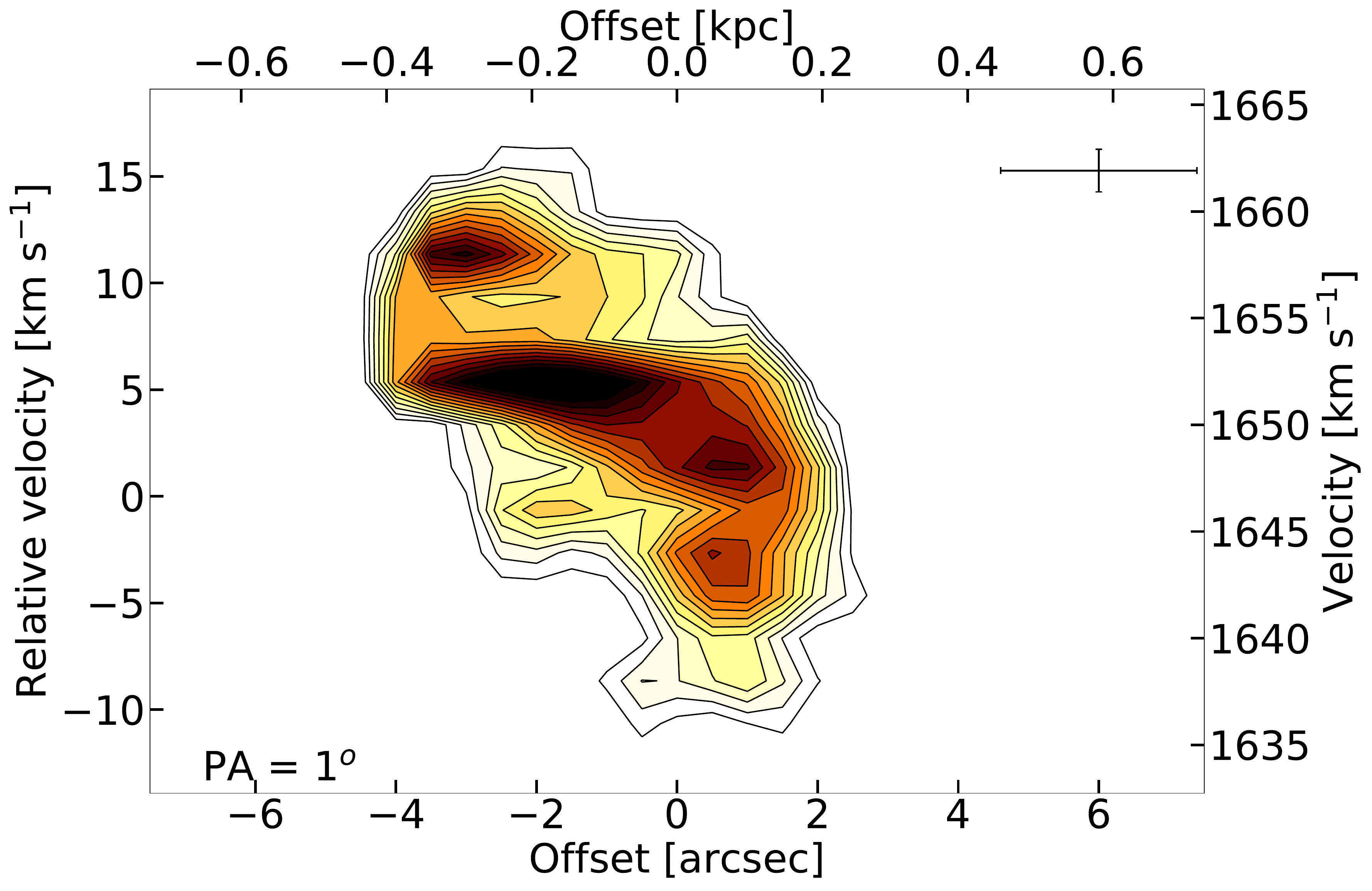}}	
	\hspace{6mm}
	\subfloat[]
		{\includegraphics[height=0.355\textwidth]{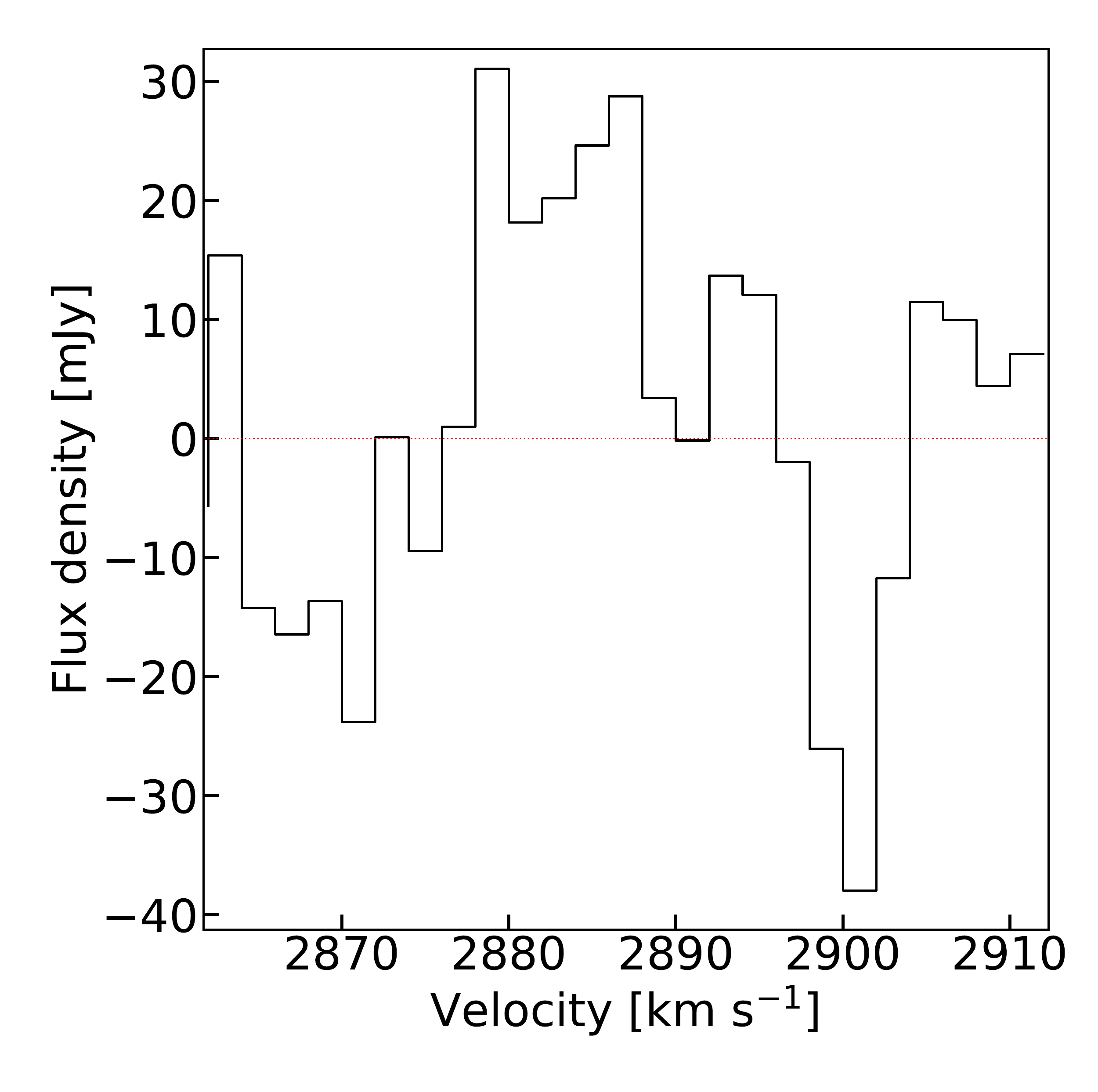}}
		
		
	\caption{FCCFCC261, similar to Figure \ref{fig:NGC1351A}, except that the velocity channels are 2 \kms wide.}
	\label{fig:FCCFCC261}
\end{figure*}

FCC261 (Figures \ref{fig:FCCFCC261} and \ref{subfig:overplot_FCC261}, FCC282 (Figures \ref{fig:FCC282} and \ref{subfig:overplot_FCC207}), and FCC332 (Figures \ref{fig:FCC332} and \ref{subfig:overplot_FCC332}) are small galaxies with molecular gas that is located off-centre (this is especially evident in FCC282). In FCC261 and FCC282 some rotation can be identified, although the kinematics of the molecular gas in all three galaxies are disturbed. In these galaxies the disturbance is reflected in the optical images as well. The stellar body of FCC261 is asymmetric, and there appears to be some colour gradient. The peak of the CO emission is located next to the peak of the optical emission. Something similar is seen in FCC282, where the CO emission peaks at the south-east of the galaxy, which appears bluer in the three-colour image and possibly shows the presence of dust. The stellar body of FCC332 looks slightly more regular, and its CO velocity map more disturbed. 

\subsubsection{NGC1436 (FCC290)}
\label{subsub:NGC1436}
NGC1436 (Figures \ref{fig:NGC1436} and \ref{subfig:overplot_NGC1436}) is a good example of a late-type galaxy. The patchy nature of the CO emission is very clear, showing the resolved GMAs in this galaxy. The gas rotates regularly. Because of the fragmentary distribution of the molecular gas, the PVD in this case was obtained using the full width of the galaxy.

\subsubsection{NGC1437B (FCC308)}
\label{subsub:NGC1437B}
Although the morphology and kinematics of NGC1437B (Figure \ref{fig:NGC1437B} and \ref{subfig:overplot_NGC1437B}) are disturbed, a rotation in the molecular gas can still be identified, and it is the least H$_2$ deficient of the disturbed galaxies (see Figure \ref{fig:gas_fraction}). This galaxy is discussed in more detail in \Section \ref{sub:RPS}.

\subsubsection{PGC013571}
\label{subsub:PGC013571}
\begin{figure*}

	\centering

	\subfloat[]
	{\hspace{-5mm}\includegraphics[height=0.35\textwidth ]{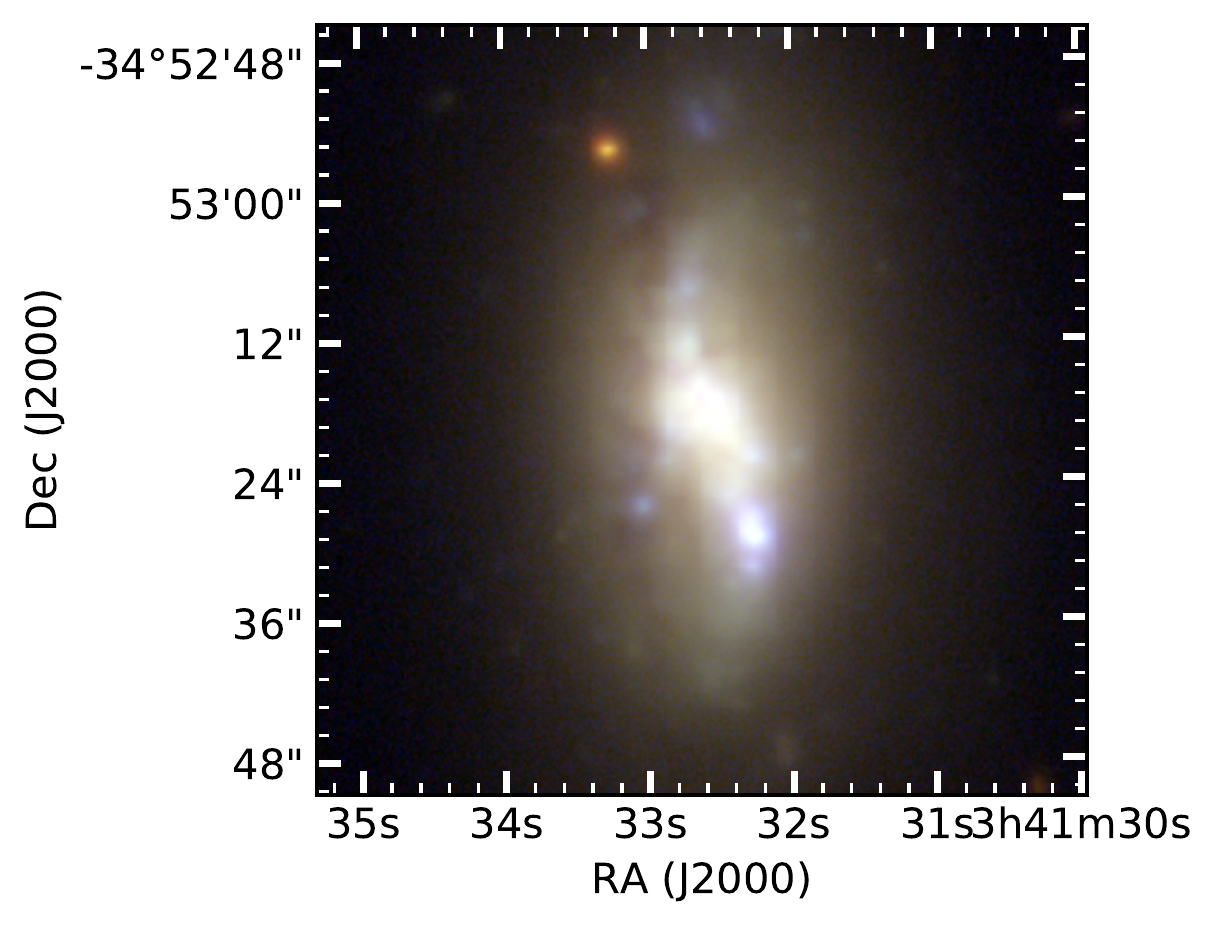}}	
	\hspace{-4mm}
	\subfloat[]
		{\includegraphics[height=0.35\textwidth]{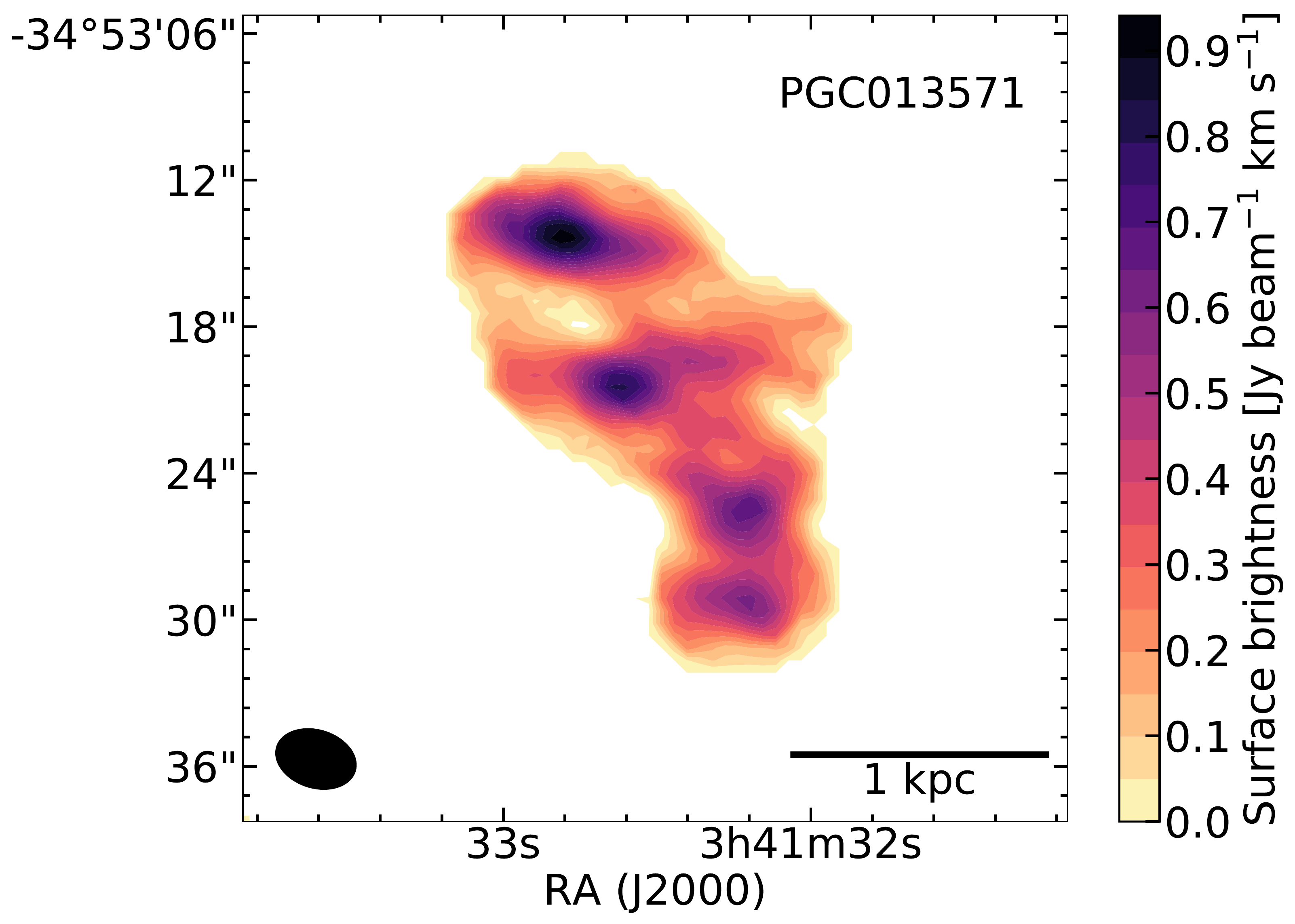}}	
	
	
	\subfloat[]
		{\hspace{-5mm}\includegraphics[height=0.35\textwidth]{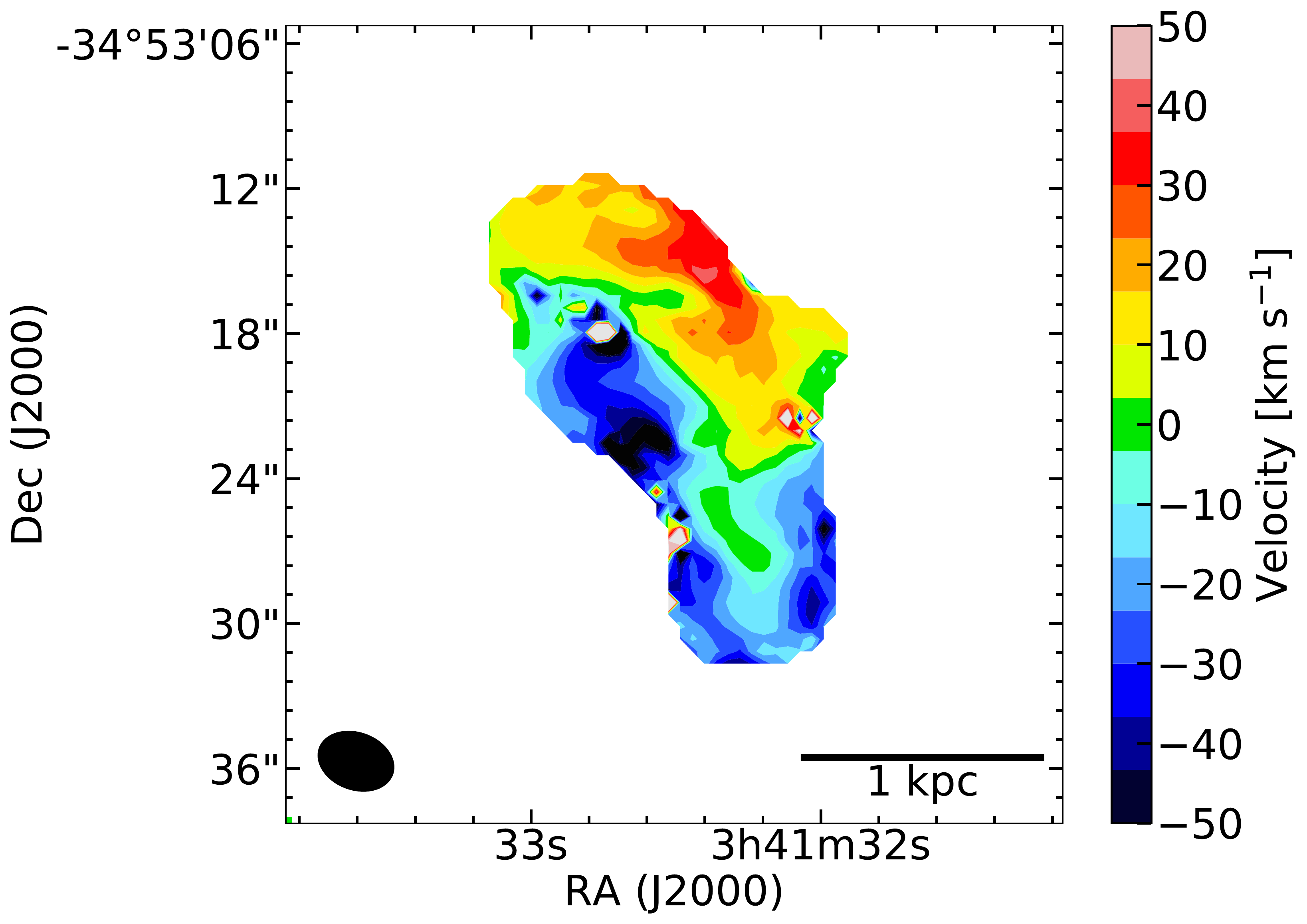}}
	\hspace{7mm}
	\subfloat[]
		{\includegraphics[height=0.35\textwidth]{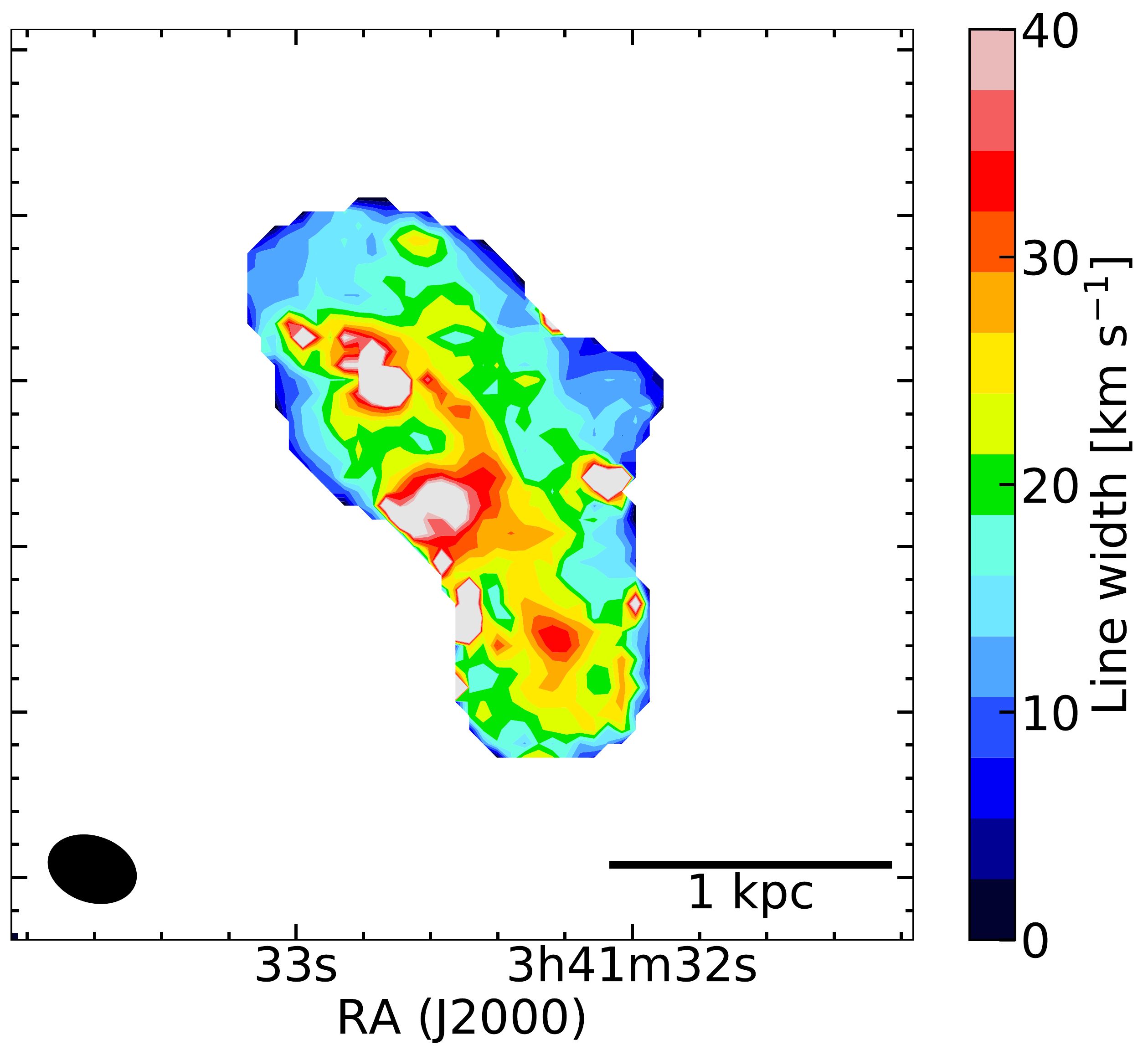}}
		
		
	\subfloat[]
		{\includegraphics[height=0.39\textwidth]{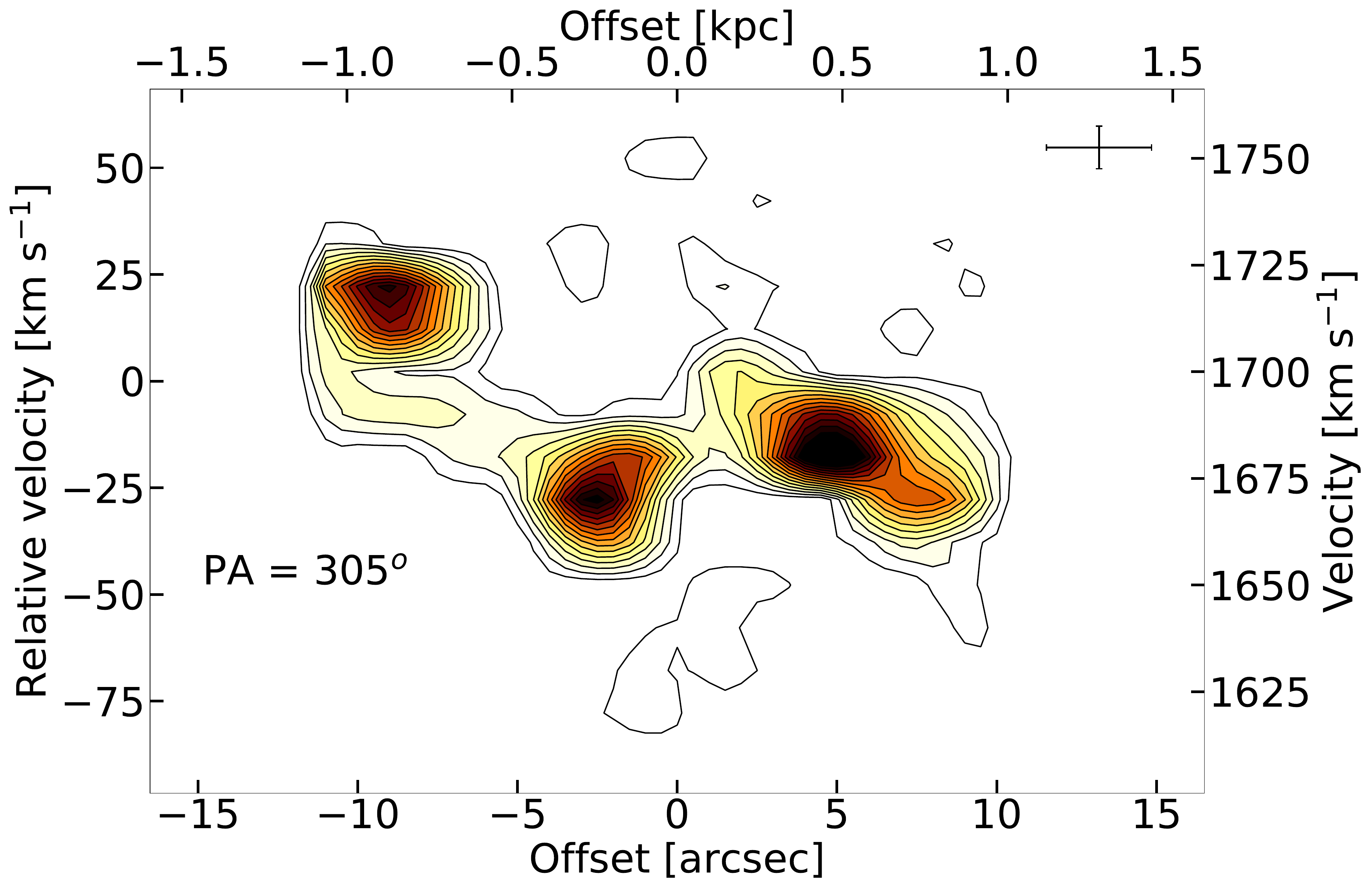}}	
	\hspace{0mm}
	\subfloat[]
		{\includegraphics[height=0.355\textwidth]{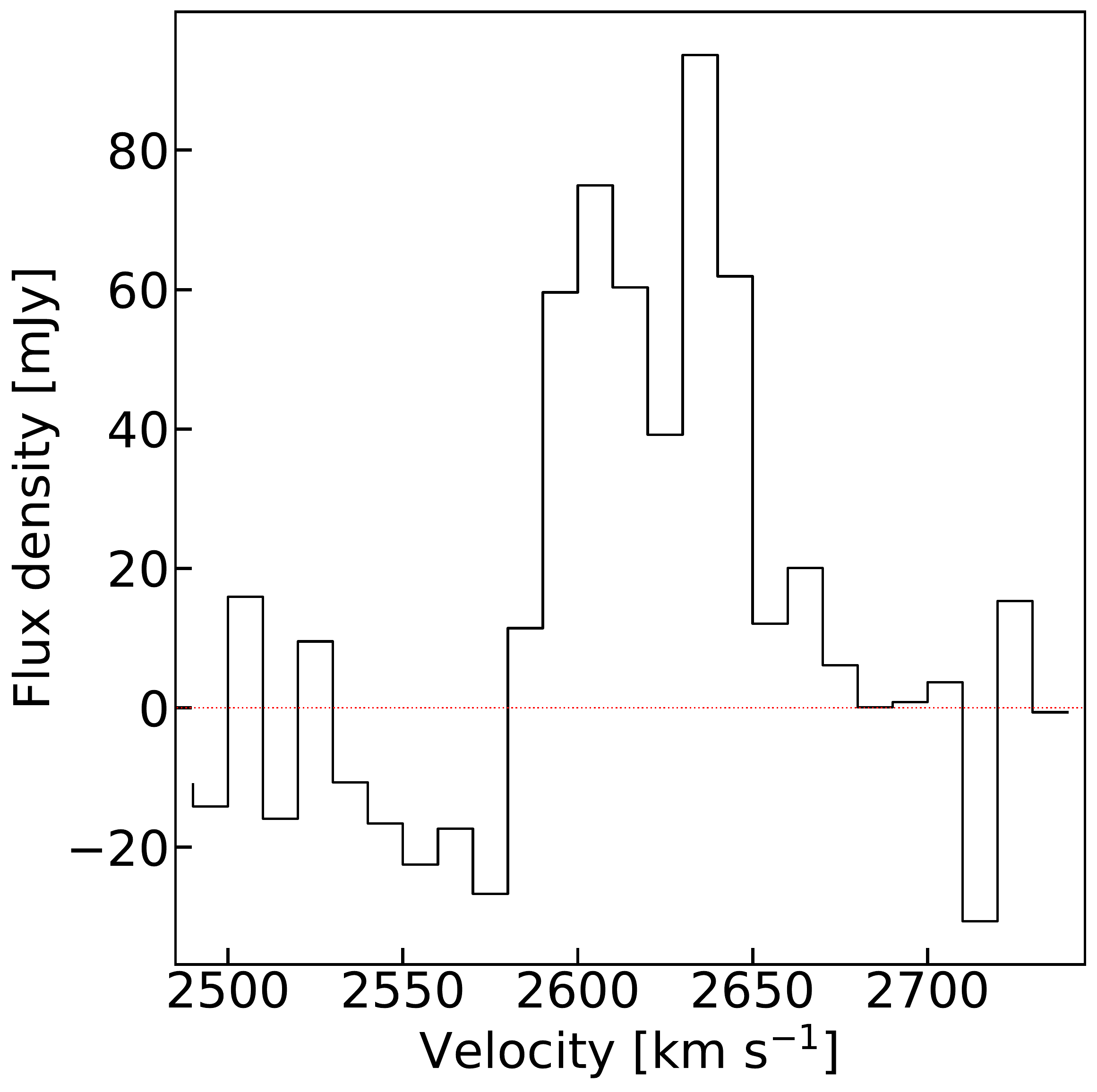}}
		
		
	\caption{PGC013571, similar to Figure \ref{fig:NGC1351A}.}
	\label{fig:PGC013571}
\end{figure*}

\begin{figure*}

	\centering

	\subfloat[]
	{\hspace{-7mm}\includegraphics[height=0.35\textwidth ]{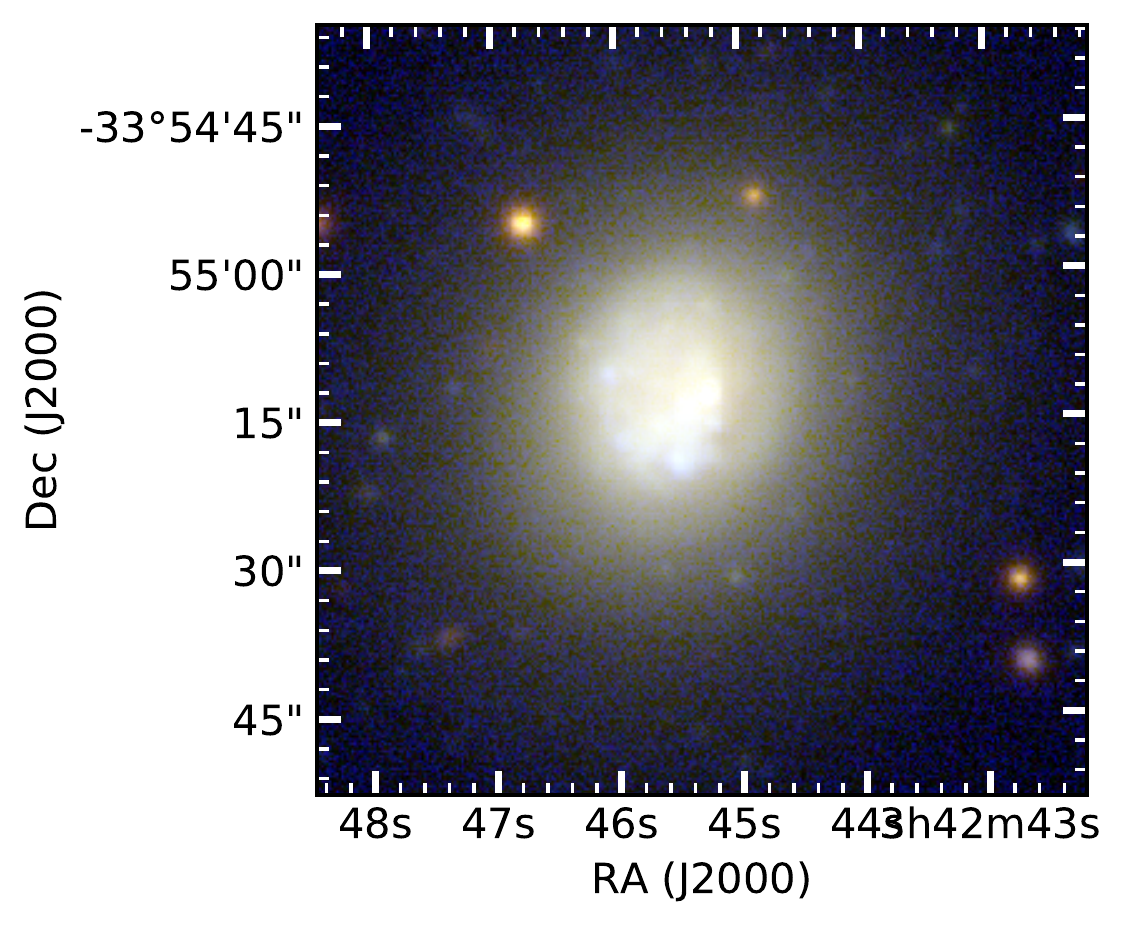}}	
	\hspace{0mm}
	\subfloat[]
		{\includegraphics[height=0.35\textwidth]{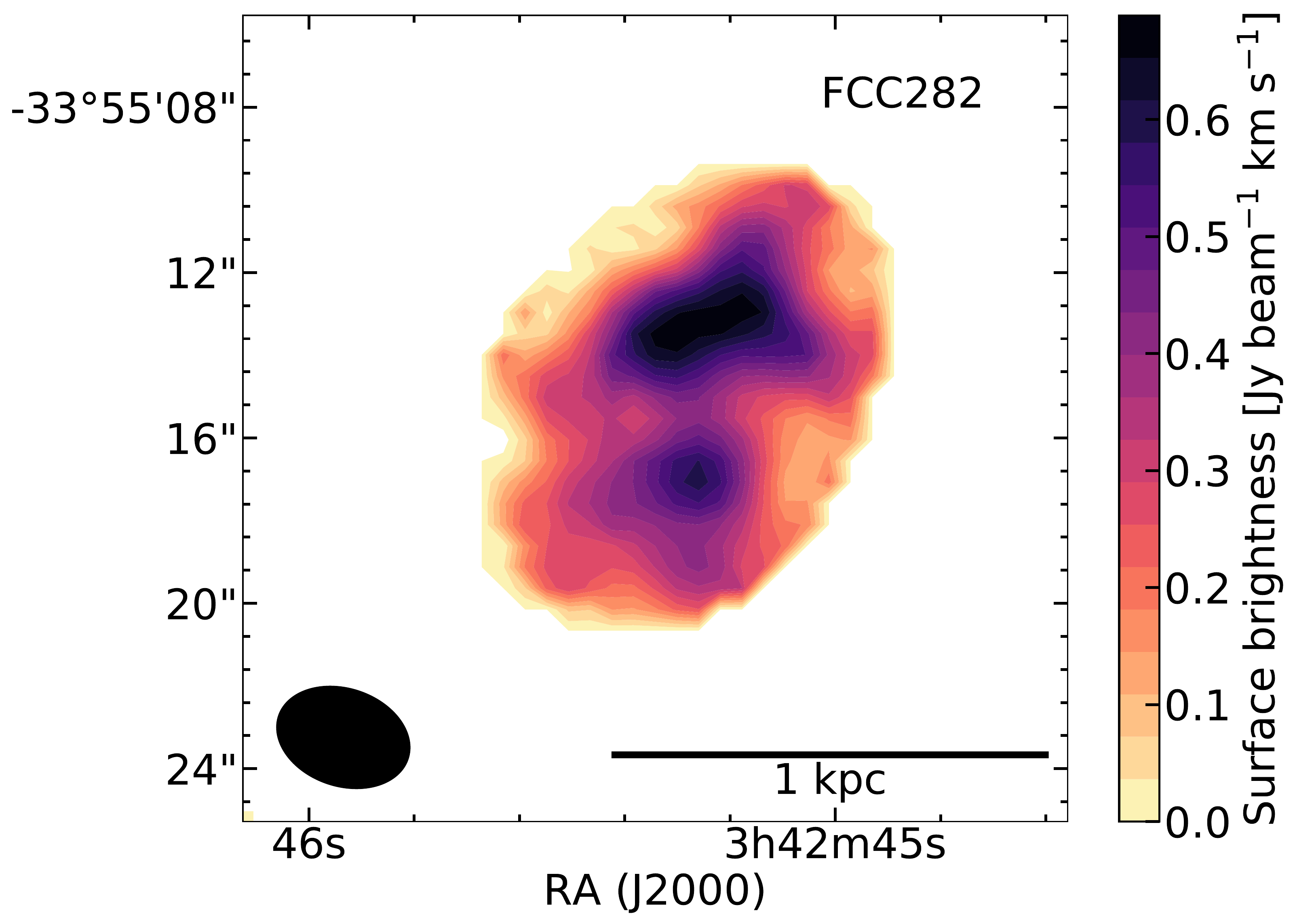}}	
	
	
	\subfloat[]
		{\hspace{-5mm}\includegraphics[height=0.35\textwidth]{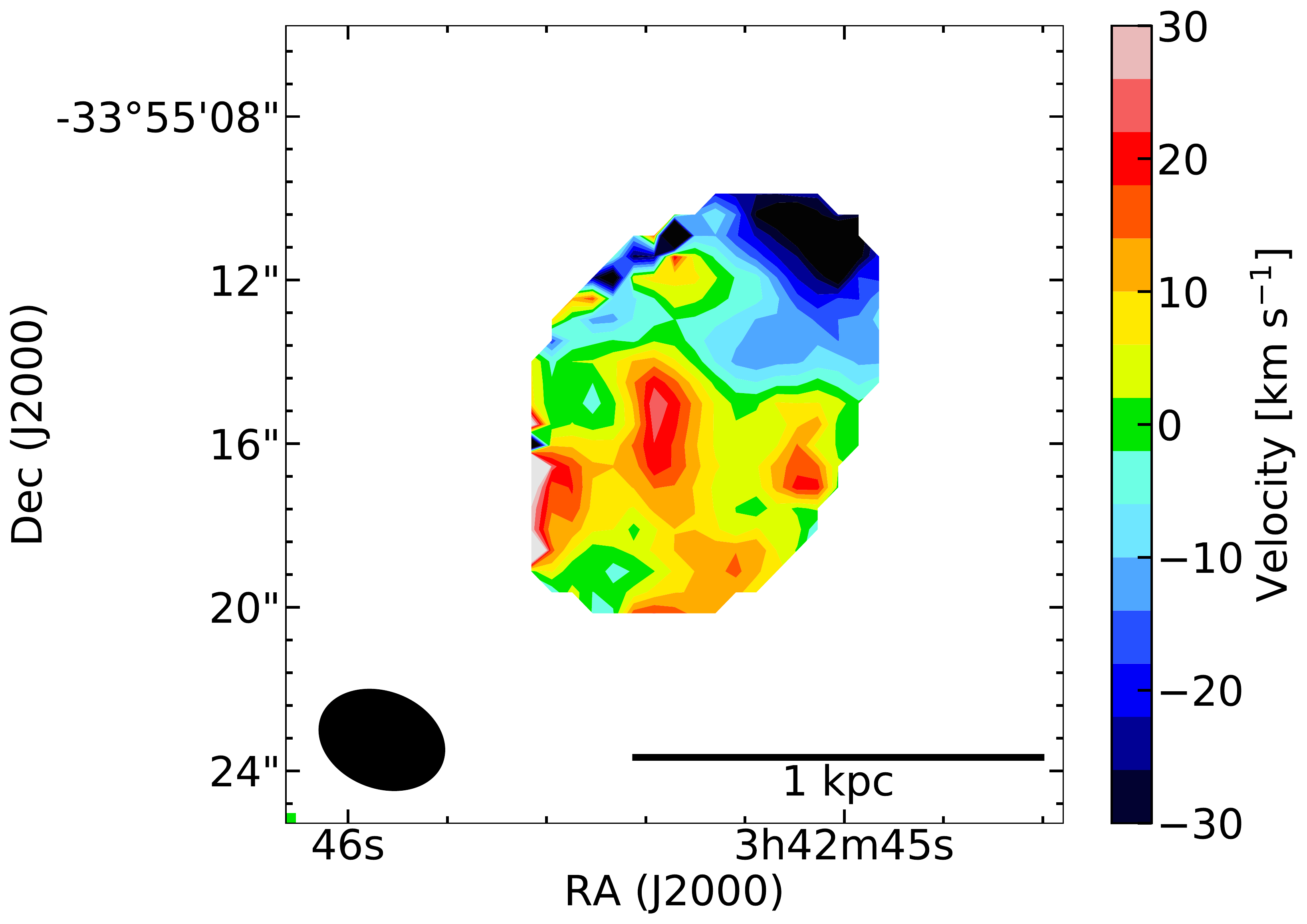}}
	\hspace{7mm}
	\subfloat[]
		{\includegraphics[height=0.35\textwidth]{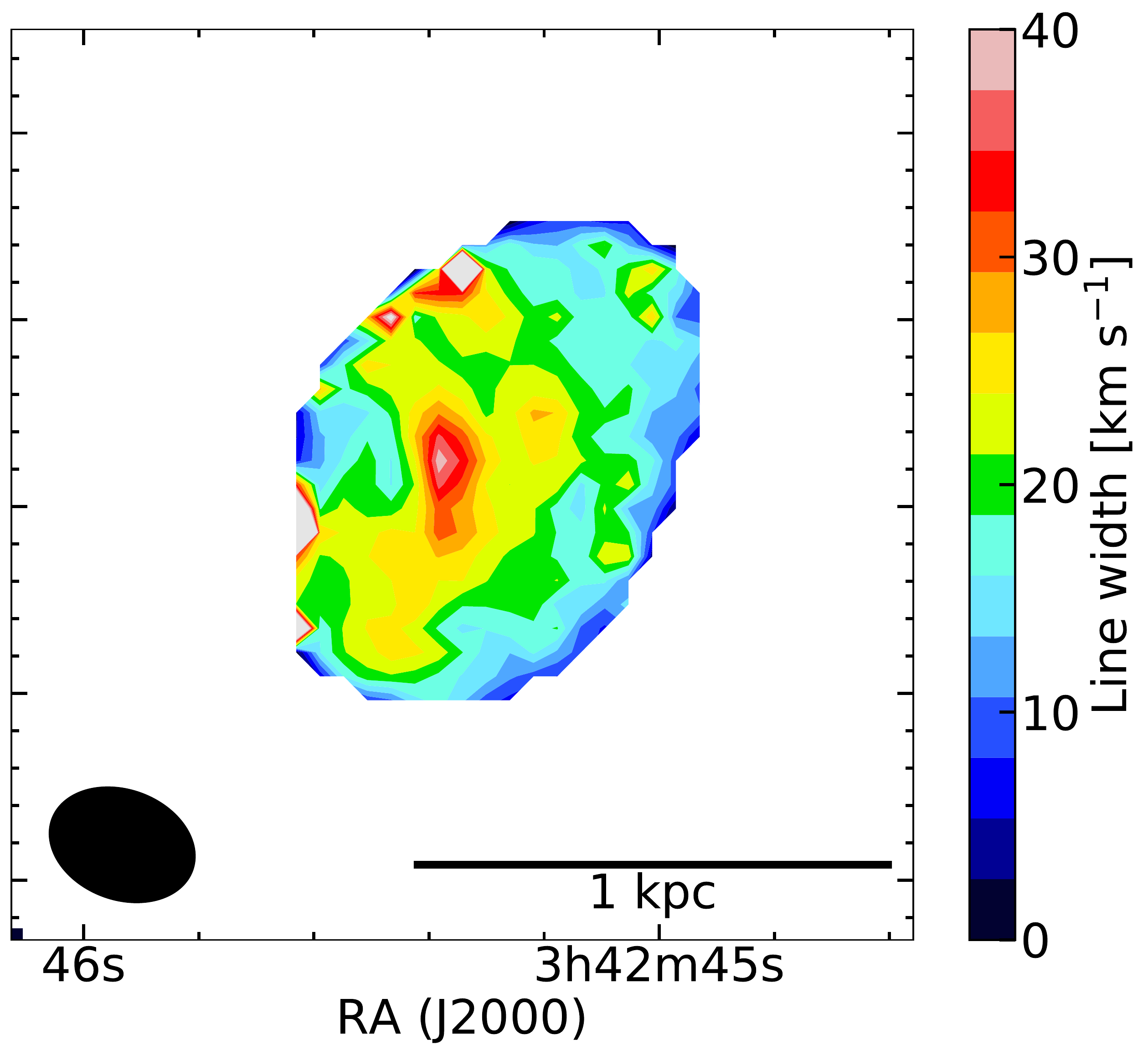}}
		
		
	\subfloat[]
		{\includegraphics[height=0.39\textwidth]{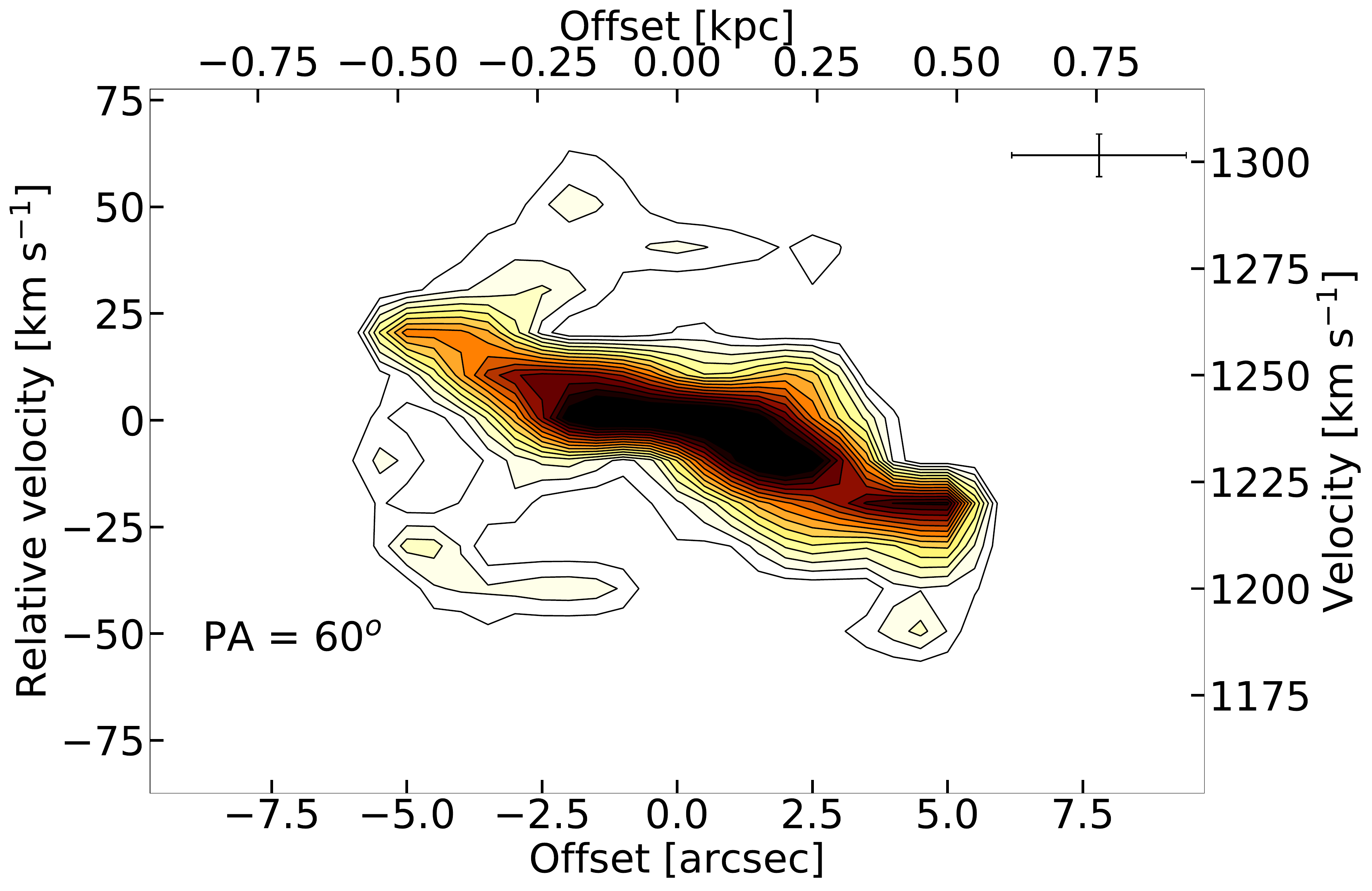}}	
	\hspace{6mm}
	\subfloat[]
		{\includegraphics[height=0.355\textwidth]{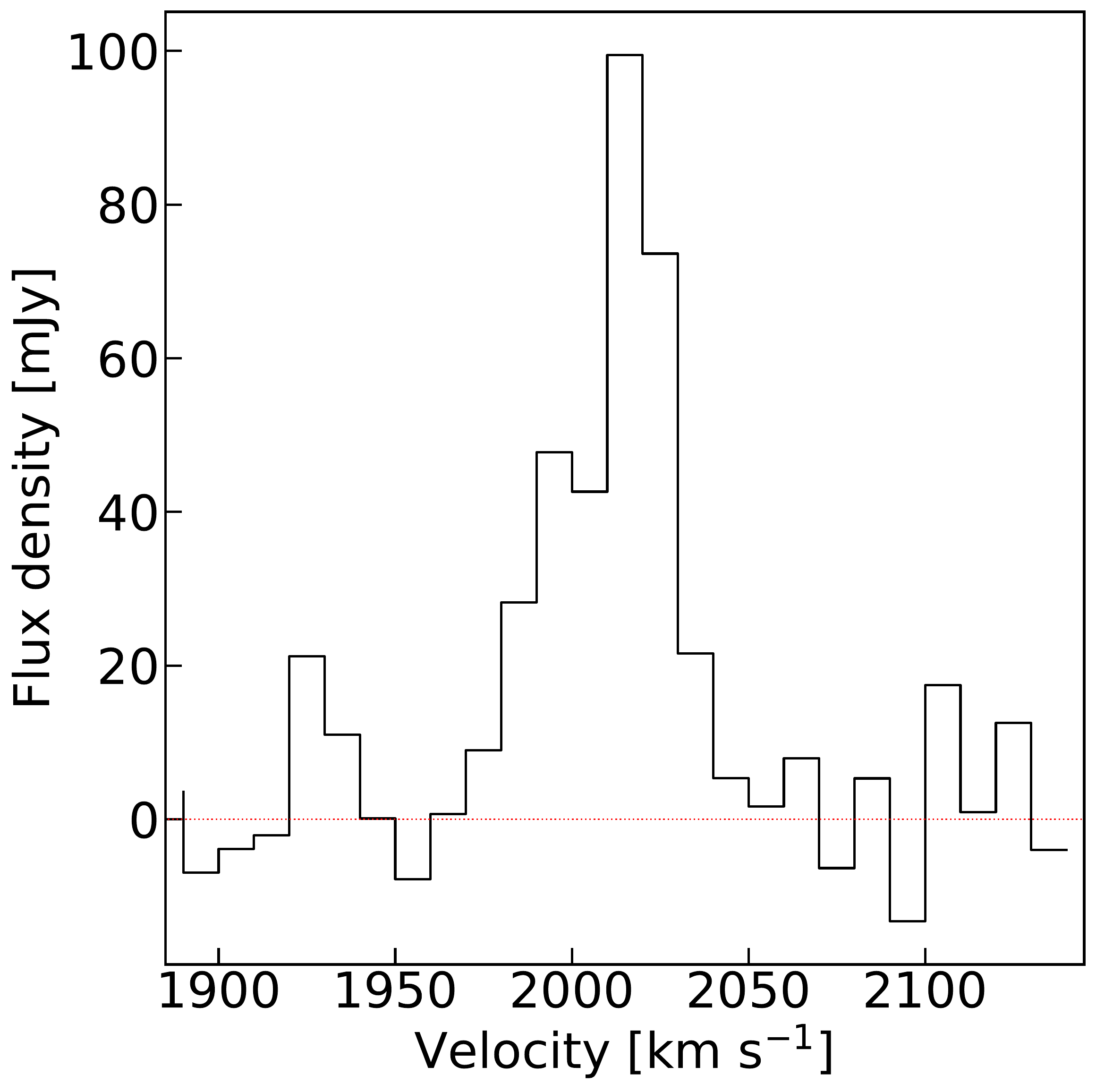}}
		
		
	\caption{FCC282, similar to Figure \ref{fig:NGC1351A}.}
	\label{fig:FCC282}
\end{figure*}

\begin{figure*}

	\centering

	\subfloat[]
	{\hspace{-5mm}\includegraphics[height=0.35\textwidth ]{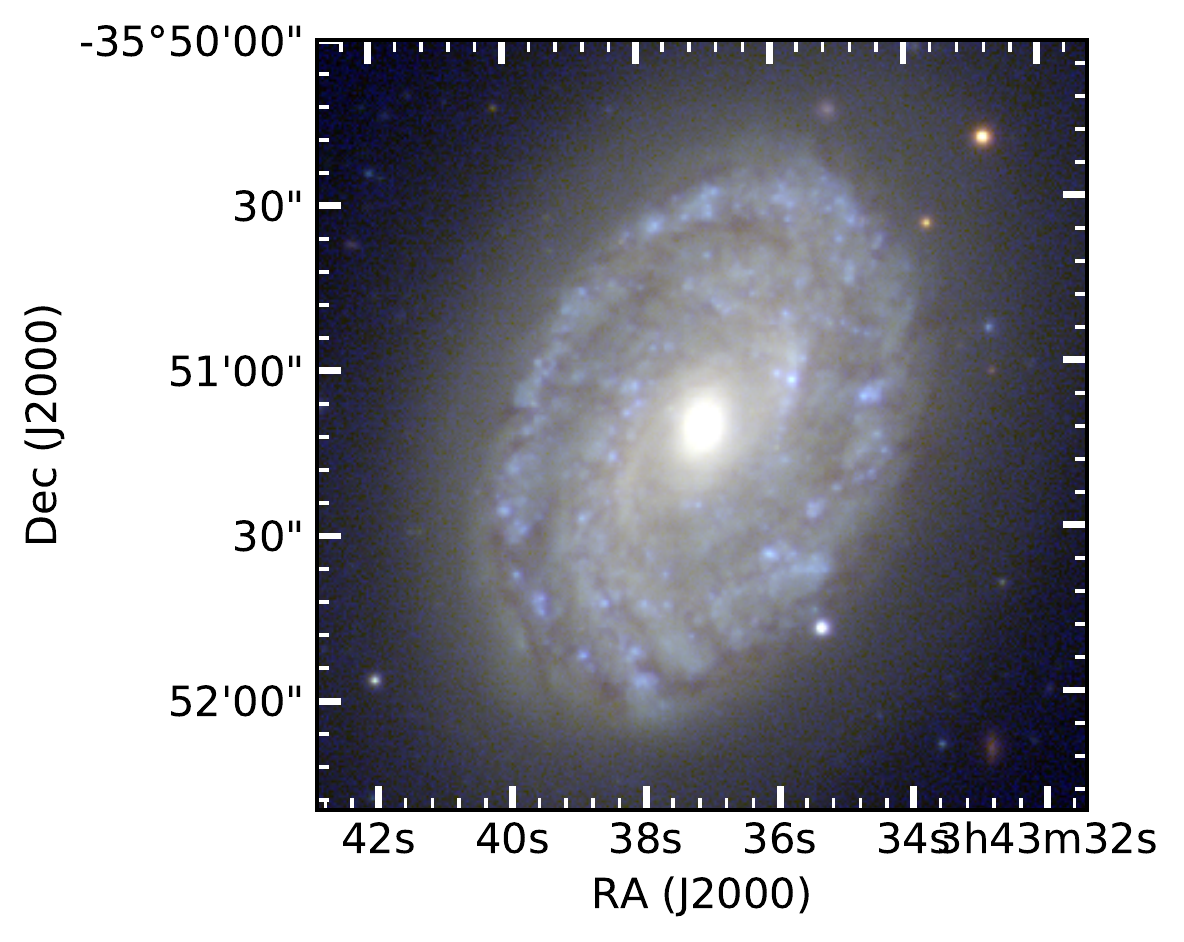}}	
	\hspace{0mm}
	\subfloat[]
		{\includegraphics[height=0.35\textwidth]{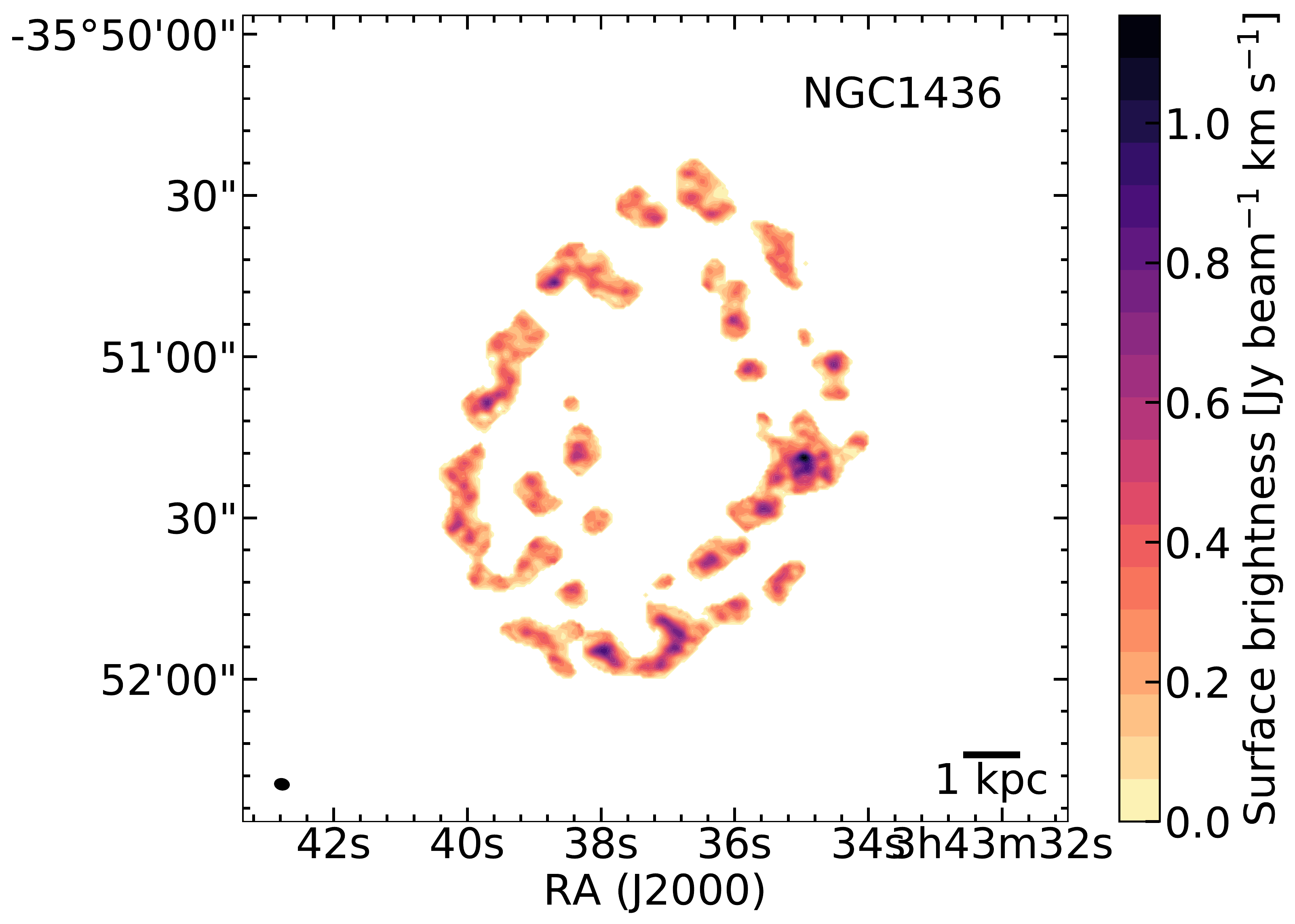}}	
	
	
	\subfloat[]
		{\hspace{-5mm}\includegraphics[height=0.35\textwidth]{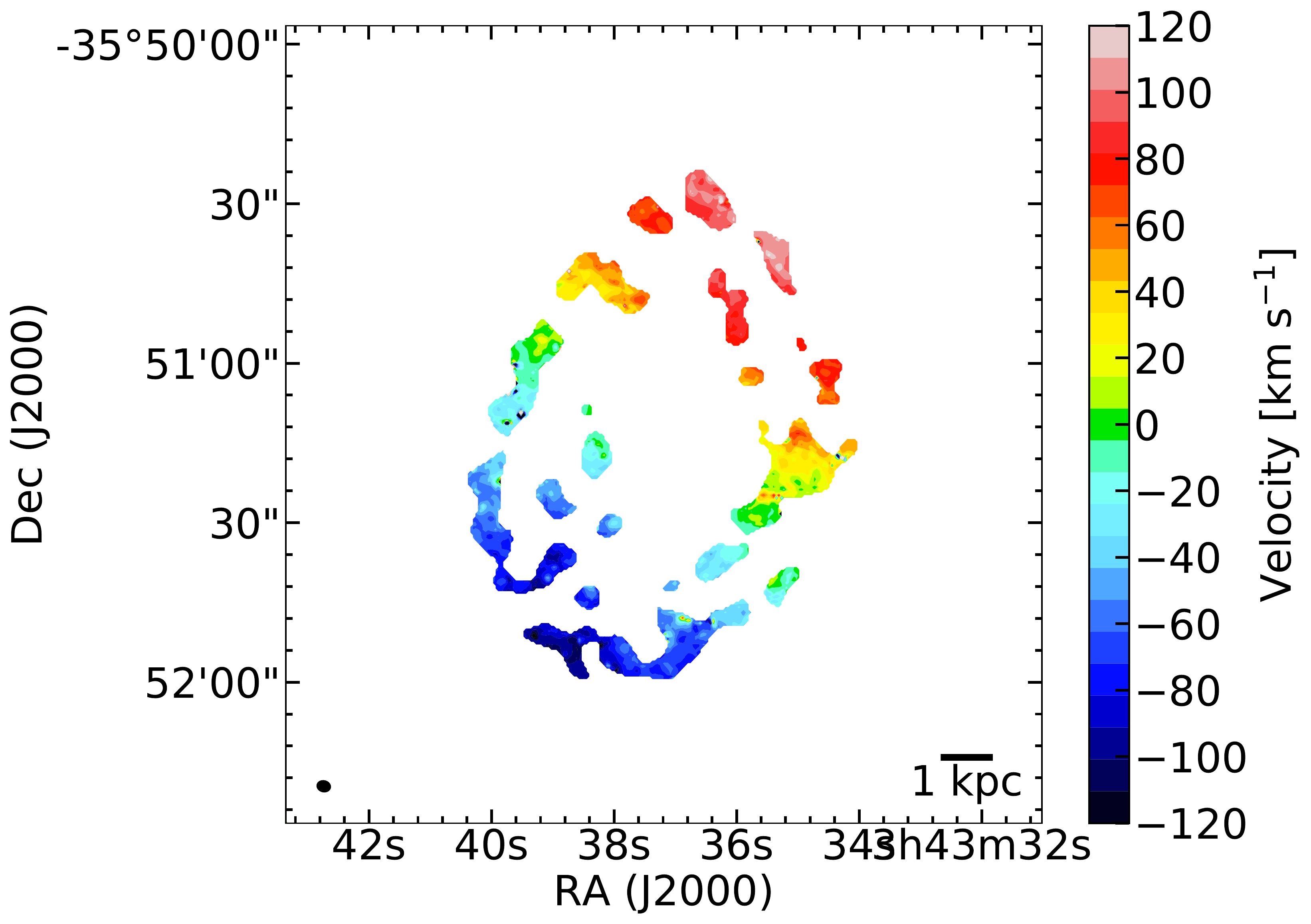}}
	\hspace{7mm}
	\subfloat[]
		{\includegraphics[height=0.35\textwidth]{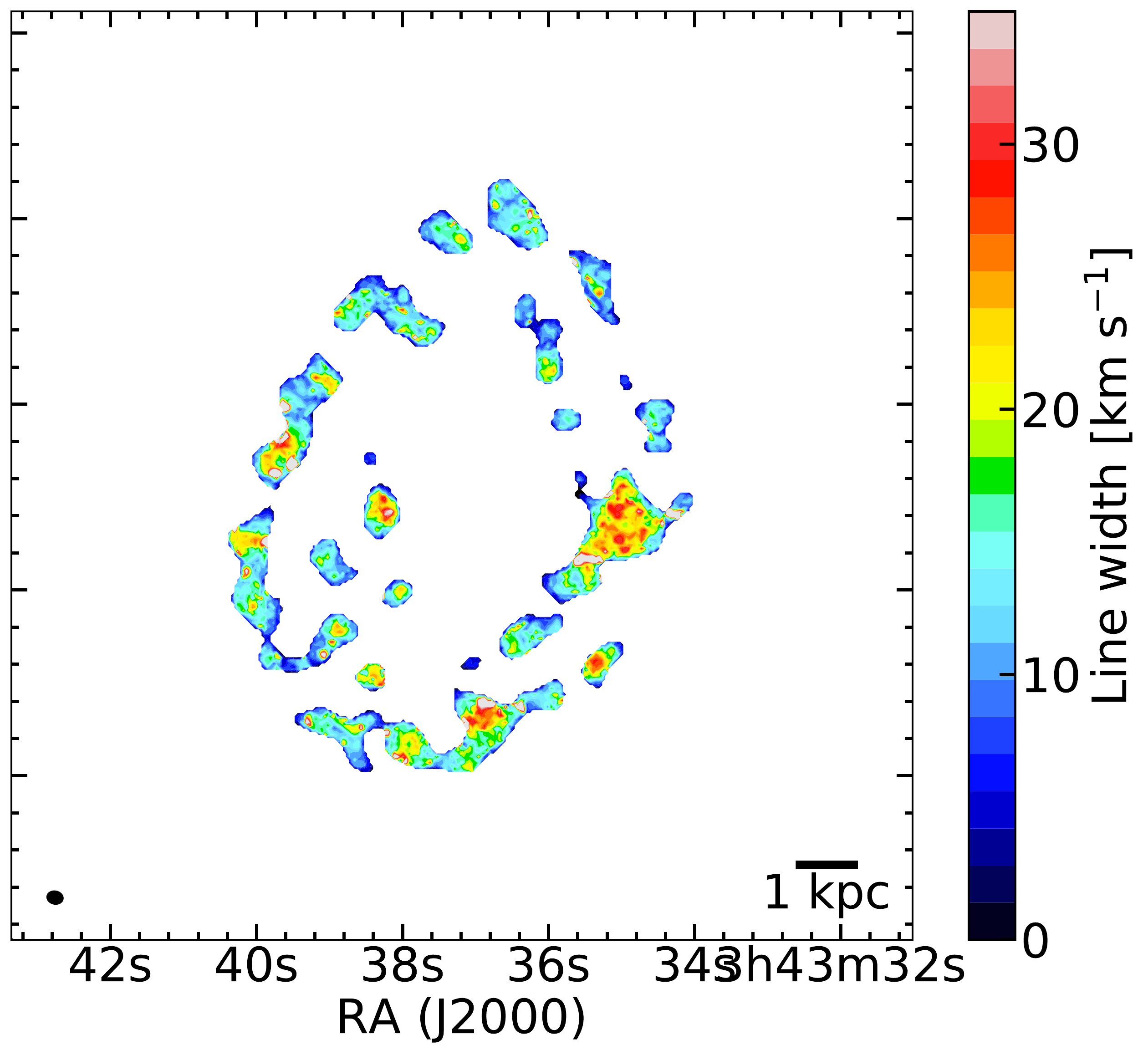}}
		
		
	\subfloat[]
		{\includegraphics[height=0.39\textwidth]{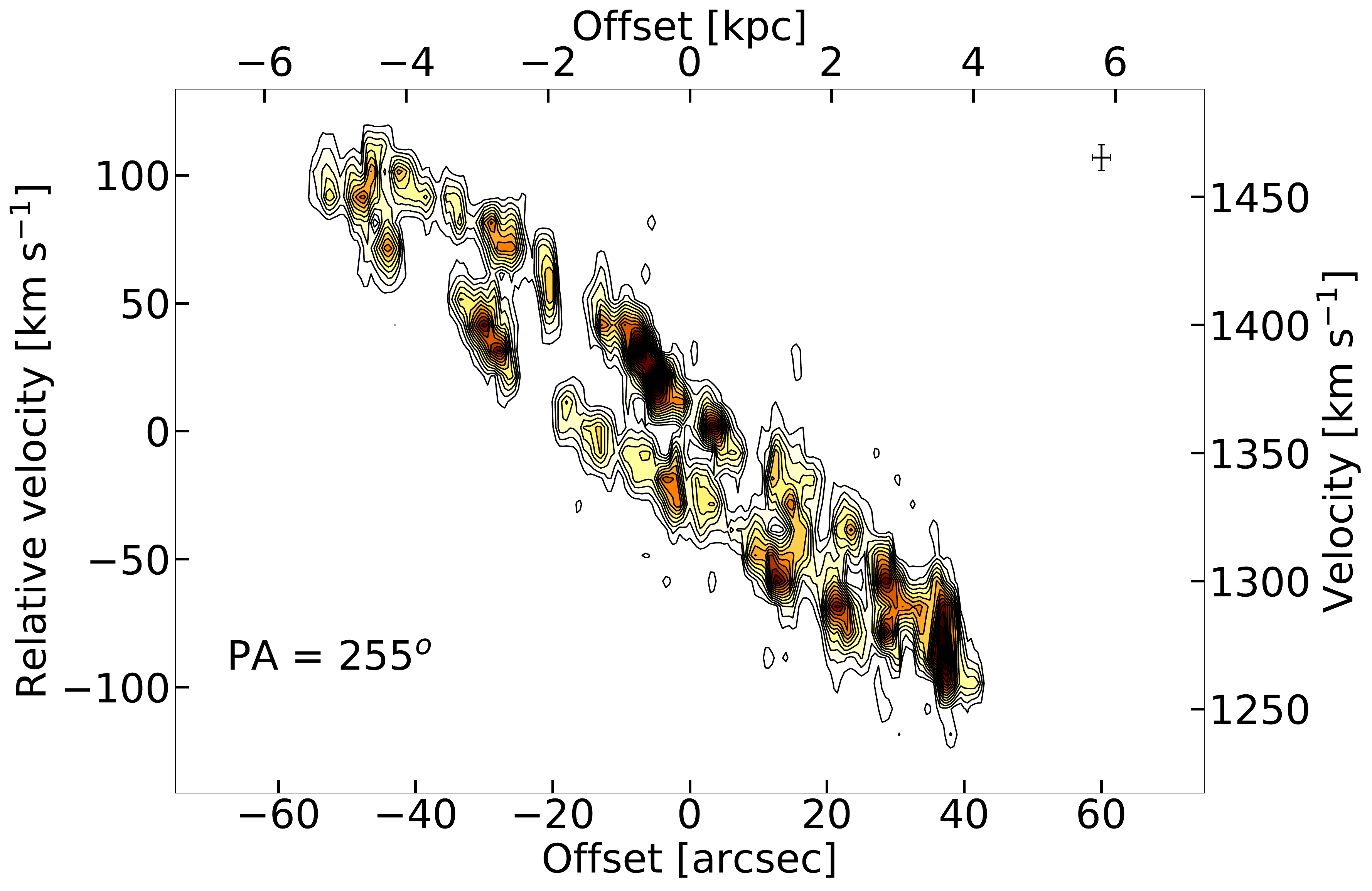}}	
	\hspace{6mm}
	\subfloat[]
		{\includegraphics[height=0.355\textwidth]{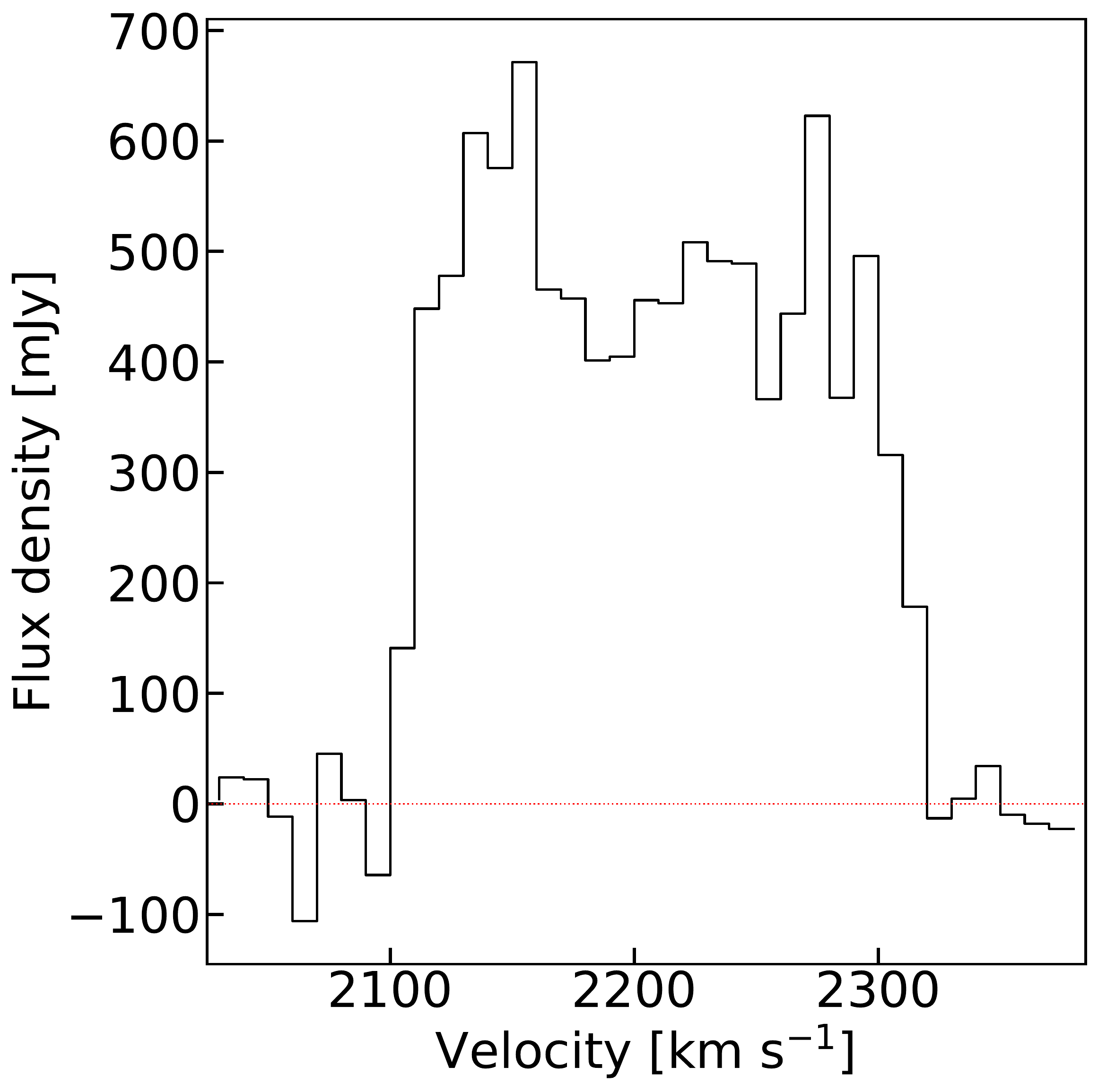}}
		
		
	\caption{NGC1436, similar to Figure \ref{fig:NGC1351A}.}
	\label{fig:NGC1436}
\end{figure*}

\begin{figure*}

	\centering

	\subfloat[]
	{\hspace{-7mm}\includegraphics[height=0.35\textwidth ]{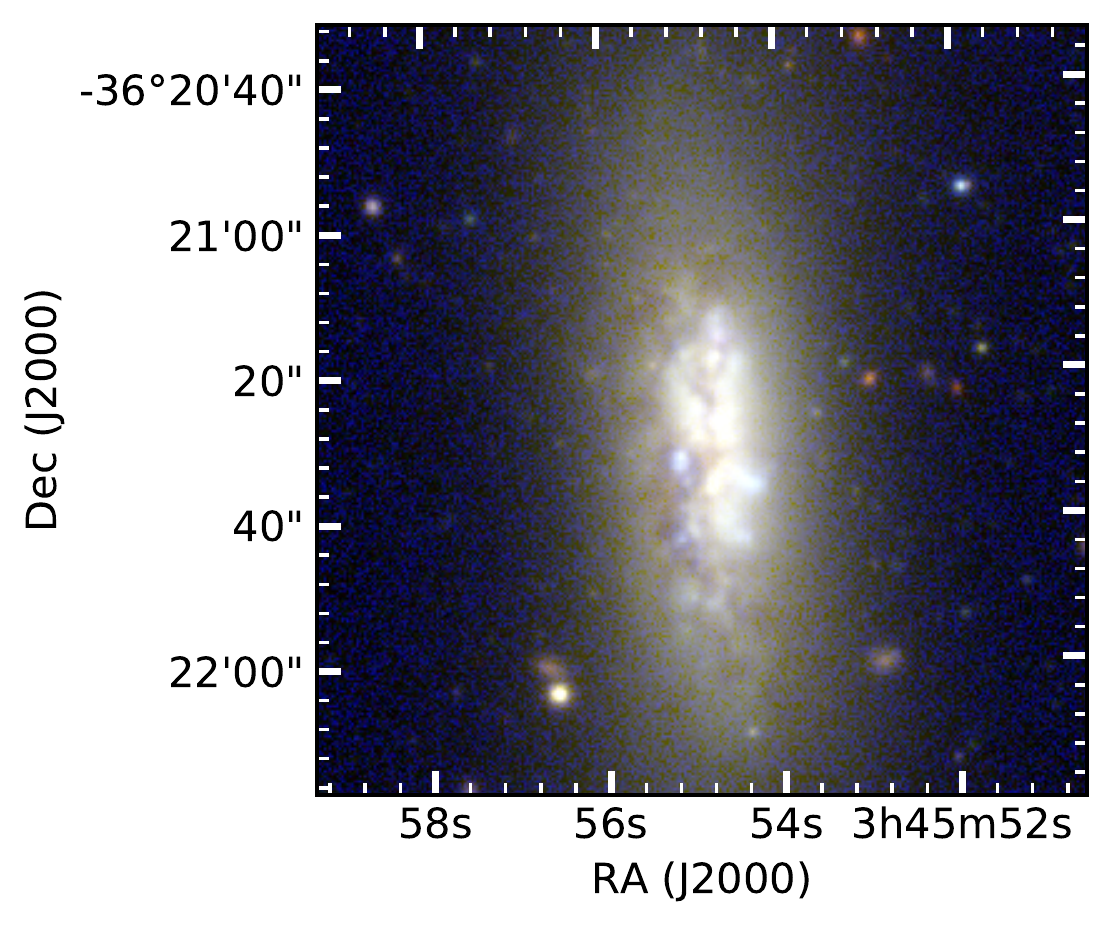}}	
	\hspace{4mm}
	\subfloat[]
		{\includegraphics[height=0.35\textwidth]{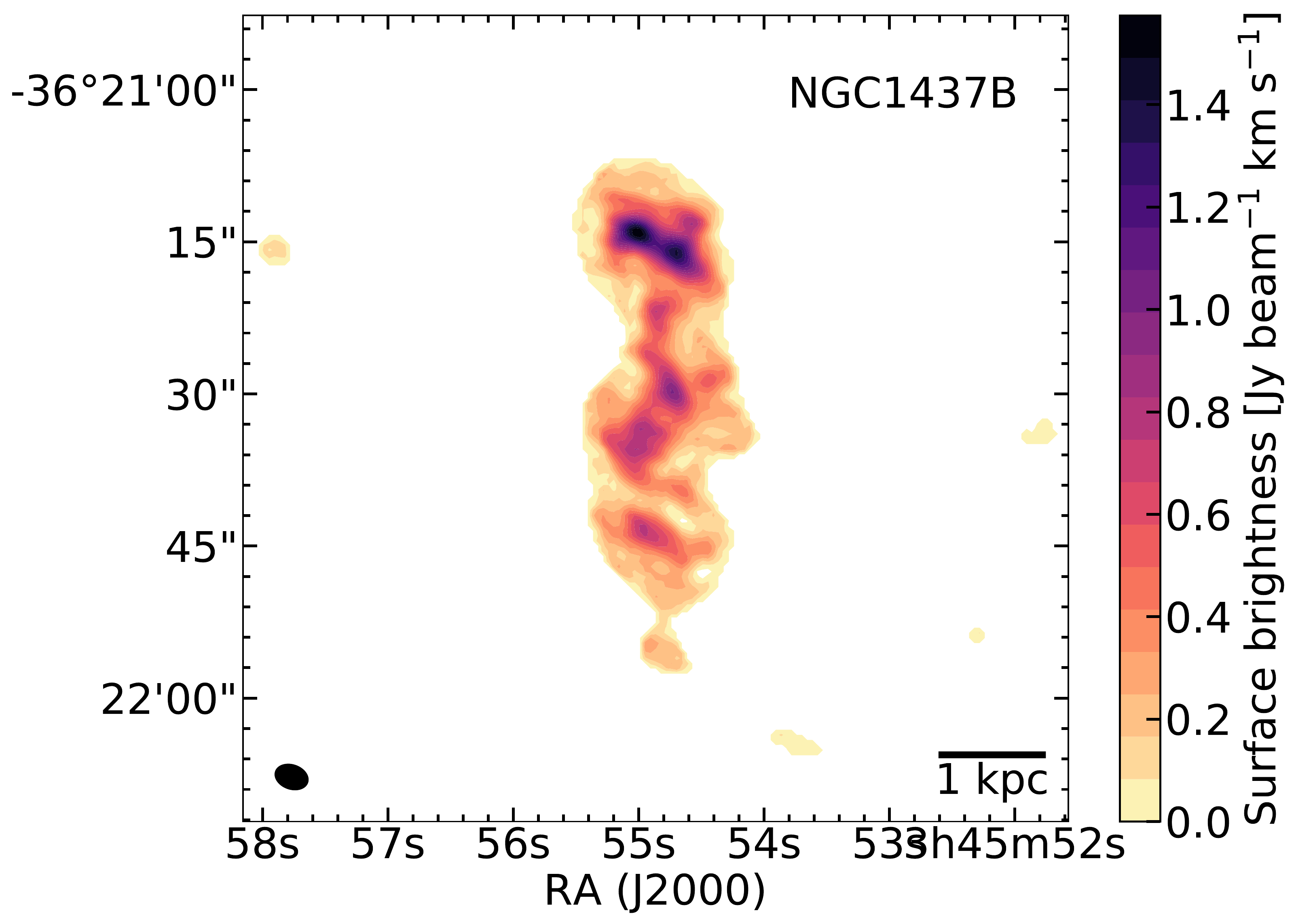}}	
	
	
	\subfloat[]
		{\hspace{-5mm}\includegraphics[height=0.35\textwidth]{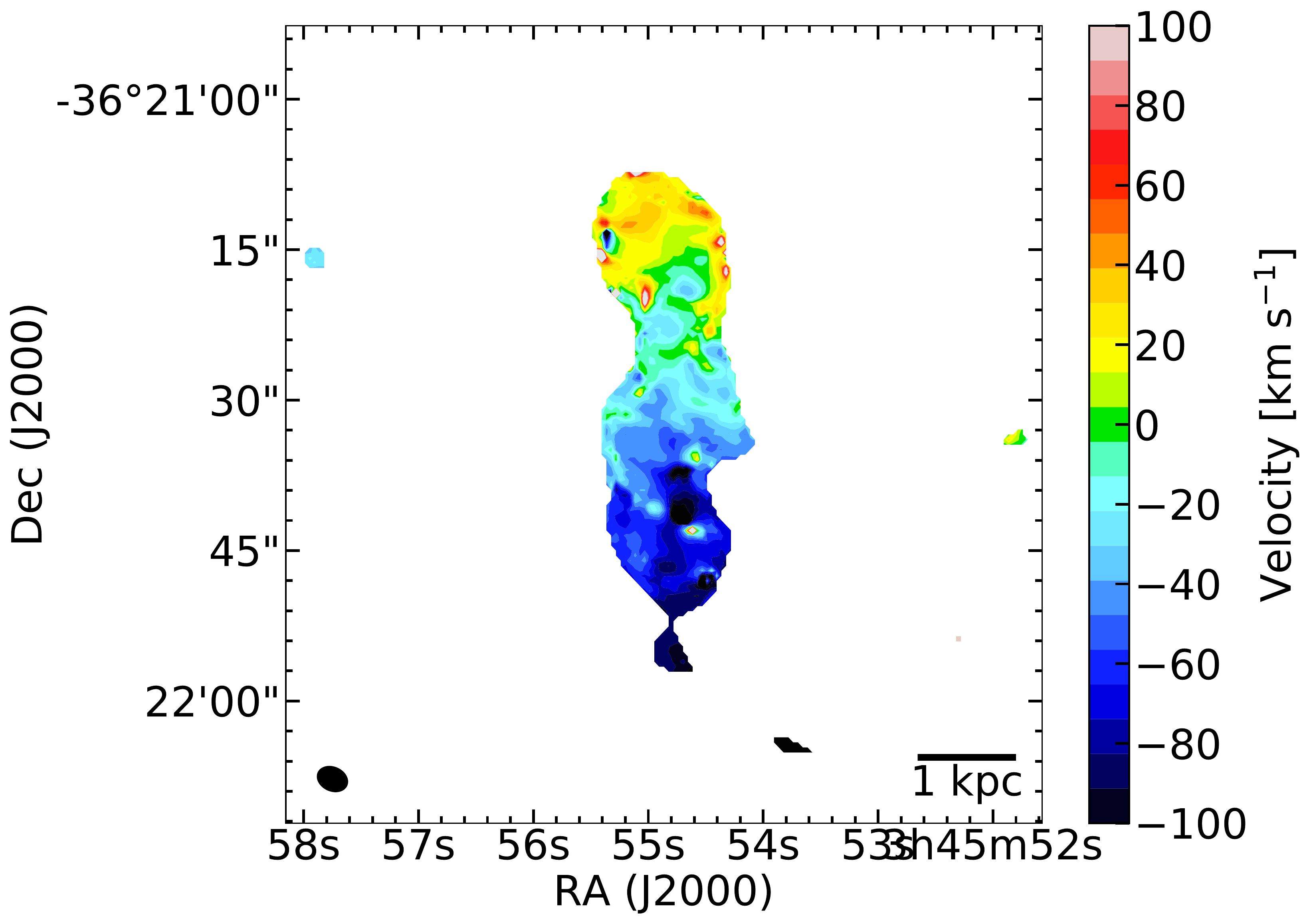}}
	\hspace{7mm}
	\subfloat[]
		{\includegraphics[height=0.35\textwidth]{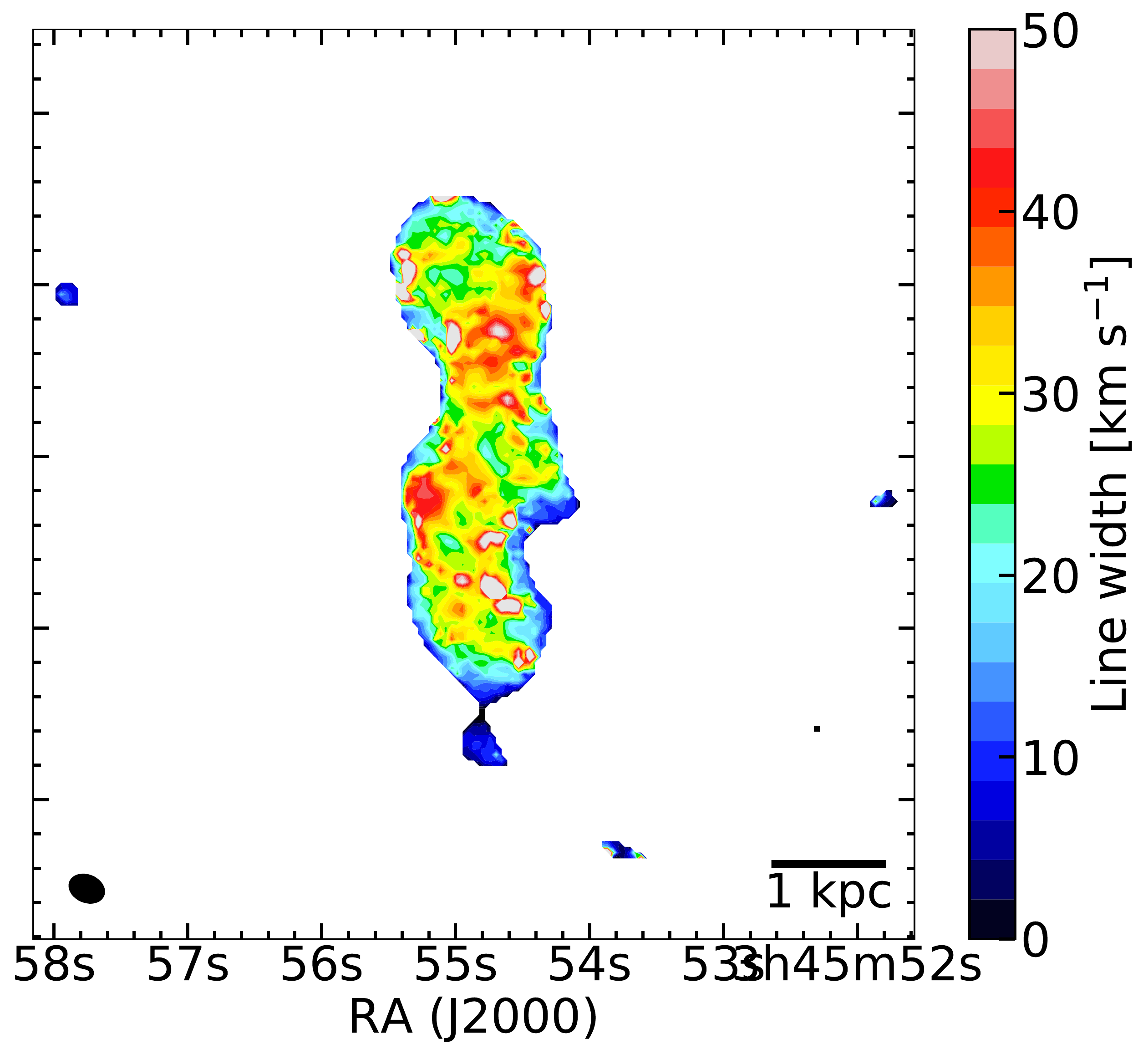}}
		
		
	\subfloat[]
		{\includegraphics[height=0.39\textwidth]{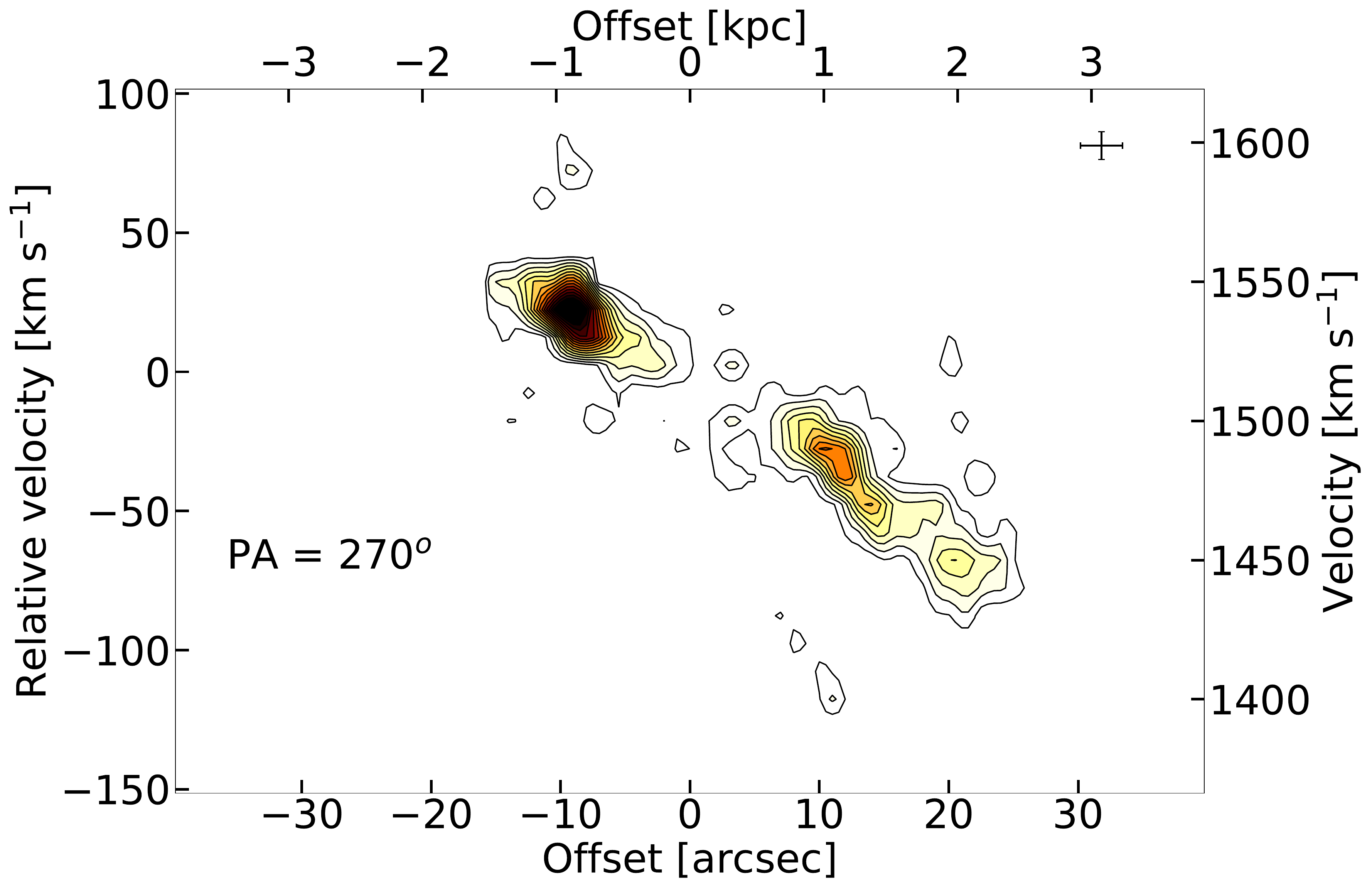}}	
	\hspace{6mm}
	\subfloat[]
		{\includegraphics[height=0.355\textwidth]{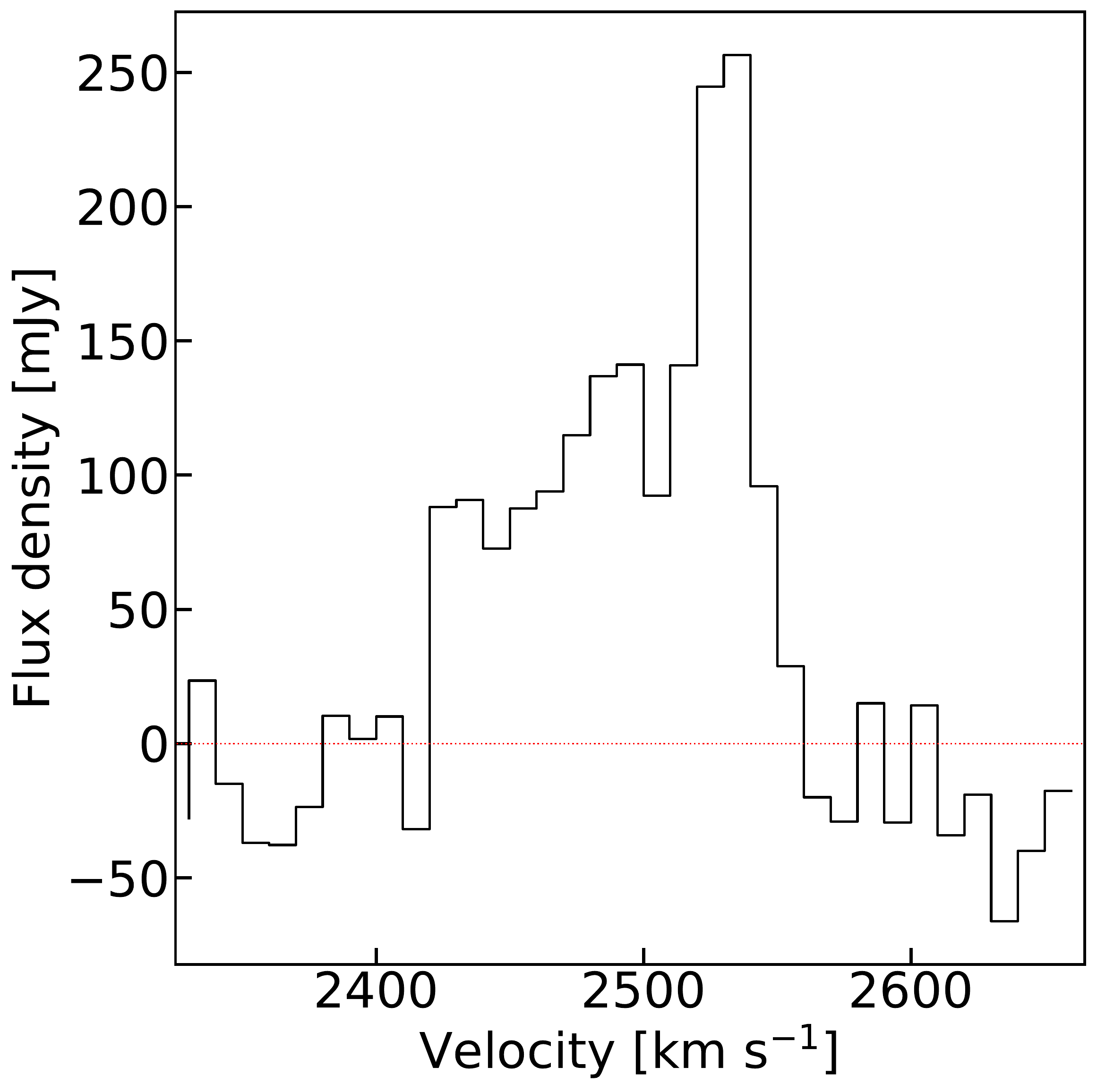}}
		
		
	\caption{NGC1437B, similar to Figure \ref{fig:NGC1351A}.}
	\label{fig:NGC1437B}
\end{figure*}

The velocity map of PGC013571 (see Figure \ref{fig:PGC013571}) seems to consist of more than one rotation around different axes. In the intensity map there are multiple peaks. The PVD of this galaxy consists of multiple velocity structures. There are two possible explanations for these observations. First, they could be an indication that this galaxy has a strong bar/spiral arms. The pattern of the contours in the velocity map and the two components near the centre of the PVD indicate strong bar-like non-circular streaming motions, which is consistent with this interpretation. Similar kinematics are seen in the Virgo cluster spiral NGC4064 \citep{Cortes2006}, which indeed has a bar and two spiral arms, resulting in strong non-circular streaming motions (see Figure 11 in \citealt{Cortes2006}). The irregularity, both in the molecular gas and the stellar body, could have been caused by a tidal interaction or minor merger that caused the strong two-arm spiral disturbance.

Alternatively, it is possible that this galaxy has recently undergone a merging event and is still in the process of relaxing. The velocity map could be interpreted as two different rotating bodies: one rotating along a horizontal axis, and one around a more diagonal axis. In the three-colour image of this galaxy we can see a split in the optical emission at the south side of the galaxy, with two blue ends. Usually mergers in cluster environments are rare, however they can occur in the outskirts.

\clearpage

\section{Moment maps}
\label{app:moment_maps}
This appendix shows three-colour images, moment maps, position-velocity diagrams, and the CO(\textit{J} = 1-0) line for all galaxies detected here.

\begin{figure*}

	\centering

	\subfloat[]
	{\hspace{-6mm}\includegraphics[height=0.35\textwidth ]{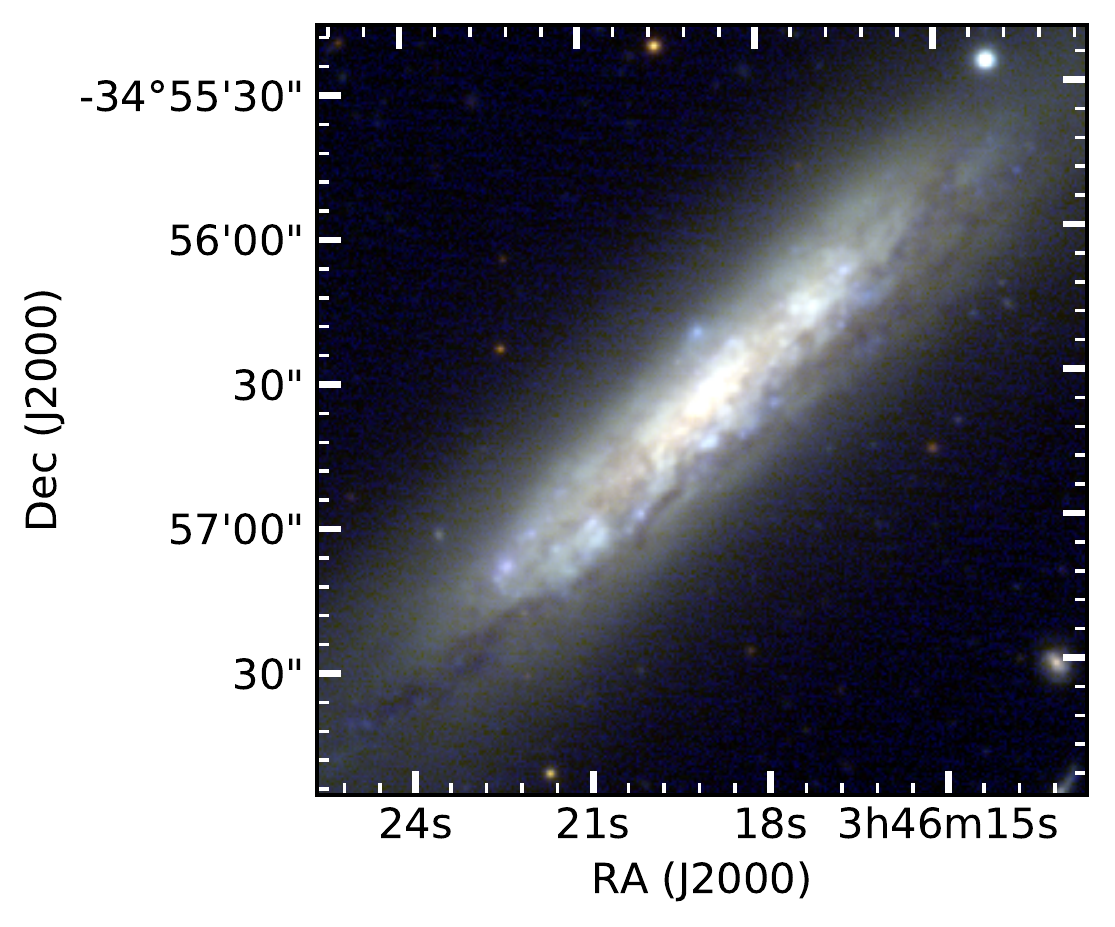}}	
	\hspace{2mm}
	\subfloat[]
		{\includegraphics[height=0.35\textwidth]{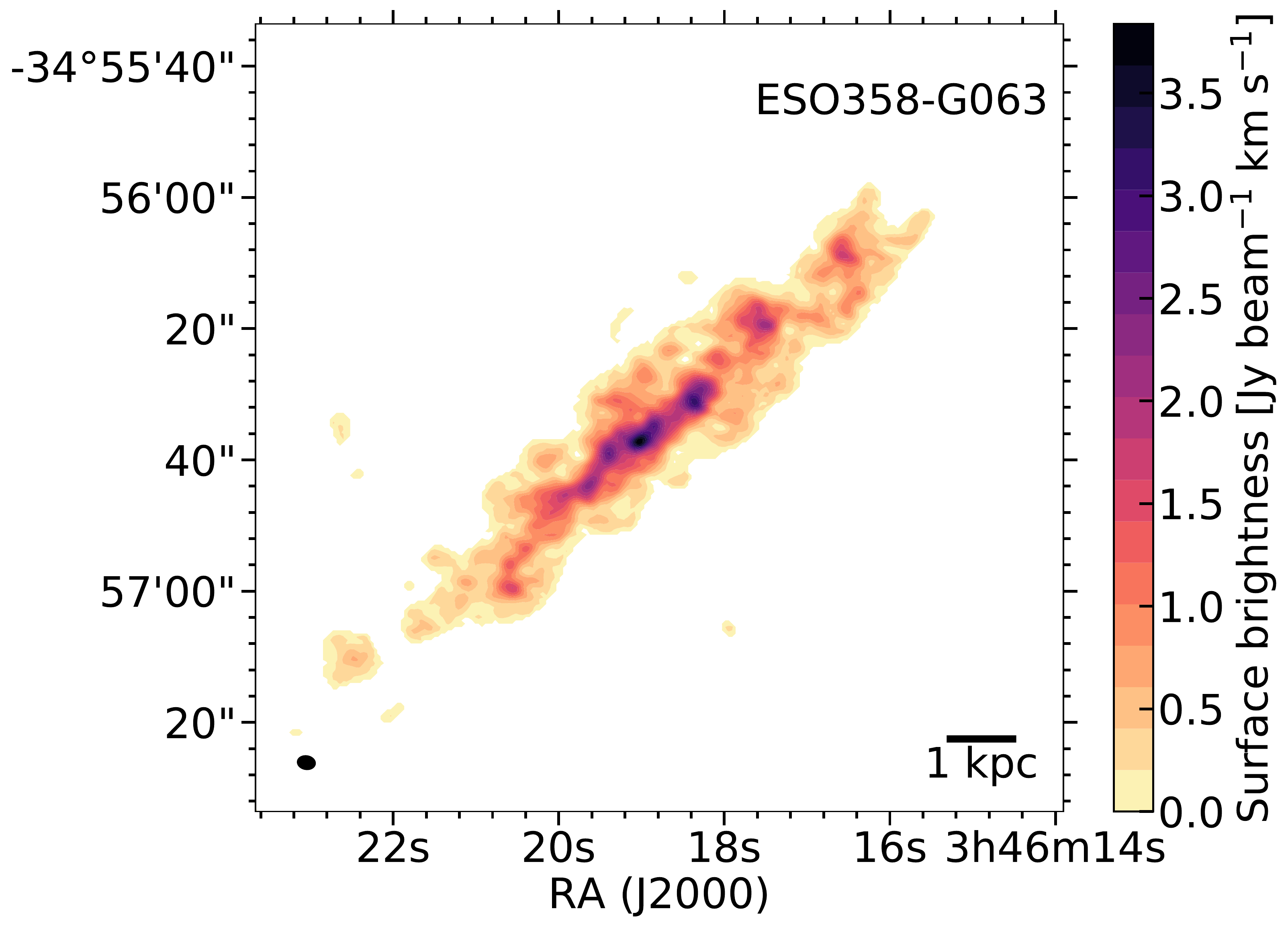}}	
	
	
	\subfloat[\label{subfig:063_mom1}]
		{\hspace{-5mm}\includegraphics[height=0.35\textwidth]{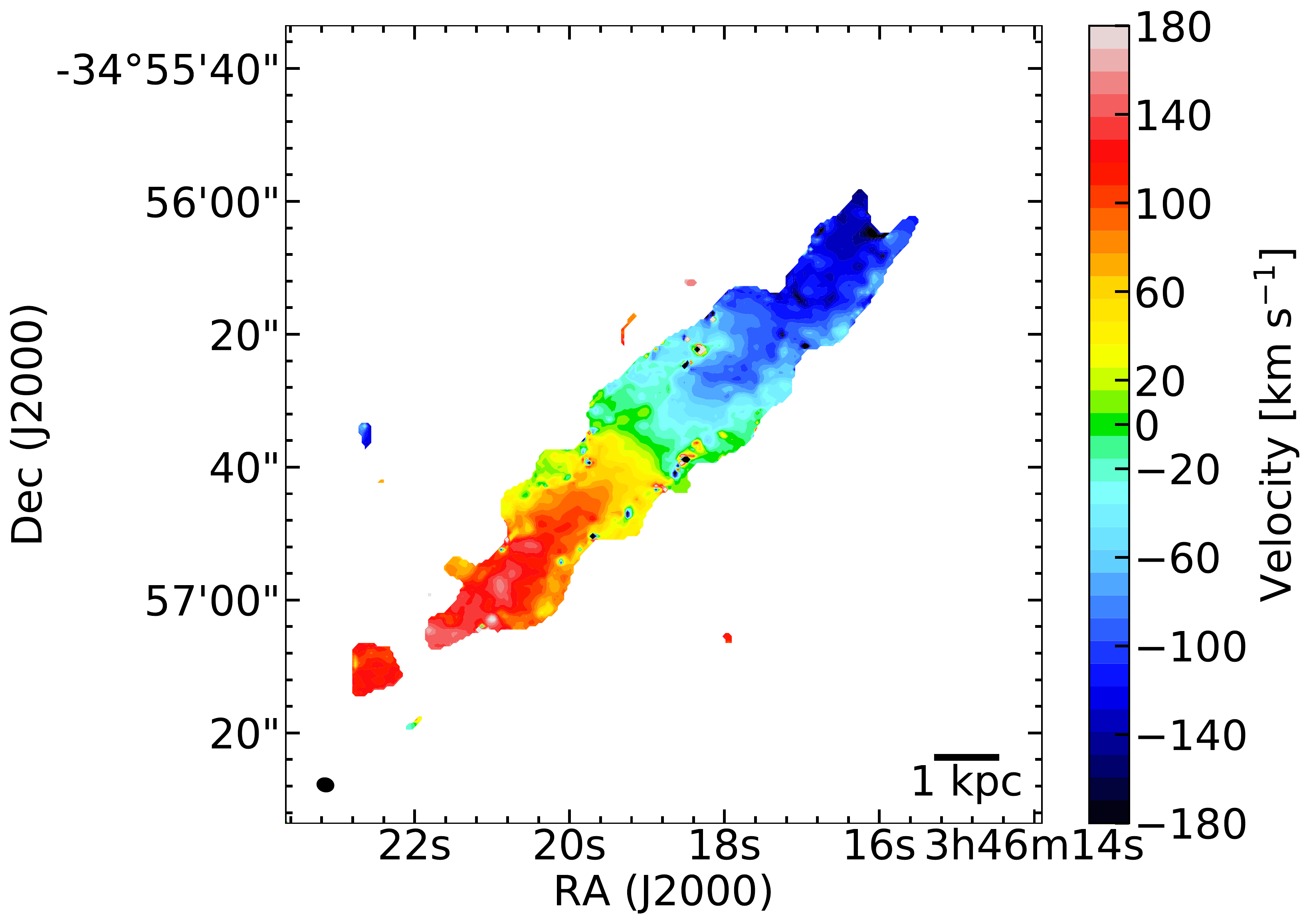}}
	\hspace{7mm}
	\subfloat[\label{subfig:063_mom2}]
		{\includegraphics[height=0.35\textwidth]{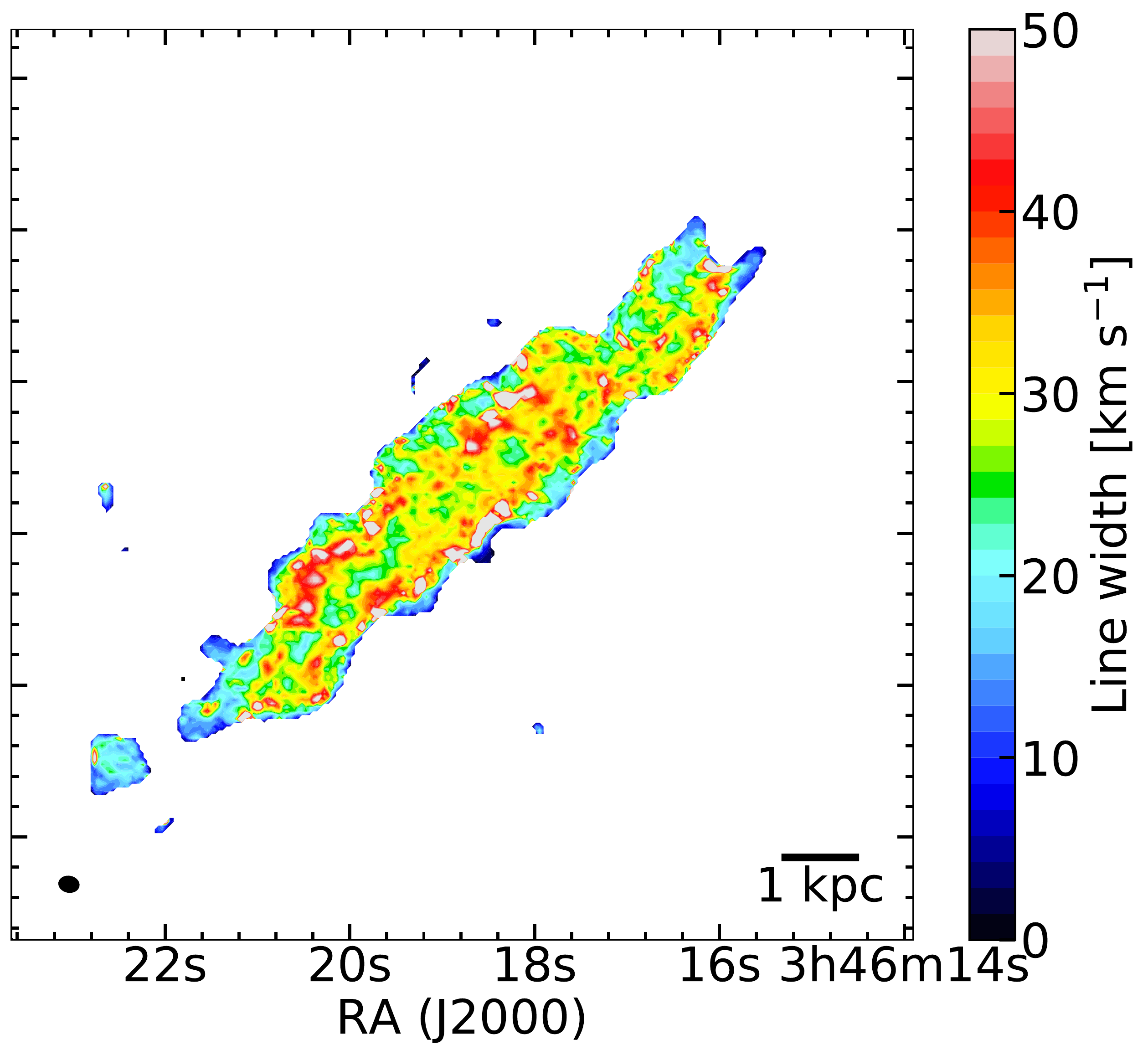}}
		
		
	\subfloat[\label{subfig:063_PVD}]
		{\includegraphics[height=0.39\textwidth]{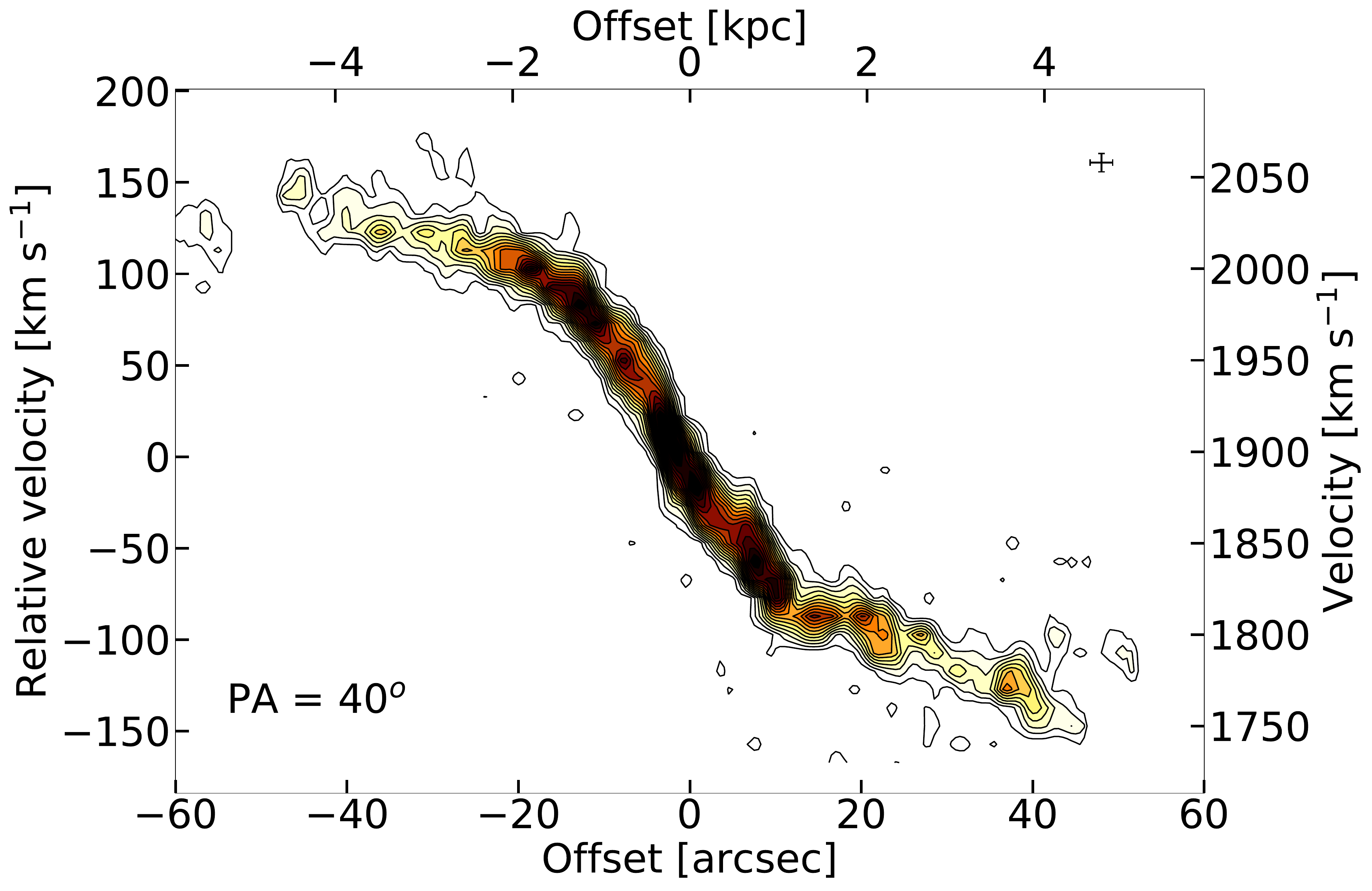}}	
	\hspace{6mm}
	\subfloat[\label{subfig:063_spectrum}]
		{\includegraphics[height=0.355\textwidth]{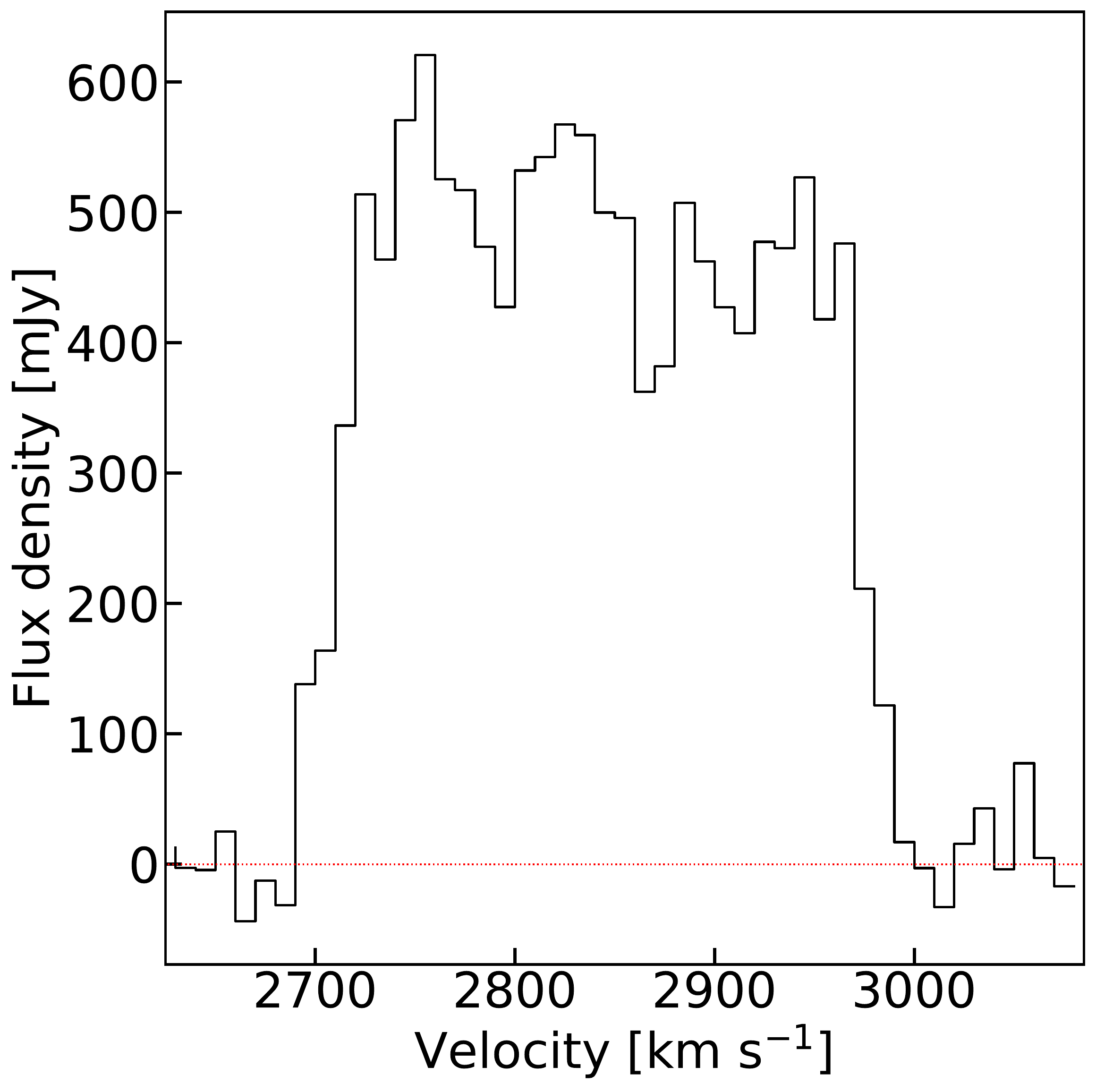}}
		
		
	\caption{ESO358-G063, similar to Figure \ref{fig:NGC1351A}.}
	\label{fig:ESO358-G063}
\end{figure*}

\begin{figure*}

	\centering

	\subfloat[]
	{\hspace{-5mm}\includegraphics[height=0.35\textwidth ]{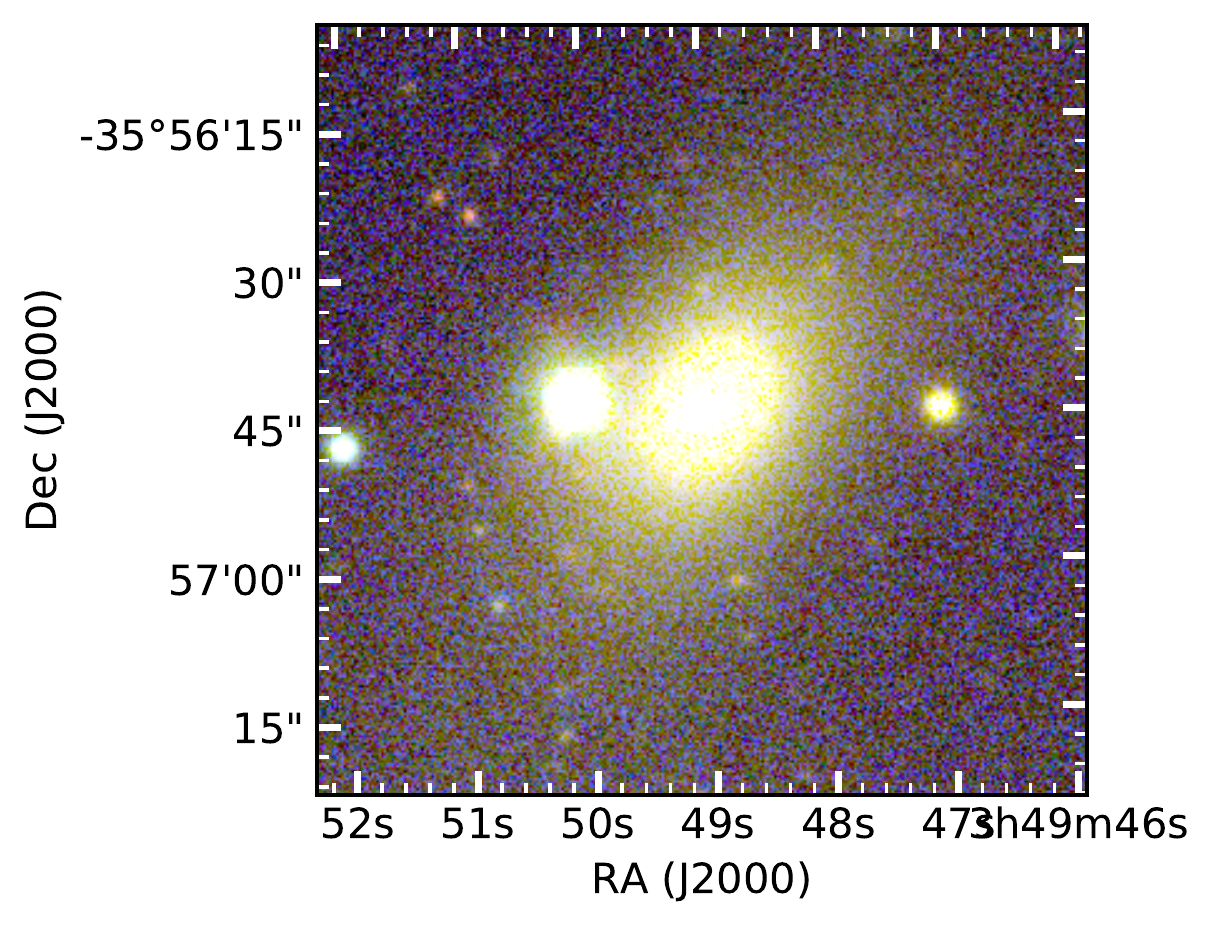}}	
	\hspace{-2mm}
	\subfloat[]
		{\includegraphics[height=0.35\textwidth]{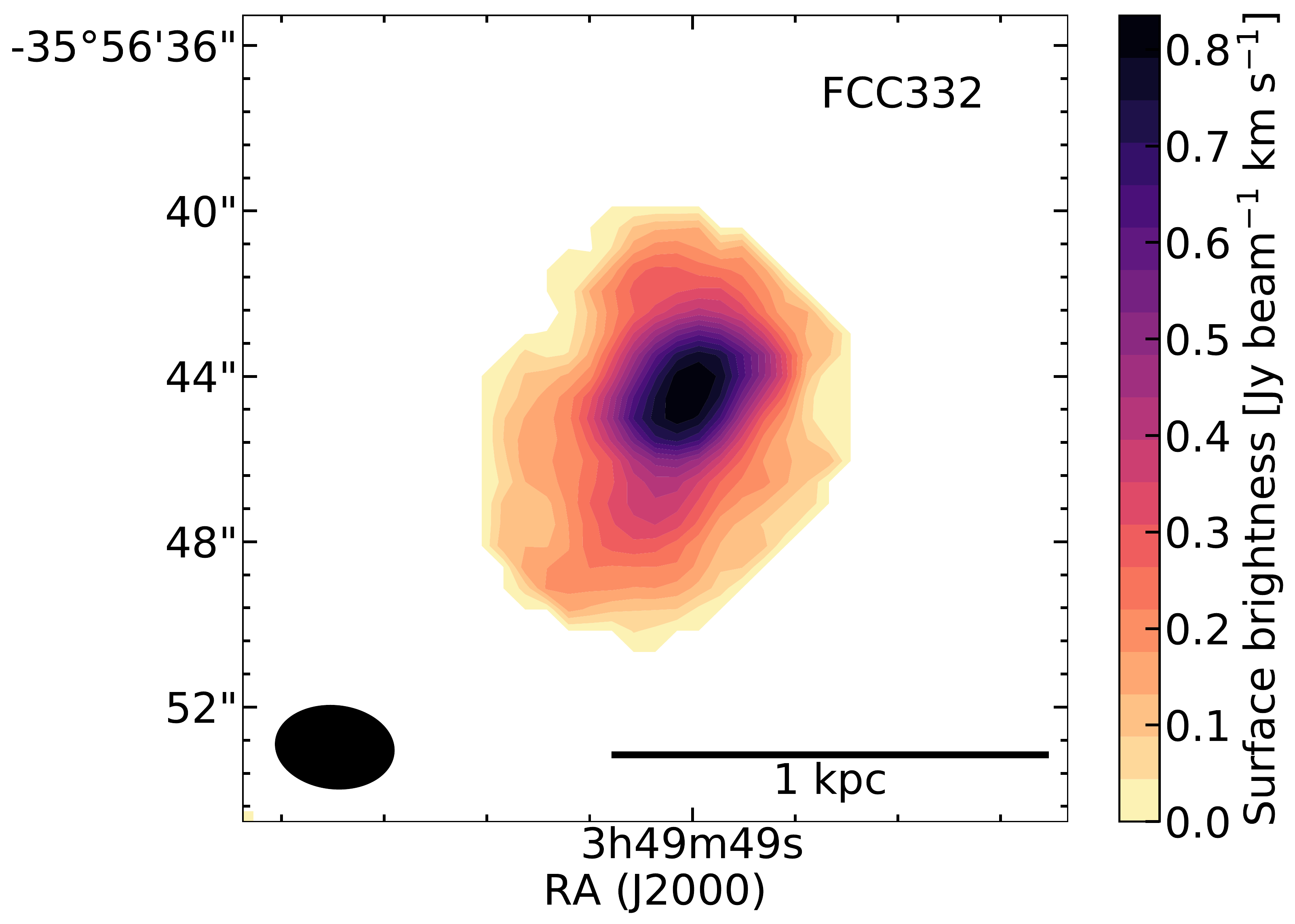}}	
	
	
	\subfloat[]
		{\hspace{-5mm}\includegraphics[height=0.35\textwidth]{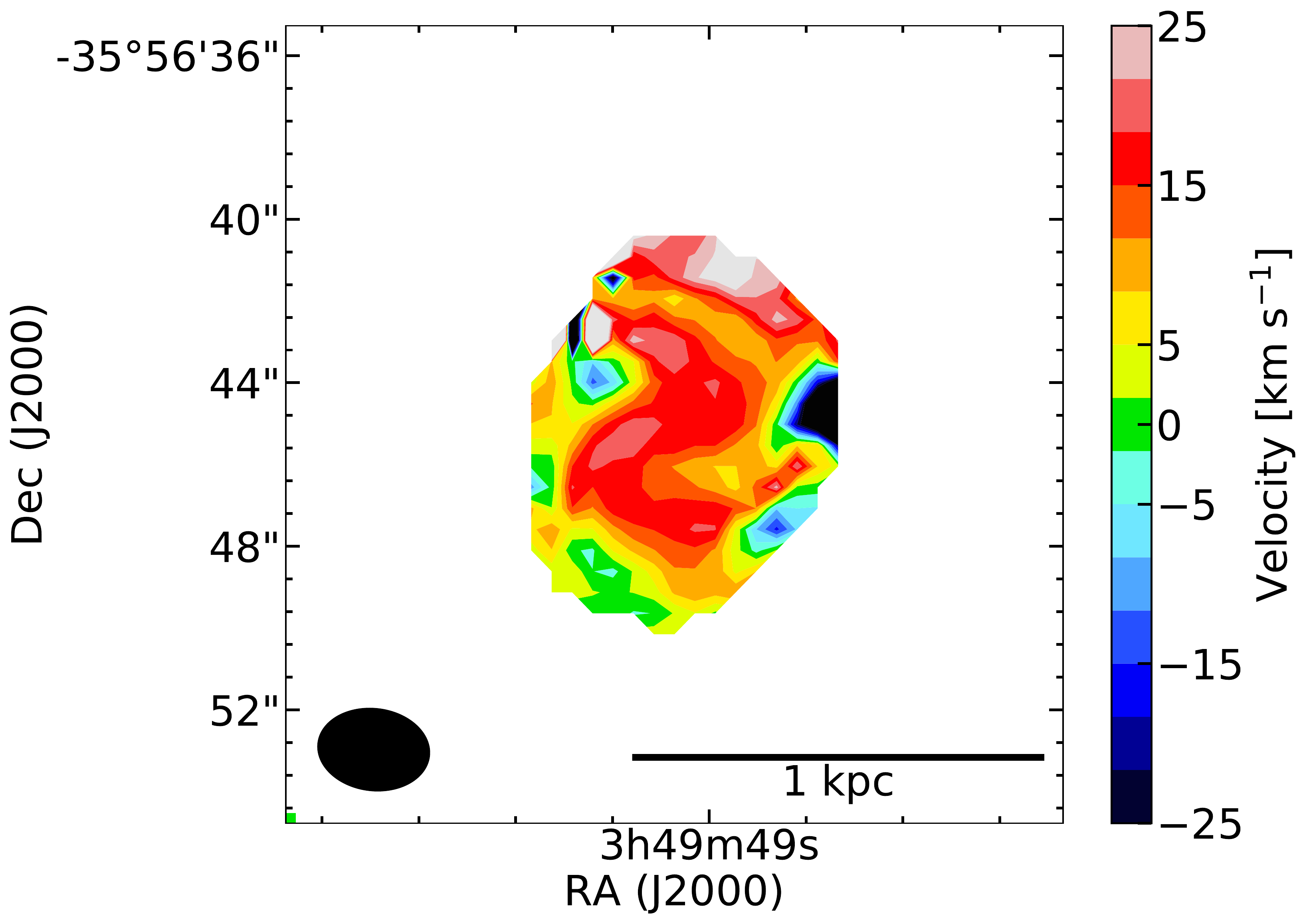}}
	\hspace{8mm}
	\subfloat[]
		{\includegraphics[height=0.35\textwidth]{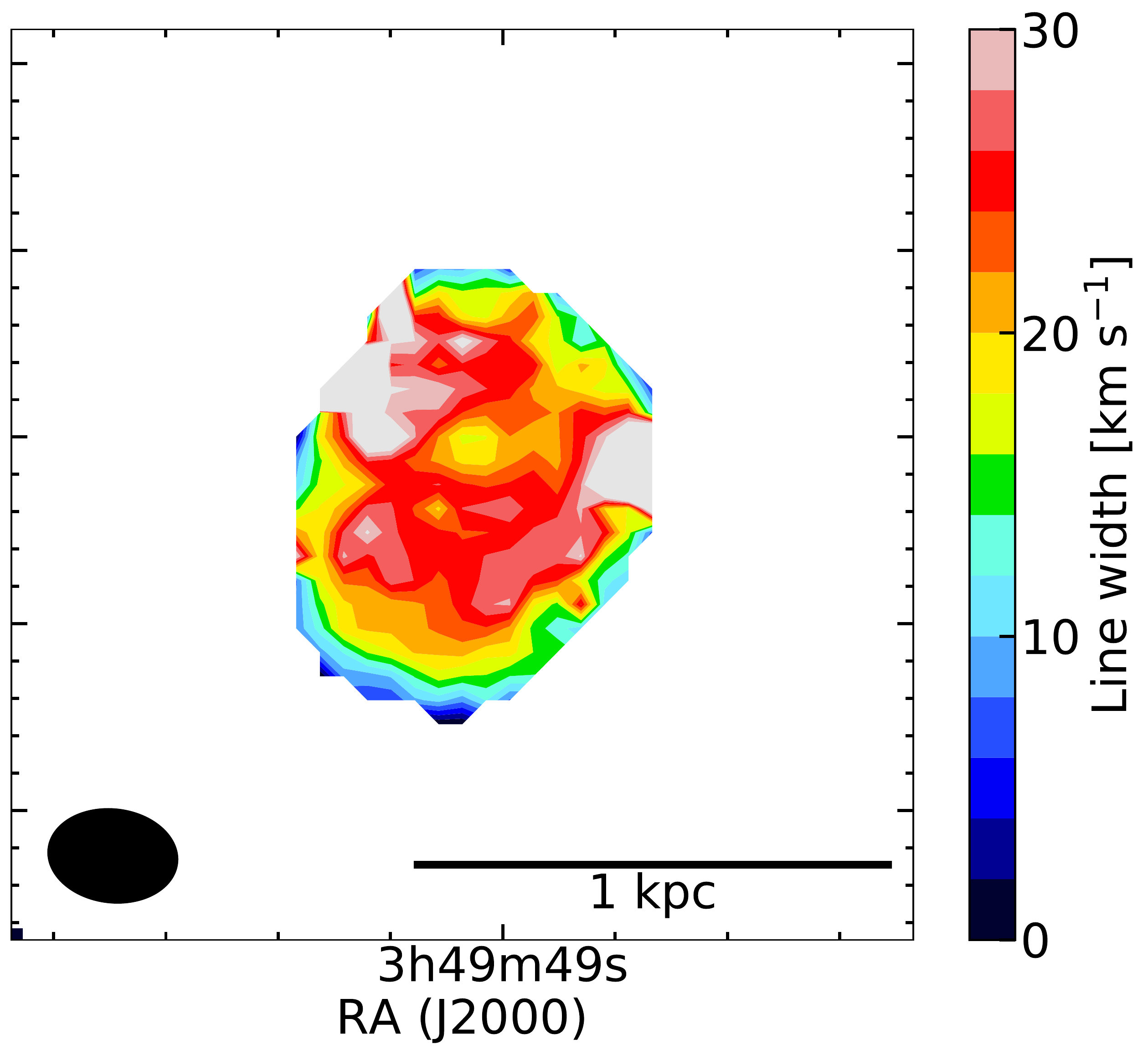}}
		
		
	\subfloat[]
	{\includegraphics[height=0.41\textwidth]{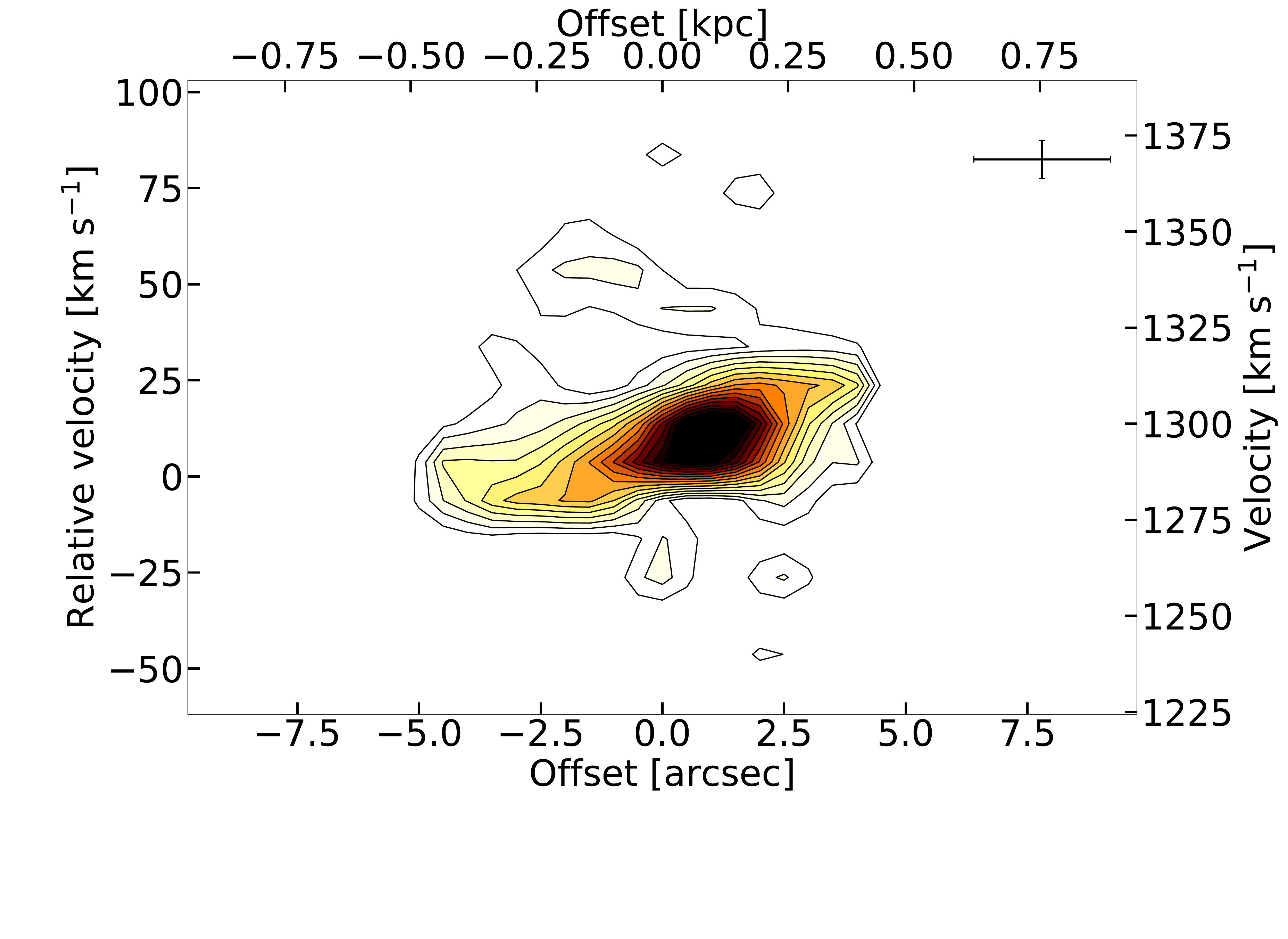}}	
	\hspace{0mm}
	\subfloat[]
		{\includegraphics[height=0.355\textwidth]{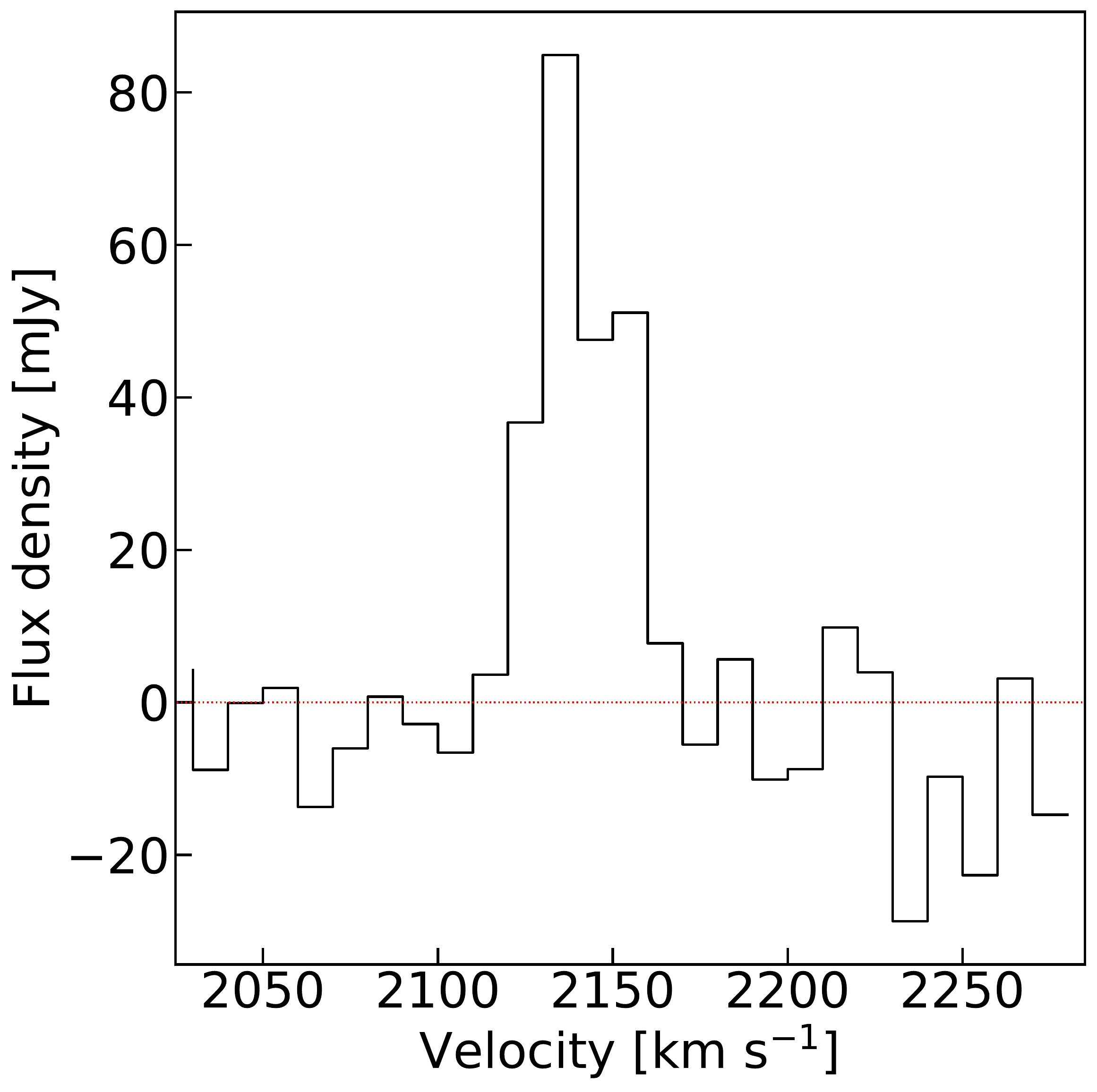}}
		
		
	\caption{FCC332, similar to Figure \ref{fig:NGC1351A}.}
	\label{fig:FCC332}
\end{figure*}

\begin{figure*}

	\centering

	\subfloat[]
	{\hspace{-5mm}\includegraphics[height=0.35\textwidth ]{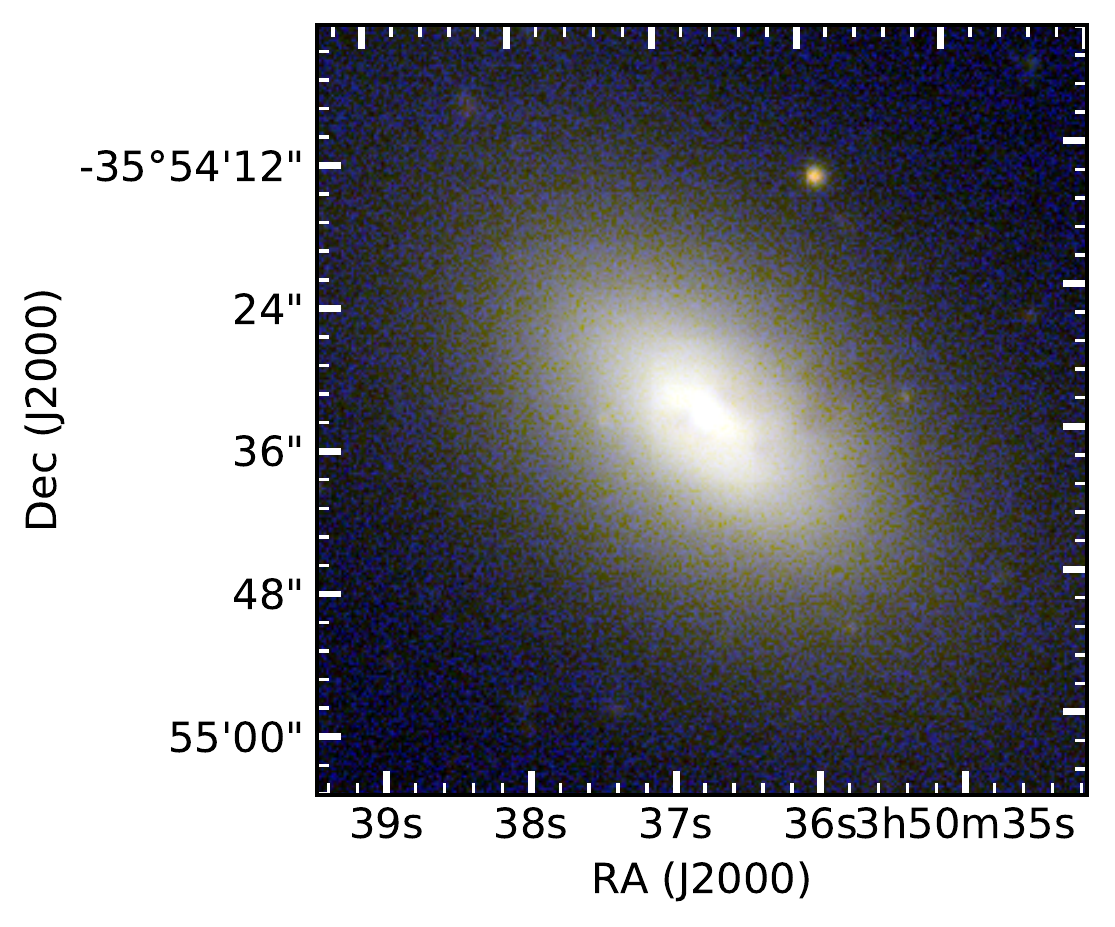}}	
	\hspace{2mm}
	\subfloat[]
		{\includegraphics[height=0.35\textwidth]{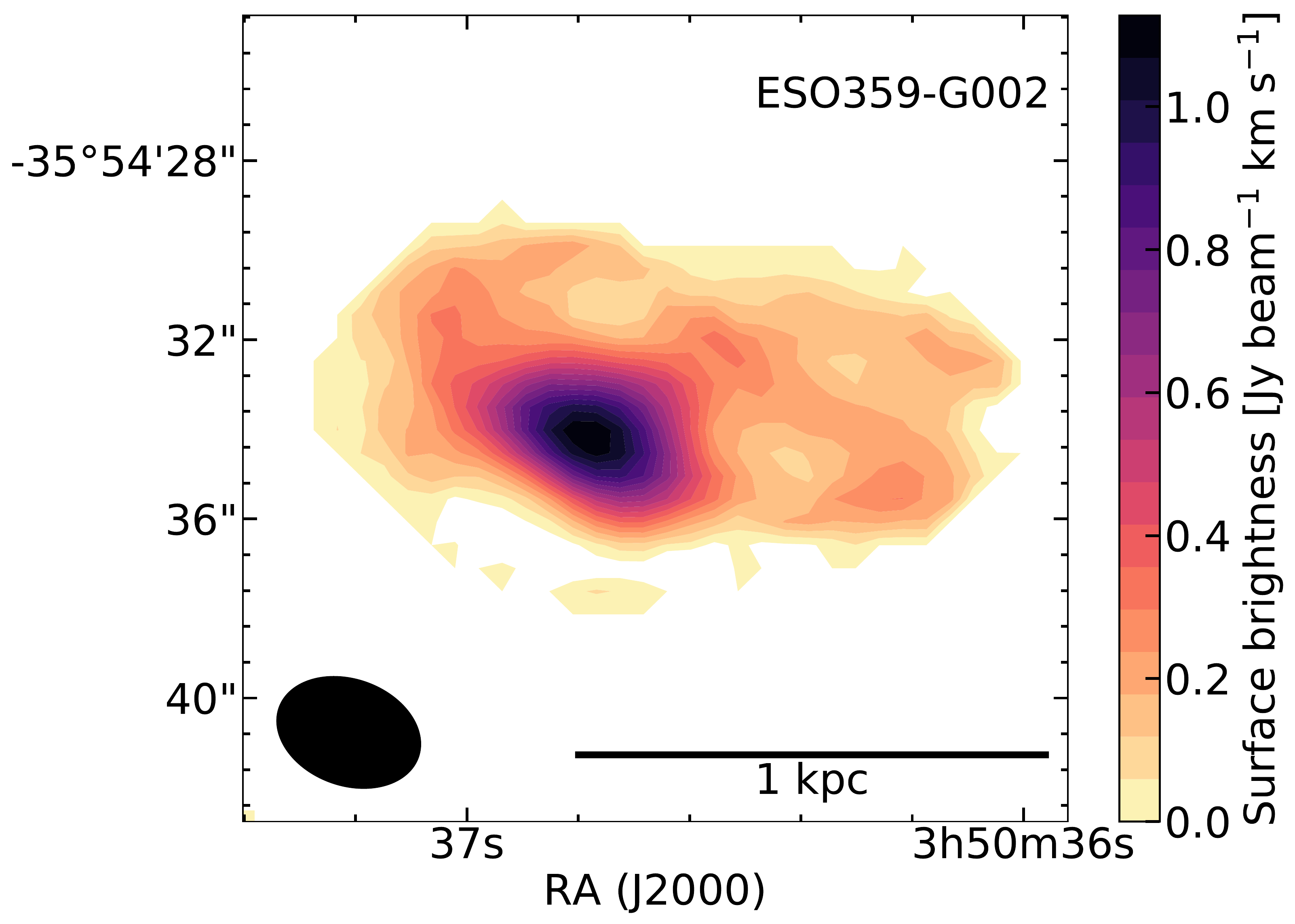}}	
	
	
	\subfloat[]
		{\hspace{-5mm}\includegraphics[height=0.35\textwidth]{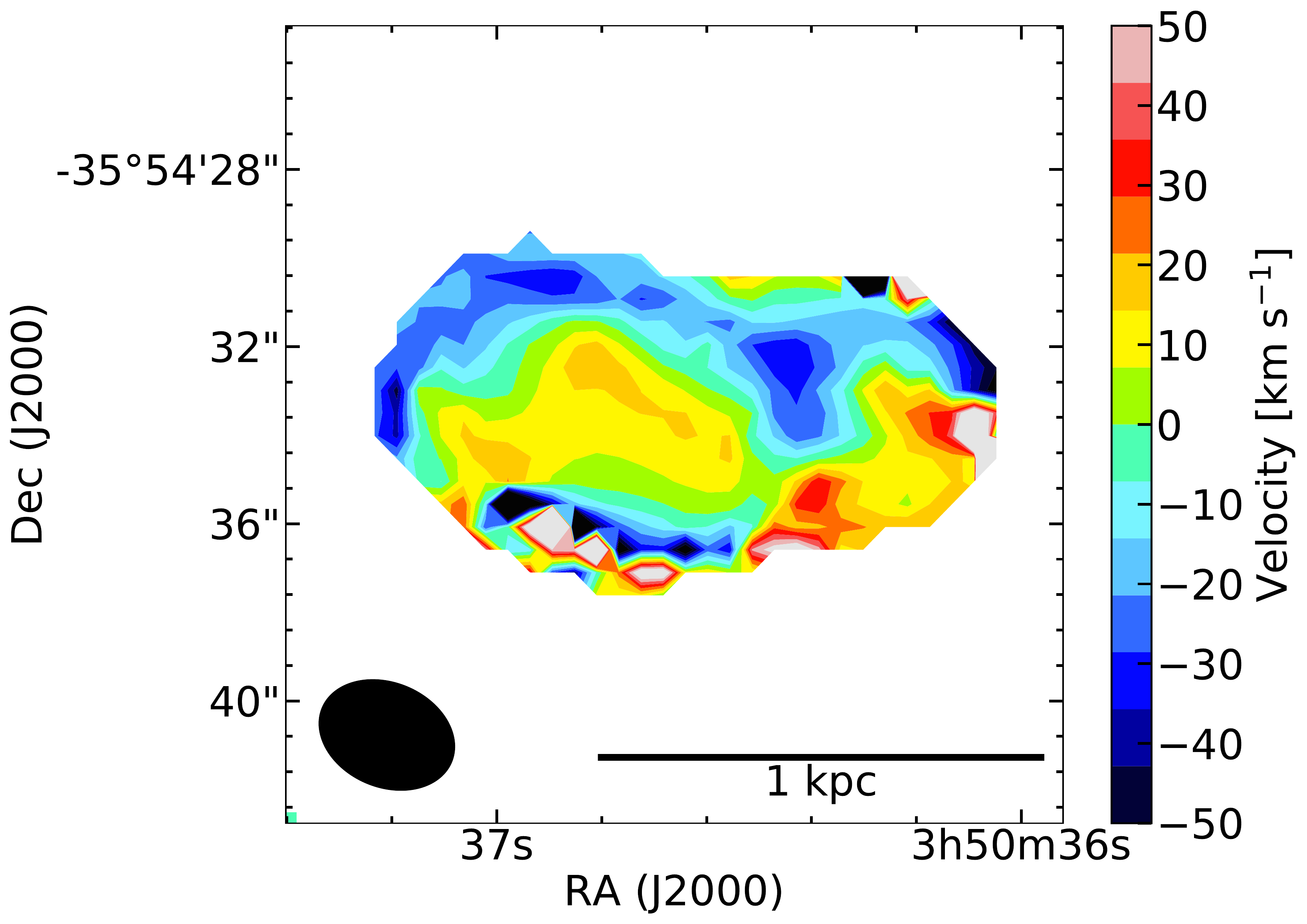}}
	\hspace{9mm}
	\subfloat[]
		{\includegraphics[height=0.35\textwidth]{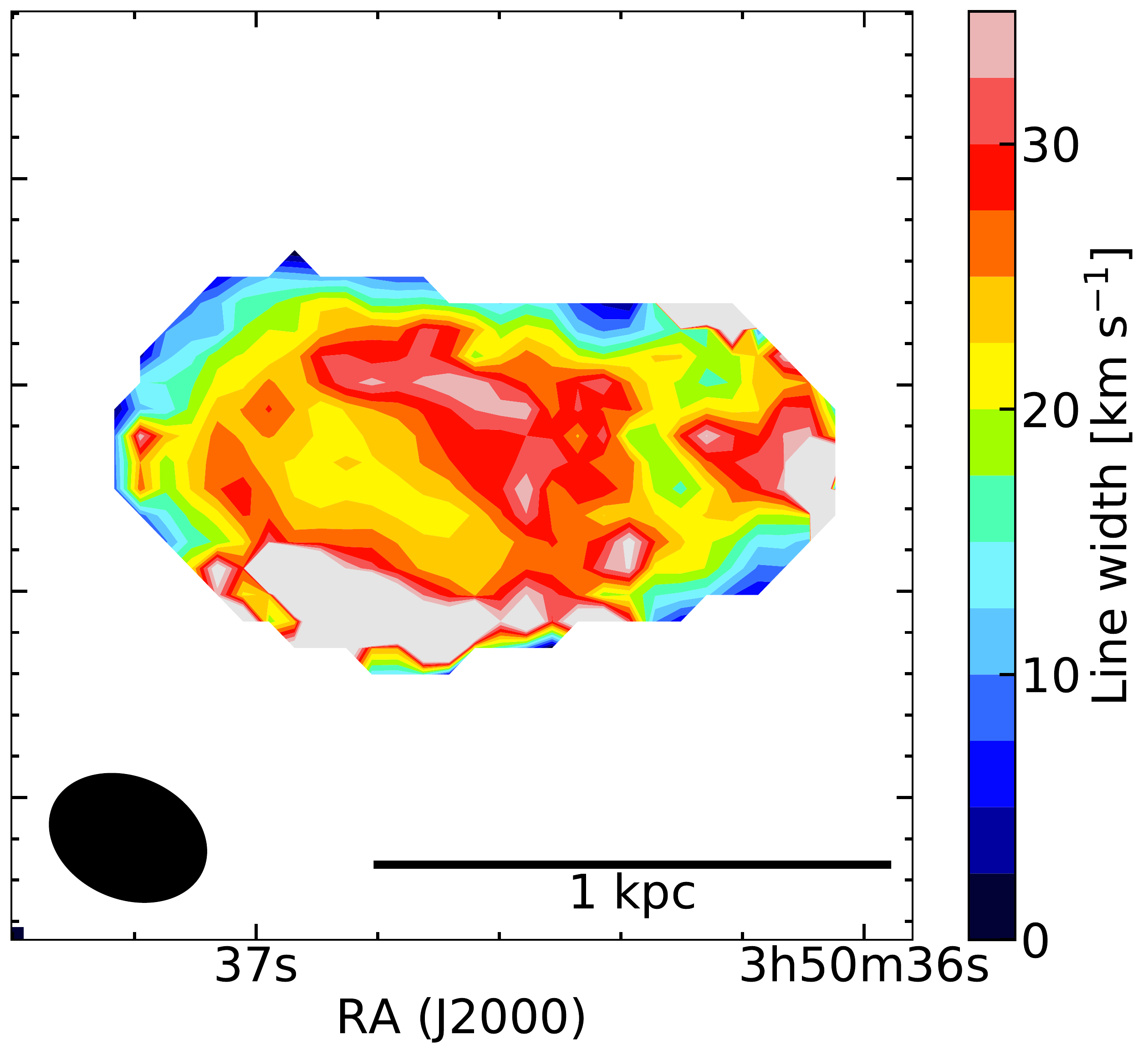}}
		
		
	\subfloat[]
		{\includegraphics[height=0.39\textwidth]{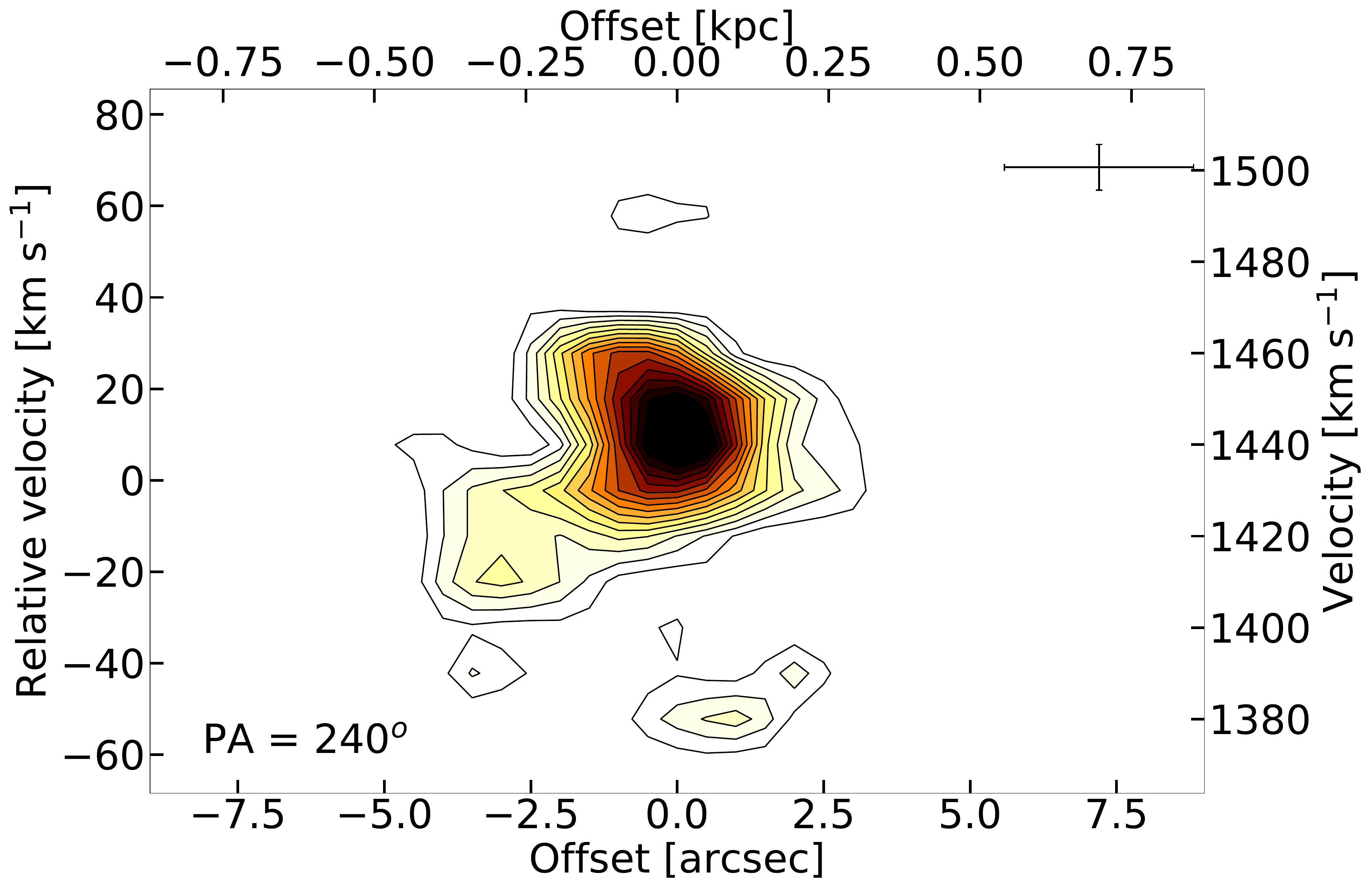}}	
	\hspace{6mm}
	\subfloat[]
		{\includegraphics[height=0.355\textwidth]{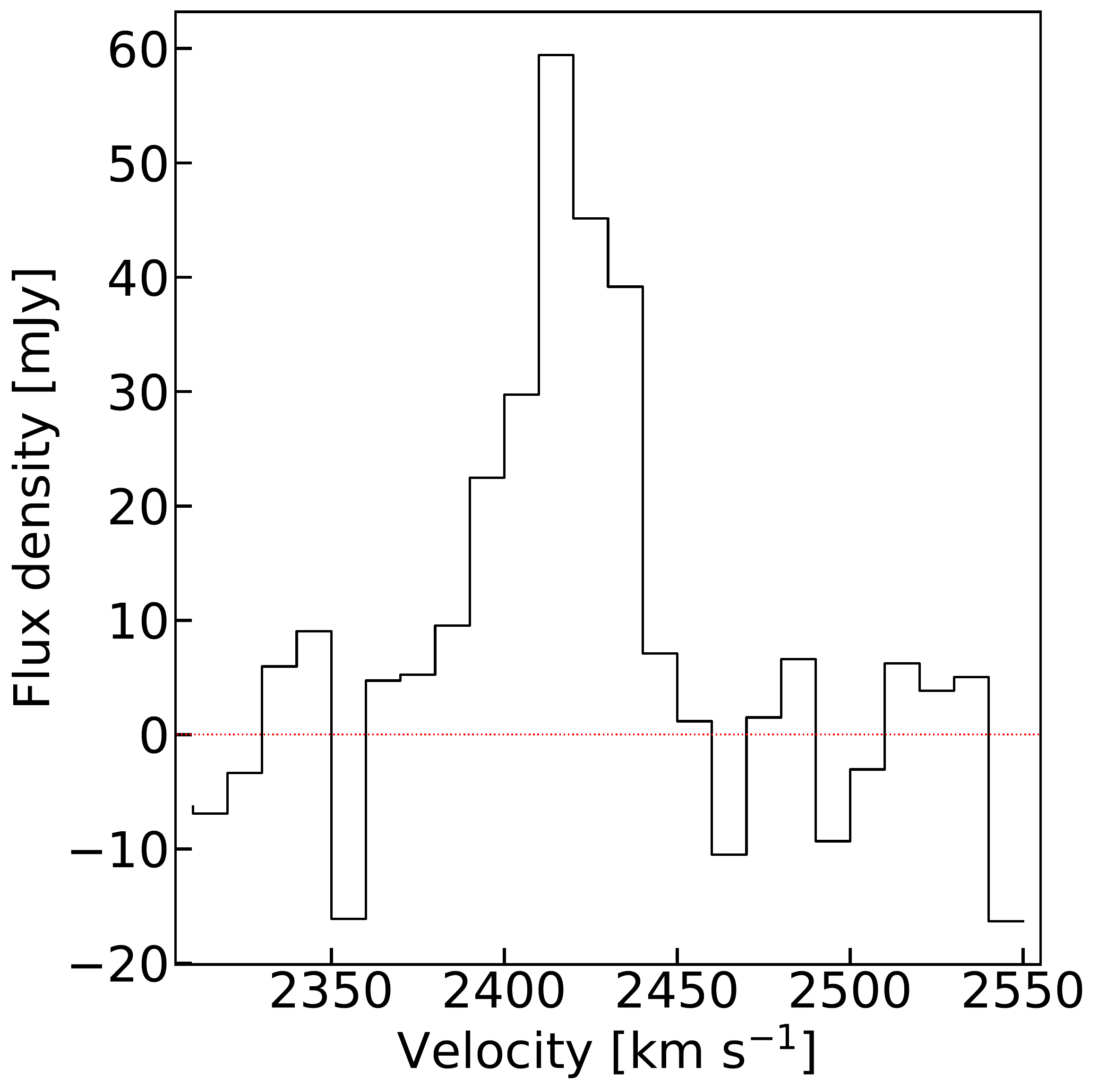}}
		
		
	\caption{ESO359-G002, similar to Figure \ref{fig:NGC1351A}.}
	\label{fig:ESO359-G002}
\end{figure*}

\clearpage

\section{Comparison with optical images}
\label{app:overplots}
Images in this appendix show the contours of the CO emission overplotted on optical (\textit{g}-band) images from the Fornax Deep Survey (\FDS). Additional information about these figures can be found in the caption of Figure \ref{fig:overplots}.

\begin{figure*}

	\centering
	
	\subfloat[\label{subfig:overplot_NGC1351A}]
	{\includegraphics[height=0.4\textwidth]{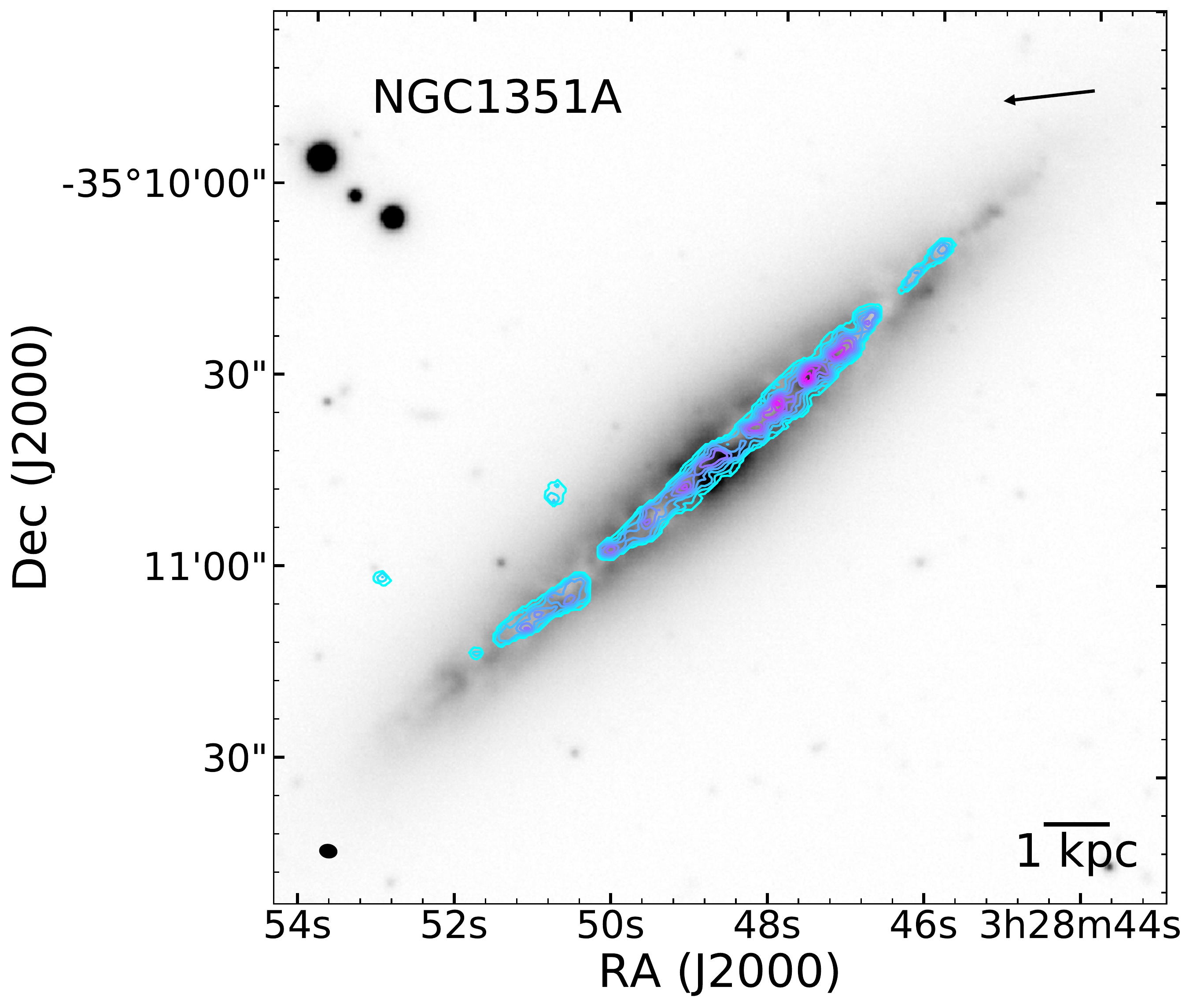}}	
	\subfloat[\label{subfig:overplot_MCG-06-08-024}]
	{\includegraphics[height=0.4\textwidth]{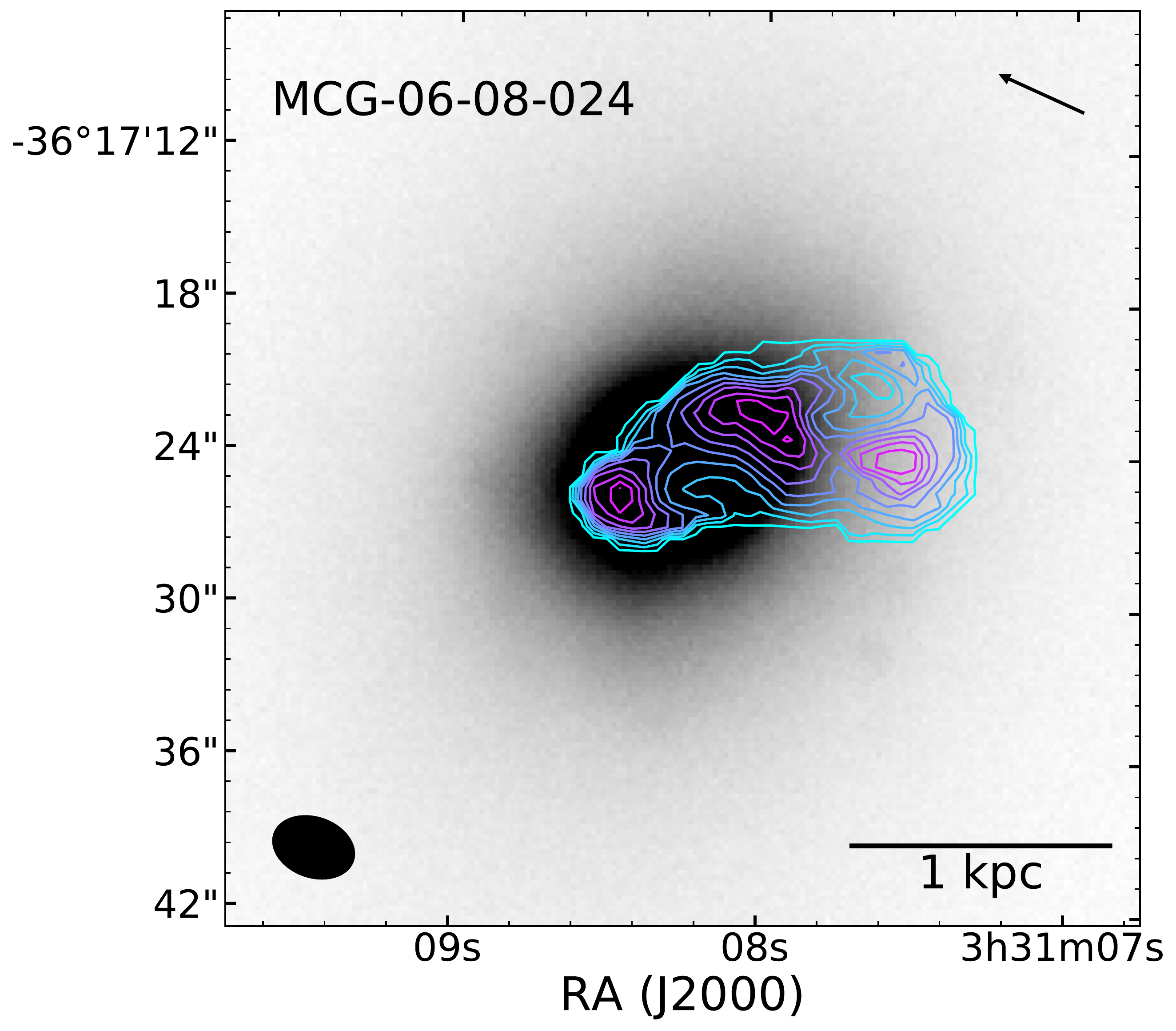}}	
	
	\subfloat[\label{subfig:overplot_NGC1365}]
	{\includegraphics[height=0.4\textwidth]{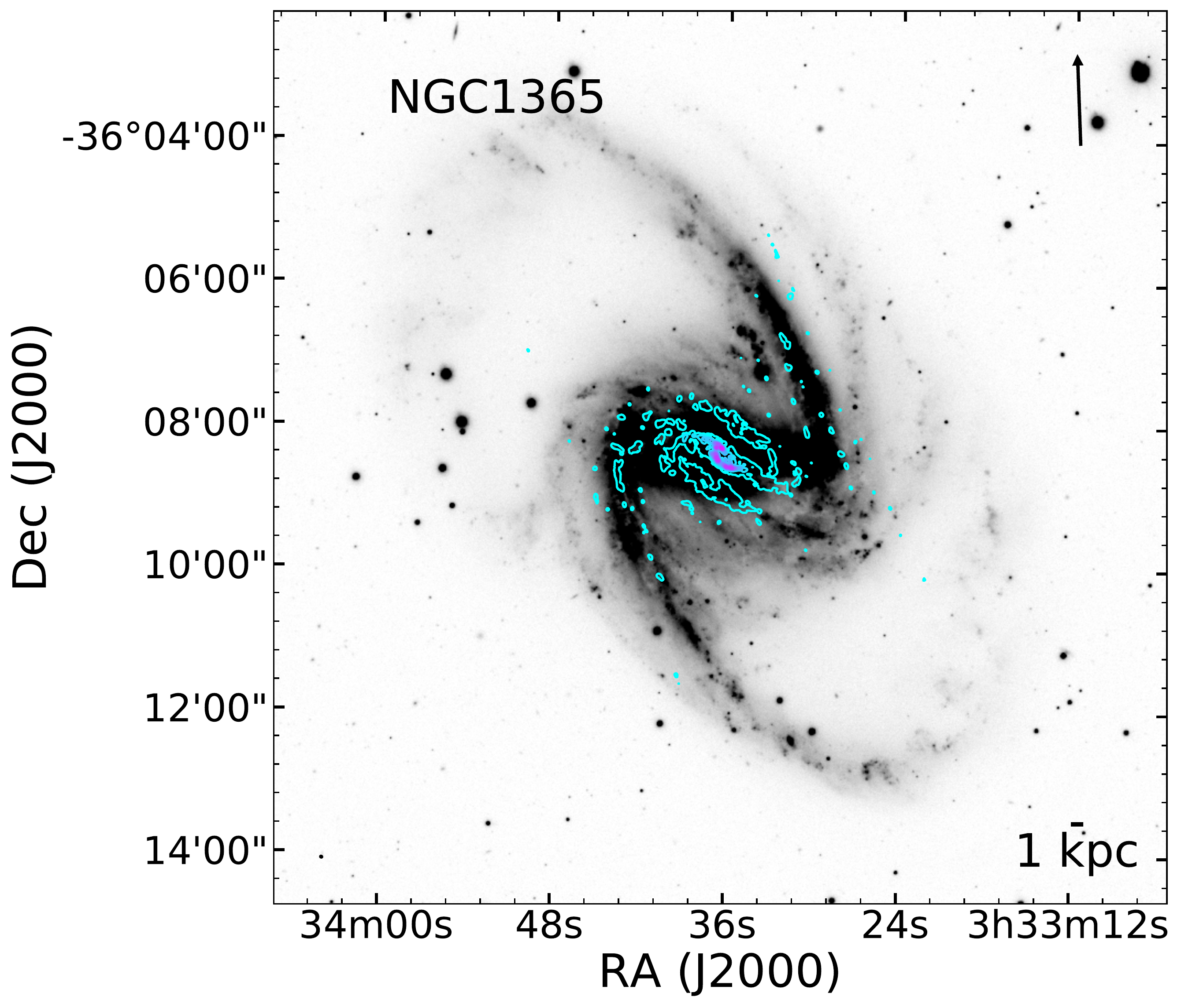}}	
	\subfloat[\label{subfig:overplot_NGC1380}]
	{\includegraphics[height=0.4\textwidth]{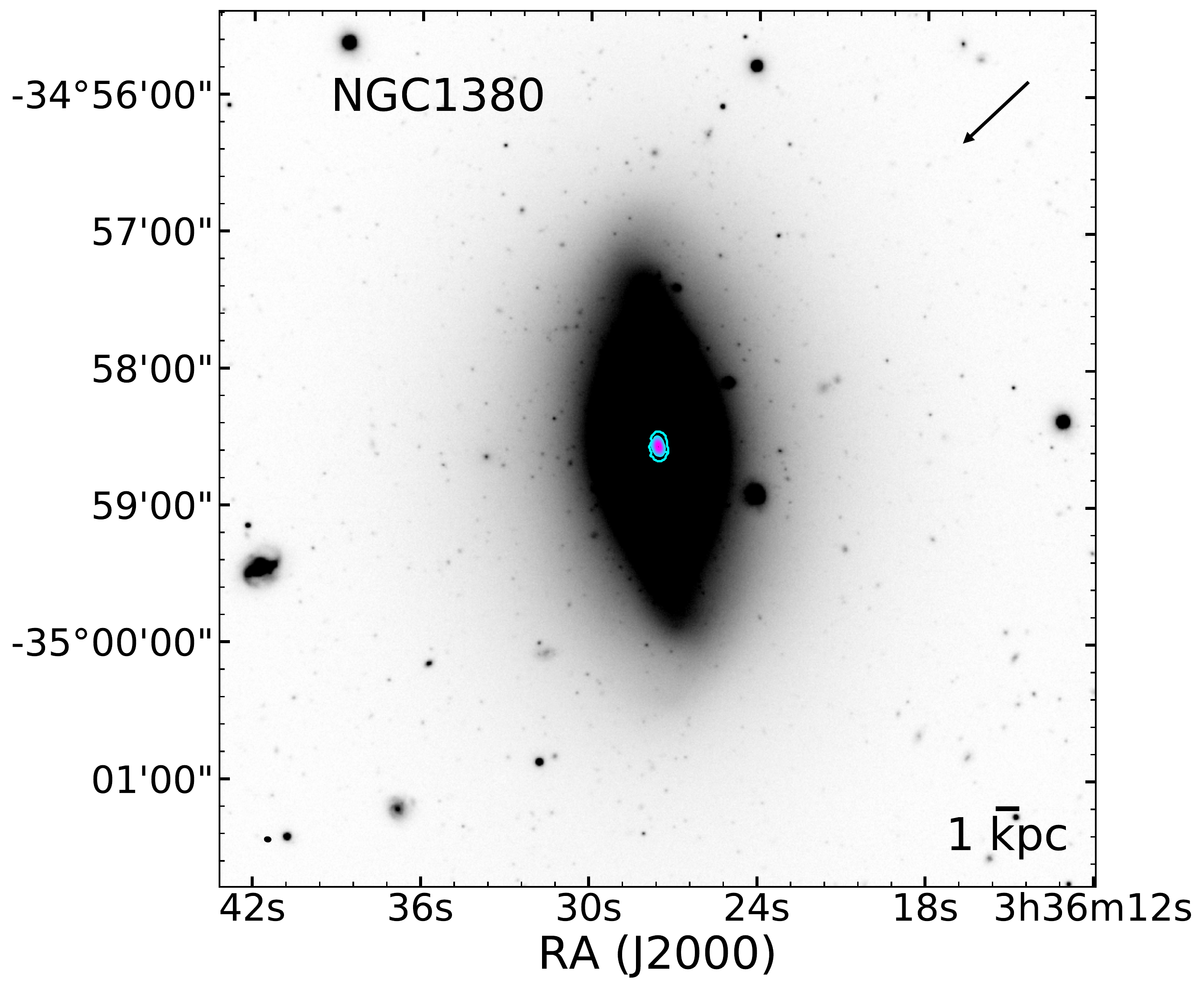}}	
	
	\subfloat[\label{subfig:overplot_NGC1386}]
	{\includegraphics[height=0.4\textwidth]{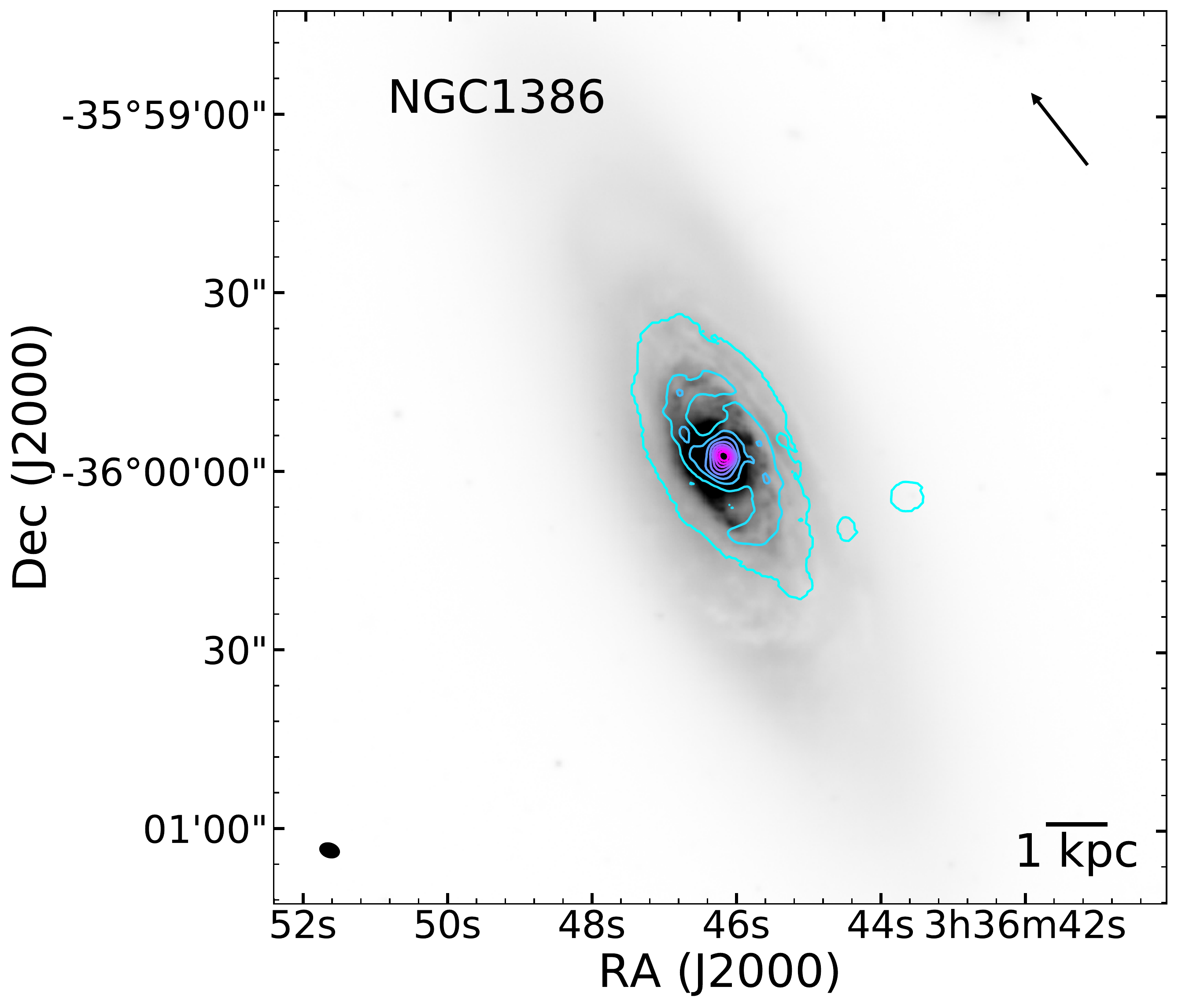}}
	\subfloat[\label{subfig:overplot_NGC1387}]
	{\includegraphics[height=0.4\textwidth]{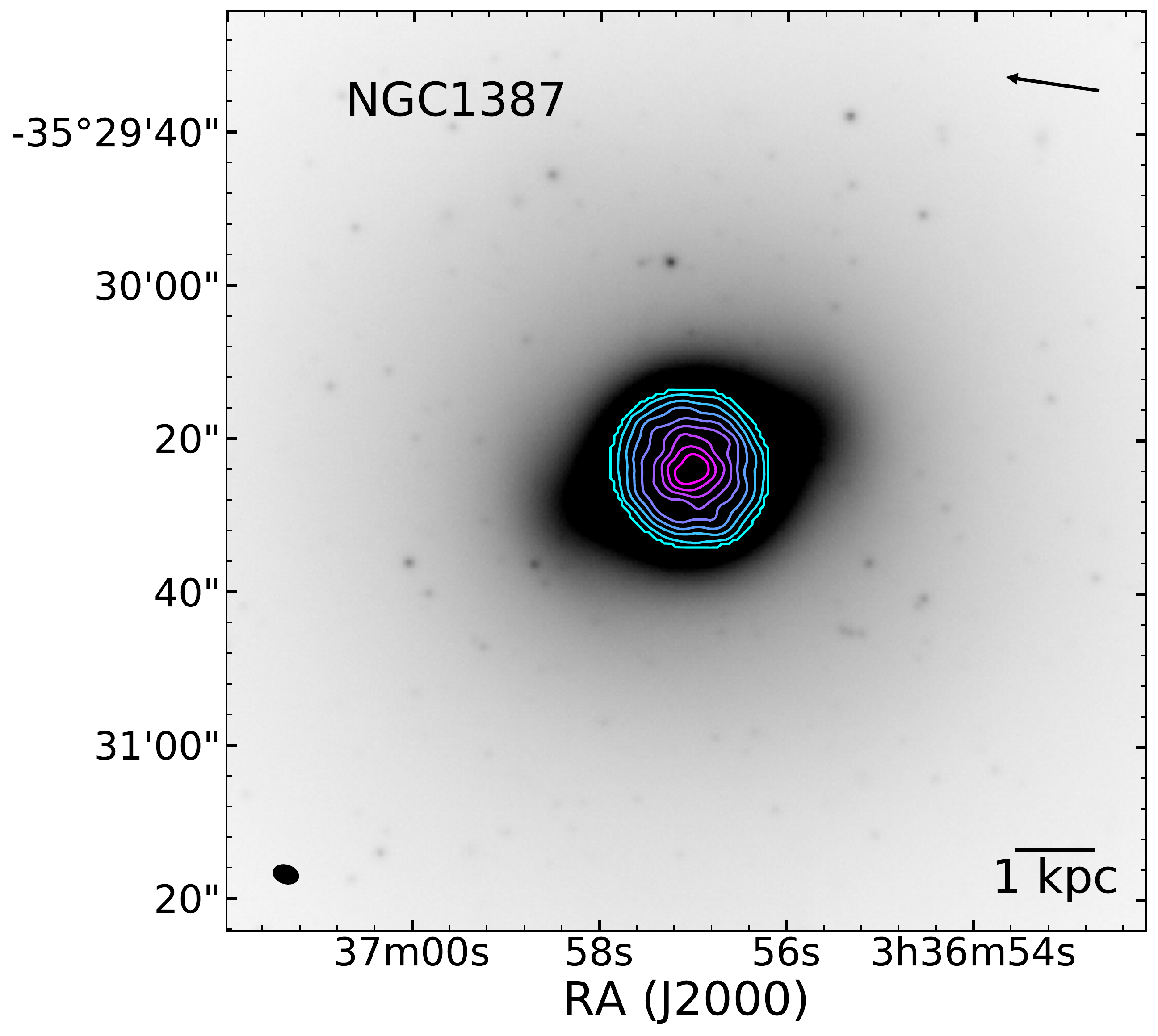}}
	
\end{figure*}

\begin{figure*}
	
	\ContinuedFloat
	\centering

	\subfloat[\label{subfig:overplot_FCC207}]
	{\includegraphics[height=0.4\textwidth]{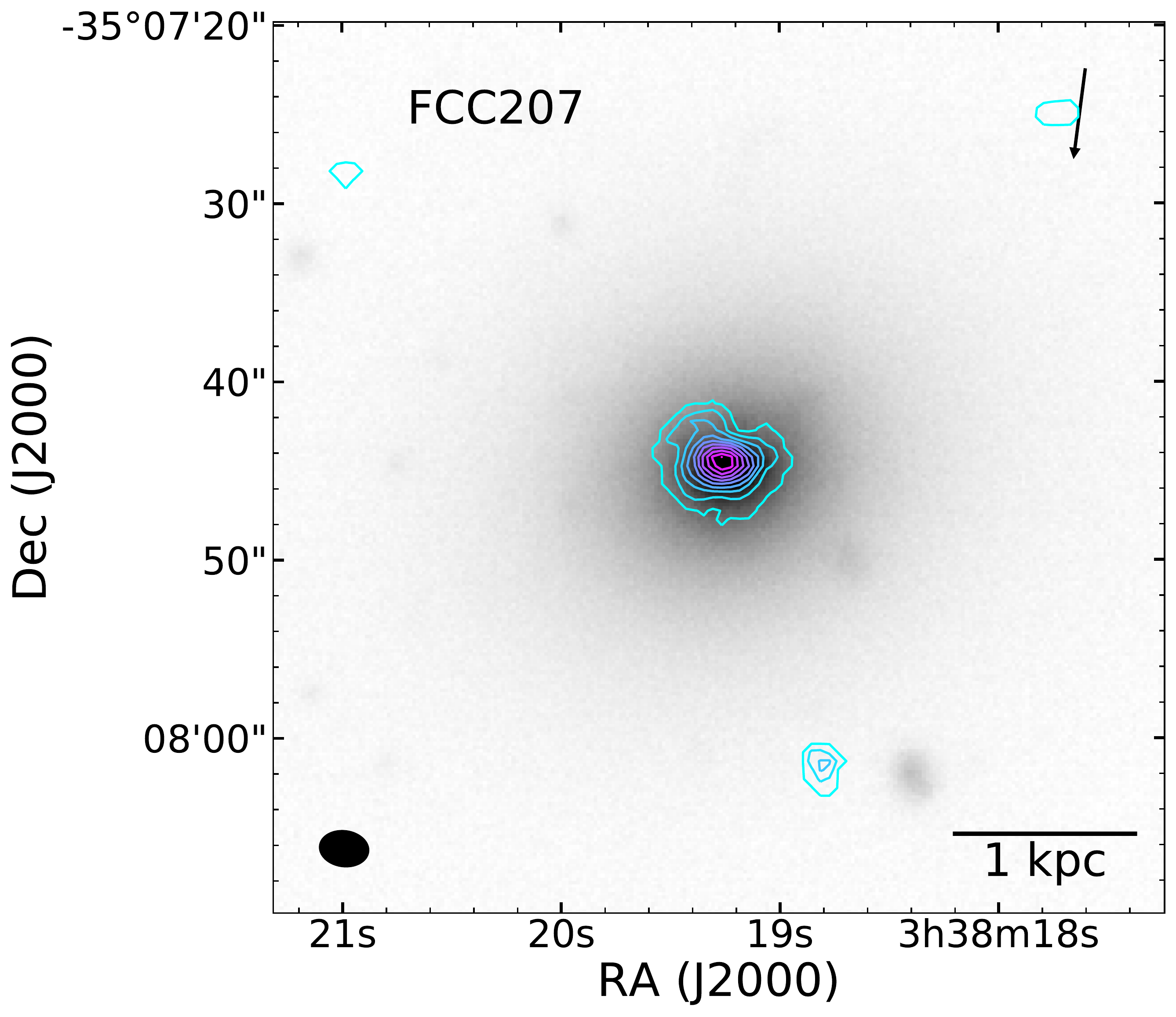}}
	\subfloat[\label{subfig:overplot_FCC261}]
	{\includegraphics[height=0.4\textwidth]{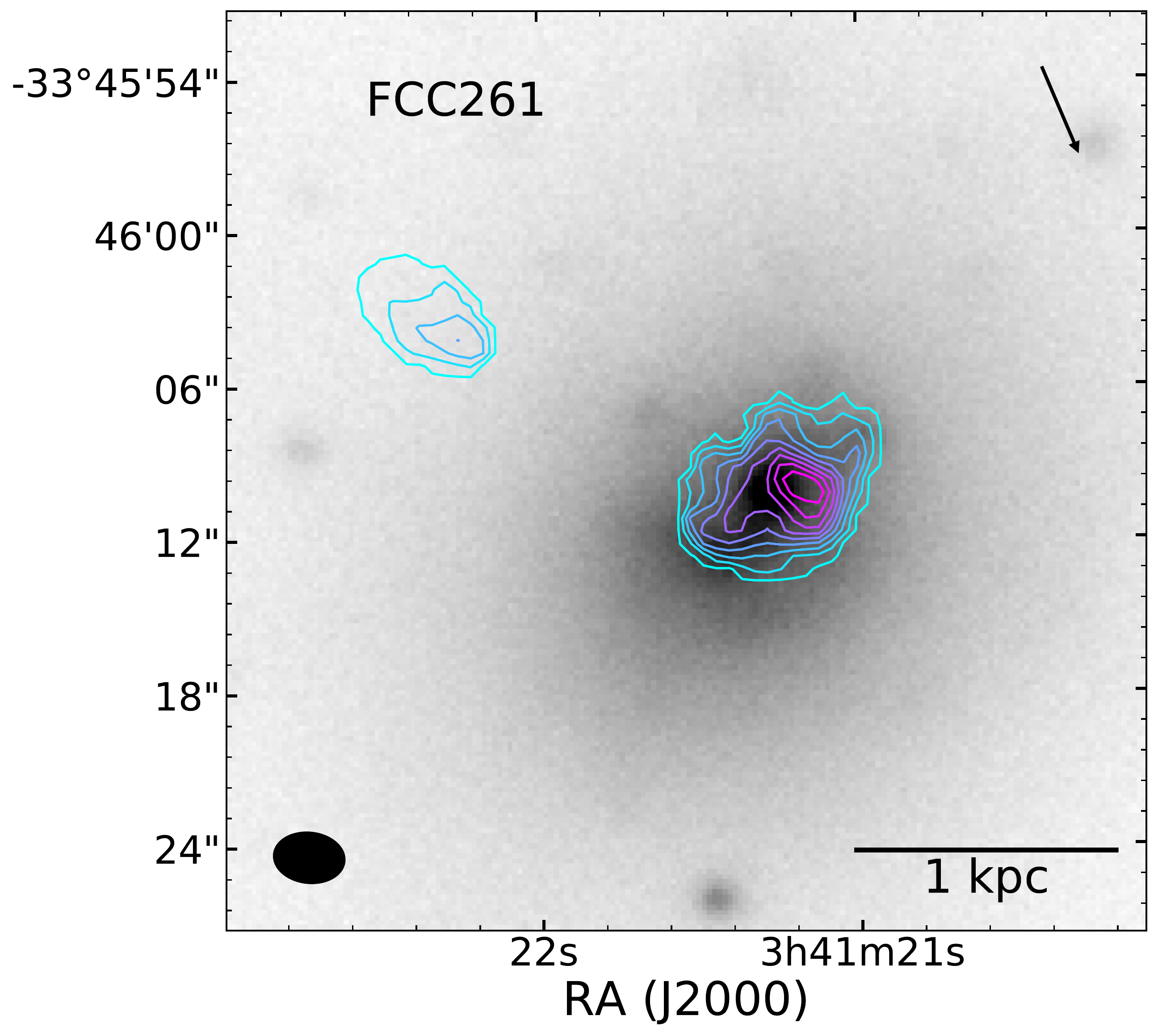}}	
	
	\subfloat[\label{subfig:overplot_PGC013571}]
	{\includegraphics[height=0.4\textwidth]{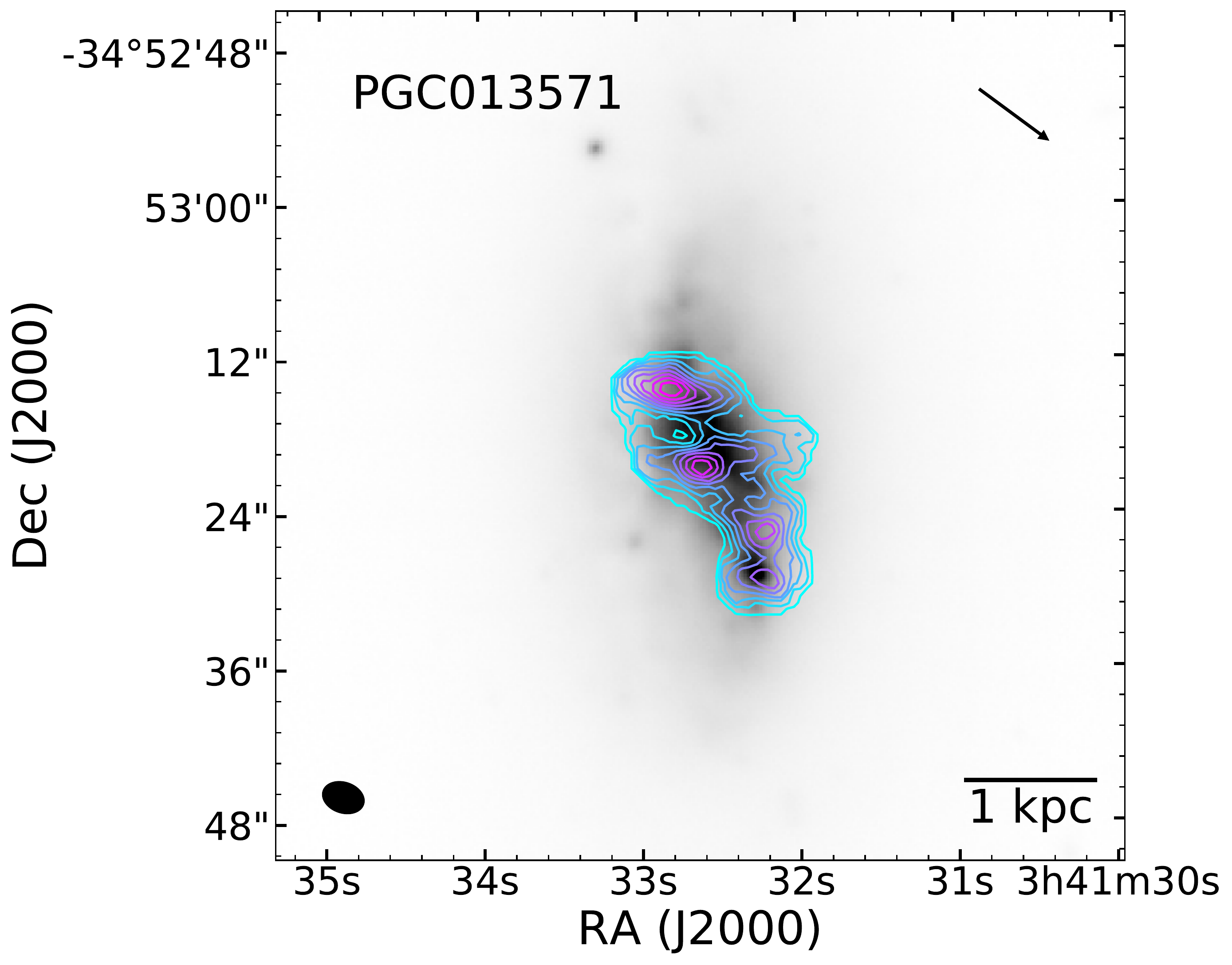}}
	\subfloat[\label{subfig:overplot_FCC282}]
	{\includegraphics[height=0.4\textwidth]{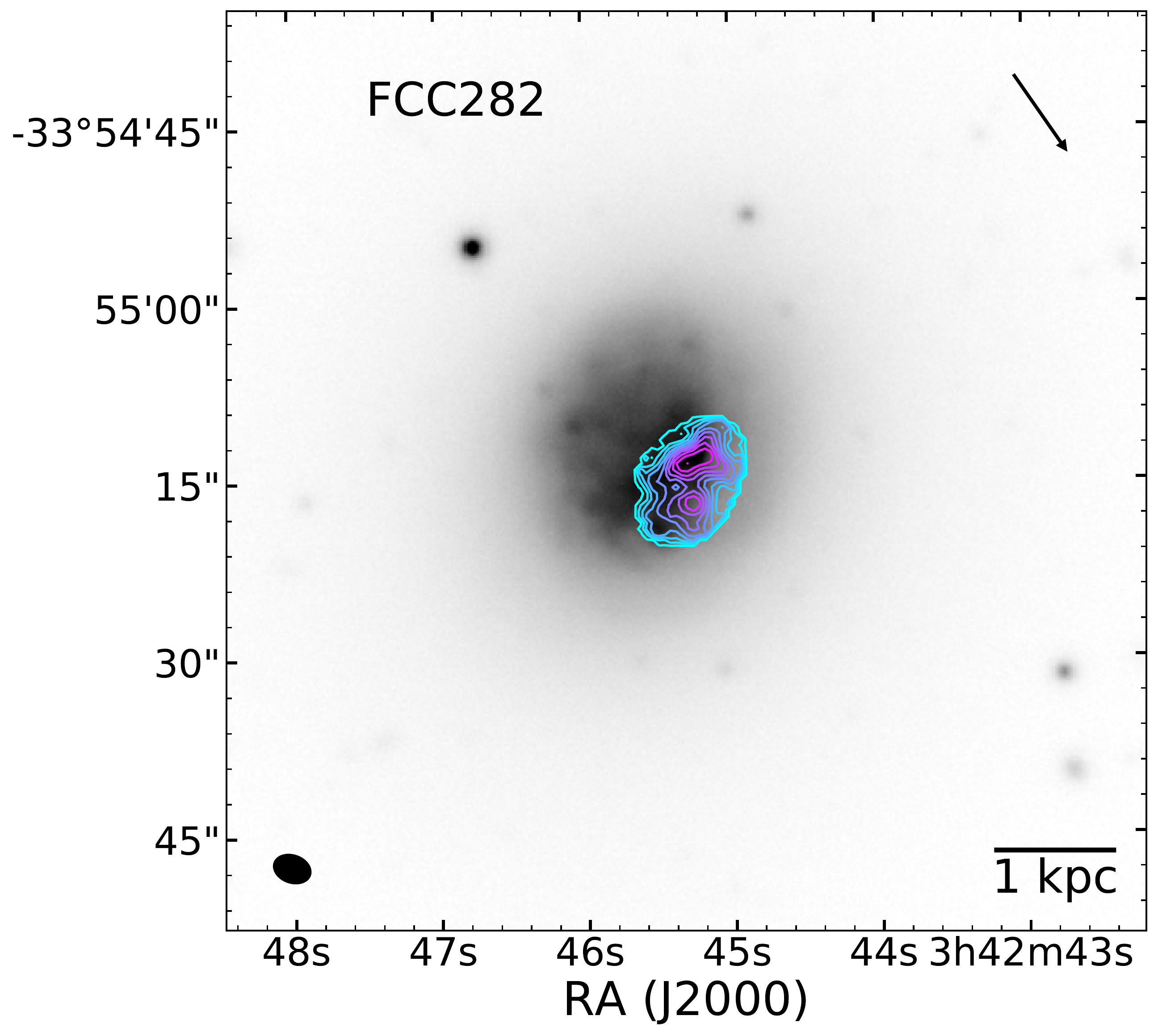}}
	
	\subfloat[\label{subfig:overplot_NGC1436}]
	{\includegraphics[height=0.4\textwidth]{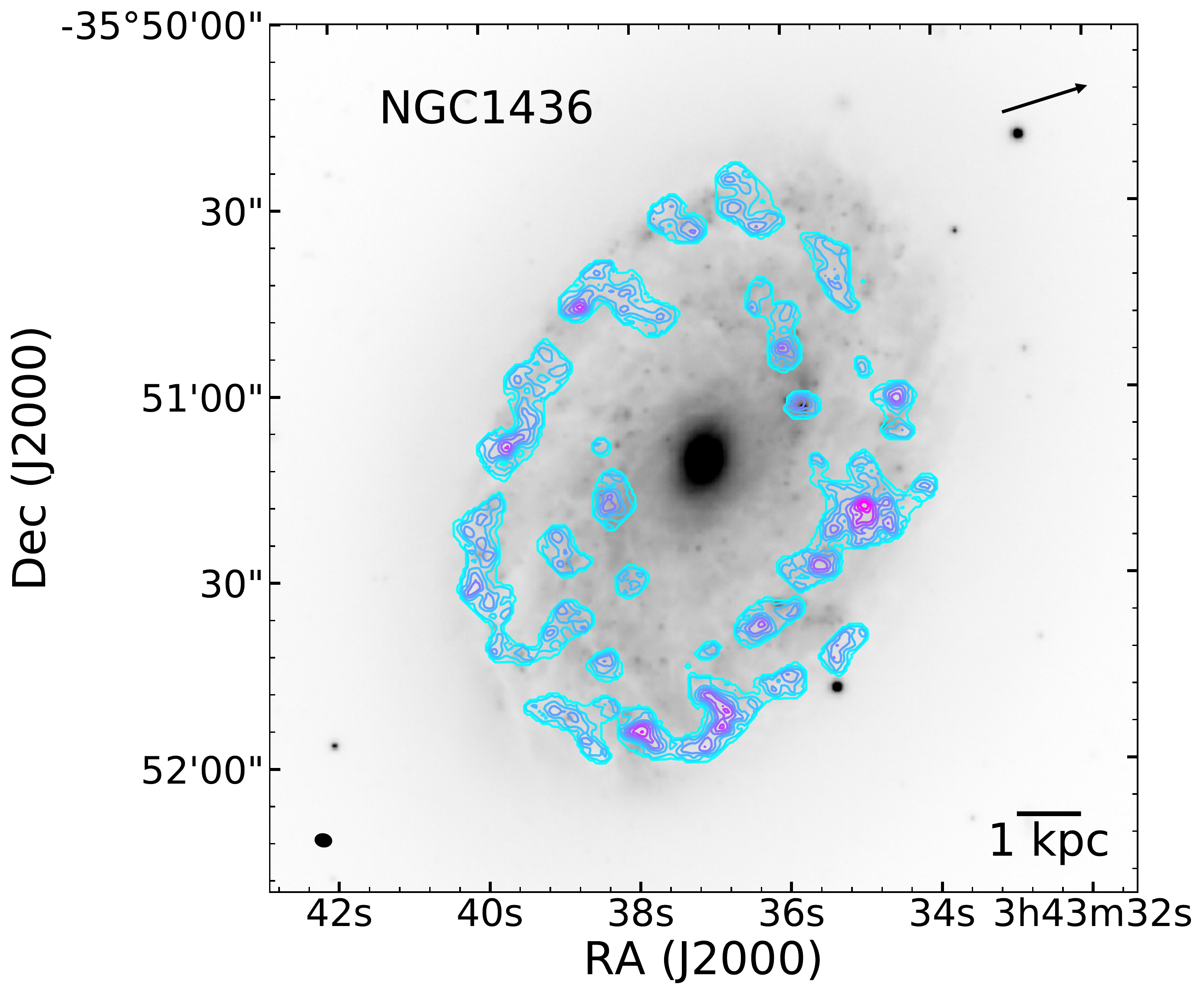}}
	\subfloat[\label{subfig:overplot_NGC1437B}]
	{\includegraphics[height=0.4\textwidth]{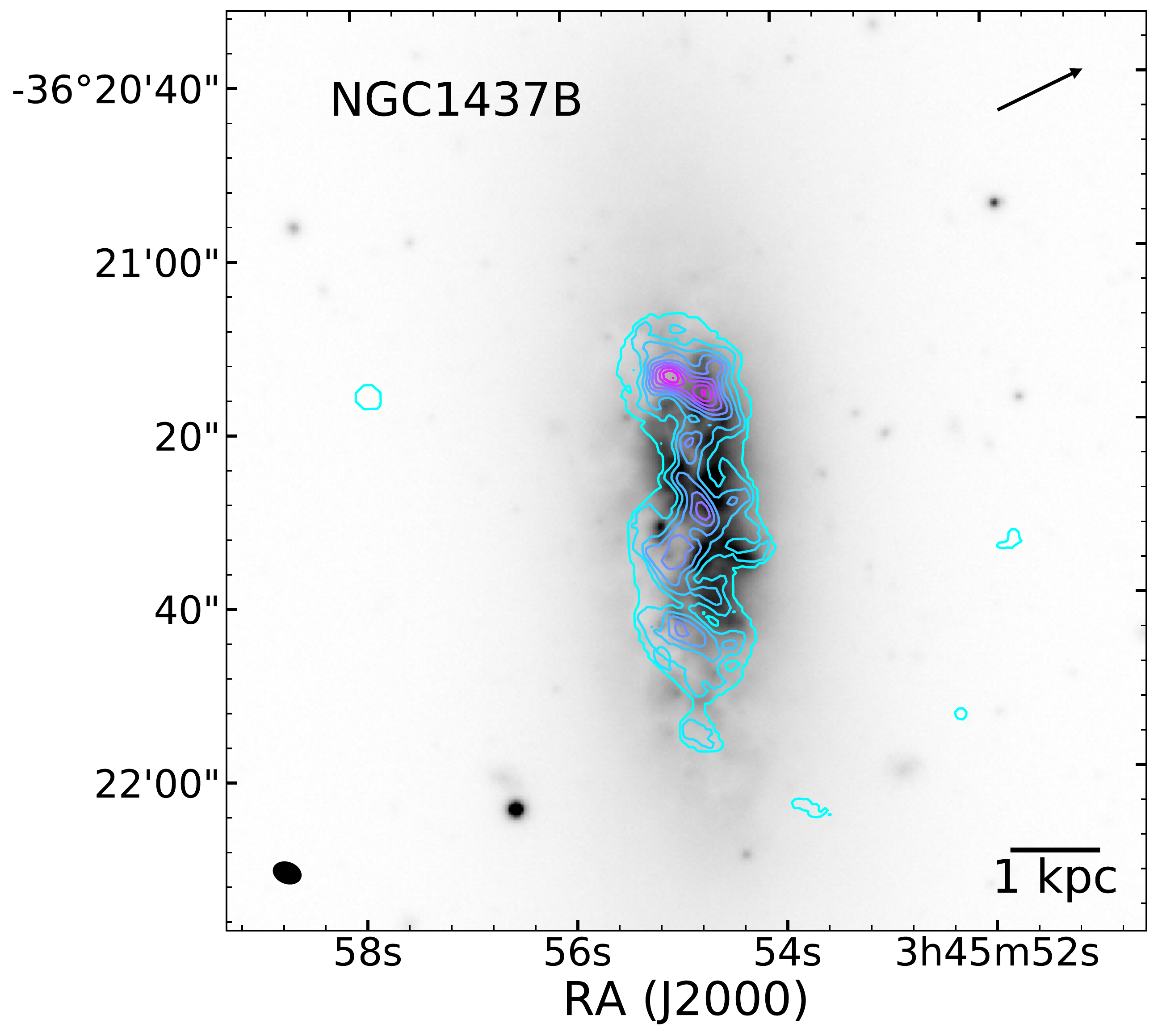}}	
	
\end{figure*}

\begin{figure*}

	\ContinuedFloat
	\centering
	
	\subfloat[\label{subfig:overplot_ESO358-G063}]
	{\includegraphics[height=0.4\textwidth]{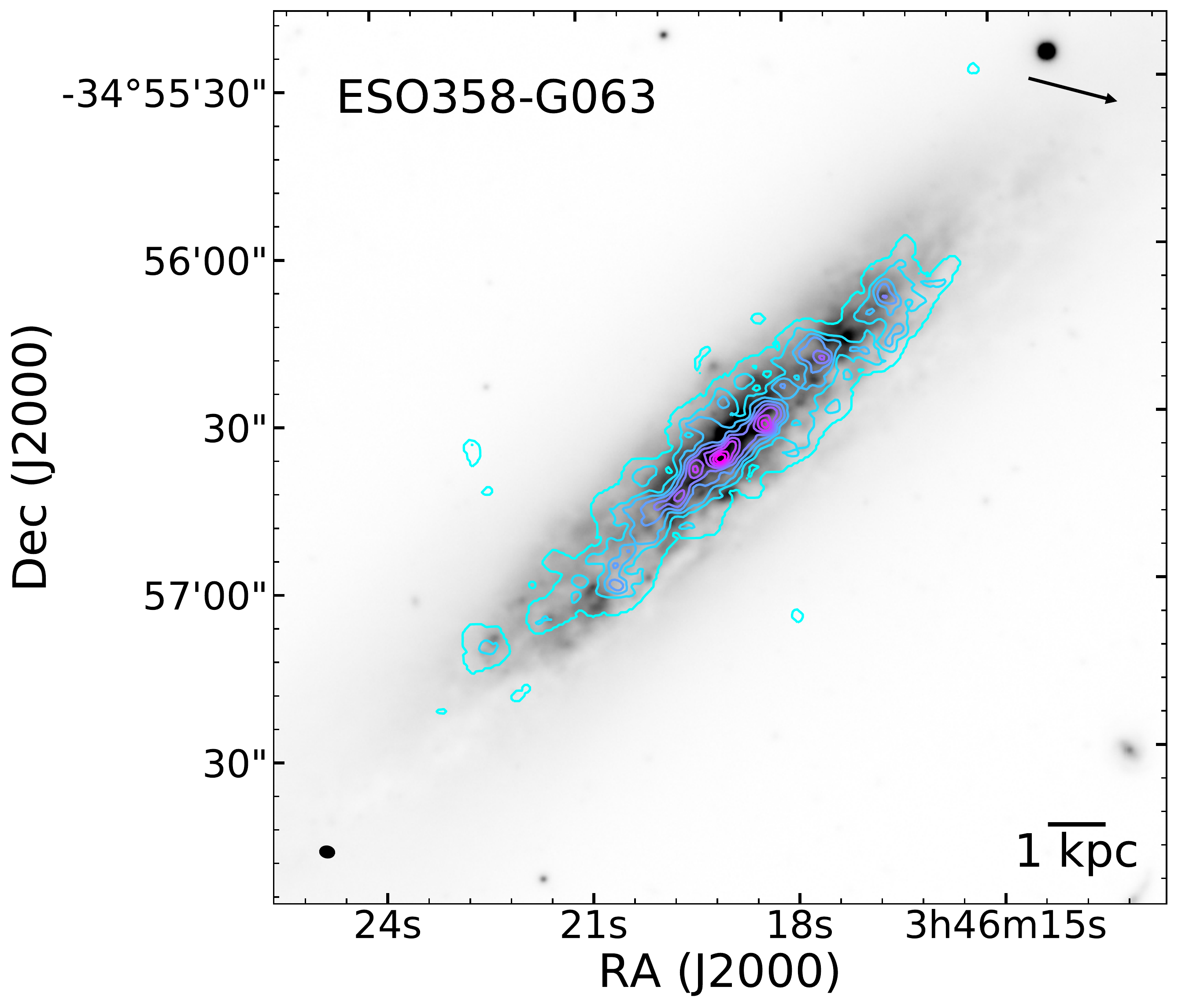}}		
	\subfloat[\label{subfig:overplot_FCC332}]
	{\includegraphics[height=0.4\textwidth]{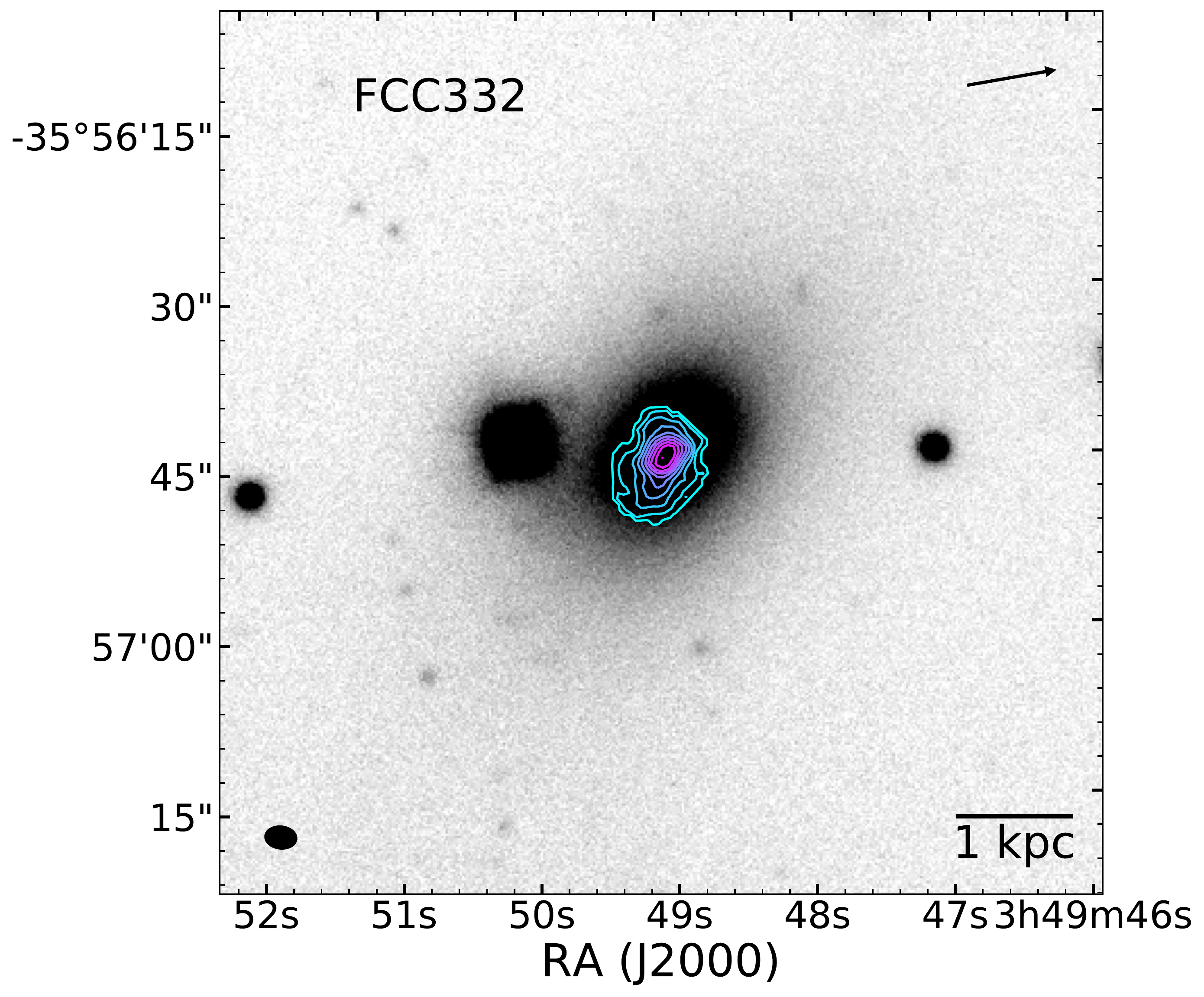}}
	
	\subfloat[\label{subfig:overplot_ESO359-G002}]
	{\includegraphics[height=0.4\textwidth]{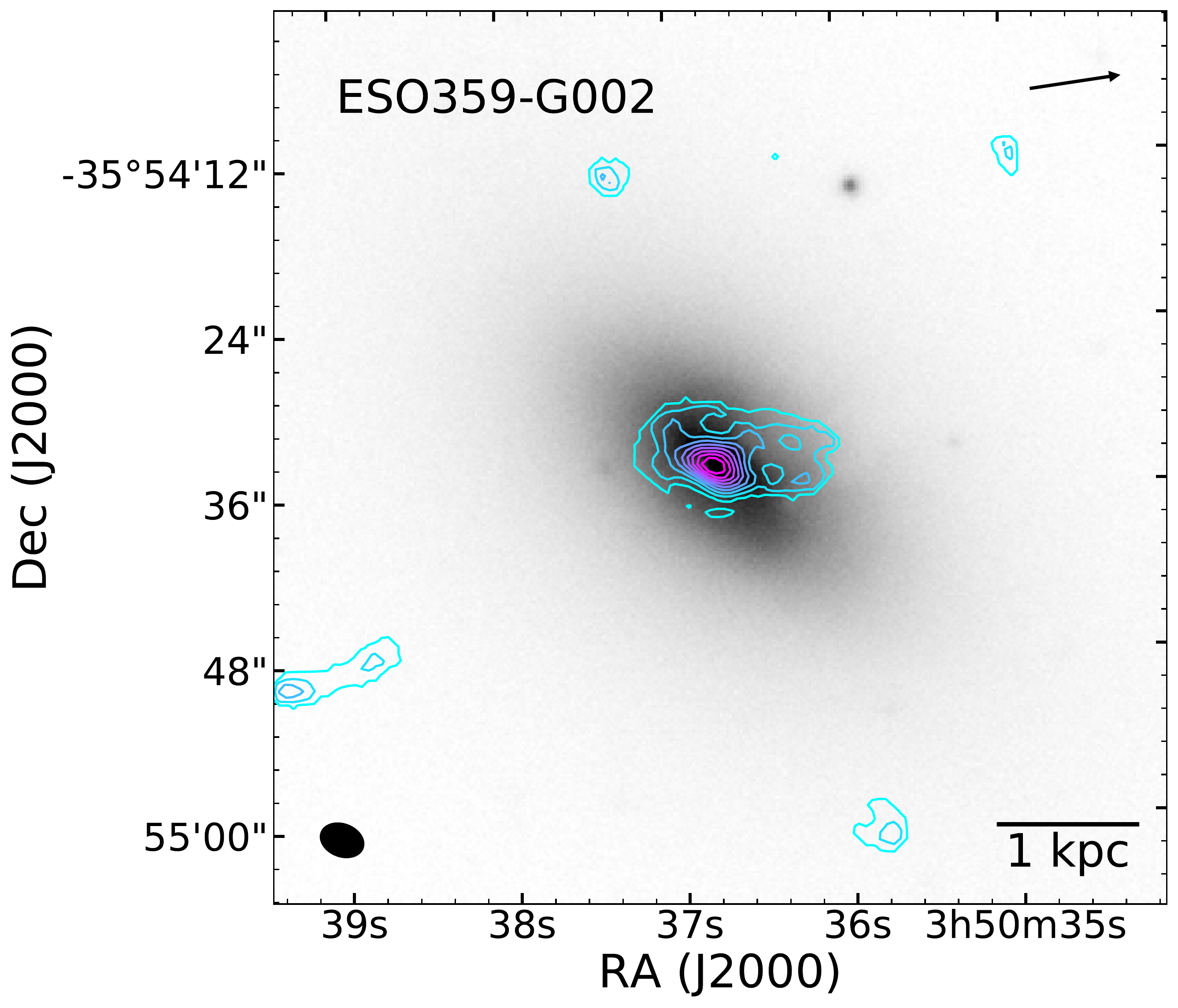}}

	\caption{CO integrated intensity contours overplotted on optical (\textit{g}-band) images of the galaxies (see \Section \ref{subsub:optical}).}
	\label{fig:overplots_app}
\end{figure*}


\bsp	
\label{lastpage}
\end{document}